\documentclass[a4paper,11pt]{article}
\usepackage{jheppub} % for details on the use of the package, please
\usepackage{autobreak}
\usepackage{graphicx}
\usepackage{subfigure}
\usepackage{color}
\graphicspath{{figures/}{fig/}}
\usepackage{amsmath}
\usepackage{bm}
\usepackage{slashed}
\usepackage{epsfig}
\usepackage{amsfonts}
\usepackage{epstopdf}
\usepackage{hyperref}
\usepackage{booktabs}
\usepackage{multirow}
\usepackage{float}
\usepackage{makecell}
\usepackage{graphicx}
\allowdisplaybreaks

\begin{document}

\title{NLO results with operator mixing for fully heavy tetraquarks in QCD sum rules}

\author[a]{Ren-Hua Wu}

\author[a]{Yu-Sheng Zuo}

\author[b]{Chen-Yu Wang}

\author[a]{Ce Meng}

\author[a,c,d]{Yan-Qing Ma}

\author[a,c]{Kuang-Ta Chao}

\affiliation[a]{School of Physics and State Key Laboratory of Nuclear Physics and Technology, Peking University,\\Beijing 100871, China}
\affiliation[b]{Institute for Theoretical Particle Physics, KIT, \\Karlsruhe, Germany}
\affiliation[c]{Center for High Energy Physics, Peking University,\\Beijing 100871, China}
\affiliation[d]{Collaborative Innovation Center of Quantum Matter,\\Beijing 100871, China}

%\affiliation{
%$^{1}$School of Physics and State Key Laboratory of Nuclear Physics and Technology, Peking University, Beijing 100871, China\\
%$^{2}$Institute for Theoretical Particle Physics, KIT, Karlsruhe, Germany\\
%$^{3}$Center for High Energy Physics,
%Peking University, Beijing 100871, China\\
%$^{4}$Collaborative Innovation Center of Quantum Matter,
%Beijing 100871, China
%}%

\emailAdd{renhuawu@pku.edu.cn}

\emailAdd{1801210125@pku.edu.cn}

\emailAdd{chen-yu.wang@kit.edu}

\emailAdd{mengce75@pku.edu.cn}

\emailAdd{yqma@pku.edu.cn}

\emailAdd{ktchao@pku.edu.cn}

\date{\today}

\preprint{TTP-22-006, P3H-22-013}

\abstract{We study the mass spectra of $\bar{Q}Q\bar{Q}Q\ (Q=c,b)$ systems in QCD sum rules with the complete next-to-leading order (NLO) contribution to the perturbative QCD part of the correlation functions. Instead of  meson-meson or diquark-antidiquark currents, we use diagonalized currents under operator renormalization.
	  We find that differing from conventional mesons $\bar qq$ and baryons $qqq$, a unique feature of the multiquark  systems like $\bar{Q}Q\bar{Q}Q$ is the operator mixing or color configuration mixing induced by NLO corrections, which is crucial to understand the color structure of the states. Our numerical results show that the NLO corrections are very important for the $\bar{Q}Q\bar{Q}Q$ system, because they not only give significant contributions but also reduce the scheme and scale dependence and make Borel platform more distinct, especially for the $\bar{b}b\bar{b}b$ in the $\overline{\rm{MS}}$ scheme. We use currents that have good  perturbation convergence in our phenomenological analysis.
	With the $\overline{\rm{MS}}$ scheme, we get three  $J^{PC}=0^{++}$ states, with masses  $6.35^{+0.20}_{-0.17}$~GeV, $6.56^{+0.18}_{-0.20}$~GeV and $6.95^{+0.21}_{-0.31}$~GeV, respectively. The first two seem to agree with the broad structure around $6.2\sim6.8$~GeV measured by the LHCb collaboration in the $J/\psi J/\psi$ spectrum, and the third seems to agree with the narrow resonance $X(6900)$.  For the $2^{++}$ states we find one with mass $7.03^{+0.22}_{-0.26}$~GeV, which is also close to that of $X(6900)$, and another one around $7.25^{+0.21}_{-0.35}$~GeV, which has good scale dependence but slightly large scheme dependence.}
%%%%%%%%%%%%%%%%%%%%%%%%%%%%%%%%%%%%%%%%%%%%%%%
%\linenumbersx
%%%%%%%%%%%%%%%%%%%%%%%%%%%%%%%%%%%%%%%%%%%%%%%

\maketitle
\flushbottom

%%%%%%%%%%%%%%%%%%%%%%%%%%%%%%%%%%%%%%%%%%%%%%%%%
\section{Introduction}

In recent years, a large number of new hadronic states containing heavy quarks (the charm quark $c$ or bottom quark $b$) have been observed at hadron colliders and $e^+e^-$ colliders~\cite{Tanabashi:2018oca}. They are expected to be candidates of tetraquark states, pentaquark states, and baryons which contain two heavy quarks~\cite{Chen:2016qju,Liu:2019zoy,Brambilla:2019esw}. These findings have opened up a new stage for the study of hadron physics and QCD. Lately, the LHCb collaboration has discovered a narrow resonance X(6900) and a broad structure around $6.2 \sim 6.8$~GeV in the double-$J/\psi$ spectrum~\cite{LHCb:2020bwg}, where the X(6900) may be a $\bar{c}c\bar{c}c$ resonance.

Fully heavy tetraquark $\bar{Q}Q\bar{Q}Q$ system is a good platform for studying QCD and exotic states because the system has a strong symmetry in structure and avoids pollution from light quarks. Since 1975, there have been many theoretical studies of fully heavy tetraquark systems using potential models~\cite{Iwasaki:1975pv,Chao:1980dv,Ader:1981db,Ballot:1983iv,Heller:1985cb,Lloyd:2003yc,SilvestreBrac:1992mv,SilvestreBrac:1993ss,Barnea:2006sd,Karliner:2016zzc,Wu:2016vtq,Anwar:2017toa,Richard:2017vry,Debastiani:2017msn,Liu:2019zuc,Jin:2020jfc,Wang:2021kfv}, QCD Sum Rules~\cite{Chen:2016jxd,Wang:2017jtz, Wang:2020ols, Wang:2018poa, Albuquerque:2020hio,Yang:2020wkh,Zhang:2020xtb} and other techniques~\cite{Heupel:2012ua,Guo:2020pvt,Dong:2020nwy,Tiwari:2021tmz}. But it is still under debate whether there exist compact bound states below di-heavy-quarkonium threshold, e.g.\ di-$\eta_c$, di-$J/\psi$, di-$\eta_b$, di-$\Upsilon(1S)$ and so on.
Some works imply that there is no stable state below the corresponding threshold~\cite{Ader:1981db,Wu:2016vtq,Richard:2017vry,Hughes:2017xie,Liu:2019zuc,Ke:2021iyh,Zhao:2020zjh,Jin:2020jfc}, while some other works have opposite conclusion~\cite{SilvestreBrac:1993ss,Lloyd:2003yc,Barnea:2006sd,Heupel:2012ua,Berezhnoy:2011xn,Bai:2016int,Anwar:2017toa,Debastiani:2017msn,Karliner:2017qhf,Esposito:2018cwh,Wu:2016vtq,
Bedolla:2019zwg,Lundhammar:2020xvw,Zhu:2020xni}.
Moreover, there are different interpretations of the nature of X(6900) state, e.g.\ tetraquark~\cite{Albuquerque:2020hio,liu:2020eha,Jin:2020jfc,Lu:2020cns,Giron:2020wpx,Huang:2020dci,Faustov:2021hjs,Wang:2021kfv,Li:2021ygk,Sonnenschein:2020nwn,Tiwari:2021tmz,Ke:2021iyh,Zhao:2020zjh}, gluonic tetracharm~\cite{Wan:2020fsk}, or coupled channel effect~\cite{Guo:2020pvt,Dong:2020nwy,Cao:2020gul}. Therefore, further study of $\bar{Q}Q\bar{Q}Q$ system is still needed.

The QCD Sum Rule~\cite{Shifman:1978bx,Shifman:1978by} approach is a powerful tool to study hadronic properties~\cite{Colangelo:2000dp,Narison:2010wb,Narison:2014wqa,Albuquerque:2018jkn}. Currently, there have been many leading order (LO) in $\alpha_s$ calculations of the $\bar{Q}Q\bar{Q}Q$ system~\cite{Chen:2016jxd,Wang:2017jtz, Wang:2020ols, Wang:2018poa, Yang:2020wkh,Zhang:2020xtb}, which however results in  different conclusions. The importance of purely perturbative part, denoted as $C_1$, at the next-to-leading order (NLO) in $\alpha_s$ has been emphasized in many works, e.g.\ for the proton~\cite{Ovchinnikov:1991mu,Groote:2008hz}, singly heavy baryons~\cite{Groote:2008dx}, the doubly heavy baryon $\Xi_{cc}^{++}$~\cite{Wang:2017qvg}, and fully heavy baryons $\Omega_{ccc}^{++}$ and $\Omega_{bbb}^{-}$~\cite{Wu:2021tzo}. Our previous work on $\Omega_{QQQ}\ (Q=c,b)$~\cite{Wu:2021tzo} shows that the NLO contribution of fully heavy quark system can not only lead to a large correction, but also reduce  parameters dependence, which makes the Borel platform more distinct. Especially for $\Omega_{bbb}$ in the $\overline{\text{MS}}$ scheme,  the platform appears only at NLO but not at LO. Therefore, it is reasonable to expect that the NLO corrections are also sizable and important for the $\bar{Q}Q\bar{Q}Q$ system.

Partial NLO contributions of $C_1$  for the $\bar{Q}Q\bar{Q}Q$ system, originated from the so-called factorized diagrams, have been considered in Ref.~\cite{Albuquerque:2020hio}. However, to further reduce theoritical uncertainties, it is necessary to perform a complete NLO corrections to  $C_1$, which will be presented in this paper. The rest of the paper is organized as the following. In Sec.~\ref{sec:QCD sum rules},  sum rules for calculation of the mass of $\bar{Q}Q\bar{Q}Q$ are given. In Sec.~\ref{sec:Calculation of C1 and CGG}, we present our methods to calculate perturbative coefficients. Phenomenological results and discussions are given in Sec.~\ref{sec:Phenomenology}. Some details of our calculations and results are given in Apps.~\ref{sec:matrix}, \ref{sec:charm} and \ref{sec:bottom}.

%%%%%%%%%%%%%%%%%%%%%%%%%%%%%%%%%%%%%%%%%%%%%%%%%
\section{QCD Sum Rule}\label{sec:QCD sum rules}

In this section, we briefly review the framework of the QCD sum rules used to calculate the mass of the tetraquark ground state. See Ref.~\cite{Colangelo:2000dp} for more details. We start with a two-point correlation function
\begin{align}\label{eq:corrFun}
\Pi(q^2) &= i \int {\mathrm{d}^D x e^{iq \cdot x} \langle \Omega|T[J(x) J^\dagger(0)] |\Omega\rangle} ,
\end{align}
where $D$ denotes the spacetime dimension, $\Omega$ denotes the QCD vacuum and $J$ is the (pseudo-)scalar tetraquark current to be defined later.

On the one hand, the correlation function $\Pi(q^2)$ can be related to  the phenomenological spectrum by the  K\"{a}ll\'{e}n-Lehmann representation~\cite{Colangelo:2000dp},
\begin{align}\label{eq:K-LSR}
\Pi(q^2) &=  \int {\mathrm{d} s \frac{\rho(s)}{s-q^2-i\epsilon}}\,,
\end{align}
where $\rho(s)$ denotes the physical spectrum density. Taking the narrow resonance approximation for the physical ground state, one can parametrize the spectrum density as a pole plus a continuum part
\begin{align}\label{eq:SP}
\rho(s) = \lambda_H \delta(s-M_H^2)+ \rho_{\text{cont}}(s) \theta(s-s_{h})\,,
\end{align}
where $M_H$ and $\lambda_H$ denote the mass of the ground state and pole residue, respectively. $\rho_{\text{cont}}(s)$ denotes the continuum spectrum density, which could also contain information of higher resonances.
$s_{h}$ is the threshold of the continuum spectrum.

On the other hand, in the region where $-q^2=Q^2\gg\Lambda^2_{\text{QCD}}$, one can calculate correlation function $\Pi(q^2)$ using the operator product expansion (OPE), which reads

\begin{align}\label{eq:OPE}
\begin{split}
\Pi(q^2) &=C_1(q^2) +\sum_i C_i(q^2) \langle O_i \rangle \, ,
\end{split}
\end{align}
where $C_{1}$ and $C_{i}$ are perturbatively calculable Wilson coefficients, and $\langle O_i  \rangle$ is a shorthand of the vacuum condensate $\langle \Omega|O_i |\Omega \rangle $, which is a nonperturbative but universal quantity. The relative importance of the vacuum condensate is power suppressed by the dimension of the operator $O_i$. In our calculations, we will only keep the relevant vacuum condensates up to dimension four, which gives the approximated expression of the OPE as
\begin{align}\label{eq:DisInt}
\begin{split}
\Pi(q^2) & = C_1(q^2)+ C_{GG}(q^2) \langle g_s^2 \hat{G}\hat{G} \rangle \, ,
\end{split}
\end{align}
where $\langle g_s^2 \hat{G}\hat{G} \rangle$ denotes the gluon-gluon ($GG$) condensate $\langle \Omega|g_s^2 \hat{G}\hat{G} |\Omega\rangle$.

According to Eq.~(\ref{eq:K-LSR}), one can relate the physical spectrum density to the imaginary part of $\Pi(q^2)$ in Eq.~(\ref{eq:DisInt}) using the dispersion relation, which gives
\begin{align}\label{eq:KL-DisInt}
\begin{split}
\Pi(q^2) & =  \int {\mathrm{d} s \frac{\rho(s)}{s-q^2-i\epsilon}} \, \\
&= \frac{1}{\pi} \int_{s_{\text{th}}}^\infty \mathrm{d} s \frac{\text{Im} C_1(s)+ \text{Im} C_{GG}(s)\langle g_s^2 \hat{G}\hat{G} \rangle }{s-q^2-i\epsilon} \,,
\end{split}
\end{align}
where $s_{\text{th}}=16 m_Q^2$ is the QCD threshold for the $\bar{Q}Q\bar{Q}Q$ system, and the integral in the second line has been assumed to be convergent.
Then by employing the quark-hadron duality and Borel transformation~\cite{Colangelo:2000dp}, we obtain a sum rule for $\Pi(q^2)$,
\begin{align}\label{eq:Borel-QHD}
\begin{split}
\lambda_H  e^{-\frac{M_H^2}{M_B^2}} =&  \int_{s_{\text{th}}}^{s_0} \mathrm{d} s \frac{1}{\pi}{\text{Im}} C_1(s)\, e^{-\frac{s}{M_B^2}}  \,+\int_{s_{\text{th}}}^\infty \mathrm{d} s \frac{1}{\pi}{\text{Im}} C_{GG}(s) e^{-\frac{s}{M_B^2}}\langle g_s^2 \hat{G}\hat{G} \rangle\,,
\end{split}
\end{align}
where $s_0$ is the threshold parameter and $M_B$ is the Borel parameter. They are introduced into the formula due to the qurak-hadron duality and Borel transformation, respectively.
By differentiating both sides of Eq.~(\ref{eq:Borel-QHD}) with respect to $-\frac{1}{M_B^2}$, one can get
\begin{align}\label{eq:diff}
\begin{split}
\lambda_H \ M_H^2 \ e^{-\frac{M_H^2}{M_B^2}} =&  \int_{s_{\text{th}}}^{s_0} \mathrm{d} s \frac{1}{\pi}{\text{Im}} C_1(s)\, e^{-\frac{s}{M_B^2}} s +\int_{s_{\text{th}}}^\infty \mathrm{d} s \frac{1}{\pi}{\text{Im}} C_{GG}(s) e^{-\frac{s}{M_B^2}} s\langle g_s^2 \hat{G}\hat{G} \rangle
\,.
\end{split}
\end{align}
Finally, one can solve $M_H$ according to Eq.~(\ref{eq:Borel-QHD}) and (\ref{eq:diff}),
\begin{align}\label{eq:MH}
\begin{split}
M_H^2 &= \frac{\int_{s_{\text{th}}}^{s_0} \mathrm{d} s\, \, s \ \rho_{1}(s)\  e^{-\frac{s}{M_B^2}} +\int_{s_{\text{th}}}^\infty \mathrm{d} s\  s\  \rho_{GG}(s)\  e^{-\frac{s}{M_B^2}} \langle g_s^2 \hat{G}\hat{G} \rangle}{\int_{s_{\text{th}}}^{s_0} \mathrm{d} s\, \, \rho_{1}(s)\ e^{-\frac{s}{M_B^2}}+\int_{s_{\text{th}}}^\infty \mathrm{d} s \ \rho_{GG}(s)\  e^{-\frac{s}{M_B^2}}\langle g_s^2 \hat{G}\hat{G} \rangle}\,,
\end{split}
\end{align}
where $\rho_1=\frac{1}{\pi}{\text{Im}} C_1$ and $\rho_{GG}=\frac{1}{\pi}{\text{Im}} C_{GG}$.

Similar to Eq.~(\ref{eq:corrFun}), for the (axial-)vector and tensor tetraquark currents $J_\mu$ and $J_{\mu\nu}$ (to be defined later), one can introduce two-point correlation functions as
\begin{align}\label{eq:corrFun2}
\Pi^{V(A)}_{\mu \nu}(q^2) &= i \int {\mathrm{d}^{D} x e^{iq \cdot x} \langle \Omega|T[J_\mu (x) J^\dagger_\nu(0)] |\Omega\rangle} \,,\\
\Pi^T_{\mu \nu,\rho\sigma}(q^2) &= i \int {\mathrm{d}^{D} x e^{iq \cdot x} \langle \Omega|T[J_{\mu\rho} (x) J^\dagger_{\nu\sigma}(0)] |\Omega\rangle} \,.
\end{align}
For $ J^P=1^- $ vector particle and $ J^P=1^+ $ axial vector particle, the correlation function $\Pi_{\mu \nu}^V$ and $\Pi_{\mu \nu}^A$ can be decomposed as
\begin{align}\label{eq:VAcorrFun}
	\begin{split}
		\Pi_{\mu \nu}^{V}(q^2) &= \left( -g_{\mu \nu}+\frac{q_\mu q_\nu}{q^2} \right) \Pi^{V}_{1}(q^2)\,,\\
		\Pi_{\mu \nu}^A(q^2) &= \left( -g_{\mu \nu}+\frac{q_\mu q_\nu}{q^2} \right) \Pi^A_{1}(q^2)+\frac{q_\mu q_\nu}{q^2} \Pi^A_2(q^2)\,.
	\end{split}
\end{align}
While for $ J^P=2^+ $ tensor particle, the correlation function $\Pi_{\mu \nu,\rho \sigma}^T$ can be decomposed as
\begin{align}\label{eq:TAcorrFun}
\begin{split}
\Pi_{\mu \nu,\rho \sigma}^T(q^2) =& \left( \frac{\theta_{\mu\rho} \theta_{\nu\sigma}+\theta_{\mu\sigma} \theta_{\nu\rho}}{2}-\frac{\theta_{\mu\nu} \theta_{\rho\sigma}}{D-1} \right) \Pi_1^T(q^2)+
\frac{\theta_{\mu\rho} \omega_{\nu\sigma}+ \theta_{\mu\sigma} \omega_{\nu\rho}+ \omega_{\mu\rho} \theta_{\nu\sigma}+\omega_{\mu\sigma} \theta_{\nu\rho}}{2} \Pi_2^T(q^2)\\
& + \frac{\theta_{\mu\nu} \theta_{\rho\sigma}}{D-1} \Pi_3^T(q^2)+ \omega_{\mu\nu} \omega_{\rho\sigma} \Pi_4^T(q^2) +
\frac{\theta_{\mu\nu} \omega_{\rho\sigma}}{\sqrt{D-1}} \Pi_5^T(q^2)+ \frac{\omega_{\mu\nu} \theta_{\rho\sigma}}{\sqrt{D-1}} \Pi_6^T(q^2)\\
& + \frac{\theta_{\mu\rho} \omega_{\nu\sigma}- \theta_{\mu\sigma} \omega_{\nu\rho}+ \omega_{\mu\rho} \theta_{\nu\sigma}-\omega_{\mu\sigma} \theta_{\nu\rho}}{2} \Pi_7^T(q^2)+
\frac{\theta_{\mu\rho} \theta_{\nu\sigma}-\theta_{\mu\sigma} \theta_{\nu\rho}}{2} \Pi_8^T(q^2)\\
& + \frac{\theta_{\mu\rho} \omega_{\nu\sigma}- \theta_{\mu\sigma} \omega_{\nu\rho}- \omega_{\mu\rho} \theta_{\nu\sigma}+\omega_{\mu\sigma} \theta_{\nu\rho}}{2} \Pi_9^T(q^2)\\
& + \frac{\theta_{\mu\rho} \omega_{\nu\sigma}+ \theta_{\mu\sigma} \omega_{\nu\rho}- \omega_{\mu\rho} \theta_{\nu\sigma}-\omega_{\mu\sigma} \theta_{\nu\rho}}{2} \Pi_{10}^T(q^2)\,,
\end{split}
\end{align}
where $\theta_{\mu\nu}=g_{\mu\nu}-\frac{q_\mu q_\nu}{q^2}$ and $\omega_{\mu\nu}=\frac{q_\mu q_\nu}{q^2}$.
In this paper we use $\Pi^{V(A)}_{1}$ and $\Pi_1^T$ to construct sum rules, as they project out the spin-1 and spin-2 degrees of freedom we are interested in.
The calculation of the corresponding ground state masses is similar to that in Eq.~(\ref{eq:MH}).

 %%%%%%%%%%%%%%%%%%%%%%%%%%%%%%%%%%%%%%%%%%%%%%%%%
\section{Calculation of $C_1$ and $C_{GG}$ }\label{sec:Calculation of C1 and CGG}

In QCD Sum Rules, there are two kinds of expansions: the OPE and the perturbative expansion in $\alpha_s$. For the OPE, we only consider the most important contributions, the purely perturbative term $C_1$ and the $GG$ condensate term $C_{GG} \langle g_s^2 \hat{G}\hat{G} \rangle$, because other higher dimensional operators are power suppressed in the OPE. According to Eq.~(\ref{eq:MH}), we need to calculate the imaginary parts of $C_1$ and $C_{GG}$ perturbatively. We can expect that the LO contribution of $C_1$ is the dominant one, and the next important contribution can be the NLO corrections for $C_1$ or the LO contribution of $C_{GG}$. Therefore, the NLO corrections to $C_1$ need to be considered in the calculation in order to reduce  theoretical uncertainties. For convenience, we will call the sum of the LO of $C_1$ and $C_{GG}$ as the LO contribution and the NLO corrections to $C_1$ as the NLO contribution in the following.

\begin{figure}[htb]
\centering
\subfigure[$C_1$-LO]{
\includegraphics[scale=0.9]{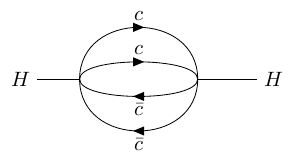}
}
\subfigure[$C_{GG}$-LO]{
\includegraphics[scale=0.9]{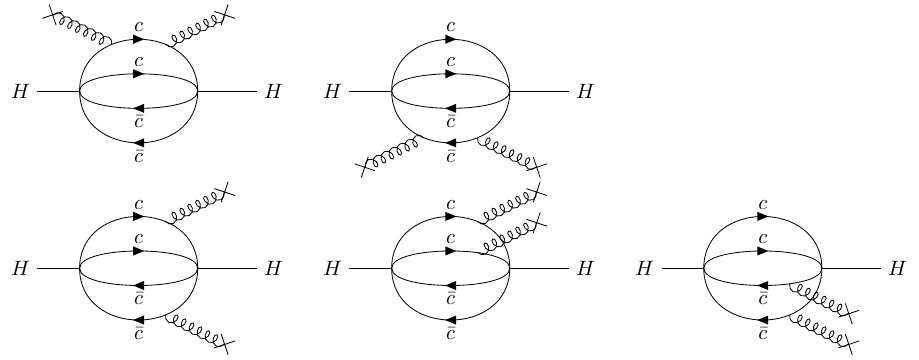}
}
\caption{\label{fig:FeynmanDiagrams-LO}
  LO Feynman diagrams of $C_1$ and $C_{GG}$. $H$ denotes the interpolating current.}
\end{figure}

\begin{figure}[htb]
\centering
\subfigure[$C_1$-NLO]{
\includegraphics[scale=1]{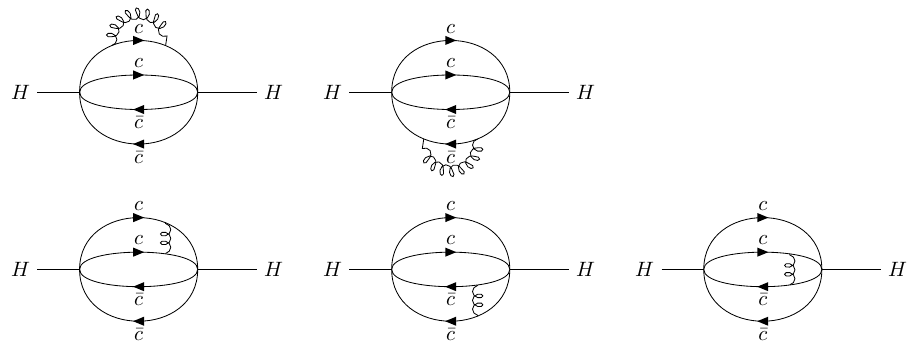}
}
\subfigure[$C_1$-NLOct]{
\includegraphics[scale=1]{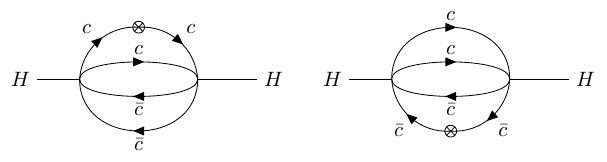}
}
\caption{\label{fig:FeynmanDiagrams-NLO}
  NLO and counter term Feynman diagrams of $C_1$. $H$ denotes the interpolating current.}
\end{figure}

We use \texttt{FeynArts}~\cite{Kublbeck:1990xc,Hahn:2000kx} to generate Feynman diagrams and Feynman amplitudes of $C_1$ and $C_{GG}$. Some representative Feynman diagrams at the LO and the NLO are shown in Fig.~\ref{fig:FeynmanDiagrams-LO} and Fig.~\ref{fig:FeynmanDiagrams-NLO}, respectively.

The calculation procedure for $C_1$ and $C_{GG}$ are summarized below:
\begin{itemize}
\item
  1. We use \texttt{FeynCalc}~\cite{Mertig:1990an,Shtabovenko:2016sxi} to simplify spinor structures of Feynman amplitudes with the Naive-$\gamma_5$ scheme~\cite{Korner:1991sx}.
\item
  2. We use \texttt{Reduze}~\cite{vonManteuffel:2012np} to reduce all loop integrals to linear combinations of a set of simpler integrals,  which are called master integrals (MIs).

 \item
  3. We set up differential equations for MIs~\cite{Kotikov:1990kg,Bern:1992em,Remiddi:1997ny,Gehrmann:1999as} and solve them numerically~\cite{Liu:2022chg}, with boundary conditions obtained via auxiliary mass flow~\cite{Liu:2017jxz}.
  MIs and thus $C_1$ and $C_{GG}$  are expressed as general series expansion.

 \item
  4. Renormalization. There are no infrared divergences in the NLO amplitude of $C_1$. After performing wave-function and mass renormalization of quarks ($m_Q$ is renormalized in either the $\overline{\text{MS}}$ scheme or the on-shell scheme), the remaining ultraviolet divergences can be removed by the renomalization of the current operators. When there are more than one current operator share the same quantum number $J^{PC}$, they are usually mixed with each others under the renormalization. We get operator renormalization matrices for different $J^{PC}$  in the $\overline{\text{MS}}$ scheme, which are shown explicitly in Appendix~\ref{sec:matrix}.
\end{itemize}
Because the expressions of $C_1$ and $C_{GG}$ are too complicated to be shown in the paper, we attach the imaginary part of $C_1$ and $C_{GG}$, which are the only needed information in phenomenological study, as ancillary files.

\section{Current operators}\label{sec:Current operators}

\subsection{$J^{P} = 0^{+}$}

For the $J^{PC}=0^{++}$ scalar $\bar{Q}Q\bar{Q}Q$ system, there are five independent interpolating currents. The operator basis, in the color-singlet meson-meson type  currents, can be chosen as
\begin{align}\label{eq:0++meson}
\begin{split}
J_{\text{S},1}^{\text{M-M}}&= (\bar{Q}_a \gamma^\mu Q_a)(\bar{Q}_b \gamma_\mu Q_b)\,, \\
J_{\text{S},2}^{\text{M-M}}&= (\bar{Q}_a \gamma^\mu \gamma^5 Q_a)(\bar{Q}_b \gamma_\mu \gamma^5 Q_b)\,, \\
J_{\text{S},3}^{\text{M-M}}&= (\bar{Q}_a  Q_a)(\bar{Q}_b Q_b)\,, \\
J_{\text{S},4}^{\text{M-M}}&= (\bar{Q}_a i\gamma^5 Q_a)(\bar{Q}_b i\gamma^5 Q_b)\,, \\
J_{\text{S},5}^{\text{M-M}}&= (\bar{Q}_a \sigma^{\mu\nu} Q_a)(\bar{Q}_b \sigma_{\mu\nu} Q_b)\,,
\end{split}
\end{align}
 where $a$ and $b$ represent color indices. Alternatively, one can choose the diquark-antidiquark type currents as the basis like those in Ref.~\cite{Chen:2016jxd}, which are given by
 \begin{align}\label{eq:0++Diquark}
\begin{split}
J_{\text{S},1}^{\text{Di-Di}}&= (Q^T_a \hat{C}\gamma^\mu Q_b)(\bar{Q}_a \gamma_\mu \hat{C} \bar{Q}^T_b)\,, \\
J_{\text{S},2}^{\text{Di-Di}}&= (Q^T_a \hat{C}\gamma^\mu\gamma^5 Q_b)(\bar{Q}_a \gamma_\mu\gamma^5 \hat{C} \bar{Q}^T_b)\,, \\
J_{\text{S},3}^{\text{Di-Di}}&= (Q^T_a \hat{C} Q_b)(\bar{Q}_a  \hat{C} \bar{Q}^T_b)\,, \\
J_{\text{S},4}^{\text{Di-Di}}&= (Q^T_a \hat{C}i\gamma^5 Q_b)(\bar{Q}_a i\gamma^5 \hat{C} \bar{Q}^T_b)\,, \\
J_{\text{S},5}^{\text{Di-Di}}&= (Q^T_a \hat{C}\sigma^{\mu\nu} Q_b)(\bar{Q}_a \sigma_{\mu\nu} \hat{C} \bar{Q}^T_b)\,, \\
\end{split}
\end{align}
where $\hat{C}$ is the charge-conjugation matrix. The two types of bases can be associated with each other by the Fierz transformation in 4 dimension,
\begin{equation}\label{eq:0++FierzTrans}
\vec{J}_{\text{S}}^{\text{Di-Di}}=\frac{1}{8}\begin{pmatrix}
  4 &  -4 & 8&8&0 \\
 -4 &  4 & 8&8&0 \\
 2 &  2 & -2 &2& 1 \\
 2 &  2 & 2 &-2& -1 \\
  0 &  0 & 24 &-24& 4 \\
\end{pmatrix} \cdot \vec{J}_{\text{S}}^{\text{M-M}}\,,
\end{equation}
where we use the column vector $\vec{J}$ to represent the basis in Eq.~(\ref{eq:0++meson}) or (\ref{eq:0++Diquark}).

A physical state can well be a mixture of all possible currents that share the same quantum numbers. The operator mixing has significant effects on the QCD sum rules calculations, say, for the heavy baryon spectrum~\cite{Wang:2017qvg}.  However, there are no natural standards to pin down the mixing scheme only based on the LO calculation of $C_1$ and $C_{GG}$. Thanks to the NLO calculations, the currents are mixed with each other naturally under the renormalization. If one choose the basis which diagonalizes the anomalous dimension matrix, then the operators in the basis have universal anomalous dimensions, separately. Thus, inserting these operators into the calculations in QCD sum rules, the dependence on the renormalization scale $\mu$ tends to be cancelled out in the righthand side of Eq.~(\ref{eq:MH}), which is desirable since the left-hand side $M_H^2$ is a physical quantity.

For $J^{PC}=0^{++}$ state, if we choose the currents in Eq.~(\ref{eq:0++meson}) as the operator basis, the operator anomalous dimension matrix $\mathcal{A}_{\text{S}}^{\text{M-M}}$ is given by
\begin{equation}\label{eq:0++meson-AnoDim}
\mathcal{A}_{\text{S}}^{\text{M-M}}=\delta \begin{pmatrix}
  -6 &  -2 & -12&-12&0 \\
 -2 &  -6 & 12&12&0 \\
 0 &  0 & 26 &6& \frac{1}{3} \\
 0 &  0 & 6& 26 &-\frac{1}{3} \\
   0 &  0 & -40&40&-\frac{68}{3} \\
\end{pmatrix}\,,
\end{equation}
where $\delta=-\frac{\alpha_s}{16 \pi }$.
To diagonalize the matrix in Eq.~(\ref{eq:0++meson-AnoDim}), one needs the following transformation matrix
\begin{equation}\label{eq:0++meson-Diagonal}
\mathcal{T}_{\text{S}}^{\text{Dia}}=\begin{pmatrix}
  \frac{1}{2} &  \frac{1}{2} &0&0&0 \\
  -\frac{1}{2} &  \frac{1}{2} & -\frac{1}{3}&-\frac{1}{3}&0 \\
 0 &  0 & \frac{1}{2}&\frac{1}{2}& 0 \\
  0 &  0 & -\frac{15}{\sqrt{241}}&\frac{15}{\sqrt{241}}&\frac{1}{2}-\frac{8}{\sqrt{241}} \\
 0 &  0 & \frac{15}{\sqrt{241}}& -\frac{15}{\sqrt{241}}&\frac{1}{2}+\frac{8}{\sqrt{241}} \\
\end{pmatrix}\,.
\end{equation}
So, we get the new basis
\begin{equation}\label{eq:0++basis-mixing}
\vec{J}_{\text{S}}^{\text{Dia}}=\mathcal{T}_{\text{S}}^{\text{Dia}}\cdot \vec{J}_{\text{S}}^{\text{M-M}}\,.
\end{equation}
The anomalous dimension matrix of $\vec{J}_{\text{S}}^{\text{Dia}}$ is diagonal, which is given by
\begin{equation}\label{eq:0++Diagonal-AnoDim}
\begin{split}
\mathcal{A}_{\text{S}}^{\text{Dia}}& =\mathcal{T}_{\text{S}}^{\text{Dia}}\cdot \mathcal{A}_{\text{S}}^{\text{M-M}}\cdot (\mathcal{T}_{\text{S}}^{\text{Dia}})^{-1}\,\\
&=\frac{4}{3} \delta \begin{pmatrix}
  -6 &  0 & 0&0&0 \\
   0 & -3 & 0&0&0 \\
      0 &  0 & 24&0&0 \\
         0 &  0 & 0&-1+\sqrt{241}&0 \\
            0 &  0 & 0&0&-1-\sqrt{241} \\
\end{pmatrix}\,.
\end{split}
\end{equation}
Because the eigenvalues of anomalous dimension matrix do not degenerate, the transformation matrix $\mathcal{T}_{\text{S}}^{\text{Dia}}$ is unique, and thus the basis $\vec{J}_{\text{S}}^{\text{Dia}}$ in Eq.~(\ref{eq:0++basis-mixing}) is unique.

\subsection{$J^{P} = 0^{-}$}

For the $J^{P}=0^{-}$ pseudoscalar system, there are three independent interpolating currents. The operator basis, in the color-singlet meson-meson type  currents, can be chosen as
\begin{align}\label{eq:0--meson}
\begin{split}
J_{\text{P},1}^{\text{M-M}}&= (\bar{Q}_a \gamma^\mu Q_a)(\bar{Q}_b \gamma_\mu\gamma^5 Q_b)\,, \\
J_{\text{P},2}^{\text{M-M}}&= (\bar{Q}_a  Q_a)(\bar{Q}_b i \gamma^5 Q_b)\,, \\
J_{\text{P},3}^{\text{M-M}}&= (\bar{Q}_a \sigma^{\mu\nu} Q_a)(\bar{Q}_b\sigma_{\mu\nu}i\gamma^5  Q_b)\,,
\end{split}
\end{align}
where $J_{\text{P},1}^{\text{M-M}}$ couples to the state with $J^{PC}=0^{--}$, while $J_{\text{P},2}^{\text{M-M}}$ and $J_{\text{P},3}^{\text{M-M}}$ couple to the state with $J^{PC}=0^{-+}$. Of course, one can choose the diquark-antidiquark type currents~\cite{Chen:2016jxd} as the basis, which are given by
 \begin{align}\label{eq:0--Diquark}
\begin{split}
J_{\text{P},1}^{\text{Di-Di}}&= (Q^T_a \hat{C} Q_b)(\bar{Q}_a i\gamma^5 \hat{C} \bar{Q}^T_b)-(Q^T_a \hat{C} i\gamma^5 Q_b)(\bar{Q}_a  \hat{C} \bar{Q}^T_b)\,, \\
J_{\text{P},2}^{\text{Di-Di}}&= (Q^T_a \hat{C} Q_b)(\bar{Q}_a i\gamma^5 \hat{C} \bar{Q}^T_b)+(Q^T_a \hat{C} i\gamma^5 Q_b)(\bar{Q}_a  \hat{C} \bar{Q}^T_b)\,, \\
J_{\text{P},3}^{\text{Di-Di}}&= (Q^T_a \hat{C}\sigma^{\mu\nu} Q_b)(\bar{Q}_a \sigma_{\mu\nu}i\gamma^5 \hat{C} \bar{Q}^T_b)\,, \\
\end{split}
\end{align}
where $J_{\text{P},1}^{\text{Di-Di}}$  couples to the state with $J^{PC}=0^{--}$, while $J_{\text{P},2}^{\text{Di-Di}}$  and $J_{\text{P},3}^{\text{Di-Di}}$ couple to the state with $J^{PC}=0^{-+}$.
The two types bases can be associated with each other by the Fierz transformation in 4 dimension, which is given as
\begin{equation}\label{eq:0--FierzTrans}
\vec{J}_{\text{P}}^{\text{Di-Di}}=\frac{1}{4}\begin{pmatrix}
 -4\, i &  0 & 0 \\
 0 &  -4 & 1 \\
 0 &  24 & 2
\end{pmatrix} \cdot \vec{J}_{\text{P}}^{\text{M-M}}\,.
\end{equation}

Choosing the currents in Eq.~(\ref{eq:0--meson}) as the operator basis, one can get the anomalous dimension matrix
\begin{equation}\label{eq:0--meson-AnoDim}
\mathcal{A}_{\text{P}}^{\text{M-M}}=\frac{\delta}{3}\begin{pmatrix}
 -24 &  0 &0 \\
  0 &   60&1 \\
   0  &-240 &-68 \\
\end{pmatrix}\,.
\end{equation}
To diagonalize the matrix in Eq.~(\ref{eq:0--meson-AnoDim}), one needs the transformation matrix
\begin{equation}\label{eq:0--meson-Diagonal}
\mathcal{T}_{\text{P}}^{\text{Dia}}=\begin{pmatrix}
  1 & 0 & 0 \\
   0 &  -\frac{30}{\sqrt{241}} &\frac{1}{2}-\frac{8}{\sqrt{241}}\\
 0 &  \frac{30}{\sqrt{241}} &\frac{1}{2}+\frac{8}{\sqrt{241}}
\end{pmatrix}\,.
\end{equation}
And one can get a unique set of the diagonalized currents
\begin{equation}\label{eq:0-basis-mixing}
\vec{J}_{\text{P}}^{\text{Dia}}=\mathcal{T}_{\text{P}}^{\text{Dia}}\cdot \vec{J}_{\text{P}}^{\text{M-M}}\,.
\end{equation}
The anomalous dimension matrix of $\vec{J}_{\text{P}}^{\text{Dia}}$ is diagonal, which is given by
\begin{equation}\label{eq:0--Diagonal-AnoDim}
\begin{split}
\mathcal{A}_{\text{P}}^{\text{Dia}}& =\mathcal{T}_{\text{P}}^{\text{Dia}}\cdot\mathcal{A}_{\text{P}}^{\text{M-M}}\cdot (\mathcal{T}_{\text{P}}^{\text{Dia}})^{-1}\,\\
&=\frac{4}{3} \delta \begin{pmatrix}
 -6 &  0 & 0 \\
   0 &  -1+\sqrt{241} & 0 \\
      0 &  0 & -1-\sqrt{241} \\
\end{pmatrix}
\,.
\end{split}
\end{equation}

\subsection{$J^{P} = 1^{+}$}

For the $J^{P}=1^{+}$ axial vector system, there are four independent interpolating currents. The operator basis, in the color singlet meson-meson type  currents, can be chosen as
\begin{align}\label{eq:1+-meson}
\begin{split}
J_{\text{A},1}^{\text{M-M}}&= (\bar{Q}_a  Q_a)(\bar{Q}_b \gamma^\mu\gamma^5 Q_b)\,, \\
J_{\text{A},2}^{\text{M-M}}&= (\bar{Q}_a \sigma^{\mu\nu}i\gamma^5 Q_a)(\bar{Q}_b \gamma_\nu Q_b)\,, \\
J_{\text{A},3}^{\text{M-M}}&= (\bar{Q}_a i\gamma^5 Q_a)(\bar{Q}_b \gamma^\mu  Q_b)\,,\\
J_{\text{A},4}^{\text{M-M}}&= (\bar{Q}_a \sigma^{\mu\nu} Q_a)(\bar{Q}_b\gamma_\nu \gamma^5  Q_b)\,,
\end{split}
\end{align}
where $J_{\text{A},1}^{\text{M-M}}$ and $J_{\text{A},2}^{\text{M-M}}$ couple to states with $J^{PC}=1^{++}$, while $J_{\text{A},3}^{\text{M-M}}$ and $J_{\text{A},4}^{\text{M-M}}$  couple to states with $J^{PC}=1^{+-}$.
Alternatively, one can choose the diquark-antidiquark type currents~\cite{Chen:2016jxd} as the basis, which are given by
 \begin{align}\label{eq:1+-Diquark}
\begin{split}
J_{\text{A},1}^{\text{Di-Di}}&= (Q^T_a \hat{C} \gamma^\mu\gamma^5 Q_b)(\bar{Q}_a  \hat{C} \bar{Q}^T_b)+(Q^T_a \hat{C} Q_b)(\bar{Q}_a \gamma^\mu\gamma^5 \hat{C} \bar{Q}^T_b)\,, \\
J_{\text{A},2}^{\text{Di-Di}}&= (Q^T_a \hat{C}\sigma^{\mu\nu}i\gamma^5 Q_b)(\bar{Q}_a \gamma_\nu \hat{C} \bar{Q}^T_b) +(Q^T_a \hat{C} \gamma_\nu Q_b)(\bar{Q}_a \sigma^{\mu\nu}i\gamma^5 \hat{C} \bar{Q}^T_b)\,,  \\
J_{\text{A},3}^{\text{Di-Di}}&=(Q^T_a \hat{C} \gamma^\mu\gamma^5 Q_b)(\bar{Q}_a  \hat{C} \bar{Q}^T_b)-(Q^T_a \hat{C} Q_b)(\bar{Q}_a \gamma^\mu\gamma^5 \hat{C} \bar{Q}^T_b) \,, \\
J_{\text{A},4}^{\text{Di-Di}}&= (Q^T_a \hat{C}\sigma^{\mu\nu}i\gamma^5 Q_b)(\bar{Q}_a \gamma_\nu \hat{C} \bar{Q}^T_b)-(Q^T_a \hat{C} \gamma_\nu Q_b)(\bar{Q}_a \sigma^{\mu\nu}i\gamma^5 \hat{C} \bar{Q}^T_b)\,,
\end{split}
\end{align}
where $J_{\text{A},1}^{\text{Di-Di}}$ and $J_{\text{A},2}^{\text{Di-Di}}$ couple to states with $J^{PC}=1^{++}$, while $J_{\text{A},3}^{\text{Di-Di}}$ and $J_{\text{A},4}^{\text{Di-Di}}$  couple to  states with $J^{PC}=1^{+-}$.
In the calculation, $J_{\text{A}}^{\text{Di-Di}}$ can be associated with $J_{\text{A}}^{\text{M-M}}$ by the Fierz transformation in 4 dimension.
\begin{equation}\label{eq:1+-FierzTrans}
\vec{J}_{\text{A}}^{\text{Di-Di}}=\begin{pmatrix}
  -1 &  1 & 0 &0 \\
  3 &  1 & 0 &0 \\
  0 &  0 & i &i \\
  0 &  0 & -3\, i &i \\
\end{pmatrix} \cdot \vec{J}_{\text{A}}^{\text{M-M}}\,.
\end{equation}

Choosing the currents in Eq.~(\ref{eq:1+-meson}) as the operator basis, one can get the anomalous dimension matrix
\begin{equation}\label{eq:1+-meson-AnoDim}
\mathcal{A}_{\text{A}}^{\text{M-M}}= \frac{\delta}{3}\begin{pmatrix}
  30 & 2& 0 &0 \\
  -30 &  -34&0&0\\
  0& 0  &30 &-2 \\
    0& 0  &30 &-34 \\
\end{pmatrix}\,.
\end{equation}
To diagonalize the matrix in Eq.~(\ref{eq:1+-meson-AnoDim}), one needs the transformation matrix
\begin{equation}\label{eq:1+-meson-Diagonal}
\mathcal{T}_{\text{A}}^{\text{Dia}}=\frac{1}{2\sqrt{241}}\begin{pmatrix}
15 &  \sqrt{241}+16 &0&0 \\
-15 &  \sqrt{241}-16 &0&0 \\
  0 & 0 & -15&\sqrt{241}+16 \\
  0 & 0 &15& \sqrt{241}-16 \\
\end{pmatrix}\,.
\end{equation}
And one can get a unique set of the diagonalized currents
\begin{equation}\label{eq:1+basis-mixing}
\vec{J}_{\text{A}}^{\text{Dia}}=\mathcal{T}_{\text{A}}^{\text{Dia}}\cdot \vec{J}_{\text{A}}^{\text{M-M}}\,.
\end{equation}
The anomalous dimension matrix of $\vec{J}_{\text{A}}^{\text{Dia}}$ is diagonal, which is given by

\begin{equation}\label{eq:1+-Diagonal-AnoDim}
\begin{split}
\mathcal{A}_{\text{A}}^{\text{Dia}} =  -\frac{2}{3}\delta\begin{pmatrix}
  1+\sqrt{241} &  0 & 0&0 \\
   0 &   1-\sqrt{241} &  0 & 0 \\
      0 &  0 &  1+\sqrt{241} &  0 \\
            0 &  0 & 0& 1-\sqrt{241} \\
\end{pmatrix}\,.
\end{split}
\end{equation}

\subsection{$J^{P} = 1^{-} $}

For the $J^{P}=1^{-}$ vector system, there are four independent interpolating currents. The operator basis, in the color singlet meson-meson type  currents, can be chosen as
\begin{align}\label{eq:1--meson}
\begin{split}
J_{\text{V},1}^{\text{M-M}}&= (\bar{Q}_a  Q_a)(\bar{Q}_b \gamma^\mu Q_b)\,, \\
J_{\text{V},2}^{\text{M-M}}&= (\bar{Q}_a \sigma^{\mu\nu}i\gamma^5 Q_a)(\bar{Q}_b \gamma_\nu \gamma^5 Q_b)\,, \\
J_{\text{V},3}^{\text{M-M}}&= (\bar{Q}_a i\gamma^5 Q_a)(\bar{Q}_b \gamma^\mu \gamma^5  Q_b)\,,\\
J_{\text{V},4}^{\text{M-M}}&= (\bar{Q}_a \sigma^{\mu\nu} Q_a)(\bar{Q}_b\gamma_\nu  Q_b)\,,
\end{split}
\end{align}
where $J_{\text{V},1}^{\text{M-M}}$ and $J_{\text{V},2}^{\text{M-M}}$ couple to  states with $J^{PC}=1^{--}$, while $J_{\text{V},3}^{\text{M-M}}$ and $J_{\text{V},4}^{\text{M-M}}$  couple to  states with $J^{PC}=1^{-+}$.
Of course, one can choose the diquark-antidiquark type currents~\cite{Chen:2016jxd} as the basis, which are given by
 \begin{align}\label{eq:1--Diquark}
\begin{split}
J_{\text{V},1}^{\text{Di-Di}}&= (Q^T_a \hat{C} \gamma^\mu\gamma^5 Q_b)(\bar{Q}_a i\gamma^5 \hat{C} \bar{Q}^T_b)-(Q^T_a \hat{C}i\gamma^5 Q_b)(\bar{Q}_a \gamma^\mu\gamma^5 \hat{C} \bar{Q}^T_b)\,,  \\
J_{\text{V},2}^{\text{Di-Di}}&= (Q^T_a \hat{C}\sigma^{\mu\nu} Q_b)(\bar{Q}_a \gamma_\nu \hat{C} \bar{Q}^T_b)-(Q^T_a \hat{C} \gamma_\nu Q_b)(\bar{Q}_a \sigma^{\mu\nu} \hat{C} \bar{Q}^T_b)\,,\\
J_{\text{V},3}^{\text{Di-Di}}&= (Q^T_a \hat{C} \gamma^\mu\gamma^5 Q_b)(\bar{Q}_a i\gamma^5 \hat{C} \bar{Q}^T_b)+(Q^T_a \hat{C} i\gamma^5 Q_b)(\bar{Q}_a \gamma^\mu\gamma^5 \hat{C} \bar{Q}^T_b)\,, \\
J_{\text{V},4}^{\text{Di-Di}}&= (Q^T_a \hat{C}\sigma^{\mu\nu} Q_b)(\bar{Q}_a \gamma_\nu \hat{C} \bar{Q}^T_b)+(Q^T_a \hat{C} \gamma_\nu Q_b)(\bar{Q}_a \sigma^{\mu\nu} \hat{C} \bar{Q}^T_b)\,,
\end{split}
\end{align}
where $J_{\text{V},1}^{\text{Di-Di}}$ and $J_{\text{V},2}^{\text{Di-Di}}$ couple to states with $J^{PC}=1^{--}$, while $J_{\text{V},3}^{\text{Di-Di}}$ and $J_{\text{V},4}^{\text{Di-Di}}$ couple to  states with $J^{PC}=1^{-+}$.

In the calculation, $\vec{J}_{\text{V}}^{\text{Di-Di}}$ can be associated with $\vec{J}_{\text{V}}^{\text{M-M}}$ by Fierz Transformation in 4 dimension, which is given by
\begin{equation}\label{eq:1--FierzTrans}
\vec{J}_{\text{V}}^{\text{Di-Di}}=\begin{pmatrix}
 -i &  i & 0 &0 \\
 -3\, i &-i & 0 &  0  \\
  0 &  0 & -1 &-1 \\
   0 &0 &-3 &  1 \\

\end{pmatrix} \cdot \vec{J}_{\text{V}}^{\text{M-M}}\,.
\end{equation}

 If we choose  the currents in Eq.~(\ref{eq:1--meson}) as the operator basis, the anomalous dimension matrix $\mathcal{A}_{\text{V}}^{\text{M-M}}$ is the same as $\mathcal{A}_{\text{A}}^{\text{M-M}}$ shown in  Eq.~(\ref{eq:1+-meson-AnoDim}). Thus, similar to axial vector ($J^{P}=1^{+}$) system, one can get a unique set of diagonalized currents
 \begin{equation}\label{eq:1--Mixed-currents}
 \begin{split}
\vec{J}_{\text{V}}^{\text{Di-Di}}&=\mathcal{T}_{\text{A}}^{\text{Dia}}\cdot \vec{J}_{\text{V}}^{\text{M-M}}\\
         &=\frac{1}{2\sqrt{241}}\begin{pmatrix}
15 &  \sqrt{241}+16 &0&0 \\
-15 &  \sqrt{241}-16 &0&0 \\
  0 & 0 & -15&\sqrt{241}+16 \\
  0 & 0 &15& \sqrt{241}-16 \\
\end{pmatrix} \cdot \vec{J}_{\text{V}}^{\text{M-M}}\,,
\end{split}
\end{equation}
which make the anomalous dimension matrix $\mathcal{A}_{\text{V}}^{\text{M-M}}$ diagonal.

\subsection{$J^{P} = 2^{+}$}

For the $J^{PC}=2^{++}$ tensor system, there are three independent interpolating currents. The operator basis, in the color-singlet meson-meson type  currents, can be chosen as
\begin{align}\label{eq:2++meson}
\begin{split}
J_{\text{T},1}^{\text{M-M}}&= (\bar{Q}_a \gamma^\mu Q_a)(\bar{Q}_b \gamma^\nu Q_b)\,, \\
J_{\text{T},2}^{\text{M-M}}&= (\bar{Q}_a \gamma^\mu\gamma^5 Q_a)(\bar{Q}_b \gamma^\nu\gamma^5 Q_b)\,, \\
J_{\text{T},3}^{\text{M-M}}&= (\bar{Q}_a \sigma^{\mu\alpha} Q_a)(\bar{Q}_b\sigma^{\nu\alpha} Q_b)\,.
\end{split}
\end{align}
One could also construct the following operators
\begin{align}
	\begin{split}
		J_{\text{T},4}^{\text{M-M}}&= g^{\mu \nu} (\bar{Q}_a Q_a)(\bar{Q}_b Q_b)\,, \\
		J_{\text{T},5}^{\text{M-M}}&= g^{\mu \nu} (\bar{Q}_a i \gamma^{5} Q_a)(\bar{Q}_b i \gamma^{5} Q_b)\,, \\
		J_{\text{T},6}^{\text{M-M}}&= (\bar{Q}_a Q_a)(\bar{Q}_b \sigma^{\mu \nu} Q_b)\,, \\
	\end{split}
\end{align}
but they won't contribute to $\Pi^{T}_{1}$.
This suggests that these operators can not correspond to a tensor particle and we discard them from our analysis.

Of course, one can choose the diquark-antidiquark type currents~\cite{Chen:2016jxd} as the basis, which are given by
 \begin{align}\label{eq:2++Diquark}
\begin{split}
J_{\text{T},1}^{\text{Di-Di}}&= (Q^T_a \hat{C} \gamma^\mu Q_b)(\bar{Q}_a \gamma^\nu \hat{C} \bar{Q}^T_b)+(Q^T_a \hat{C} \gamma^\nu Q_b)(\bar{Q}_a  \gamma^\mu \hat{C} \bar{Q}^T_b)\,, \\
J_{\text{T},2}^{\text{Di-Di}}&= (Q^T_a \hat{C} \gamma^\mu\gamma^5 Q_b)(\bar{Q}_a \gamma^\nu\gamma^5 \hat{C} \bar{Q}^T_b)+(Q^T_a \hat{C} \gamma^\nu\gamma^5 Q_b)(\bar{Q}_a \gamma^\mu\gamma^5 \hat{C} \bar{Q}^T_b)\,, \\
J_{\text{T},3}^{\text{Di-Di}}&= (Q^T_a \hat{C} \sigma^{\mu\alpha} Q_b)(\bar{Q}_a \sigma^{\nu\alpha} \hat{C} \bar{Q}^T_b)+(Q^T_a \hat{C} \sigma^{\nu\alpha} Q_b)(\bar{Q}_a \sigma^{\mu\alpha} \hat{C} \bar{Q}^T_b)\,. \\
\end{split}
\end{align}
In the calculation, $\vec{J}_{\text{T}}^{\text{Di-Di}}$ can associate with $\vec{J}_{\text{T}}^{\text{M-M}}$ by the Fierz Transformation in 4 dimension, which is given by
\begin{equation}\label{eq:2++FierzTrans}
\vec{J}_{\text{T}}^{\text{Di-Di}}=-\frac{1}{2}\begin{pmatrix}
  -1 &  1 & 1 \\
 -1 &  1 & -1 \\
 2 &  2 & 0
\end{pmatrix} \cdot \vec{J}_{\text{T}}^{\text{M-M}}\,.
\end{equation}

Choosing the currents in Eq.~(\ref{eq:2++meson}) as the operator basis, one can get the anomalous dimension matrix
\begin{equation}\label{eq:2++meson-AnoDim}
\mathcal{A}_{\text{T}}^{\text{M-M}}=-\frac{2}{3}\delta\begin{pmatrix}
  3 & 5 & 3 \\
   5 &  3 &-3\\
 0 &  0 &16
\end{pmatrix}\,.
\end{equation}
To diagonalize the matrix in Eq.~(\ref{eq:2++meson-AnoDim}), one needs the transformation matrix
\begin{equation}\label{eq:2++meson-Diagonal}
\mathcal{T}_{\text{T}}^{\text{Dia}}=\frac{1}{6}\begin{pmatrix}
  -3 & 3 & 1 \\
   3 &  3 &0\\
 0 &  0 &6
\end{pmatrix}
\,.
\end{equation}
And one can get a unique set of the diagonalized currents
\begin{equation}\label{eq:2++basis-mixing}
\vec{J}_{\text{T}}^{\text{Dia}}=\mathcal{T}_{\text{T}}^{\text{Dia}}\cdot \vec{J}_{\text{T}}^{\text{M-M}}\,.
\end{equation}
The anomalous dimension matrix of $\vec{J}_{\text{T}}^{\text{Dia}}$ is diagonal, which is given by
\begin{equation}\label{eq:2++Diagonal-AnoDim}
\begin{split}
\mathcal{A}_{\text{T}}^{\text{Dia}}& =\mathcal{T}_{\text{T}}^{\text{Dia}}\cdot\mathcal{A}_{\text{T}}^{\text{M-M}}\cdot (\mathcal{T}_{\text{T}}^{\text{Dia}})^{-1}\,\\
&=-\frac{4}{3}\delta \begin{pmatrix}
 -1 &  0 & 0 \\
   0 &  4 & 0 \\
      0 &  0 & 8 \\
\end{pmatrix}
\,.
\end{split}
\end{equation}
Thus, all diagonalized currents can be determined uniquely.
%%%%%%%%%%%%%%%%%%%%%%%%%%%%%%%%%%%%%%%%%%%%%%%%%

\section{Phenomenology}\label{sec:Phenomenology}
In our numerical analysis, we choose the following parameters~\cite{Bagan:1992za,Dominguez:1994ce,Dominguez:2014pga,Aoki:2016frl,Wang:2017qvg},
\begin{align}\label{eq:parameters}
\begin{split}
m_c^{\overline{\text{MS}}}(m_c)&=1.27 \pm 0.03\, \,  {\text{GeV}}\,,\\
m_c^{\text{OS}}&=1.46 \pm 0.07\, \,  {\text{GeV,}}\,\\
m_b^{\overline{\text{MS}}}(m_b)&=4.18 \pm 0.03\, \,  {\text{GeV}}\,,\\
m_b^{\text{OS}}&=4.65 \pm 0.05\, \,  {\text{GeV}}\,,\\
\langle g_s^2 \hat{G}\hat{G} \rangle &= 4\pi^2(0.037\pm0.015) \, \, {\text{GeV}}^4 \,,\\
\alpha_s(m_Z&=91.1876 ~{\text{GeV}})=0.1181\,.
\end{split}
\end{align}
It is worth emphasizing that $\alpha_s (\mu)$ and the heavy quark mass $m_Q^{\overline{\text{MS}}}(\mu)$ are obtained through two-loop running. Note that we don't need to consider the running of $\langle g_s^2 \hat{G}\hat{G} \rangle$ for the LO $GG$ condensate contribution, as its anomalous dimension vanishes up to this order. As a typical choice, we set $\mu=M_B$ in our phenomenological analysis~\cite{Shifman:1978bx,Bertlmann:1981he}, but the renormalization scale dependence will also be discussed.
On-Shell (OS) masses $m_c^{\text{OS}}$ and $m_b^{\text{OS}}$ are extracted from the QCD sum rules analysis of the $J/\psi$ and $\Upsilon(1S)$ spectrum, respectively, in which the mass renormalization scheme and truncation order of $\alpha_s$ are the same as this paper.

According to Eq.~(\ref{eq:MH}), numerical result $M_H$ also depends on other two parameters: $s_0$ and $M_B$. However, the  physical value of $M_H$ should be independent of any artificial parameters. So a credible result should be obtained from an appropriate region where the dependence of $s_0$ and $M_B$ is weak. On the other hand, the choice of $M_B$ and $s_0$ should ensure the validity of the OPE and ground-state contribution dominance, which constrain the two parameters to be the so-called ``Borel window''. Within the Borel window, one should find the region, the so-called ``Borel platform'', in which $M_H$ depends on $s_0$ and $M_B$  weakly.

To search for the Borel window, we define the relative contributions of the condensate and continuum as
\begin{align}\label{eq:borelwindow}
\begin{split}
r_{GG} &=\frac{\langle g_s^2 \hat{G}\hat{G} \rangle\int_{s_{\text th}}^{\infty} \mathrm{d} s\, \, \rho_{GG}(s)\, e^{-\frac{s}{M_B^2}}}{\int_{s_{\text th}}^{\infty} \mathrm{d} s\, \, \rho_{1}(s)\, e^{-\frac{s}{M_B^2}}}\,,\\
r_{\text{cont}} &=\frac{\int_{s_{0}}^{\infty} \mathrm{d} s\, \, \rho_{1}(s)\, e^{-\frac{s}{M_B^2}}}{\int_{s_{\text th}}^{\infty} \mathrm{d} s\, \, \rho_{1}(s)\, e^{-\frac{s}{M_B^2}}}\,,
\end{split}
\end{align}
and impose the following constraints:
\begin{align}\label{eq:BWcondition}
|r_{GG}|\leq 30\%, \, \quad \, |r_{\text{cont}}|\leq 30\%\,.
\end{align}
The two constraints guarantee the validity of OPE and the ground-state contribution dominance, respectively.  In addition to the conditions given in Eq.~(\ref{eq:BWcondition}), we also impose the following constrain on $s_0$:
\begin{equation}\label{eq:S0condition}
s_0<(M_H+1 \ \text{GeV})^2,
\end{equation}
since, roughly speaking, $s_0$ denotes the energy scale where the continuum spectrum begins to contribute and that the binding energy in a purely heavy hadron is usually smaller than 1~GeV.
To find the Borel platform, we search for the point where the parameter dependence of $M_H$ is weakest within the Borel window.
More explicitly, we choose the variables as $x=s_0$ and $y=M_B^2$ and define the function
\begin{align}\label{eq:findPlatform}
\Delta(x,y)=\left( \frac{\partial M_H}{\partial x} \right)^2+ \left( \frac{\partial M_H}{\partial y}\right)^2\,.
\end{align}
By minimizing the function $\Delta(x,y)$ within the Borel window and with the constrain Eq.~(\ref{eq:S0condition}), we get a point ($x_0,y_0$), which will be used to calculate the central value of $M_H$. To estimate errors of $M_H$, we vary the values of $s_0$ and $M_B^2$ around the point ($x_0,y_0$) up to 10\% in magnitude. It should be emphasized that the central point ($x_0,y_0$) may lies on the margin of the Borel window in some cases. Therefore, the parameter space used to estimate errors of $M_H$ may exceed the Borel window, and also, the upper and the lower errors are usually asymmetric.

\subsection{Numerical results and discussions for the $\bar{c}c\bar{c}c$ system}

\begin{figure}[htb]
	\vspace{-0.3cm}
	\setlength{\abovecaptionskip}{-0.8cm}
	\begin{center}
		\includegraphics[scale=0.5]{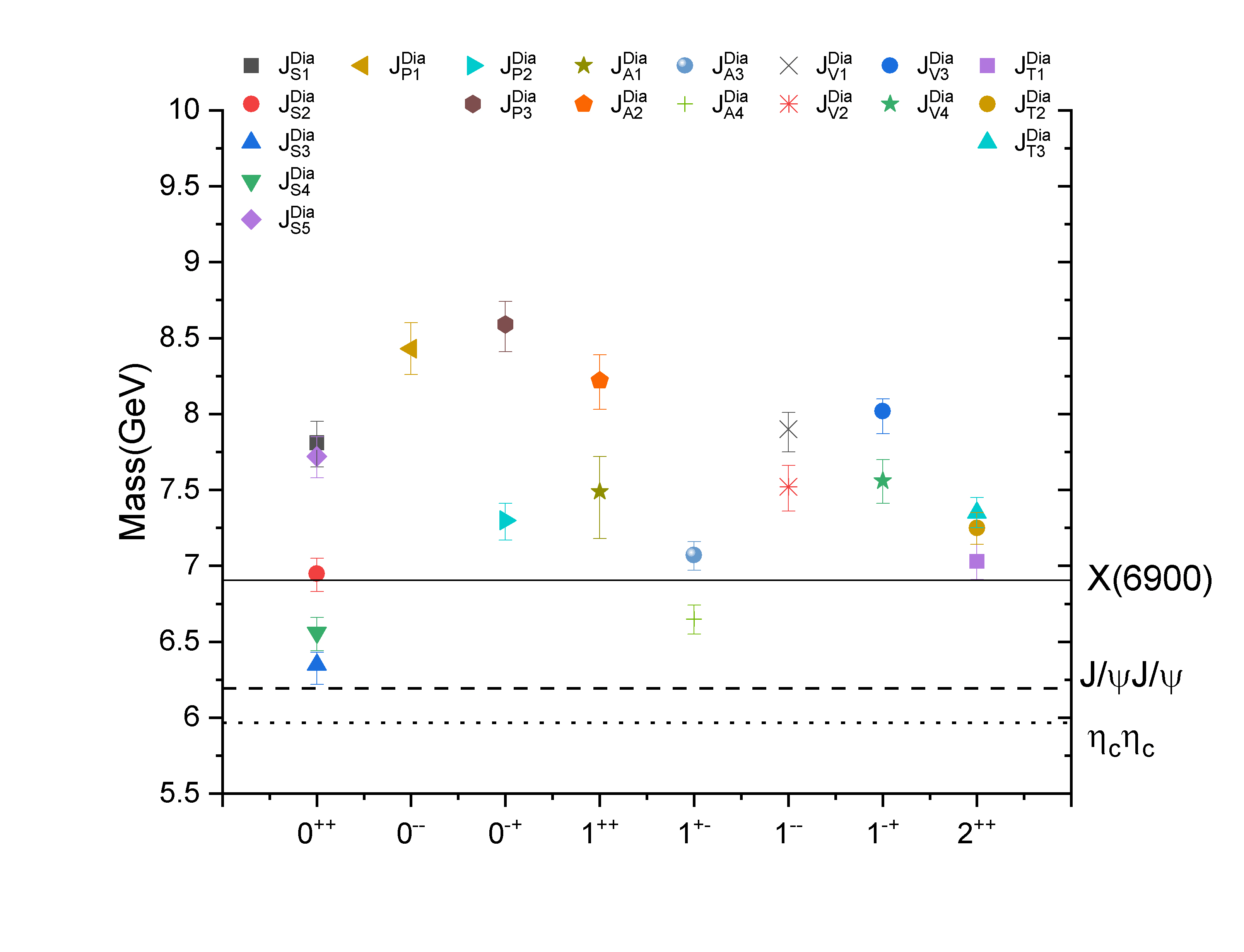}
		\caption{\label{fig:4c-Mass-Spectrum}
			The NLO mass spectra of the $\bar{c}c\bar{c}c$ system in the $\overline{\text{MS}}$ scheme. The errors of masses shown in this figure just come from the the parameter dependence on $s_0$ and $M_B^2$.}
	\end{center}
\end{figure}

Our main results are shown in Fig.~\ref{fig:4c-Mass-Spectrum}, where the 19 diagonalized currents, which should be more reasonable to be used in the QCD sum rules, are clustered by different quantum numbers. We set  $\mu=M_B$ and choose the $\overline{\text{MS}}$ renormalization scheme, and errors of $M_H$ include only that originated from uncertainties of $s_0$ and $M_B^2$. In the plot we also indicate the mass of $X(6900)$ and the double $J/\psi$ threshold.

The most comprehensive results are listed in Tabs.~\ref{tab:S-NLOresult-MSbar}--\ref{tab:T-NLOresult-OS} in Appendix~\ref{sec:charm}, where we include  both LO and NLO,  both $\overline{\text{MS}}$ scheme and on-shell scheme, and all currents of meson-meson types, diquark-antidiquark types, and also diagonalized ones. Again in these tables we set $\mu=M_B$ and thus errors of $M_H$ are due to choices of $s_0$ and $M_B^2$.
Further information of $s_0$ and $M_B^2$ dependence is shown in   Figs.~\ref{fig:0+-Mixed1-NLO-MSbar-OS}--\ref{fig:2+-Mixed3-NLO-MSbar-OS} in Appendix~\ref{sec:charm}, where only results of the more reasonable diagonalized currents are shown.
In these plots, a black dot denotes the central point ($x_0,y_0$), and shadows denote the Borel window determined by Eq.~(\ref{eq:BWcondition}).

Let us first emphasize the importance of the NLO corrections. On one hand,  NLO corrections to hadron masses are significant, which are larger than 0.5~GeV in both  $\overline{\text{MS}}$ and on-shell schemes for almost all the currents involved in Tabs.~\ref{tab:S-NLOresult-MSbar}--\ref{tab:T-NLOresult-OS}. On the other hand, with the NLO corrections, the quark mass scheme dependence of $M_H$ tends to be reduced, especially for some diagonalized currents. To see this, we examine the difference of the predicted hadron masses between the two schemes,
\begin{equation}\label{eq:DeltaMH}
\Delta M_H=M_H^{\text{OS}}-M_H^{\overline{\text{MS}}}\,.
\end{equation}
From Tabs.~\ref{tab:S-NLOresult-MSbar}--\ref{tab:T-NLOresult-OS}, for almost all the currents, one can find that the mass difference at LO is about $\Delta M_H^{\text{LO}}\approx1.2$~GeV, which implies a roughly linear dependence of $\Delta M_H^{\text{LO}}$ on the quark mass difference between the two schemes. One can also find that the NLO corrections to $M_H^{\overline{\text{MS}}}$ are positive while those to $M_H^{\text{OS}}$ are negative, therefore, the scheme dependence of $M_H$ tends to be reduced with the NLO corrections. Taking the $J^{P}=0^+$ system as an example (see Tab.~\ref{tab:S-NLOresult-MSbar} and \ref{tab:S-NLOresult-OS}), for the three diagonalized currents $J_{S,2,3,4}^{\text{Dia}}$, the NLO mass difference $|\Delta M_H^{\text{NLO}}|<0.4$~GeV, which is explicitly smaller than that at LO. As for the currents $J_{S,1}^{\text{Dia}}$ and $J_{S,5}^{\text{Dia}}$, the NLO corrections to $M_H^{\overline{\text{MS}}}$ are larger than 1~GeV, which implies that there are genuine large corrections other than the quark mass renormalization effects and the perturbation convergence may be bad for these currents.

The convergence of perturbation can also be explored by the $\mu$ dependence of the NLO results. One would expect that the $\mu$ dependence of the NLO result will be significantly reduced in comparison with the LO one for the current for which the perturbation convergence is good, since the truncation of the perturbation series up to NLO has weak effect on the result. On the other hand, the large $\mu$ dependence of the NLO result may imply that the perturbation convergence is bad, say, the next-to-next-to-leading-order (NNLO) corrections should be important in this case. In Tabs.~\ref{tab:S-NLOresult-MSbar}--\ref{tab:T-NLOresult-OS}, we have chose $\mu=M_B$ in the $\overline{\text{MS}}$ scheme. To study the $\mu$ dependence, we vary $\mu=k \ M_B$ with $k \in (0.8, 2.0)$, where the range  is chosen with the requirement that the Borel platform can be achieved and the perturbative expansion is under good control.
We investigate the $\mu$ dependence for the diagonalized operators with $J^{PC}=0^{++}$, which are shown in Fig.~\ref{4c-mu-dependence-S-1}--\ref{4c-mu-dependence-S-5} in Appendix~\ref{sec:charm} for  the LO and the NLO results.
From these plots, one can see that the $\mu$ dependence of the results for $J_{S,2,3,4}^{\text{Dia}}$ are improved significantly after including the NLO contributions, especially for the last two operators $J_{S,3}^{\text{Dia}}$ and $J_{S,4}^{\text{Dia}}$, which implies that those operators may have good perturbation convergence. While the NLO results of $J_{S,1}^{\text{Dia}}$ and $J_{S,5}^{\text{Dia}}$ are still very sensitive to the renormalization scale $\mu$, which implies that those currents may have bad perturbation convergence. This may indicate that the mass difference $\Delta M_H^{\text{NLO}}$ between the two schemes and the $\mu$ dependence of the NLO $M_H^{\overline{\text{MS}}}$ are correlated. That is, when there is a good perturbation convergence, we should expect a small $\Delta M_H^{\text{NLO}}$ and weak $\mu$ dependence at NLO.

As for states other than $J^{PC}=0^{++}$, there are only three diagonalized operators that satisfy $\Delta M_H^{\text{NLO}}<0.5$~GeV. They are $J_{P,2}^{\text{Dia}}$ with $J^{PC}=0^{-+}$, $J_{A,4}^{\text{Dia}}$ with $J^{PC}=1^{+-}$ and $J_{T,1}^{\text{Dia}}$ with $J^{PC}=2^{++}$. The $\mu$ dependence of the LO and NLO $\overline{\text{MS}}$ masses for these three diagonalized operators are shown in Fig.~\ref{4c-mu-dependence-P-2}--\ref{4c-mu-dependence-T-1} in Appendix~\ref{sec:charm}. Just as one would expect, the $\mu$ dependence of NLO results are improved significantly, compared with that of the LO one.

In cases where convergence of perturbation  is bad, uncertainties from higher order corrections should be large. As higher order corrections, such as the NNLO ones for $C_1$, are beyond the scope of this paper, we will only choose diagonalized operators that have good perturbation convergence in the following analysis. For the results of the diagonalized operators $J_{S,2,3,4}^{\text{Dia}}$ with $J^{PC}=0^{++}$, $J_{P,2}^{\text{Dia}}$ with $J^{PC}=0^{-+}$, $J_{A,4}^{\text{Dia}}$ with $J^{PC}=1^{+-}$ and $J_{T,1}^{\text{Dia}}$ with $J^{PC}=2^{++}$, we also estimate the uncertainties coming from errors of the quark masses in Eq.~(\ref{eq:parameters}), and the uncertainties are shown in Tabs.~\ref{tab:R2-NLOresult-MSbar-OS}--\ref{tab:T1-NLOresult-MSbar-OS}.

\begin{table}[H]
	\renewcommand\arraystretch{1.7}
\begin{center}
	\setlength{\tabcolsep}{1.2 mm}
		\begin{tabular}{cccccccccc}
			\hline\hline
			Current& Order&   $M_H$ (GeV)   &  $s_0$ (${\text{GeV}}^2$)  & $M_B^2$ (${\text{GeV}}^2$)  & \makecell{Error from \\$s_0$ and $M_B^2$ } & \makecell{Error from \\$m_Q$ } & \makecell{Error from \\$\mu$} \\ \hline
			
			\multirow{4}{*}{$J_{S,2}^{\text{Dia}}$} & LO($\overline{\text{MS}}$) &    $6.19^{+0.26}_{-0.23}$   &    $51(\pm 10\%)$   &   $3.50(\pm 10\%)$   &  $^{+0.07}_{-0.12}$ &  $^{+0.11}_{-0.14}$ &  $^{+0.22}_{-0.23}$ \\
			
			&NLO($\overline{\text{MS}}$) &   $6.95^{+0.21}_{-0.31}$   &    $61(\pm 10\%)$   &   $5.00(\pm 10\%)$    &  $^{+0.10}_{-0.12}$ &  $^{+0.15}_{-0.13}$ &  $^{+0.11}_{-0.26}$ \\
			
			&LO(OS)     &         $7.31^{+0.29}_{-0.24}$         &      $64(\pm 10\%)$    &   $3.75(\pm 10\%)$    & $^{+0.08}_{-0.12}$ &  $^{+0.28}_{-0.21}$ & \\
			
			&NLO(OS)     &         $6.58^{+0.28}_{-0.29}$           &   $48(\pm 10\%)$     &   $2.00(\pm 10\%)$    & $^{+0.08}_{-0.11}$ &  $^{+0.27}_{-0.27}$ &\\
			\hline\hline
		\end{tabular}
		\caption{The LO and NLO results for the mass of $J_{S,2}^{\text{Dia}}$ in $\overline{\text{MS}}$ and On-Shell schemes. Here the errors for $M_H$ are from $s_0, M_B$, the charm quark mass, and the renormalization scale $\mu$ with $\mu=kM_B$ and $k \in (0.8, 1.2)$ (the central values correspond to $\mu=M_B$ ).}
		\label{tab:R2-NLOresult-MSbar-OS}
	\end{center}
\end{table}

\begin{table}[H]
	\renewcommand\arraystretch{1.7}
\begin{center}
	\setlength{\tabcolsep}{1.2 mm}
		\begin{tabular}{cccccccccc}
			\hline\hline
			Current& Order&   $M_H$ (GeV)   &  $s_0$ (${\text{GeV}}^2$)  & $M_B^2$ (${\text{GeV}}^2$)  & \makecell{Error from \\$s_0$ and $M_B^2$ } & \makecell{Error from \\$m_Q$ } & \makecell{Error from \\$\mu$} \\ \hline
			
			\multirow{4}{*}{$J_{S,3}^{\text{Dia}}$} & LO($\overline{\text{MS}}$) &    $5.93^{+0.31}_{-0.26}$   &    $45(\pm 10\%)$   &   $3.00(\pm 10\%)$   &  $^{+0.07}_{-0.10}$ &  $^{+0.18}_{-0.09}$ &  $^{+0.24}_{-0.22}$ \\
			
			&NLO($\overline{\text{MS}}$) &   $6.35^{+0.20}_{-0.17}$   &    $51(\pm 10\%)$   &   $3.50(\pm 10\%)$    &  $^{+0.08}_{-0.13}$ &  $^{+0.18}_{-0.11}$ &  $^{+0.00}_{-0.03}$ \\
			
			&LO(OS)     &         $7.06^{+0.32}_{-0.26}$         &      $60(\pm 10\%)$    &   $3.00(\pm 10\%)$    & $^{+0.07}_{-0.10}$ &  $^{+0.31}_{-0.24}$ & \\
			
			&NLO(OS)     &         $6.47^{+0.29}_{-0.30}$           &   $46(\pm 10\%)$     &   $1.75(\pm 10\%)$    & $^{+0.08}_{-0.10}$ &  $^{+0.28}_{-0.28}$ &\\
			\hline\hline
		\end{tabular}
		\caption{The LO and NLO results for the mass of $J_{S,3}^{\text{Dia}}$ in $\overline{\text{MS}}$ and On-Shell schemes. Here the errors for $M_H$ are from $s_0, M_B$, the charm quark mass, and the renormalization scale $\mu$ with $\mu=kM_B$ and $k \in (0.8, 1.2)$ (the central values correspond to $\mu=M_B$ ).}
		\label{tab:R3-NLOresult-MSbar-OS}
	\end{center}
\end{table}

\begin{table}[H]
	\renewcommand\arraystretch{1.7}
	\begin{center}
		\setlength{\tabcolsep}{1.2 mm}
		\begin{tabular}{cccccccccc}
			\hline\hline
			Current& Order&   $M_H$ (GeV)   &  $s_0$ (${\text{GeV}}^2$)  & $M_B^2$ (${\text{GeV}}^2$)  & \makecell{Error from \\$s_0$ and $M_B^2$ } & \makecell{Error from \\$m_Q$ } & \makecell{Error from \\$\mu$} \\ \hline
			
			\multirow{4}{*}{$J_{S,4}^{\text{Dia}}$} & LO($\overline{\text{MS}}$) &    $6.02^{+0.24}_{-0.28}$   &    $49(\pm 10\%)$   &   $3.00(\pm 10\%)$   &  $^{+0.05}_{-0.06}$ &  $^{+0.09}_{-0.14}$ &  $^{+0.22}_{-0.23}$\\
			
			&NLO($\overline{\text{MS}}$) &   $6.56^{+0.18}_{-0.20}$   &    $55(\pm 10\%)$   &   $4.00(\pm 10\%)$    &  $^{+0.10}_{-0.12}$ &  $^{+0.15}_{-0.13}$ & $^{+0.03}_{-0.10}$\\
			
			&LO(OS)     &         $7.16^{+0.24}_{-0.30}$         &      $66(\pm 10\%)$    &   $3.00(\pm 10\%)$    & $^{+0.04}_{-0.05}$ &  $^{+0.24}_{-0.30}$ & \\
			
			&NLO(OS)     &         $6.49^{+0.29}_{-0.30}$           &   $46(\pm 10\%)$     &   $1.75(\pm 10\%)$    & $^{+0.07}_{-0.10}$ &  $^{+0.28}_{-0.28}$ &\\
			\hline\hline
		\end{tabular}
		\caption{The LO and NLO results for the mass of $J_{S,4}^{\text{Dia}}$ in $\overline{\text{MS}}$ and On-Shell schemes. Here the errors for $M_H$ are from $s_0, M_B$, the charm quark mass, and the renormalization scale $\mu$ with $\mu=kM_B$ and $k \in (0.8, 1.2)$ (the central values correspond to $\mu=M_B$ ).}
		\label{tab:R4-NLOresult-MSbar-OS}
	\end{center}
\vspace{-0.4cm}
\end{table}

\begin{table}[H]
	\renewcommand\arraystretch{1.7}
	\begin{center}
		\setlength{\tabcolsep}{1.2 mm}
		\begin{tabular}{cccccccccc}
			\hline\hline
			Current& Order&   $M_H$ (GeV)   &  $s_0$ (${\text{GeV}}^2$)  & $M_B^2$ (${\text{GeV}}^2$)  & \makecell{Error from \\$s_0$ and $M_B^2$ } & \makecell{Error from \\$m_Q$ } & \makecell{Error from \\$\mu$} \\ \hline
			
			\multirow{4}{*}{$J_{P,2}^{\text{Dia}}$} & LO($\overline{\text{MS}}$) &    $6.53^{+0.29}_{-0.27}$ &$56.(\pm 10\%)$ &$4.25(\pm 10\%)$    &  $^{+0.12}_{-0.14}$ &  $^{+0.12}_{-0.16}$ &  $^{+0.24}_{-0.16}$ \\
			
			&NLO($\overline{\text{MS}}$) &    $7.30^{+0.19}_{-0.20}$ &$68.(\pm 10\%)$ &$6.00(\pm 10\%)$   & $^{+0.11}_{-0.13}$ &  $^{+0.14}_{-0.15}$ &  $^{+0.00}_{-0.03}$\\
			
			&LO(OS)     & $7.79^{+0.31}_{-0.27}$ &$74.(\pm 10\%)$ &$4.50(\pm 10\%)$  & $^{+0.10}_{-0.15}$  & $^{+0.29}_{-0.23}$     \\
			
			&NLO(OS)     &   $6.89^{+0.32}_{-0.34}$ &$52.(\pm 10\%)$ &$2.75(\pm 10\%)$  &   $^{+0.14}_{-0.23}$ & $^{+0.29}_{-0.25}$ \\
			\hline\hline
		\end{tabular}
		\caption{The LO and NLO results for the mass of $J_{P,2}^{\text{Dia}}$ in $\overline{\text{MS}}$ and On-Shell schemes. Here the errors for $M_H$ are from $s_0, M_B$, the charm quark mass, and the renormalization scale $\mu$ with $\mu=kM_B$ and $k \in (0.8, 1.2)$ (the central values correspond to $\mu=M_B$ ).}
		\label{tab:P2-NLOresult-MSbar-OS}
	\end{center}
\vspace{-0.4cm}
\end{table}

\begin{table}[H]
	\renewcommand\arraystretch{1.7}
	\begin{center}
		\setlength{\tabcolsep}{1.2 mm}
		\begin{tabular}{cccccccccc}
			\hline\hline
			Current& Order&   $M_H$ (GeV)   &  $s_0$ (${\text{GeV}}^2$)  & $M_B^2$ (${\text{GeV}}^2$)  & \makecell{Error from \\$s_0$ and $M_B^2$ } & \makecell{Error from \\$m_Q$ } & \makecell{Error from \\$\mu$} \\ \hline
			
			\multirow{4}{*}{$J_{A,4}^{\text{Dia}}$} & LO($\overline{\text{MS}}$) &   $6.04^{+0.26}_{-0.26}$ &$48.(\pm 10\%)$ &$3.25(\pm 10\%)$     &  $^{+0.06}_{-0.08}$ &  $^{+0.10}_{-0.13}$ &  $^{+0.23}_{-0.21}$ \\
			
			&NLO($\overline{\text{MS}}$) & $6.65^{+0.18}_{-0.23}$ &$58.(\pm 10\%)$ &$4.25(\pm 10\%)$   & $^{+0.09}_{-0.10}$ &  $^{+0.15}_{-0.17}$ &  $^{+0.01}_{-0.11}$ \\
			
			&LO(OS)     &$7.23^{+0.24}_{-0.31}$ &$67.(\pm 10\%)$ &$3.25(\pm 10\%)$  & $^{+0.04}_{-0.05}$  & $^{+0.24}_{-0.31}$      \\
			
			&NLO(OS)    &$6.53^{+0.27}_{-0.28}$ &$47.(\pm 10\%)$ &$2.00(\pm 10\%)$&   $^{+0.08}_{-0.11}$ & $^{+0.26}_{-0.26}$ \\
			\hline\hline
		\end{tabular}
		\caption{The LO and NLO results for the mass of $J_{A,4}^{\text{Dia}}$ in $\overline{\text{MS}}$ and On-Shell schemes. Here the errors for $M_H$ are from $s_0, M_B$, the charm quark mass, and the renormalization scale $\mu$ with $\mu=kM_B$ and $k \in (0.8, 1.2)$ (the central values correspond to $\mu=M_B$ ).}
		\label{tab:A4-NLOresult-MSbar-OS}
	\end{center}
\vspace{-0.4cm}
\end{table}

\begin{table}[H]
	\renewcommand\arraystretch{1.7}
\begin{center}
	\setlength{\tabcolsep}{1.2 mm}
		\begin{tabular}{cccccccccc}
			\hline\hline
			Current& Order&   $M_H$ (GeV)   &  $s_0$ (${\text{GeV}}^2$)  & $M_B^2$ (${\text{GeV}}^2$)  & \makecell{Error from \\$s_0$ and $M_B^2$ } & \makecell{Error from \\$m_Q$ } & \makecell{Error from \\$\mu$} \\ \hline
			
			\multirow{4}{*}{$J_{T,1}^{\text{Dia}}$} & LO($\overline{\text{MS}}$) &    $6.14^{+0.25}_{-0.29}$   &    $51(\pm 10\%)$   &   $3.50(\pm 10\%)$   &  $^{+0.07}_{-0.11}$ &  $^{+0.07}_{-0.15}$ &  $^{+0.23}_{-0.22}$ \\
			
			&NLO($\overline{\text{MS}}$) &   $7.03^{+0.22}_{-0.26}$   &    $63(\pm 10\%)$   &   $5.50(\pm 10\%)$    &  $^{+0.11}_{-0.12}$ &  $^{+0.17}_{-0.14}$ & $^{+0.08}_{-0.18}$ \\
			
			&LO(OS)     &         $7.31^{+0.25}_{-0.30}$         &      $68(\pm 10\%)$    &   $3.50(\pm 10\%)$    & $^{+0.05}_{-0.08}$ &  $^{+0.24}_{-0.29}$ & \\
			
			&NLO(OS)     &         $6.56^{+0.28}_{-0.32}$           &   $47(\pm 10\%)$     &   $2.00(\pm 10\%)$    & $^{+0.09}_{-0.15}$ &  $^{+0.27}_{-0.28}$ &\\
			\hline\hline
		\end{tabular}
		\caption{The LO and NLO results for the mass of $J_{T,1}^{\text{Dia}}$ in $\overline{\text{MS}}$ and On-Shell schemes. Here the errors for $M_H$ are from $s_0, M_B$, the charm quark mass, and the renormalization scale $\mu$ with $\mu=kM_B$ and $k \in (0.8, 1.2)$ (the central values correspond to $\mu=M_B$ ).}
		\label{tab:T1-NLOresult-MSbar-OS}
	\end{center}
\end{table}

For $J_{T,2}^{\text{Dia}}$, although $\Delta M_H^{\text{NLO}}\simeq 0.7$~GeV, the $\mu$ dependence of this current is good, as shown in Fig.~\ref{4c-mu-dependence-T-2}, so we also meticulously estimate the uncertainties, which are shown in Tab.~\ref{tab:T2-NLOresult-MSbar-OS}. ($J_{T,3}^{\text{Dia}}$ will not be further considered, since it has large $\mu$ dependence.)
\begin{table}[H]
	\renewcommand\arraystretch{1.7}
\begin{center}
	\setlength{\tabcolsep}{1.2 mm}
		\begin{tabular}{cccccccccc}
			\hline\hline
			Current& Order&   $M_H$ (GeV)   &  $s_0$ (${\text{GeV}}^2$)  & $M_B^2$ (${\text{GeV}}^2$)  & \makecell{Error from \\$s_0$ and $M_B^2$ } & \makecell{Error from \\$m_Q$ } & \makecell{Error from \\$\mu$} \\ \hline
			
			\multirow{4}{*}{$J_{T,2}^{\text{Dia}}$} & LO($\overline{\text{MS}}$) &    $6.15^{+0.32}_{-0.21}$   &    $49(\pm 10\%)$   &   $3.75(\pm 10\%)$   &  $^{+0.08}_{-0.10}$ &  $^{+0.13}_{-0.09}$ &  $^{+0.28}_{-0.16}$ \\
			
			&NLO($\overline{\text{MS}}$) &   $7.25^{+0.21}_{-0.35}$   &    $67(\pm 10\%)$   &   $5.75(\pm 10\%)$    &  $^{+0.10}_{-0.11}$ &  $^{+0.14}_{-0.17}$ & $^{+0.12}_{-0.29}$ \\
			
			&LO(OS)     &         $7.32^{+0.29}_{-0.26}$         &      $65(\pm 10\%)$    &   $3.75(\pm 10\%)$    & $^{+0.07}_{-0.13}$ &  $^{+0.28}_{-0.22}$ & \\
			
			&NLO(OS)     &         $6.57^{+0.30}_{-0.37}$           &   $47(\pm 10\%)$     &   $2.25(\pm 10\%)$    & $^{+0.12}_{-0.28}$ &  $^{+0.27}_{-0.24}$ &\\
			\hline\hline
		\end{tabular}
		\caption{The LO and NLO results for the mass of $J_{T,2}^{\text{Dia}}$ in $\overline{\text{MS}}$ and On-Shell schemes. Here the errors for $M_H$ are from $s_0, M_B$, the charm quark mass, and the renormalization scale $\mu$ with $\mu=kM_B$ and $k \in (0.8, 1.2)$ (the central values correspond to $\mu=M_B$ ).}
		\label{tab:T2-NLOresult-MSbar-OS}
	\end{center}
\end{table}

Phenomenologically, it is interesting to compare our calculations with the LHCb measurements of the possible $\bar{c}c\bar{c}c$ tetraquark states in the $J/\psi J/\psi$ spectrum~\cite{LHCb:2020bwg}. The most likely quantum numbers $J^{PC}$ for the tetraquark states are $0^{++}$ and $2^{++}$, since they can  couple to $J/\psi J/\psi$ in S-wave. The predicted NLO $\overline{\text{MS}}$ masses for the two operators $J_{S,3}^{\text{Dia}}$ and $J_{S,4}^{\text{Dia}}$ are $6.35^{+0.20}_{-0.17}$~GeV and $6.56^{+0.18}_{-0.20}$~GeV, respectively, which might account for the broad structure around $6.2\sim6.8$~GeV measured by the LHCb collaboration~\cite{LHCb:2020bwg}.  As for the narrow resonance $X(6900)$~\cite{LHCb:2020bwg}, the central value of the mass is consistent with the NLO $\overline{\text{MS}}$ mass for the operator $J_{S,2}^{\text{Dia}}$, which gives $6.95^{+0.21}_{-0.31}$~GeV. Moreover, the predicted NLO $\overline{\text{MS}}$ mass, $7.03^{+0.22}_{-0.26}$~GeV, for the operator $J_{T,1}^{\text{Dia}}$ with $J^{PC}=2^{++}$ is also close to that of $X(6900)$, so we can not assert that the quantum number of X(6900) is $0^{++}$, while $2^{++}$ may also be possible. Since the quality of the Borel platform in On-Shell scheme is worse than that in $\overline{\text{MS}}$ one, which can be seen from Figs.~\ref{fig:0+-Mixed1-NLO-MSbar-OS}--\ref{fig:2+-Mixed3-NLO-MSbar-OS}, we only use the corresponding $\overline{\text{MS}}$ masses in the above analysis. As for the NLO On-Shell masses of $J_{S,2,3,4}^{\text{Dia}}$ and $J_{T,1}^{\text{Dia}}$ (see Table~\ref{tab:R2-NLOresult-MSbar-OS}--\ref{tab:R4-NLOresult-MSbar-OS} and \ref{tab:T1-NLOresult-MSbar-OS}), they all lie on the the broad structure around $6.2\sim6.8$~GeV measured by the LHCb collaboration~\cite{LHCb:2020bwg}.

Since the above $\bar{c}c\bar{c}c$ states are pure heavy-quark systems, their non-relativistic (NR) attributes should be important to understand them. In our calculations, although the amplitude are calculated in full QCD for the covariant operators given in Sec.~\ref{sec:Current operators}, the results still exhibit some NR features. Taking  the $J^{PC}=0^{++}$ states as examples, both the diagonalized operators $J_{S,3}^{\text{Dia}}$ and $J_{S,4}^{\text{Dia}}$ are roughly mixing of the meson-meson type ones  $J_{S,3}^{\text{M-M}}$ and $J_{S,4}^{\text{M-M}}$ with the same weight, which can be seen from the transition matrix given in Eq.~(\ref{eq:0++meson-Diagonal}) (the $J_{S,5}^{\text{M-M}}$ component of $J_{S,4}^{\text{Dia}}$ can be neglected).  However, in the NR limit, the operator $J_{S,4}^{\text{M-M}}$ leads to dimension 6 operator $\psi^\dag\chi\,\chi^\dag\psi+{\rm h.c.}$ ($\psi$ and $\chi$ are two-component Pauli spinors),  while $J_{S,3}^{\text{M-M}}$ leads to dimension 8 one $\psi^\dag({\bf \sigma}\cdot\overleftrightarrow{\bf D})\chi\,\chi^\dag({\bf \sigma}\cdot\overleftrightarrow{\bf D})\psi+{\rm h.c.}$, where $\sigma$ is Pauli matrix and $\overleftrightarrow{\bf D}=\overrightarrow{\bf D}-\overleftarrow{\bf D}$ is the NR covariant derivative operator. Thus, one can expected that the state for $J_{S,3}^{\text{Dia}}$ and $J_{S,4}^{\text{Dia}}$ are dominated by the same $J_{S,4}^{\text{M-M}}$ component, and should be degenerated in the NR limit. This is roughly the case in our result, the NLO $\overline{\text{MS}}$ masses of the two states are roughly equal, and from Tab.~\ref{tab:S-NLOresult-MSbar}, one can see that they are both close to the NLO $\overline{\text{MS}}$ mass for $J_{S,4}^{\text{M-M}}$ ($6.36^{+0.06}_{-0.10}$ GeV), and are not consistent with that for $J_{S,3}^{\text{M-M}}$ ($7.91^{+0.16}_{-0.19}$ GeV). Similarly, in the NR limit, the state for $J_{T,2}^{\text{Dia}}$ is dominated by its $J_{T,1}^{\text{M-M}}$ component (see Eq.~(\ref{eq:2++meson-Diagonal})), which leads to dimension 6 operator and can survive in the limit. Correspondingly, from Tab.~\ref{tab:T-NLOresult-MSbar} one can see that the mass of $J_{T,2}^{\text{Dia}}$ is close to that of $J_{T,1}^{\text{M-M}}$, and is not consistent with that of $J_{T,2}^{\text{M-M}}$, which leads to dimension 8 operator in the NR limit.

\begin{figure}[htb]
	
	\begin{center}
		\subfigure[$J_{S,4}^{\text{M-M}}$]{
			\includegraphics[scale=0.23]{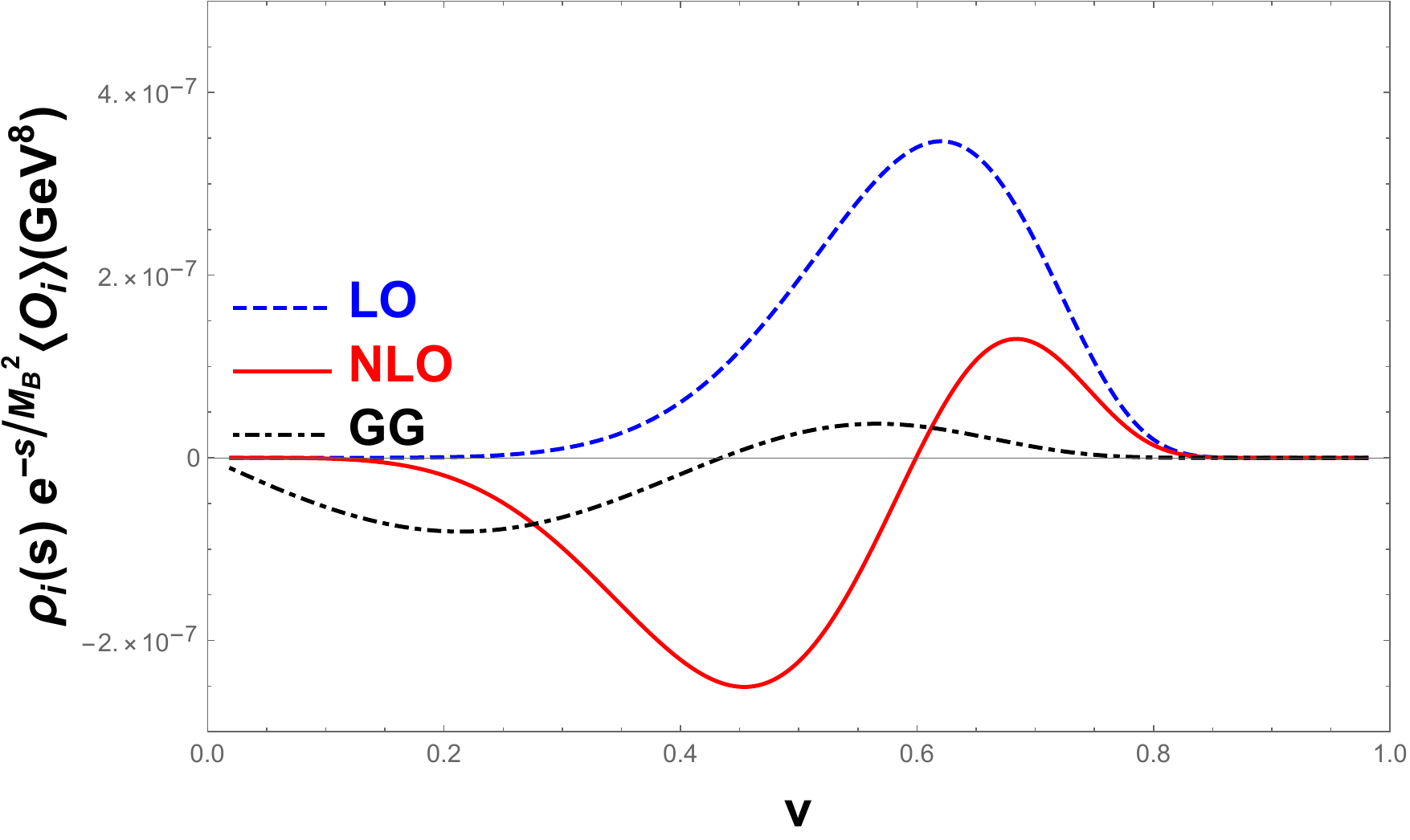}}
		\quad
		\subfigure[$J_{S,2}^{\text{Dia}}$]{
			\includegraphics[scale=0.23]{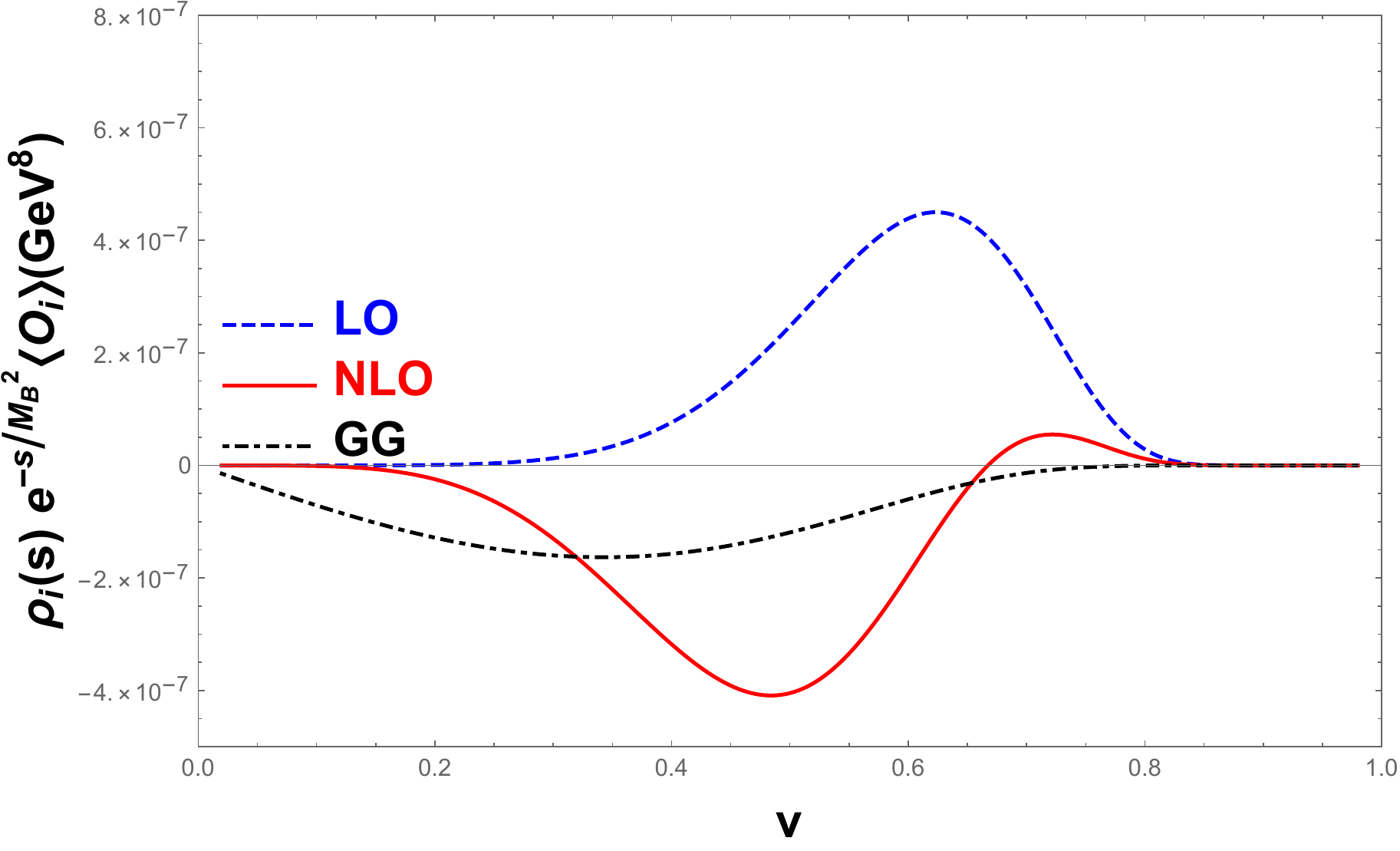}}\\
		\subfigure[$J_{S,3}^{\text{Dia}}$]{
			\includegraphics[scale=0.23]{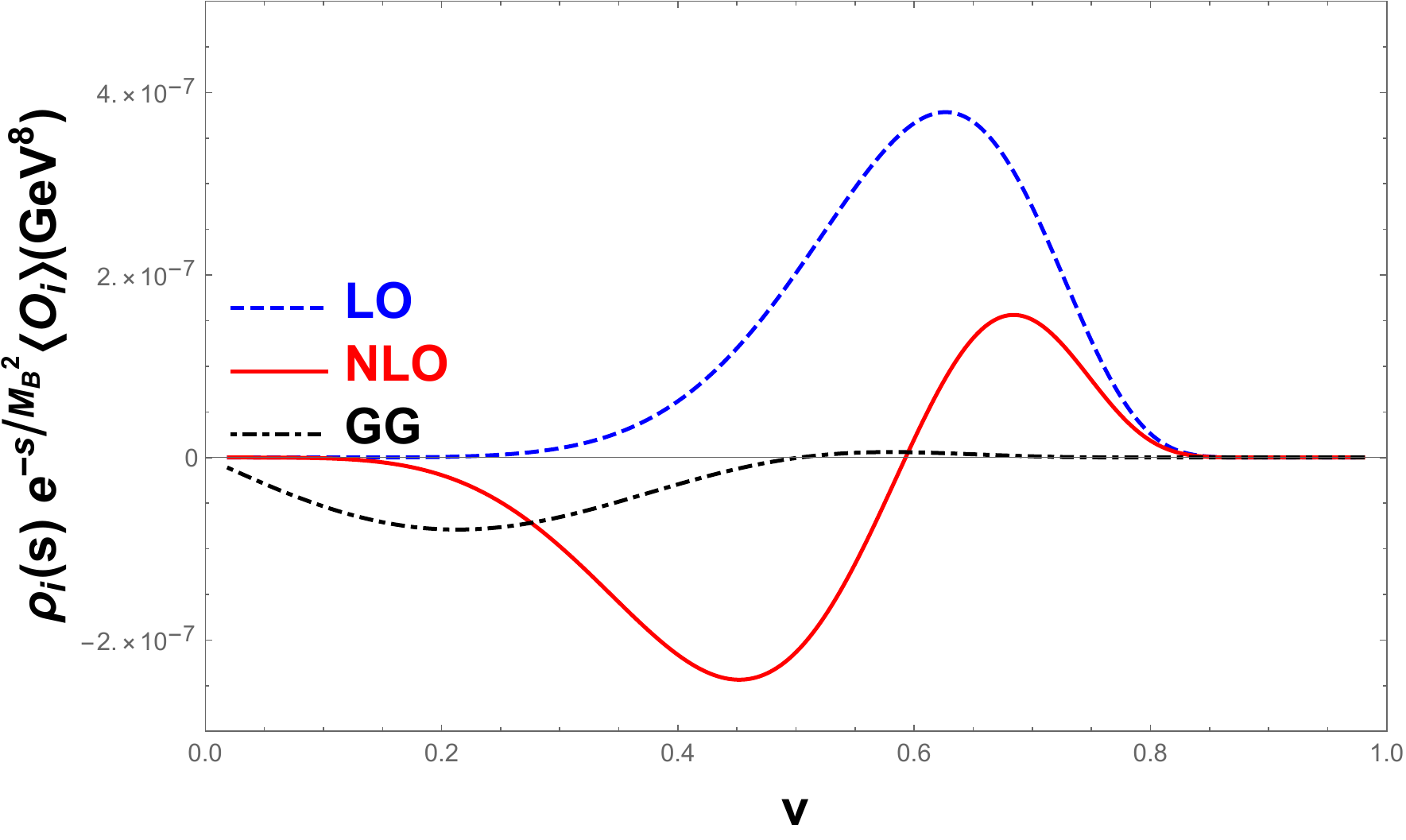}}
		\quad
		\subfigure[$J_{S,4}^{\text{Dia}}$]{
			\includegraphics[scale=0.23]{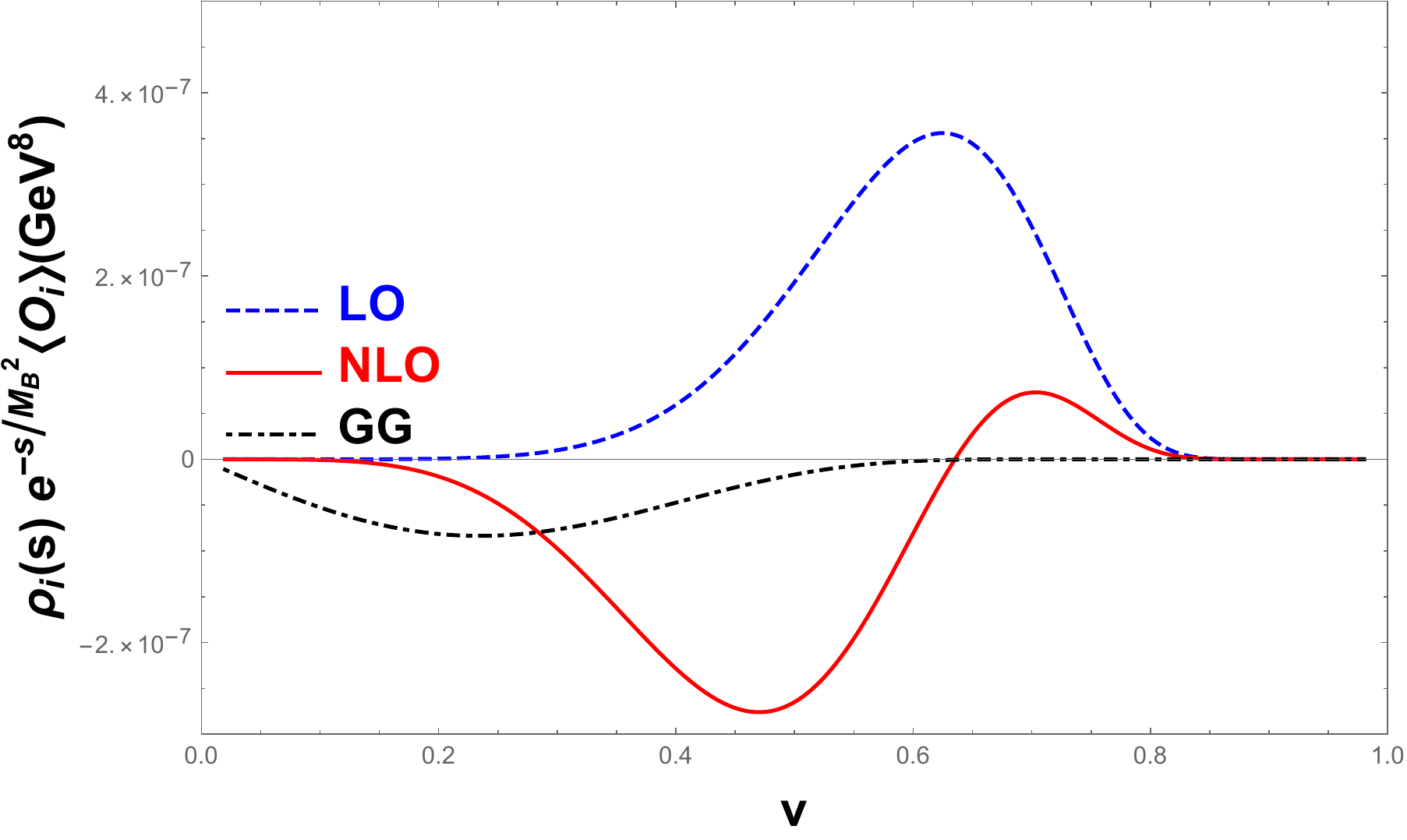}}\\
		\caption{\label{fig:4c-ampCurve-S}
			The $v$-dependence of integrand function $\rho_i(s) e^{-s/M_B^2} \left \langle O_i   \right \rangle $ for $J_{S,4}^{\text{M-M}}$, $J_{S,2}^{\text{Dia}}$, $J_{S,3}^{\text{Dia}}$ and $J_{S,4}^{\text{Dia}}$ of the $\bar{c}c\bar{c}c$ system in $\overline{\text{MS}}$ scheme ($\mu = M_B =\sqrt{3.5}\,  \rm{GeV}$, $v=\sqrt{1-16 m_c^2/s}$)}
	\end{center}
	
\end{figure}

To see the NR behaviors of the amplitudes more explicitly, we define $v=\sqrt{1-\frac{16 m^2_Q}{s}}$ (here, $Q=c$), and show the $v$-dependence of the integrands in Eq.~(\ref{eq:MH}) in the $\overline{\text{MS}}$ scheme for $J_{S,2}^{\text{Dia}}$, $J_{S,3}^{\text{Dia}}$, $J_{S,4}^{\text{Dia}}$ and $J_{S,4}^{\text{M-M}}$ in Fig.~\ref{fig:4c-ampCurve-S}, where the dashed, solid and dot-dashed lines denote $\rho_{1}^{\rm LO} e^{-s/M_B^2}$, $\rho_{1}^{\rm NLO} e^{-s/M_B^2}$ and $\rho_{GG} \langle g_s^2 \hat{G}\hat{G} \rangle e^{-s/M_B^2}$, respectively. For comparison, we set $\mu = M_B =\sqrt{3.5}\,  \rm{GeV}$ for all the four operators in Fig.~\ref{fig:4c-ampCurve-S}. As we have mentioned, the NR behaviors of the operators  $J_{S,3}^{\text{Dia}}$ and $J_{S,4}^{\text{Dia}}$ are dominated by the same $J_{S,4}^{\text{M-M}}$ component. To see this, we have enlarged the integrands for $J_{S,3}^{\text{Dia}}$ and $J_{S,4}^{\text{Dia}}$, respectively, by factor $4$ and $\frac{241}{225}$ to balance the coefficients in the the transition matrix given in Eq.~(\ref{eq:0++meson-Diagonal}). As one can expected, the integrands for $J_{S,3}^{\text{Dia}}$, $J_{S,4}^{\text{Dia}}$ and $J_{S,4}^{\text{M-M}}$ exhibit similar behaviors, especially in the near threshold region, where $v$ is small. For the above four operators, more explicitly analysis indicates that the near threshold behaviors of $\rho_{1}^{\rm LO}$, $\rho_{1}^{\rm NLO}$ and $\rho_{GG}$  in $\overline{\text{MS}}$ scheme are of $\mathcal{O}(v^7)$, $\mathcal{O}(v^5)$ and $\mathcal{O}(v)$, respectively, which can be roughly seen in Fig.~\ref{fig:4c-ampCurve-S}. This is to say that $\rho_{1}^{\rm NLO}$ is enhanced by a factor of $v^{-2}$ with respect to $\rho_{1}^{\rm LO}$ in the near threshold region. Because of the exponential suppression of the lager $v$ region and the threshold parameter $s_0$, the dominant domain of the integration in Eq.~(\ref{eq:MH}) corresponding to $v=0.4\sim 0.7$ for the $\bar{c}c\bar{c}c$ system, where the NLO contributions are comparable with the LO ones. This indicates the importance of the NLO corrections to the QCD sum rules for the  $\bar{c}c\bar{c}c$ system.

Finally, let's compare our predictions with those given in other works~\cite{Chen:2016jxd,Albuquerque:2020hio,Zhang:2020xtb} within the framework of QCD sum rules. The predicted masses of the $\bar{c}c\bar{c}c$ tetraquark states are listed in table~\ref{tab:4c-DiffWorks}.
The authors of Ref.~\cite{Chen:2016jxd} adopt momentum sum rules rather than Laplace sum rules (i.e. \ Borel transformation) applied here. Thus, it is difficult to compare the results between theirs and ours. The Laplace sum rules were applied in Ref.~\cite{Albuquerque:2020hio,Zhang:2020xtb}. However, the parameters (such as the renormalization scale $\mu$) and the scheme to determine the Borel platform in Ref.~\cite{Albuquerque:2020hio,Zhang:2020xtb} are different from ours. Moreover, only partial NLO contributions are considered in Ref.~\cite{Albuquerque:2020hio}.

\begin{table}[H]
  \renewcommand\arraystretch{1.7}
  \begin{center}
  \setlength{\tabcolsep}{2 mm}
   \caption{ The masses of diquark-antidiquark currents obtained by different works in QCD sum rules}
\begin{tabular}{lc|c|c|c|c|c|c}
 \hline\hline
    $J^{PC}$   & Currents   & Ours(LO) & Ours(NLO) &   Ref.~\cite{Chen:2016jxd}   & Ref.~\cite{Albuquerque:2020hio}(LO) & Ref.~\cite{Albuquerque:2020hio}(NLO)&  Ref.~\cite{Zhang:2020xtb} \\ \hline
 \multirow{5}{*}{$0^{++} $}
       &$J_{\text{S},1}^{\text{Di-Di}}$ & $6.07^{+0.05}_{-0.07}$ &$6.60^{+0.09}_{-0.10}$   & $6.46\pm 0.16$ &  $6.52$ &  $6.49\pm 0.07$ & $6.46^{+0.13}_{-0.17}$\\
       &$J_{\text{S},2}^{\text{Di-Di}}$ & $6.19^{+0.07}_{-0.12}$ &$6.90^{+0.11}_{-0.12}$   & $6.59\pm 0.17$ &  $6.55$ &  $6.61\pm 0.09$ & $6.47^{+0.12}_{-0.18}$\\
       &$J_{\text{S},3}^{\text{Di-Di}}$ & $6.96^{+0.11}_{-0.14}$ &$9.25^{+0.14}_{-0.14}$  &  $6.82\pm 0.18$ &  $7.37$ &  $7.05\pm 0.07$ & $6.45^{+0.14}_{-0.16}$\\
       &$J_{\text{S},4}^{\text{Di-Di}}$ & $6.17^{+0.07}_{-0.12}$ &$7.36^{+0.10}_{-0.11}$  &  $6.44\pm 0.15$ &  $6.59$  &  $6.39\pm 0.08$ & $6.44^{+0.15}_{-0.16}$\\
       &$J_{\text{S},5}^{\text{Di-Di}}$ & $6.07^{+0.08}_{-0.10}$ &$6.69^{+0.10}_{-0.12}$   & $6.47\pm 0.16$ &  -      &  -   		 & -       \\ \hline
 \multirow{1}{*}{$0^{--} $}
       &$J_{\text{P},1}^{\text{Di-Di}}$ & $6.55^{+0.12}_{-0.14}$ & $8.43^{+0.17}_{-0.17}$   & $6.84\pm 0.18$ &  - &  - & - \\ \hline
 \multirow{2}{*}{$0^{-+} $}
        &$J_{\text{P},2}^{\text{Di-Di}}$ & $6.55^{+0.12}_{-0.14}$ & $8.08^{+0.15}_{-0.16}$   & $6.40\pm 0.19$ &  - &  - & - \\
       &$J_{\text{P},3}^{\text{Di-Di}}$ & $6.54^{+0.12}_{-0.14}$  & $7.51^{+0.12}_{-0.16}$    & $6.34\pm 0.19$ &  -  &  -& - \\ \hline
\multirow{2}{*}{$1^{++} $}
       &$J_{\text{A},1}^{\text{Di-Di}}$ & $7.00^{+0.12}_{-0.14}$ & $8.84^{+0.09}_{-0.19}$   & $6.40\pm 0.19$ &  - & - &  -\\
       &$J_{\text{A},2}^{\text{Di-Di}}$ & $7.04^{+0.13}_{-0.15}$ & $7.41^{+0.23}_{-0.30}$   & $6.34\pm 0.19$ &  - & - &  -\\ \hline
\multirow{2}{*}{$1^{+-} $}
       &$J_{\text{A},3}^{\text{Di-Di}}$ & $6.95^{+0.13}_{-0.16}$ & $8.81^{+0.08}_{-0.19}$   & $6.37\pm 0.18$ &    -  &  -& -\\
       &$J_{\text{A},4}^{\text{Di-Di}}$ & $6.08^{+0.04}_{-0.10}$ & $6.65^{+0.10}_{-0.13}$   & $6.51\pm 0.15$ &    - &  - & -\\ \hline
\multirow{2}{*}{$1^{--} $}
       &$J_{\text{V},1}^{\text{Di-Di}}$ & $6.56^{+0.11}_{-0.13}$ & $7.45^{+0.12}_{-0.14}$   & $6.84\pm 0.18$ &  - &  - & -\\
       &$J_{\text{V},2}^{\text{Di-Di}}$ &$6.61^{+0.12}_{-0.15}$ & $7.97^{+0.10}_{-0.17}$   & $6.83\pm 0.18$ &  - &  - & -\\ \hline
\multirow{2}{*}{$1^{-+} $}
       &$J_{\text{V},3}^{\text{Di-Di}}$ & $6.56^{+0.12}_{-0.15}$ & $7.52^{+0.12}_{-0.14}$   & $6.84\pm 0.18$ &    - &  - & -\\
       &$J_{\text{V},4}^{\text{Di-Di}}$ & $6.53^{+0.11}_{-0.16}$ & $8.02^{+0.08}_{-0.17}$   & $6.88\pm 0.18$ &    - &  - & -\\ \hline
\multirow{3}{*}{$2^{++} $}
       &$J_{\text{T},1}^{\text{Di-Di}}$ & $6.07^{+0.08}_{-0.10}$ & $6.98^{+0.09}_{-0.11}$   & $6.51\pm 0.15$ & - &  - & -\\
       &$J_{\text{T},2}^{\text{Di-Di}}$ & $7.02^{+0.13}_{-0.16}$ & $9.00^{+0.21}_{-0.23}$   & $6.37\pm 0.19$ &   - &  - & -\\
       &$J_{\text{T},3}^{\text{Di-Di}}$ & $6.15^{+0.08}_{-0.10}$ & $7.25^{+0.10}_{-0.11}$   & -  &  - &  - & -\\ \hline
\hline

   \end{tabular}
   \label{tab:4c-DiffWorks}
   \end{center}
   \end{table}

\subsection{Numerical results and discussions for the $\bar{b}b\bar{b}b$ system}

\begin{figure}[htb]
	\vspace{-0.3cm}
	\setlength{\abovecaptionskip}{-0.8cm}
	\begin{center}
		\includegraphics[scale=0.4]{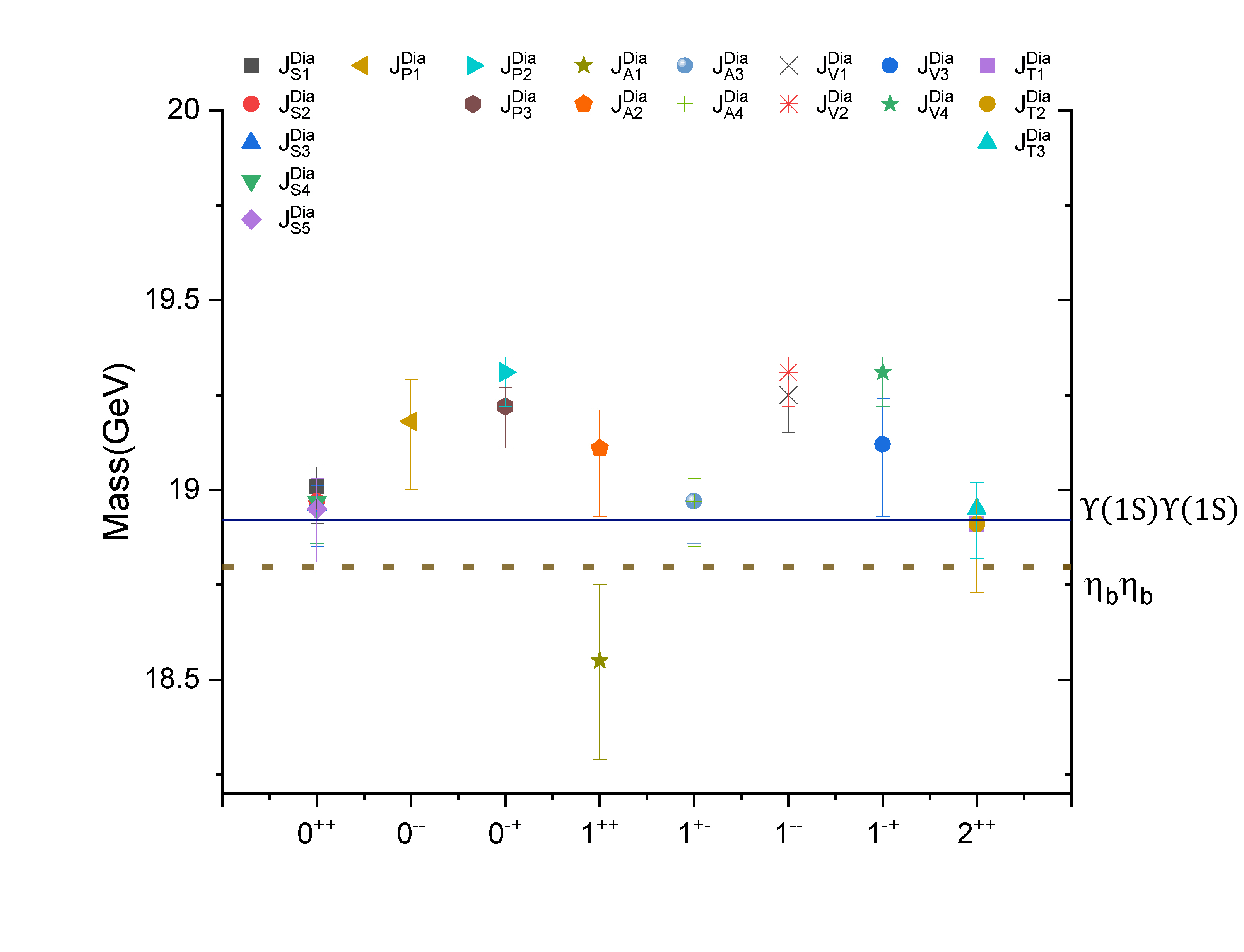}
		\caption{\label{fig:4b-Mass-Spectrum}
			The mass spectrum of $\bar{b}b\bar{b}b$ system in $\overline{\text{MS}}$ scheme. The errors of masses shown in this figure just come from the the parameter dependence on $s_0$ and $M_B^2$.}
	\end{center}
	\vspace{-0.5cm}
\end{figure}

Similar to the $\bar{c}c\bar{c}c$ system, our main results for the $\bar{b}b\bar{b}b$ system are shown in Fig.~\ref{fig:4b-Mass-Spectrum}. We set  $\mu=M_B$ and choose the $\overline{\text{MS}}$ renormalization scheme, and errors of $M_H$ include only that originated from uncertainties of $s_0$ and $M_B^2$. As references, we also plot masses of the two $\Upsilon(1S)$'s and two $\eta_b$'s.

The most comprehensive results are listed in Tabs.~\ref{tab:4b-S-NLOresult-MSbar}--\ref{tab:4b-T-NLOresult-OS} in Appendix~\ref{sec:bottom}, where we include  both LO and NLO,  both $\overline{\text{MS}}$ scheme and on-shell scheme, and all currents of meson-meson types, diquark-antidiquark types, and also diagonalized ones. Again in these tables we set $\mu=M_B$ and thus errors of $M_H$ are due to choices of $s_0$ and $M_B^2$.
Further information of $s_0$ and $M_B^2$ dependence is shown in Figs.~\ref{fig:4b-0+-Mixed1-NLO-MSbar-OS}--\ref{fig:4b-2+-Mixed3-NLO-MSbar-OS} in Appendix~\ref{sec:bottom}, where only results of the more reasonable diagonalized currents are shown.

From Figs.~\ref{fig:4b-0+-Mixed1-NLO-MSbar-OS}--\ref{fig:4b-2+-Mixed3-NLO-MSbar-OS}, one can see the qualities of the Borel platforms are improved evidently in most cases after considering the NLO contributions, especially for those in $\overline{\text{MS}}$ scheme. For example, in Fig.~\ref{fig:4b-0+-Mixed1-NLO-MSbar-OS} (a), there is no Borel platform at LO level in $\overline{\text{MS}}$ scheme, but there is a clear and distinct platform at the NLO level. Similar phenomenon was also found in the $bbb$ system~\cite{Wu:2021tzo}. We have checked that the $\bar{b}b$ system also has this phenomenon. This indicates that for the pure bottom system the NLO contribution is crucial to the formation of a stable Borel platform in the QCD sum rules.

Similar to the case of $\bar{c}c\bar{c}c$ system, from Tabs.~\ref{tab:4b-S-NLOresult-MSbar}--\ref{tab:4b-T-NLOresult-OS}, one can see that NLO contributes non-negligible corrections, with mass corrections $|M_H^{\text{NLO}}\, -\, M_H^{\text{LO}}|\, \simeq \, 0.4\mbox{-}0.6$~GeV in both $\overline{\text{MS}}$ and OS schemes. In addition, with the NLO corrections, the quark mass scheme dependence is improved significantly. The mass difference between the two schemes $|\Delta M_H|=|M_H^{\text{OS}}-M_H^{\overline{\text{MS}}}|$ is about $1.1 \sim 1.2$~GeV at LO level, while the difference is usually smaller than $0.1$~GeV at NLO level except for the $J^{PC}=1^{++}$ channel.

We choose $\mu=k\ m_B$ with $k\in (0.8, 1.2)$ to explore the renormalization scale dependence of our results, because Borel platforms can not be achieved for $k>1.2$ even with the NLO contributions. We find that the $\mu$ dependence is improved for the NLO results comparing with the LO ones, but the $\mu$ dependence of the NLO results for the $\bar{b}b\bar{b}b$ system is more sensitive than that of the $\bar{c}c\bar{c}c$ system. Typical  $\mu$ dependence at the LO and the NLO is shown in Fig.~\ref{fig:4b-mu-dependence-S-3} and \ref{fig:4b-mu-dependence-S-4}.

\begin{figure}[htb]
	
	\begin{center}
		\subfigure[$J_{S,4}^{\text{M-M}}$]{
			\includegraphics[scale=0.3]{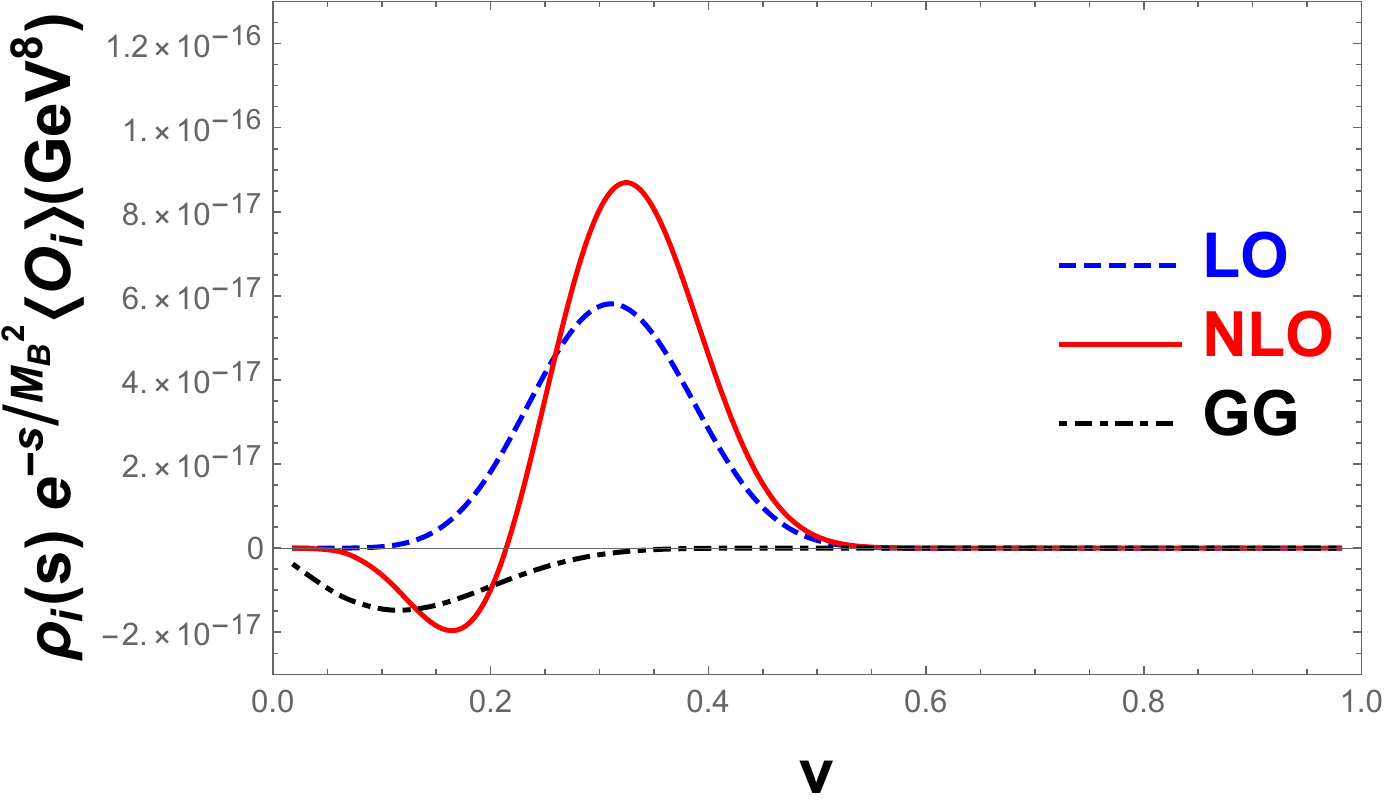}}
		\quad
		\subfigure[$J_{S,2}^{\text{Dia}}$]{
			\includegraphics[scale=0.3]{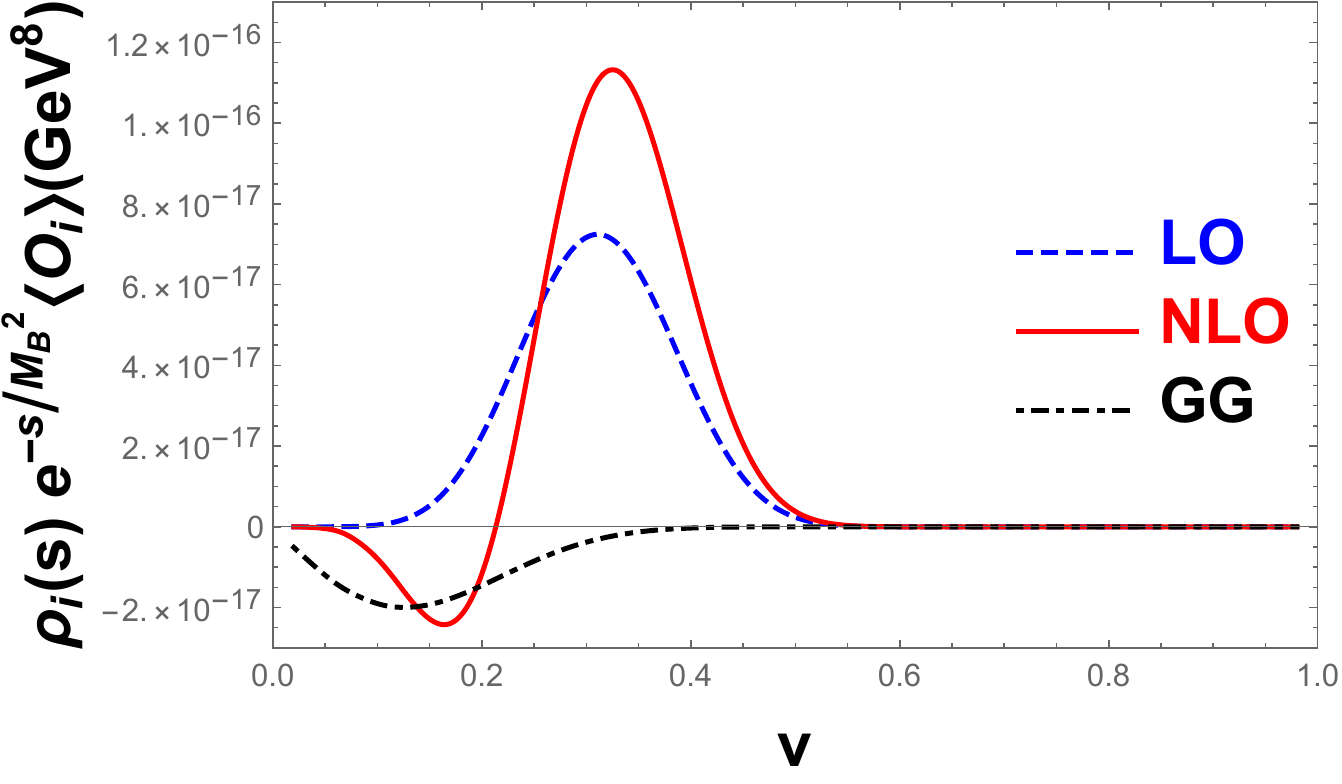}}\\
		\subfigure[$J_{S,3}^{\text{Dia}}$]{
			\includegraphics[scale=0.3]{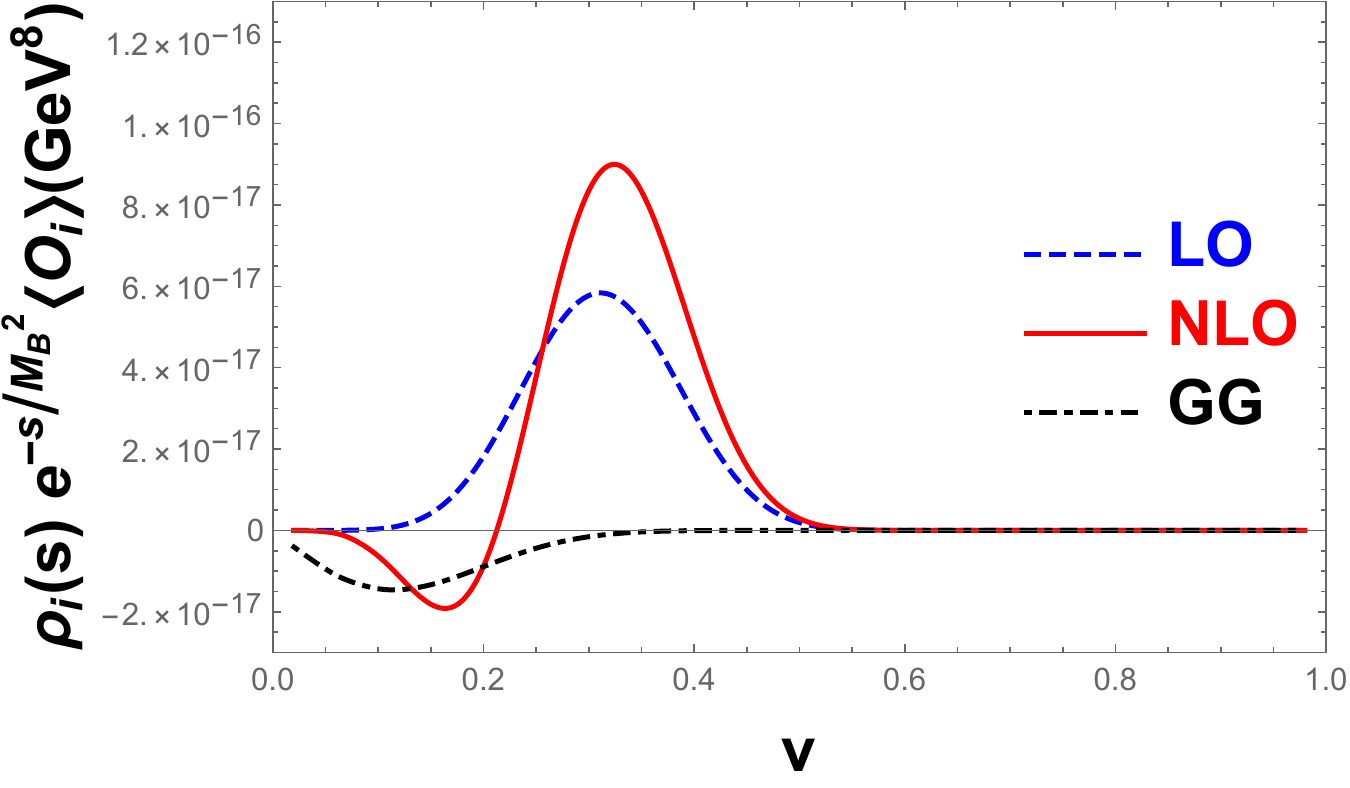}}
		\quad
		\subfigure[$J_{S,4}^{\text{Dia}}$]{
			\includegraphics[scale=0.3]{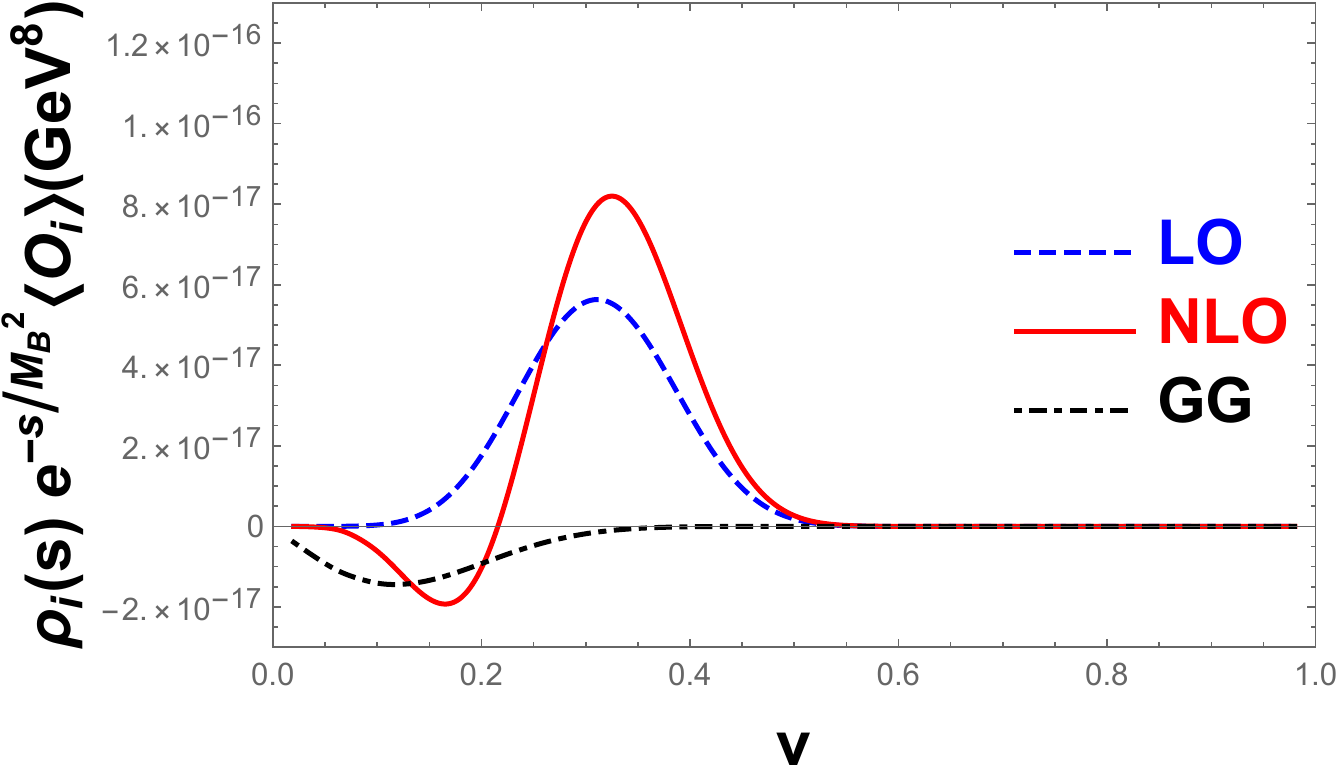}}\\
		\caption{\label{fig:4b-ampCurve-S}
			The curves of integrand function $\rho_i(s) e^{-s/M_B^2} \left \langle O_i   \right \rangle $ for $J_{S,4}^{\text{M-M}}$, $J_{S,2}^{\text{Dia}}$, $J_{S,3}^{\text{Dia}}$ and $J_{S,4}^{\text{Dia}}$ of the $\bar{b}b\bar{b}b$ system in $\overline{\text{MS}}$ scheme ($\mu = M_B =\sqrt{9.5}\,  \rm{GeV}$, $v=\sqrt{1-16 m_b^2/s}$)}
	\end{center}
	
\end{figure}

In Fig.~\ref{fig:4b-ampCurve-S}, we show the $v$-dependence (here, $v=\sqrt{1-16 m_b^2/s}$) of the integrands in Eq.~(\ref{eq:MH}) in the $\overline{\text{MS}}$ scheme for $J_{S,2}^{\text{Dia}}$, $J_{S,3}^{\text{Dia}}$, $J_{S,4}^{\text{Dia}}$ and $J_{S,4}^{\text{M-M}}$ of the $\bar{b}b\bar{b}b$ system, where we set $\mu = M_B =\sqrt{9.5}\,  \rm{GeV}$ for all the four operators and the integrands for $J_{S,3}^{\text{Dia}}$ and $J_{S,4}^{\text{Dia}}$ has been enlarged by factor $4$ and $\frac{241}{225}$, respectively, to be compared with that of $J_{S,4}^{\text{M-M}}$. Similar to the $\bar{c}c\bar{c}c$ system, the near threshold behaviors of $\rho_{1}^{\rm LO}$, $\rho_{1}^{\rm NLO}$ and $\rho_{GG}$ for the above four $\bar{b}b\bar{b}b$ operators in $\overline{\text{MS}}$ scheme are of $\mathcal{O}(v^7)$, $\mathcal{O}(v^5)$ and $\mathcal{O}(v)$, respectively, which can be roughly seen in Fig.~\ref{fig:4b-ampCurve-S}. However, the dominant domain of the integration in Eq.~(\ref{eq:MH}) corresponding to $v=0.2\sim 0.4$ for the $\bar{b}b\bar{b}b$ system, which is smaller than that for the $\bar{c}c\bar{c}c$ system. This is consistent with the general expectation that the $\bar{b}b\bar{b}b$ system is more like a NR one than the $\bar{c}c\bar{c}c$ system since, in roughly speaking, $m_b\gg m_c\gg \Lambda_{\rm QCD}$. Due to the relative enhancement of $\rho_{1}^{\rm NLO}$ with respect to $\rho_{1}^{\rm LO}$ in the near threshold region, the NLO correction to the QCD sum rules is crucial for the $\bar{b}b\bar{b}b$ system, which have been indicated by its effect on improvement of the quality of the Borel platform. On the other hand, the enhancement of the QCD perturbative correction in the near threshold region may make the perturbation convergence bad for the $\bar{b}b\bar{b}b$ system within QCD sum ruls, which have been indicated by the  $\mu$-dependence of $m_H$ up to the NLO corrections showed in Fig.~\ref{fig:4b-mu-dependence-S-3} and \ref{fig:4b-mu-dependence-S-4}. In other words, the NNLO corrections may be important and the enhancement in the near threshold region may need to be resumed for the $\bar{b}b\bar{b}b$ system within QCD sum ruls. But this is far beyond the scope of this work.

At last, we have to emphasize that there are large NLO corrections to the operator $J_{A,2}^{\text{M-M}}$. We find that the near threshold behaviors of $\rho_{1}^{\rm LO}$ and $\rho_{1}^{\rm NLO}$ for this operator are of $\mathcal{O}(v^{11})$ and $\mathcal{O}(v^7)$, respectively. This means for this operator, the $\rho_{1}^{\rm NLO}$ is enhanced by a factor of $v^{-4}$ with respect to $\rho_{1}^{\rm LO}$ in the near threshold region. This enhancement is more serious for $\bar{b}b\bar{b}b$ system than that for $\bar{c}c\bar{c}c$ system since the typical value of $v$ is smaller for the former. Thus, the  perturbation convergence is very bad for this operator, which may be indicated by the large NLO corrections and lager errors of the $\overline{\text{MS}}$ mass $m_H$ of this operator for $\bar{b}b\bar{b}b$ system (see Table~\ref{tab:4b-A-NLOresult-MSbar}). On the other hand, since the dominant component of $J_{A,1}^{\text{Dia}}$ is $J_{A,2}^{\text{M-M}}$ (see Eq.~(\ref{eq:1+-meson-Diagonal}) and (\ref{eq:1+basis-mixing})), the above analyses are basically suitable for the operator $J_{A,1}^{\text{Dia}}$.

%%%%%%%%%%%%%%%%%%%%%%%%%%%%%%%%%%%%%%%%%%%%%%%%%
\section{Summary}

In this paper, we study the NLO corrections to masses of $\bar{Q}Q\bar{Q}Q$ states within QCD sum rules. As operators with the same $J^{PC}$ can mix with each other under renormalization, we diagonalize the original operators, either in meson-meson type or diquark-antidiquark type, and use the diagonalized operators in the phenomenological study.

Numerical results show that NLO corrections are very important. On the one hand, NLO corrections to hadron masses are usually larger than 0.5~GeV in both the $\overline{\rm{MS}}$ and the on-shell schemes. On the other hand, the scheme dependence tends to be reduced with the NLO corrections. More explicitly, the LO mass difference $ M_H^{\rm{LO}\mbox{-}\rm{OS}} - M_H^{\rm{LO}\mbox{-}\overline{\rm{MS}}}=1.1\mbox{-}1.3$~GeV for all the operators, where the NLO corrections to $M_H^{\overline{\rm{MS}}}$ are positive and those to $M_H^{\rm{OS}}$ are negative, which results in the reduction of scheme dependence of the masses. Especially, for the $\bar{c}c\bar{c}c$ system, the NLO mass difference $ M_H^{\rm{NLO}\mbox{-}\rm{OS}} - M_H^{\rm{NLO}\mbox{-}\overline{\rm{MS}}}\le 0.5$~GeV  for the operators $J_{S,2,3,4}^{\rm{Dia}}$ with $J^{PC}=0^{++}$, $J_{P,2}^{\rm{Dia}}$ with $J^{PC}=0^{-+}$, $J_{A,4}^{\rm{Dia}}$ with $J^{PC}=1^{+-}$ and $J_{T,1}^{\rm{Dia}}$ with $J^{PC}=2^{++}$, which also implies that the perturbation convergence of these operators is better than that of the others.  While for the $\bar{b}b\bar{b}b$ system,  the difference is usually smaller than $0.1$~GeV at NLO except for the $J^{PC}=1^{++}$ operator. We also find that NLO corrections can significantly reduce the $\mu$ dependence.

We use currents that have good perturbative convergence in our phenomenological analysis. For the $\bar{c}c\bar{c}c$ system,
we get three  $J^{PC}=0^{++}$ states, with masses  $6.35^{+0.20}_{-0.17}$~GeV, $6.56^{+0.18}_{-0.20}$~GeV and $6.95^{+0.21}_{-0.31}$~GeV, respectively. The first two may explain the broad structure around $6.2\sim6.8$~GeV measured by the LHCb collaboration~\cite{LHCb:2020bwg}, and the third one may be assigned to the observed narrow resonance $X(6900)$.  For the $2^{++}$ states, we find one with mass $7.03^{+0.22}_{-0.26}$~GeV, which may also be a candidate for the $X(6900)$, and another one around $7.25^{+0.21}_{-0.35}$~GeV, which has good $\mu$ dependence but slightly large scheme dependence (with $\Delta M_H^{\text{NLO}}\simeq 0.7$~GeV).

As for the $\bar{b}b\bar{b}b$ system, we find that the NLO contribution improves the quality of the Borel platform evidently in $\overline{\rm{MS}}$ scheme, which is similar to the case of the $bbb$ baryon~\cite{Wu:2021tzo}.  The quark mass scheme dependence of the results are also improved significantly with the NLO contribution. However, the NLO results are still sensitive to the choice of renormalization scale $\mu$, and we find that the Borel platforms can not be achieved for $\mu>1.2m_B$.
%{\color{red} The fundamental question is bad perturbative convergence in $\bar{b}b\bar{b}b$ system because of without the near-threshold divergence resummation. But resumming near-threshold divergence for $\bar{Q}Q\bar{Q}Q$ system is a complex work , we will further refine our results in the future.}

Finally, we would like to emphasize the importance of NLO contributions, especially in the operator mixing or color configuration mixing for multiquark systems. (i) As a key NLO contribution, the one-gluon exchange is crucial even for charmonium and bottomonium states, because it provides the color Coulomb interaction between $Q$ and $\bar{Q}$, which is the most important short-range attractive force to form a heavy quarkonium.
(ii) In the fully heavy tetraquark system discussed in this paper, if one starts from a color-singlet current-current operator, the one-gluon exchange will change it to the color-octet current-current operator, therefore leads to the operator mixing. As already shown by our result, the operator mixing induced by renormalization at NLO is inevitable and has very important consequences in the QCD sum rule calculations. (iii) In the literature some works use the color-singlet current-current local operators to describe physical hadronic molecules. However, duo to the operator mixing, the color structure of the local operators must be mixed with both color-singlet and color-octet current-current configurations. It is impossible to keep the color-singlet structure unchanged if a complete NLO QCD contribution is seriously considered. In fact, a physical molecule state means that it contains two well separated color-singlet mesons at long-distances mediated by one-meson or two-meson exchanges. And a physical molecule may not be necessarily ascribed to the color-singlet current-current local operators, which only describe the very short-distance behavior of the tetraquark and are subjected to the color configuration mixing. The description for hadronic molecules needs to understand the long-distance dynamics beyond color confinement.

%%%%%%%%%%%%%%%%%%%%%%%%%%%%%%%%%%%%%%%%%%%%%%%%%
\acknowledgments

We thank Xiao Liu and Xin Guan for many useful and helpful discussions. We also thank Shi-Lin Zhu for helpful comments. K.T.C. thanks Cong-Feng Qiao for useful communications. The figures in this paper are drawn by using the Origin and Mathematica software. The work is supported in part by the National Natural Science Foundation of China (Grants No. 11875071, No. 11975029, No. 11745006), the National Key Research and Development Program of China under Contracts No. 2020YFA0406400, and the Deutsche Forschungsgemeinschaft (DFG, German Research Foundation) under grant 396021762 - TRR 257.

%%%%%%%%%%%%%%%%%%%%%%%%%%%%%%%%%%%%%%%%%%%%%%%%%
\appendix

\section{ Operator Renormalization Matrices}
\label{sec:matrix}
\subsection{Calculation Of Operator Renormalization Matrices }
We present the calculation of operator renormalization matrices of meson-meson type operators. The operator renormalization matrices of diquark-antidiquark type operators then follow from a Fierz transformation.

A general meson-meson type operators where four quarks are different flavors, are defined as
\begin{align}
\mathcal{O}_{\Gamma_1,\Gamma_2}=\left(\bar{q}_1^i \Gamma_1 q_2^j \right) \left(\bar{q}_3^k \Gamma_2 q_4^l \right)\ ,
\end{align}
which has two independent color configurations,
\begin{align}\label{eq:Operatorcolor}
\mathcal{O}_{\Gamma_1,\Gamma_2,[1]}&=\left(\bar{q}_1^i \Gamma_1 q_2^j \right) \left(\bar{q}_3^k \Gamma_2 q_4^l \right) \delta_{ij}\delta_{kl}\ ,\\
\mathcal{O}_{\Gamma_1,\Gamma_2,[2]}&=\left(\bar{q}_1^i \Gamma_1 q_2^j \right) \left(\bar{q}_3^k \Gamma_2 q_4^l \right) \delta_{il}\delta_{kj}\ ,
\end{align}
where $\mathcal{O}_{\Gamma_1,\Gamma_2,[1]}$ is called a color-singlet operator. We can also use following relation,
\begin{align}
T^a_{ij}T^a_{kl}=\frac{1}{2}\left( \delta_{il}\delta_{kj} - \frac{1}{N_c}\delta_{ij}\delta_{kl} \right)\ ,
\end{align}
to obtain color-octet operators,
\begin{align}
\mathcal{O}_{\Gamma_1,\Gamma_2,[8]}&=\left(\bar{q}_1^i \Gamma_1 T^a_{ij} q_2^j \right) \left(\bar{q}_3^k \Gamma_2 T^a_{kl} q_4^l \right)\\
&=\frac{1}{2}\left( \mathcal{O}_{\Gamma_1,\Gamma_2,[2]} - \frac{1}{N_c}\mathcal{O}_{\Gamma_1,\Gamma_2,[1]}                    \right)\,.
\end{align}
For convenience, we choose $\mathcal{O}_{\Gamma_1,\Gamma_2,[1]}$ and $\mathcal{O}_{\Gamma_1,\Gamma_2,[2]}$ as bases in our calculation.

Let us first suppress the dependence of color configuration for operators. According to our operator and definition of operator renormalization matrix,

	\begin{align}\label{eq:opReno}
		\mathcal{O}^B=\left(\sqrt{Z_2}\right)^4 \  \overline{\mathcal{O}}^{B}\, = {Z_{\mathcal{O}}}\  \mathcal{O}^{R} \ ,
	\end{align}
	where $\mathcal{O}^B=\left(\bar{q}_1 \Gamma_1 q_2 \right) \left(\bar{q}_3 \Gamma_2 q_4\right)$ denotes the bare operator, $\overline{\mathcal{O}}^{B}=\left(\bar{q}_1^R \Gamma_1 q_2^R \right) \left(\bar{q}_3^R \Gamma_2 q_4^R\right)$ denotes the bare operator replaced by renormalized fields, and $\mathcal{O}^{R}$ denotes the renormalized operator. In our NLO calculation, we directly calculate $\overline{\mathcal{O}}^{B}$, and thus quark self-energy diagrams are cancelled by counter term diagrams. The remaining diagrams can be divided into three parts,
	\begin{align}
		A=\int \frac{\mathrm{d}^{D} p}{(2\pi)^D}(A_1+A_2+A_3)\ ,
	\end{align}
	where $A_1$ denotes the contribution of gluon exchange between $q_1$ and $q_2$, $A_2$ denotes the contribution of gluon exchange between $q_3$ and $q_4$, and $A_3$ denotes others contributions e.g.\ contributions of gluon exchange between $q_1$ and $q_3$, $q_1$ and $q_4$ and so on.
	Because all infrared divergences will be cancelled, we just need to consider ultraviolet (UV) divergences therein. Therefore, the mass terms in quark propagators can be discarded. Explicitly, we have
	\begin{align}
		A_{1} & = \left[i g_s \gamma_{\mu}  \frac{i\slashed{p}}{p^{2}}\Gamma_{1}  \frac{i\slashed{p}}{p^{2}} i g \gamma^{\mu} \left(T^{a}\right)_{i^{\prime} i}\left(T^{a}\right)_{j j^{\prime}}\right]\left[\Gamma_{2} \delta_{k^{\prime} k} \delta_{l l^{\prime}}\right]  \frac{-i}{p^{2}}\ ,
	\end{align}
	\begin{align}
		A_{2} & = \left[\Gamma_{1} \delta_{i^{\prime} i} \delta_{j j^{\prime}}\right] \left[i g \gamma_{\mu}  \frac{i \slashed{p}}{p^{2}}\Gamma_{2}  \frac{i \slashed{p}}{p^{2}} i g_s \gamma^{\mu} \left(T^{a}\right)_{k^{\prime} k}\left(T^{a}\right)_{l l^{\prime}} \right] \frac{-i}{p^{2}}\ ,
	\end{align}
	\begin{align}
		A_{3} & = \left[i g_s \gamma_{\mu} \frac{i \slashed{p}}{p^{2}}\Gamma_{1} \left(T^{a}\right)_{i^{\prime} i} \delta_{j j^{\prime}} + \Gamma_{1} \frac{-i \slashed{p}}{p^{2}} i g_s \gamma_{\mu} \delta_{i^{\prime} i}\left(T^{a}\right)_{j j^{\prime}}  \right] \left[i g_s \gamma^{\mu} \frac{-i\slashed{p}}{p^{2}}\Gamma_{2} \left(T^{a}\right)_{k^{\prime} k} \delta_{l l^{\prime}}+\Gamma_{2} \frac{i \slashed{p}}{p^{2}} i g_s \gamma^{\mu} \delta_{k^{\prime} k}\left(T^{a}\right)_{l l^{\prime}}\right] \frac{-i}{p^{2}}\,.
	\end{align}
	After a simple manipulation, we get
	\begin{align}\label{eq:UV}
		A=-ig_s^2 \frac{1}{D} \int \frac{\mathrm{d}^D p}{(2\pi)^D}\ \frac{1}{(p^2)^2}\ B\ ,
	\end{align}
	where
	\begin{align}\label{eq:UVB}
		\begin{split}
			B &=\left(\gamma_{\mu} \gamma_{\nu} \Gamma_{1} \gamma^{\nu} \gamma^{\mu}\right)\left(\Gamma_{2}\right)\left(T^{a}\right)_{i^{\prime} i}\left(T^{a}\right)_{j j^{\prime}} \delta_{k^{\prime} k} \delta_{l l^{\prime}}+\left(\Gamma_{1}\right)\left(\gamma_{\mu} \gamma_{v} \Gamma_{2} \gamma^{\nu} \gamma^{\mu}\right) \delta_{i^{\prime} i} \delta_{j j^{\prime}}\left(T^{a}\right)_{k^{\prime} k}\left(T^{a}\right)_{l l^{\prime}} \\
			&+(-D)\left(\Gamma_{1}\right)\left(\Gamma_{2}\right)\left[\left(T^{a}\right)_{i^{\prime} i} \delta_{j j^{\prime}}-\delta_{i^{\prime} i}\left(T^{a}\right)_{j j^{\prime}}\right]\left[\left(T^{a}\right)_{k^{\prime} k} \delta_{l l^{\prime}}-\delta_{k^{\prime} k}\left(T^{a}\right)_{l l^{\prime}}\right] \\
			&+\frac{1}{4}\left(\left\{\sigma_{\mu \nu}, \Gamma_{1}\right\}\right)\left(\left\{\sigma_{\mu \nu}, \Gamma_{2}\right\}\right)\left[\left(T^{a}\right)_{i^{\prime} i} \delta_{j j^{\prime}}+\delta_{i^{\prime} i}\left(T^{a}\right)_{j j^{\prime}}\right]\left[\left(T^{a}\right)_{k^{\prime} k} \delta_{l l^{\prime}}+\delta_{k^{\prime} k}\left(T^{a}\right)_{l l^{\prime}}\right] \\
			&+\frac{1}{4}\left(\left[\sigma_{\mu \nu}, \Gamma_{1}\right]\right)\left(\left[\sigma_{\mu \nu}, \Gamma_{2}\right]\right)\left[\left(T^{a}\right)_{i^{\prime} i} \delta_{j j^{\prime}}-\delta_{i^{\prime} i}\left(T^{a}\right)_{j j^{\prime}}\right]\left[\left(T^{a}\right)_{k^{\prime} k} \delta_{l l^{\prime}}-\delta_{k^{\prime} k}\left(T^{a}\right)_{l l^{\prime}}\right]\,.
		\end{split}
	\end{align}
	
	According to Eq.~(\ref{eq:UV}), we get the UV divergences term
	\begin{align}
		A_{UV} & = -\left.i g_s^{2} \frac{1}{4} \frac{i}{(4 \pi)^{2}} \frac{1}{\varepsilon} B\right|_{D  = 4}  = \frac{\alpha_{s}}{\varepsilon} \frac{\left.B\right|_{D  = 4}}{16\pi} \,.
	\end{align}
	For operators with definite color configuration $\mathcal{O}_{\Gamma_1,\Gamma_2,[c]}$, we need to multiply the corresponding color configuration ($\delta_{ij}\delta_{kl}$ or $\delta_{il}\delta_{kj}$) in Eq.~(\ref{eq:UVB}).
	
	According to Eq.~(\ref{eq:opReno}), to use the renormalized operator we should multiply our result $M_{LO}+A_{UV}+\cdots$ by $Z_2^{2}Z_{\mathcal{O}}^{-1}  \approx1-\delta Z_{\mathcal{O}}+2\delta Z_2  $, where $M_{LO}=1$ is the LO amplitude and $Z_2=1+\delta Z_2$ and $Z_{\mathcal{O}}=1+\delta Z_{\mathcal{O}}$. Demanding that final results are free of UV divergences, we get
	\begin{align}
		\delta Z_{\mathcal{O}}=A_{UV}+ 2\  \delta {Z}_2\ ,
	\end{align}
	with
	\begin{align}
		\delta {Z}_2 & = -\frac{\alpha_s}{3\pi \varepsilon} \,,
	\end{align}
	in $\overline{\text{MS}}$ scheme.

\subsection{$J^P=0^+$}

Operator bases are defined as
\begin{align}\label{eq:0+-meson-renormalizationbase}
\begin{split}
J_{S,1,[1]}^{\text{M-M}}&= (\bar{Q}_a \gamma^\mu Q_a)(\bar{Q}_b \gamma_\mu Q_b)\,, \\
J_{S,2,[1]}^{\text{M-M}}&= (\bar{Q}_a \gamma^\mu \gamma^5 Q_a)(\bar{Q}_b \gamma_\mu \gamma^5 Q_b)\,, \\
J_{S,1,[2]}^{\text{M-M}}&= (\bar{Q}_a \gamma^\mu Q_b)(\bar{Q}_b \gamma_\mu Q_a)\,, \\
J_{S,2,[2]}^{\text{M-M}}&= (\bar{Q}_a \gamma^\mu \gamma^5 Q_b)(\bar{Q}_b \gamma_\mu \gamma^5 Q_a)\,, \\
J_{S,3,[1]}^{\text{M-M}}&= (\bar{Q}_a  Q_a)(\bar{Q}_b Q_b)\,, \\
J_{S,4,[1]}^{\text{M-M}}&= (\bar{Q}_a i\gamma^5 Q_a)(\bar{Q}_b i\gamma^5 Q_b)\,, \\
J_{S,5,[1]}^{\text{M-M}}&= (\bar{Q}_a \sigma^{\mu\nu} Q_a)(\bar{Q}_b \sigma_{\mu\nu} Q_b)\,, \\
J_{S,3,[2]}^{\text{M-M}}&= (\bar{Q}_a  Q_b)(\bar{Q}_b Q_a)\,, \\
J_{S,4,[2]}^{\text{M-M}}&= (\bar{Q}_a i\gamma^5 Q_b)(\bar{Q}_b i\gamma^5 Q_a)\,, \\
J_{S,5,[2]}^{\text{M-M}}&= (\bar{Q}_a \sigma^{\mu\nu} Q_b)(\bar{Q}_b \sigma_{\mu\nu} Q_a)\,.
\end{split}
\end{align}
The corresponding operator renormalization matrix is given by,

\begin{equation}\label{eq:0+-meson-renormalization}
  \delta Z_{O,S}=\frac{\alpha_s}{16 \pi} \delta_{\overline{\text{MS}}}\begin{pmatrix}
  0 &  \frac{12}{N_c} &0&-12&0&0&0&0&0&0 \\
 \frac{12}{N_c} &  0 & -12 &0&0&0&0&0&0&0 \\
 -6 &  -6 & 6 N_c & -\frac{6(N_c^2-2)}{N_c}& 0&0&0&0&0&0 \\
 -6 &  -6 & -\frac{6(N_c^2-2)}{N_c}& 6 N_c & 0&0&0&0&0&0 \\
 0&0&0&0&\frac{12(N_c^2-1)}{N_c}&0& -\frac{2}{N_c}&0&0&2\\
 0&0&0&0&0&\frac{12(N_c^2-1)}{N_c} &\frac{2}{N_c}&0&0&-2\\
 0&0&0&0&-\frac{48}{N_c}&\frac{48}{N_c}&-\frac{4(N_c^2-1)}{N_c}&48&-48&0\\
 0&0&0&0&12&0&1&-\frac{12}{N_c}&0&\frac{N_c^2-2}{N_c}\\
 0&0&0&0&0&12&-1&0&-\frac{12}{N_c}&-\frac{N_c^2-2}{N_c}\\
 0&0&0&0&24&-24&-12&\frac{24(N_c^2-2)}{N_c}&-\frac{24(N_c^2-2)}{N_c}&\frac{4(2N_c^2+1)}{N_c}
\end{pmatrix}
\,,
\end{equation}
where $\delta_{\overline{\text{MS}}}=\frac{1}{\epsilon}+\ln(4\pi)-\gamma_E$.

After renormalization, since there are identical particles in the operator of full heavy tetraquark system $\left(\bar{Q} \Gamma_1 Q \bar{Q} \Gamma_2 Q \right)$, 4-dimensional Fierz transformation can be used to related operators in different color configurations, which results in only 5 independent operators in $J^{PC}=0^{++}$ channel. We can choose any 5 independent operators to perform our phenomenological study. For example, we choose $J_{S,i,[1]}^{\text{M-M}}$ in this work. According to Fierz transformation
\begin{equation}\label{eq:0+-FierzTrans-meson12}
J_{S,i,[2]}^{\text{M-M}}=\frac{1}{8}\begin{pmatrix}
4 &  4 & -8&-8&0 \\
4 &  4 & 8&8&0 \\
-2 &  2 & -2 &2& -1 \\
-2 &  2 &2 &-2& 1 \\
0 &  0 & -24 &-24& 4 \\
\end{pmatrix} \cdot J_{S,i,[1]}^{\text{M-M}} \,,
\end{equation}
we can transform $J_{S,i,[2]}^{\text{M-M}}$ to $J_{S,i,[1]}^{\text{M-M}}$ to get the anomalous dimension Eq.~(\ref{eq:0++meson-AnoDim}),

\subsection{$J^P=0^-$}

Operator bases for $J^P=0^-$ are
\begin{align}\label{eq:0--meson-renormalizationbase}
\begin{split}
J_{P,1,[1]}^{\text{M-M}}&= (\bar{Q}_a \gamma^\mu Q_a)(\bar{Q}_b \gamma_\mu\gamma^5 Q_b)\,, \\
J_{P,1,[2]}^{\text{M-M}}&= (\bar{Q}_a \gamma^\mu Q_b)(\bar{Q}_b \gamma_\mu\gamma^5 Q_a)\,, \\
J_{P,2,[1]}^{\text{M-M}}&= (\bar{Q}_a  Q_a)(\bar{Q}_b i \gamma^5 Q_b)\,, \\
J_{P,3,[1]}^{\text{M-M}}&= (\bar{Q}_a \sigma^{\mu\nu} Q_a)(\bar{Q}_b\sigma_{\mu\nu}i\gamma^5  Q_b)\,, \\
J_{P,2,[2]}^{\text{M-M}}&= (\bar{Q}_a  Q_b)(\bar{Q}_b i \gamma^5 Q_a)\,, \\
J_{P,3,[2]}^{\text{M-M}}&= (\bar{Q}_a \sigma^{\mu\nu} Q_b)(\bar{Q}_b\sigma_{\mu\nu}i\gamma^5  Q_a)\,.
\end{split}
\end{align}
The operator renormalization matrix is
\begin{equation}\label{eq:0--meson-renormalization}
  \delta Z_{O,P}=\frac{\alpha_s}{16 \pi }\delta_{\overline{\text{MS}}}\begin{pmatrix}
 \frac{12}{N_c} &-12&0&0&0&0 \\
 -12 &\frac{12}{N_c}&0&0&0&0 \\
 0&0&\frac{12(N_c^2-1)}{N_c}&0& -\frac{2}{N_c}&2\\
 0&0&-\frac{96}{N_c}&-\frac{4(N_c^2-24 N_c-1)}{N_c} &0&0\\
 0&0&12&\frac{4(N_c^2-4)}{N_c}&\frac{4+N_c-4 N_c^2}{N_c}&\frac{N_c^2-2}{N_c}\\
 0&0&48&\frac{48(N_c^2-2)}{N_c}&-12&\frac{4(2N_c^2+1)}{N_c}
\end{pmatrix}
\,.
\end{equation}

Similay to $J^{PC}=0^{++}$, we have the Fierz transformation \begin{equation}\label{eq:0--FierzTrans-meson12}
J_{P,i,[2]}^{\text{M-M}}=\frac{1}{8}\begin{pmatrix}
8 &  0 & 0 \\
0 &  -4 & -1 \\
0 &  -64 & 4
\end{pmatrix} \cdot  J_{P,i,[1]}^{\text{M-M}}\,.
\end{equation}
which transforms $J_{P,i,[2]}^{\text{M-M}}$ to $J_{P,i,[1]}^{\text{M-M}}$ to get the anomalous dimension Eq.~(\ref{eq:0--meson-AnoDim}).

\subsection{$J^P=1^+$}

Operator bases for $J^P=1^+$ are
\begin{align}\label{eq:1+-meson-renormalizationbase}
\begin{split}
J_{A,1,[1]}^{\text{M-M}}&= (\bar{Q}_a  Q_a)(\bar{Q}_b \gamma^\mu\gamma^5 Q_b)\,, \\
J_{A,2,[1]}^{\text{M-M}}&= (\bar{Q}_a \sigma^{\mu\nu}i\gamma^5 Q_a)(\bar{Q}_b \gamma_\nu Q_b)\,, \\
J_{A,1,[2]}^{\text{M-M}}&= (\bar{Q}_a  Q_b)(\bar{Q}_b \gamma^\mu\gamma^5 Q_a)\,, \\
J_{A,2,[2]}^{\text{M-M}}&= (\bar{Q}_a \sigma^{\mu\nu}i\gamma^5 Q_b)(\bar{Q}_b \gamma_\nu Q_a)\,, \\
J_{A,3,[1]}^{\text{M-M}}&= (\bar{Q}_a i\gamma^5 Q_a)(\bar{Q}_b \gamma^\mu  Q_b)\,, \\
J_{A,4,[1]}^{\text{M-M}}&= (\bar{Q}_a \sigma^{\mu\nu} Q_a)(\bar{Q}_b\gamma_\nu \gamma^5  Q_b)\,, \\
J_{A,3,[2]}^{\text{M-M}}&= (\bar{Q}_a i\gamma^5 Q_b)(\bar{Q}_b \gamma^\mu  Q_a)\,, \\
J_{A,4,[2]}^{\text{M-M}}&= (\bar{Q}_a \sigma^{\mu\nu} Q_b)(\bar{Q}_b\gamma_\nu \gamma^5  Q_a)\,.
\end{split}
\end{align}
The operator renormalization matrix is
\begin{equation}\label{eq:1+-meson-renormalization}
  \delta Z_{O,A}=\frac{\alpha_s}{16 \pi }\delta_{\overline{\text{MS}}}\begin{pmatrix}
 \frac{6(N_c^2-1)}{N_c} &-\frac{4}{N_c}&0&4&0&0&0&0 \\
-\frac{12}{N_c} &-\frac{2(N_c^2-1)}{N_c}&12&0&0&0&0&0 \\
 6&2&-\frac{6}{N_c}&\frac{2(N_c^2-2)}{N_c}&0&0&0&0\\
 6&-6&\frac{6(N_c^2-2)}{N_c}&\frac{2(N_c^2+1)}{N_c}&0&0&0&0\\
0&0&0&0&\frac{6(N_c^2-1)}{N_c} &\frac{4}{N_c}&0&-4\\
0&0&0&0&\frac{12}{N_c} &-\frac{2(N_c^2-1)}{N_c}&-12&0\\
0&0&0&0&6&-2&-\frac{6}{N_c}&-\frac{2(N_c^2-2)}{N_c}\\
0&0&0&0&-6&-6&-\frac{6(N_c^2-2)}{N_c}&\frac{2(N_c^2+1)}{N_c}
\end{pmatrix}
\,.
\end{equation}

Similar to $J^{PC}=0^{++}$, the Fierz transformation \begin{equation}\label{eq:1+-FierzTrans-meson12}
J_{A,i,[2]}^{\text{M-M}}=\frac{1}{2}\begin{pmatrix}
-1 &  -1 & 0 &0 \\
-3 &  1 & 0 &0 \\
0 &  0 & -1 &1 \\
0 &  0 & -3 &1 \\
\end{pmatrix} \cdot J_{A,i,[1]}^{\text{M-M}}\,.
\end{equation}
transforms $J_{A,i,[2]}^{\text{M-M}}$ to $J_{A,i,[1]}^{\text{M-M}}$ and we thus get the anomalous dimension Eq.~(\ref{eq:1+-meson-AnoDim}).

\subsection{$J^P=1^-$}

Operator bases for $J^P=1^+$ are
\begin{align}\label{eq:1--meson-renormalizationbase}
\begin{split}
J_{V,1,[1]}^{\text{M-M}}&= (\bar{Q}_a  Q_a)(\bar{Q}_b \gamma^\mu Q_b)\,, \\
J_{V,2,[1]}^{\text{M-M}}&= (\bar{Q}_a \sigma^{\mu\nu}i\gamma^5 Q_a)(\bar{Q}_b \gamma_\nu \gamma^5 Q_b)\,, \\
J_{V,1,[2]}^{\text{M-M}}&= (\bar{Q}_a  Q_b)(\bar{Q}_b \gamma^\mu Q_a)\,, \\
J_{V,2,[2]}^{\text{M-M}}&= (\bar{Q}_a \sigma^{\mu\nu}i\gamma^5 Q_b)(\bar{Q}_b \gamma_\nu \gamma^5 Q_a)\,, \\
J_{V,3,[1]}^{\text{M-M}}&= (\bar{Q}_a i\gamma^5 Q_a)(\bar{Q}_b \gamma^\mu \gamma^5  Q_b)\,, \\
J_{V,4,[1]}^{\text{M-M}}&= (\bar{Q}_a \sigma^{\mu\nu} Q_a)(\bar{Q}_b\gamma_\nu  Q_b)\,, \\
J_{V,3,[2]}^{\text{M-M}}&= (\bar{Q}_a i\gamma^5 Q_b)(\bar{Q}_b \gamma^\mu \gamma^5  Q_a)\,, \\
J_{V,4,[2]}^{\text{M-M}}&= (\bar{Q}_a \sigma^{\mu\nu} Q_b)(\bar{Q}_b\gamma_\nu  Q_a)\,. \\
\end{split}
\end{align}
And the operator renormalization matrix is
\begin{equation}\label{eq:1--meson-renormalization}
  \delta Z_{O,V}=\frac{\alpha_s}{16 \pi }\delta_{\overline{\text{MS}}}\begin{pmatrix}
 \frac{6(N_c^2-1)}{N_c} &-\frac{4}{N_c}&0&4&0&0&0&0 \\
-\frac{12}{N_c} &-\frac{2(N_c^2-1)}{N_c}&12&0&0&0&0&0 \\
 6&2&-\frac{6}{N_c}&\frac{2(N_c^2-2)}{N_c}&0&0&0&0\\
 6&-6&\frac{6(N_c^2-2)}{N_c}&\frac{2(N_c^2+1)}{N_c}&0&0&0&0\\
0&0&0&0&\frac{6(N_c^2-1)}{N_c} &\frac{4}{N_c}&0&-4\\
0&0&0&0&\frac{12}{N_c} &-\frac{2(N_c^2-1)}{N_c}&-12&0\\
0&0&0&0&6&-2&-\frac{6}{N_c}&-\frac{2(N_c^2-2)}{N_c}\\
0&0&0&0&-6&-6&-\frac{6(N_c^2-2)}{N_c}&\frac{2(N_c^2+1)}{N_c}
\end{pmatrix}
\,.
\end{equation}

According to Fierz transformation
\begin{equation}\label{eq:1--FierzTrans-meson12}
J_{V,i,[2]}^{\text{M-M}}=\frac{1}{2}\begin{pmatrix}
-1 &  -1 & 0 &0 \\
-3 &  1 & 0 &0 \\
0 &  0 & -1 &1 \\
0 &  0 & -3 &1 \\
\end{pmatrix} \cdot J_{V,i,[1]}^{\text{M-M}}\,,
\end{equation}
we can transform $J_{V,i,[2]}^{\text{M-M}}$ to $J_{V,i,[1]}^{\text{M-M}}$ to get the anomalous dimension Eq.~(\ref{eq:1+-meson-AnoDim}).

\subsection{$J^P=2^+$}

Operator bases for $J^P=2^+$ are
\begin{align}\label{eq:2+-meson-renormalizationbase}
\begin{split}
J_{T,1,[1]}^{\text{M-M}}&= (\bar{Q}_a \gamma^\mu Q_a)(\bar{Q}_b \gamma^\nu Q_b)\,, \\
J_{T,1,[2]}^{\text{M-M}}&= (\bar{Q}_a \gamma^\mu Q_b)(\bar{Q}_b \gamma^\nu Q_a)\,, \\
J_{T,2,[1]}^{\text{M-M}}&= (\bar{Q}_a \gamma^\mu\gamma^5 Q_a)(\bar{Q}_b \gamma^\nu\gamma^5 Q_b)\,, \\
J_{T,2,[2]}^{\text{M-M}}&= (\bar{Q}_a \gamma^\mu\gamma^5 Q_b)(\bar{Q}_b \gamma^\nu\gamma^5 Q_a)\,, \\
J_{T,3,[1]}^{\text{M-M}}&= (\bar{Q}_a \sigma^{\mu\alpha} Q_a)(\bar{Q}_b\sigma^{\nu\alpha} Q_b)\,, \\
J_{T,3,[2]}^{\text{M-M}}&= (\bar{Q}_a \sigma^{\mu\alpha} Q_b)(\bar{Q}_b\sigma^{\nu\alpha} Q_a)\,.
\end{split}
\end{align}
And the operator renormalization matrix is
\begin{equation}\label{eq:2+-meson-renormalization}
  \delta Z_{O,T}=\frac{\alpha_s}{16 \pi}\delta_{\overline{\text{MS}}}\begin{pmatrix}
 0              &     0              &-\frac{4}{ N_c}      &    4      &  0      &0 \\
2    &    -2 N_c   &   2   &  \frac{2(N_c^2-2)}{ N_c}  &0    &   0\\
 -\frac{4}{N_c} &   4        &     0              &      0                 & 0    &   0 \\
 2    &    \frac{2(N_c^2-2)}{ N_c}   &    2   & -2 N_c     &0    &   0\\
0       &    0 &    0   &0   -&\frac{4(N_c^2-1)}{ N_c}     &   0\\
0       &    0 &    0   &0   &  -4   &    \frac{4}{N_c}\\

\end{pmatrix}
\,.
\end{equation}

According to Fierz transformation \begin{equation}\label{eq:2+-FierzTrans-meson12}
J_{T,i,[2]}^{\text{M-M}}=-\frac{1}{2}\begin{pmatrix}
1 &  1 & -1 \\
1 &  1 & 1 \\
-2 &  2 & 0
\end{pmatrix} \cdot J_{T,i,[1]}^{\text{M-M}}\,,
\end{equation}
we can transform $J_{T,i,[2]}^{\text{M-M}}$ to $J_{T,i,[1]}^{\text{M-M}}$ to get the anomalous dimension Eq.~(\ref{eq:2++meson-AnoDim}).

\section{Details for $\bar{c}c\bar{c}c$ system}
\label{sec:charm}

\subsection{Numerical Results for $J^P=0^+$ states}

\begin{table}[H]
	\vspace{-0.5cm}
	\renewcommand\arraystretch{1.8}
	\begin{center}
		\setlength{\tabcolsep}{3 mm}
		\caption{The LO and NLO Results for $J^P=0^+$ with $\bar{c}c\bar{c}c$ system in the $\overline{\text{MS}}$ scheme}
		\begin{tabular}{cccc|@{*}|ccc}
			\hline\hline
			\multirow{2}{*}{Current} &
			\multicolumn{3}{c|@{*}|}{LO}& \multicolumn{3}{c}{NLO($\overline{\text{MS}}$)} \\ \cline{2-7}
			& \makecell{$M_H$ \\ (GeV)} & \makecell{$s_0$ \\ ($\text{GeV}^2$)} & \makecell{$M_B^2$ \\ ($\text{GeV}^2$)} &  \makecell{$M_H$ \\ (GeV)} & \makecell{$s_0$ \\ ($\text{GeV}^2$)} & \makecell{$M_B^2$ \\ ($\text{GeV}^2$)} \\ \hline
			$J_{S,1}^{\text{M-M}}$ &$6.16^{+0.08}_{-0.10}$ &$49.(\pm 10\%)$ &$3.75(\pm 10\%)$    &$7.32^{+0.09}_{-0.11}$ &$69.(\pm 10\%)$ &$6.00(\pm 10\%)$\\
			$J_{S,2}^{\text{M-M}}$ &$6.38^{+0.09}_{-0.15}$ &$53.(\pm 10\%)$ &$3.75(\pm 10\%)$    &$8.33^{+0.13}_{-0.15}$ &$87.(\pm 10\%)$ &$8.00(\pm 10\%)$\\
			$J_{S,3}^{\text{M-M}}$ &$7.11^{+0.13}_{-0.15}$ &$65.(\pm 10\%)$ &$5.50(\pm 10\%)$    &$7.91^{+0.16}_{-0.19}$ &$79.(\pm 10\%)$ &$7.50(\pm 10\%)$\\
			$J_{S,4}^{\text{M-M}}$ &$5.90^{+0.06}_{-0.08}$ &$45.(\pm 10\%)$ &$3.00(\pm 10\%)$    &$6.36^{+0.06}_{-0.10}$ &$53.(\pm 10\%)$ &$3.50(\pm 10\%)$\\
			$J_{S,5}^{\text{M-M}}$ &$6.28^{+0.13}_{-0.17}$ &$51.(\pm 10\%)$ &$4.00(\pm 10\%)$    &$7.78^{+0.13}_{-0.13}$ &$77.(\pm 10\%)$ &$6.75(\pm 10\%)$\\ \hline
			$J_{S,1}^{\text{Di-Di}}$ &$6.07^{+0.05}_{-0.07}$ &$49.(\pm 10\%)$ &$3.25(\pm 10\%)$    &$6.60^{+0.09}_{-0.10}$ &$57.(\pm 10\%)$ &$4.00(\pm 10\%)$\\
			$J_{S,2}^{\text{Di-Di}}$ &$6.19^{+0.07}_{-0.12}$ &$51.(\pm 10\%)$ &$3.25(\pm 10\%)$    &$6.90^{+0.11}_{-0.12}$ &$61.(\pm 10\%)$ &$4.75(\pm 10\%)$\\
			$J_{S,3}^{\text{Di-Di}}$ &$6.96^{+0.11}_{-0.14}$ &$63.(\pm 10\%)$ &$4.75(\pm 10\%)$    &$9.25^{+0.14}_{-0.14}$ &$105.(\pm 10\%)$ &$10.00(\pm 10\%)$\\
			$J_{S,4}^{\text{Di-Di}}$ &$6.17^{+0.07}_{-0.12}$ &$51.(\pm 10\%)$ &$3.50(\pm 10\%)$    &$7.36^{+0.10}_{-0.11}$ &$69.(\pm 10\%)$ &$6.25(\pm 10\%)$\\
			$J_{S,5}^{\text{Di-Di}}$ &$6.07^{+0.08}_{-0.10}$ &$47.(\pm 10\%)$ &$3.50(\pm 10\%)$    &$6.69^{+0.10}_{-0.12}$ &$57.(\pm 10\%)$ &$4.25(\pm 10\%)$\\ \hline
			$J_{S,1}^{\text{Dia}}$ &$6.18^{+0.08}_{-0.10}$ &$49.(\pm 10\%)$ &$3.75(\pm 10\%)$    &$7.81^{+0.14}_{-0.16}$ &$77.(\pm 10\%)$ &$7.25(\pm 10\%)$\\
			$J_{S,2}^{\text{Dia}}$ &$6.19^{+0.07}_{-0.12}$ &$51.(\pm 10\%)$ &$3.50(\pm 10\%)$    &$6.95^{+0.10}_{-0.12}$ &$61.(\pm 10\%)$ &$5.00(\pm 10\%)$\\
			$J_{S,3}^{\text{Dia}}$ &$5.93^{+0.07}_{-0.10}$ &$45.(\pm 10\%)$ &$3.00(\pm 10\%)$    &$6.35^{+0.08}_{-0.13}$ &$51.(\pm 10\%)$ &$3.50(\pm 10\%)$\\
			$J_{S,4}^{\text{Dia}}$ &$6.02^{+0.05}_{-0.06}$ &$49.(\pm 10\%)$ &$3.00(\pm 10\%)$    &$6.56^{+0.10}_{-0.12}$ &$55.(\pm 10\%)$ &$4.00(\pm 10\%)$\\
			$J_{S,5}^{\text{Dia}}$ &$6.33^{+0.12}_{-0.14}$ &$53.(\pm 10\%)$ &$4.00(\pm 10\%)$    &$7.72^{+0.13}_{-0.14}$ &$75.(\pm 10\%)$ &$6.50(\pm 10\%)$\\ \hline\hline
		\end{tabular}
		
		\label{tab:S-NLOresult-MSbar}
	\end{center}
	\vspace{-0.4cm}
\end{table}

\begin{table}[H]
	\vspace{-0.4cm}
	\renewcommand\arraystretch{1.8}
	\setlength{\tabcolsep}{3 mm}
	\begin{center}
		\caption{The LO and NLO Results for $J^P=0^+$ with $\bar{c}c\bar{c}c$ system in the On-Shell scheme}
		
		\begin{tabular}{cccc|@{*}|ccc}
			\hline\hline
			\multirow{2}{*}{Current} &
			\multicolumn{3}{c|@{*}|}{LO}& \multicolumn{3}{c}{NLO(OS)} \\ \cline{2-7}
			& \makecell{$M_H$ \\ (GeV)} & \makecell{$s_0$ \\ ($\text{GeV}^2$)} & \makecell{$M_B^2$ \\ ($\text{GeV}^2$)} &  \makecell{$M_H$ \\ (GeV)} & \makecell{$s_0$ \\ ($\text{GeV}^2$)} & \makecell{$M_B^2$ \\ ($\text{GeV}^2$)} \\ \hline
			$J_{S,1}^{\text{M-M}}$ &$7.35^{+0.07}_{-0.10}$ &$66.(\pm 10\%)$ &$3.75(\pm 10\%)$    &$6.60^{+0.09}_{-0.12}$ &$48.(\pm 10\%)$ &$2.25(\pm 10\%)$\\
			$J_{S,2}^{\text{M-M}}$ &$7.44^{+0.12}_{-0.15}$ &$66.(\pm 10\%)$ &$4.00(\pm 10\%)$    &$6.60^{+0.10}_{-0.15}$ &$48.(\pm 10\%)$ &$2.25(\pm 10\%)$\\
			$J_{S,3}^{\text{M-M}}$ &$8.43^{+0.14}_{-0.18}$ &$86.(\pm 10\%)$ &$6.00(\pm 10\%)$    &$7.40^{+0.15}_{-0.21}$ &$62.(\pm 10\%)$ &$3.75(\pm 10\%)$\\
			$J_{S,4}^{\text{M-M}}$ &$7.05^{+0.06}_{-0.09}$ &$60.(\pm 10\%)$ &$3.00(\pm 10\%)$    &$6.44^{+0.08}_{-0.09}$ &$44.(\pm 10\%)$ &$1.75(\pm 10\%)$\\
			$J_{S,5}^{\text{M-M}}$ &$7.45^{+0.10}_{-0.11}$ &$68.(\pm 10\%)$ &$4.00(\pm 10\%)$    &$6.62^{+0.09}_{-0.13}$ &$48.(\pm 10\%)$ &$2.25(\pm 10\%)$\\ \hline
			$J_{S,1}^{\text{Di-Di}}$ &$7.23^{+0.04}_{-0.07}$ &$66.(\pm 10\%)$ &$3.25(\pm 10\%)$    &$6.54^{+0.06}_{-0.08}$ &$48.(\pm 10\%)$ &$1.75(\pm 10\%)$\\
			$J_{S,2}^{\text{Di-Di}}$ &$7.27^{+0.08}_{-0.11}$ &$64.(\pm 10\%)$ &$3.50(\pm 10\%)$    &$6.52^{+0.10}_{-0.14}$ &$46.(\pm 10\%)$ &$2.00(\pm 10\%)$\\
			$J_{S,3}^{\text{Di-Di}}$ &$8.17^{+0.15}_{-0.19}$ &$80.(\pm 10\%)$ &$5.25(\pm 10\%)$    &$7.19^{+0.16}_{-0.26}$ &$58.(\pm 10\%)$ &$3.25(\pm 10\%)$\\
			$J_{S,4}^{\text{Di-Di}}$ &$7.31^{+0.08}_{-0.11}$ &$64.(\pm 10\%)$ &$3.75(\pm 10\%)$    &$6.59^{+0.09}_{-0.12}$ &$48.(\pm 10\%)$ &$2.25(\pm 10\%)$\\
			$J_{S,5}^{\text{Di-Di}}$ &$7.22^{+0.08}_{-0.12}$ &$62.(\pm 10\%)$ &$3.50(\pm 10\%)$    &$6.51^{+0.09}_{-0.13}$ &$46.(\pm 10\%)$ &$2.00(\pm 10\%)$\\ \hline
			$J_{S,1}^{\text{Dia}}$ &$7.36^{+0.07}_{-0.10}$ &$66.(\pm 10\%)$ &$3.75(\pm 10\%)$    &$6.60^{+0.09}_{-0.12}$ &$48.(\pm 10\%)$ &$2.25(\pm 10\%)$\\
			$J_{S,2}^{\text{Dia}}$ &$7.31^{+0.08}_{-0.12}$ &$64.(\pm 10\%)$ &$3.75(\pm 10\%)$    &$6.58^{+0.08}_{-0.11}$ &$48.(\pm 10\%)$ &$2.00(\pm 10\%)$\\
			$J_{S,3}^{\text{Dia}}$ &$7.06^{+0.07}_{-0.10}$ &$60.(\pm 10\%)$ &$3.00(\pm 10\%)$    &$6.47^{+0.08}_{-0.10}$ &$46.(\pm 10\%)$ &$1.75(\pm 10\%)$\\
			$J_{S,4}^{\text{Dia}}$ &$7.16^{+0.04}_{-0.05}$ &$66.(\pm 10\%)$ &$3.00(\pm 10\%)$    &$6.49^{+0.07}_{-0.10}$ &$46.(\pm 10\%)$ &$1.75(\pm 10\%)$\\
			$J_{S,5}^{\text{Dia}}$ &$7.44^{+0.12}_{-0.14}$ &$66.(\pm 10\%)$ &$4.25(\pm 10\%)$    &$6.62^{+0.09}_{-0.13}$ &$48.(\pm 10\%)$ &$2.25(\pm 10\%)$\\ \hline\hline
		\end{tabular}
		
		\label{tab:S-NLOresult-OS}
	\end{center}
	\vspace{-0.8cm}
\end{table}

\begin{figure}[H]
	\vspace{-0.8cm}
	\centering
	\subfigure[$\overline{\text{MS}}$]{
		\includegraphics[scale=0.47]{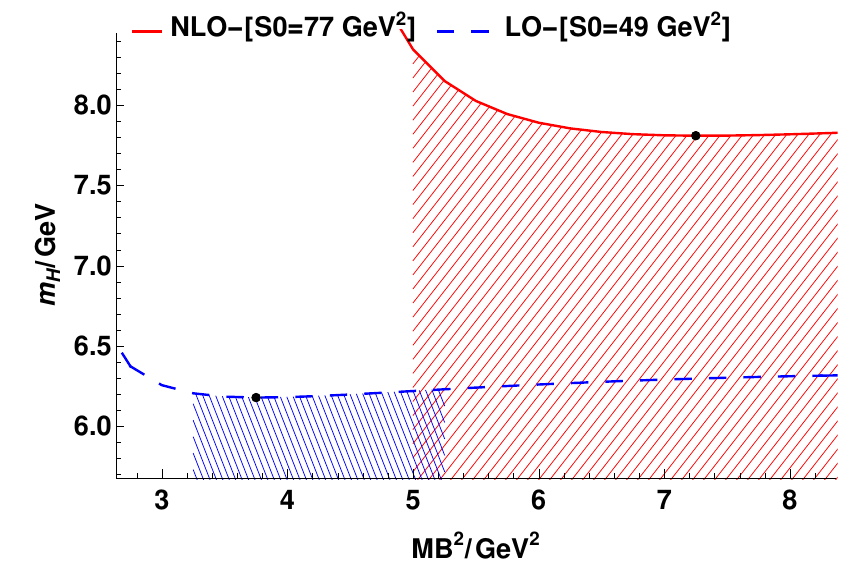}
		\includegraphics[scale=0.47]{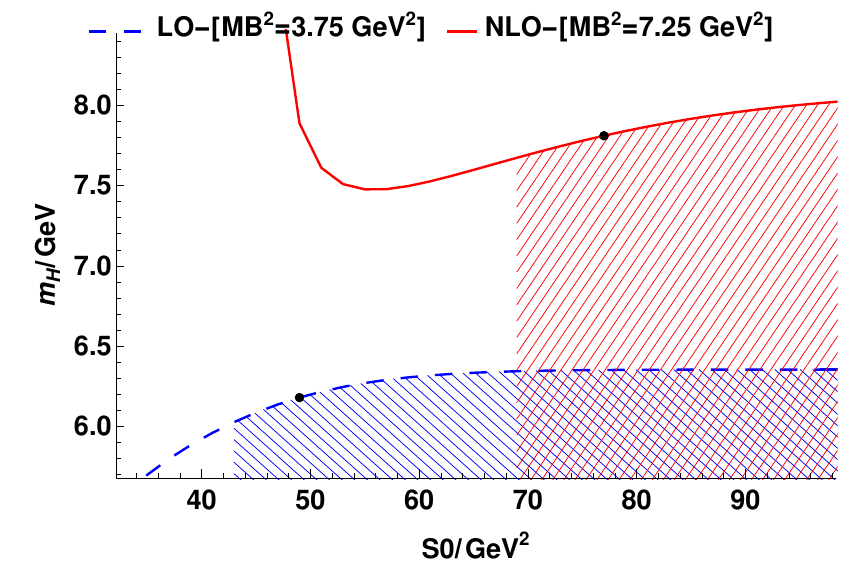}
	}\\
	\subfigure[OS]{
		\includegraphics[scale=0.47]{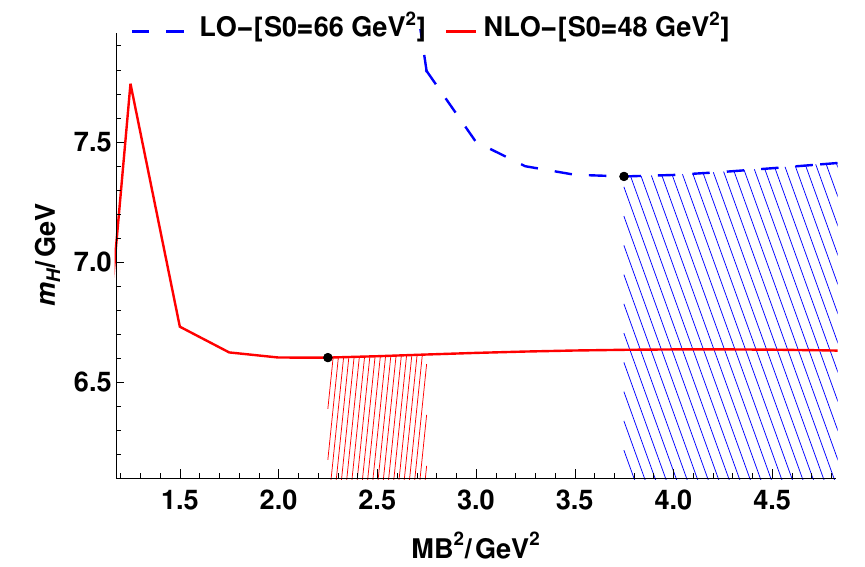}
		\includegraphics[scale=0.47]{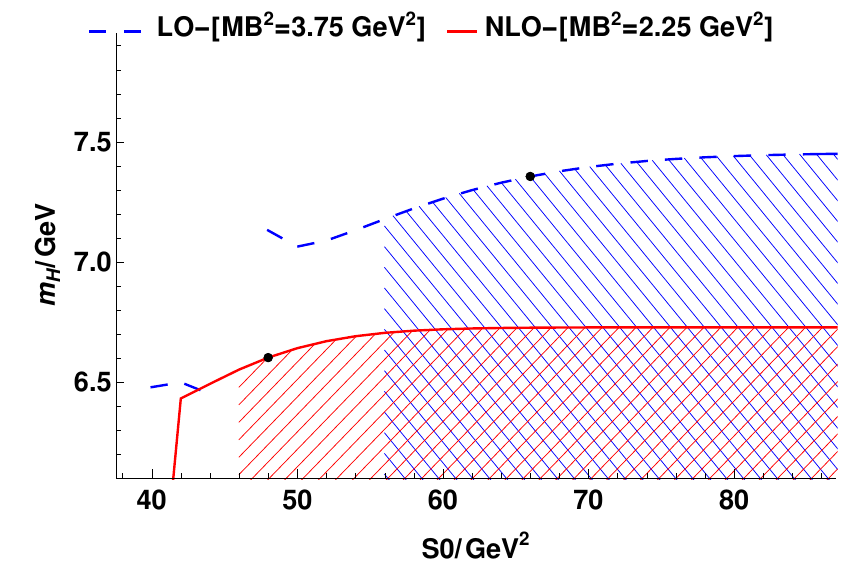}
	}
	\caption{\label{fig:0+-Mixed1-NLO-MSbar-OS}
		The Borel platform curves for $J_{S,1}^{\text{Dia}}$ with $J^{PC}=0^{++}$ in the $\overline{\text{MS}}$ and On-Shell schemes}
	\vspace{-0.5cm}
\end{figure}

\begin{figure}[H]
	\centering
	\subfigure[$\overline{\text{MS}}$]{
		\includegraphics[scale=0.47]{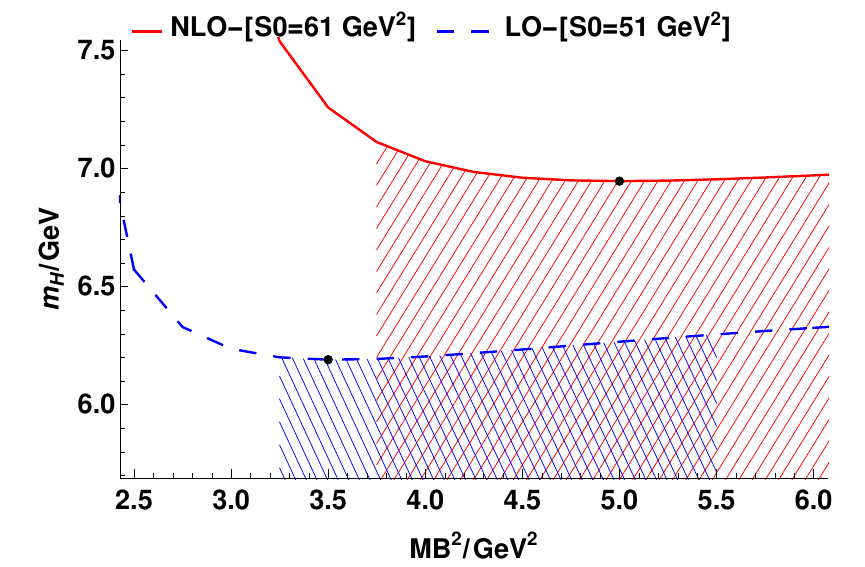}
		\includegraphics[scale=0.47]{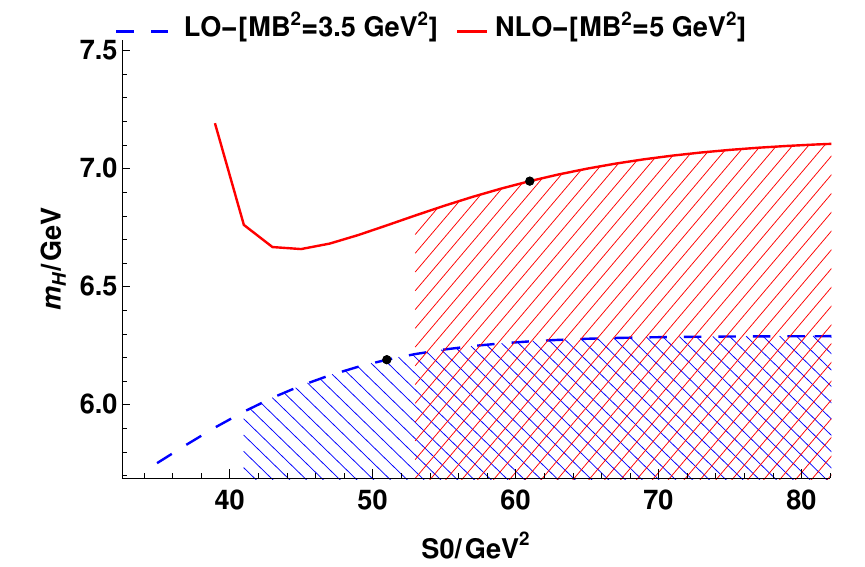}
	}\\
	\subfigure[OS]{
		\includegraphics[scale=0.47]{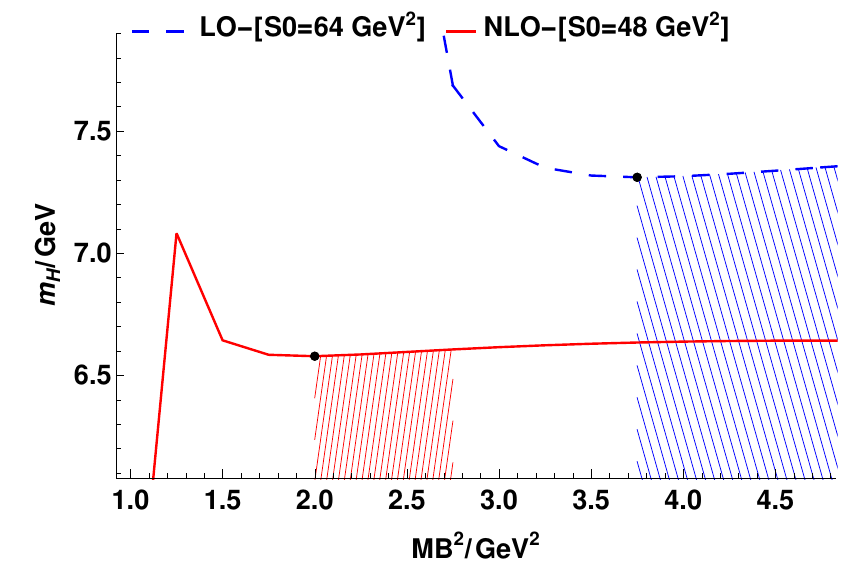}
		\includegraphics[scale=0.47]{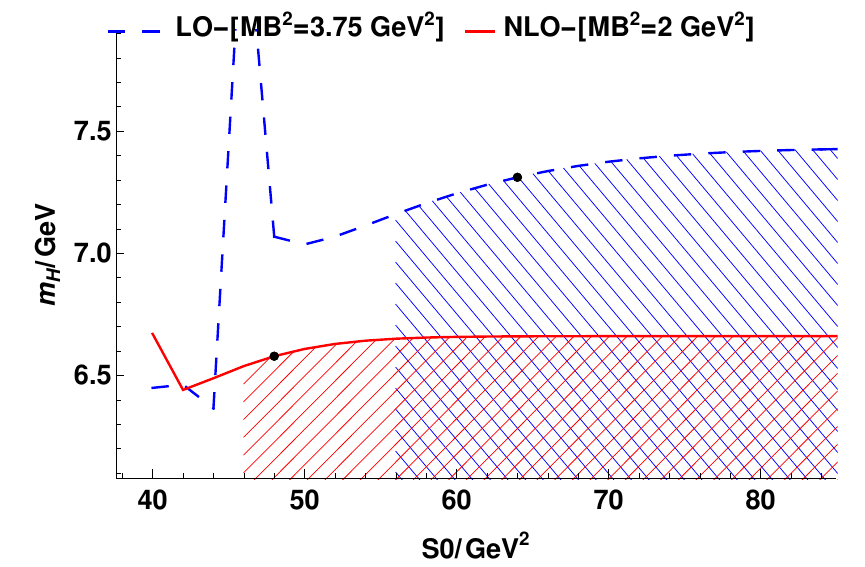}
	}
	\caption{\label{fig:0+-Mixed2-NLO-MSbar-OS}
		The Borel platform curves for $J_{S,2}^{\text{Dia}}$ with $J^{PC}=0^{++}$ in the $\overline{\text{MS}}$ and On-Shell schemes}
	\vspace{-0.8cm}
\end{figure}

\begin{figure}[H]
		\vspace{-0.8cm}
	\centering
	\subfigure[$\overline{\text{MS}}$]{
		\includegraphics[scale=0.47]{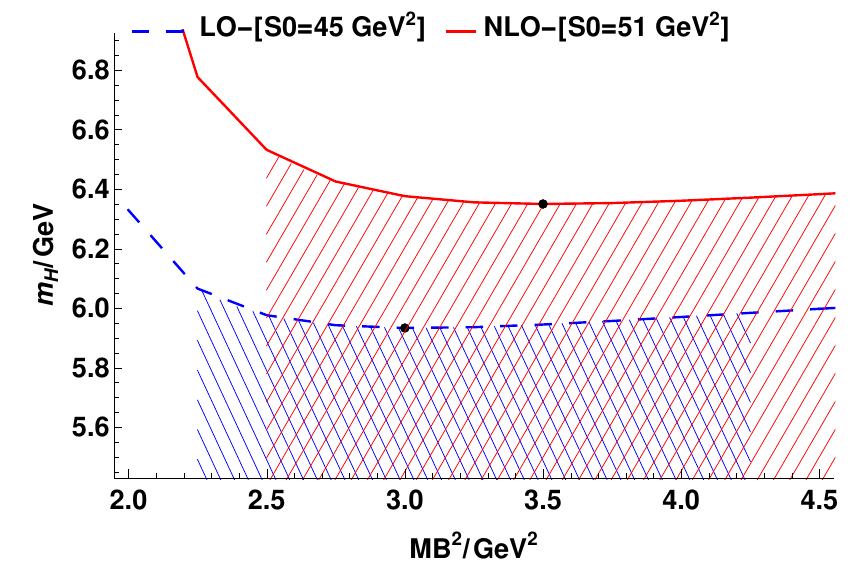}
		\includegraphics[scale=0.47]{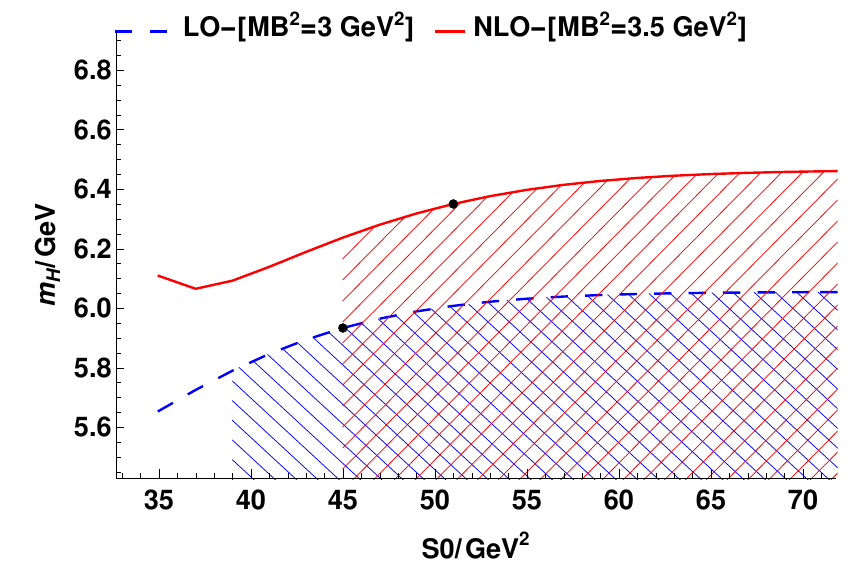}
	}\\
	\subfigure[OS]{
		\includegraphics[scale=0.47]{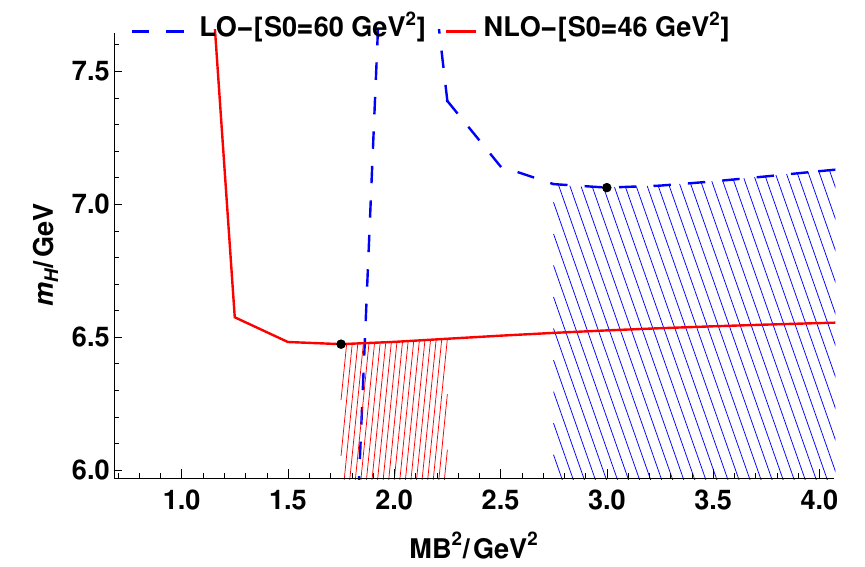}
		\includegraphics[scale=0.47]{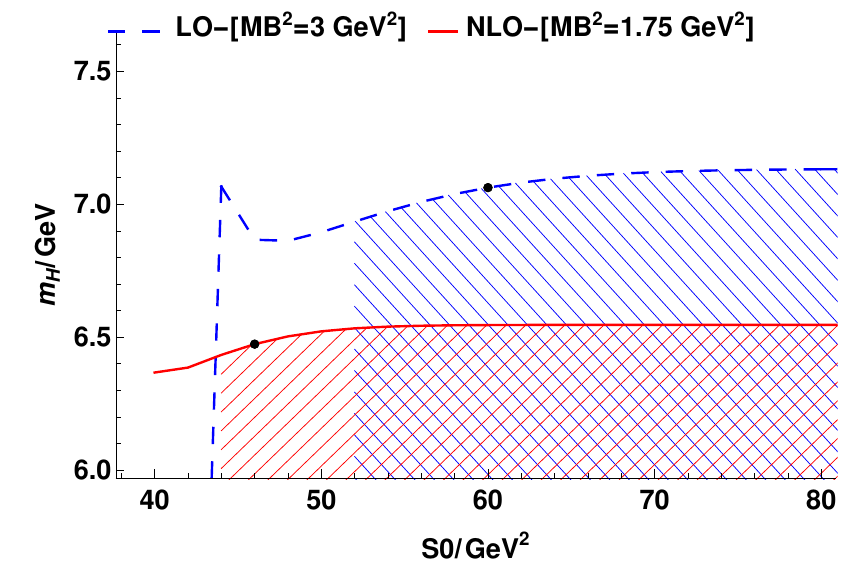}
	}
	\caption{\label{fig:0+-Mixed3-NLO-MSbar-OS}
		The Borel platform curves for $J_{S,3}^{\text{Dia}}$ with $J^{PC}=0^{++}$ in the $\overline{\text{MS}}$ and On-Shell schemes}
	\vspace{-0.5cm}
\end{figure}

\begin{figure}[H]
	\centering
	\subfigure[$\overline{\text{MS}}$]{
		\includegraphics[scale=0.47]{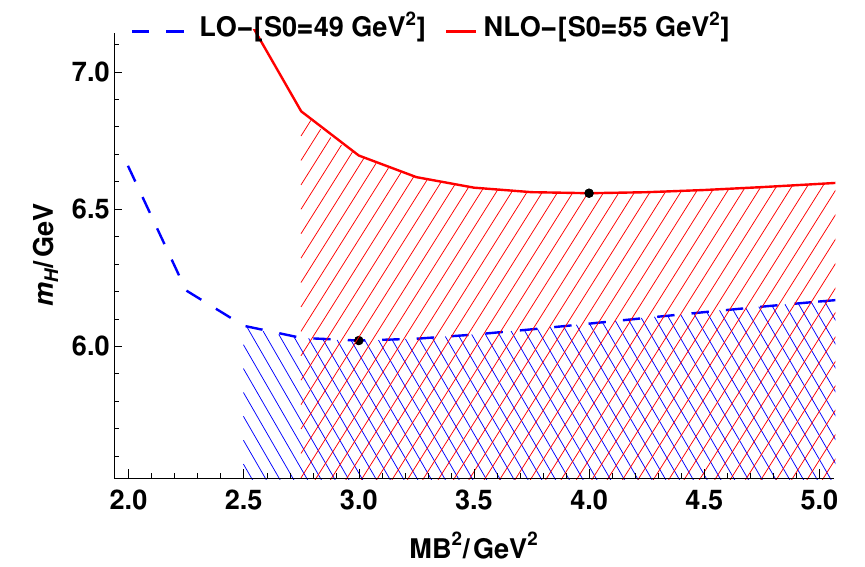}
		\includegraphics[scale=0.47]{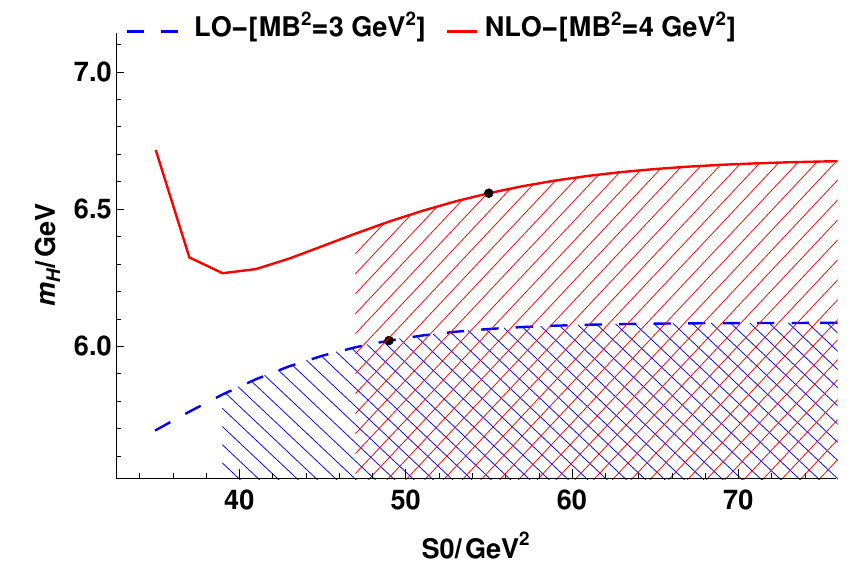}
	}\\
	\subfigure[OS]{
		\includegraphics[scale=0.47]{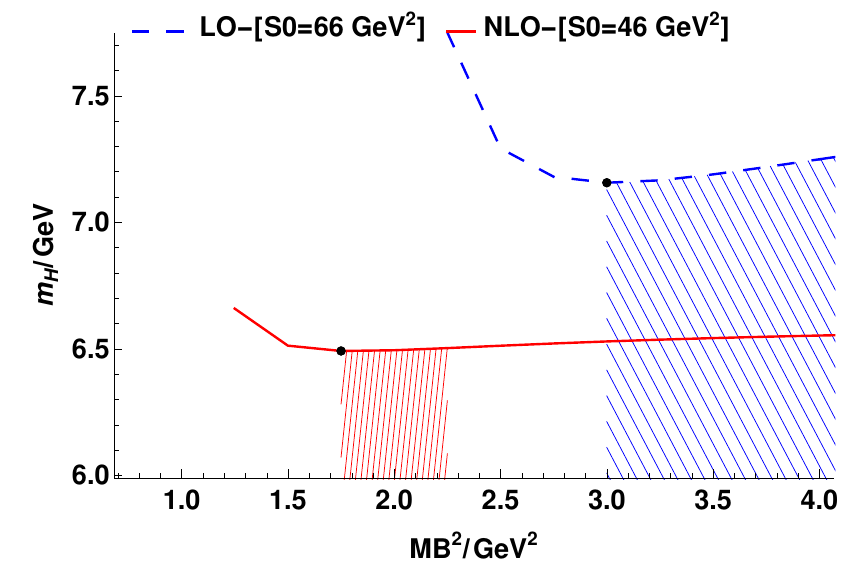}
		\includegraphics[scale=0.47]{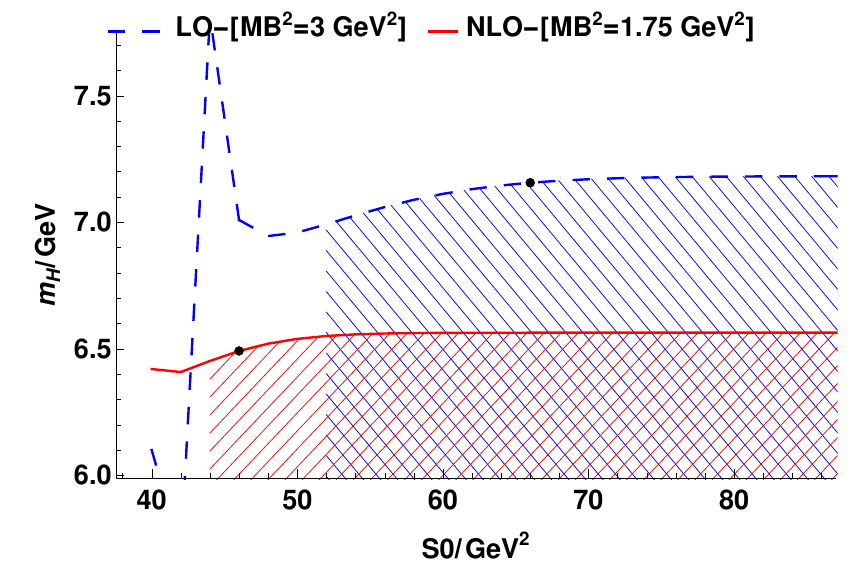}
	}
	\caption{\label{fig:0+-Mixed4-NLO-MSbar-OS}
		The Borel platform curves for $J_{S,4}^{\text{Dia}}$ with $J^{PC}=0^{++}$ in the $\overline{\text{MS}}$ and On-Shell schemes}
	\vspace{-0.8cm}
\end{figure}

\begin{figure}[H]
	\vspace{-0.9cm}
	\centering
	\subfigure[$\overline{\text{MS}}$]{
		\includegraphics[scale=0.47]{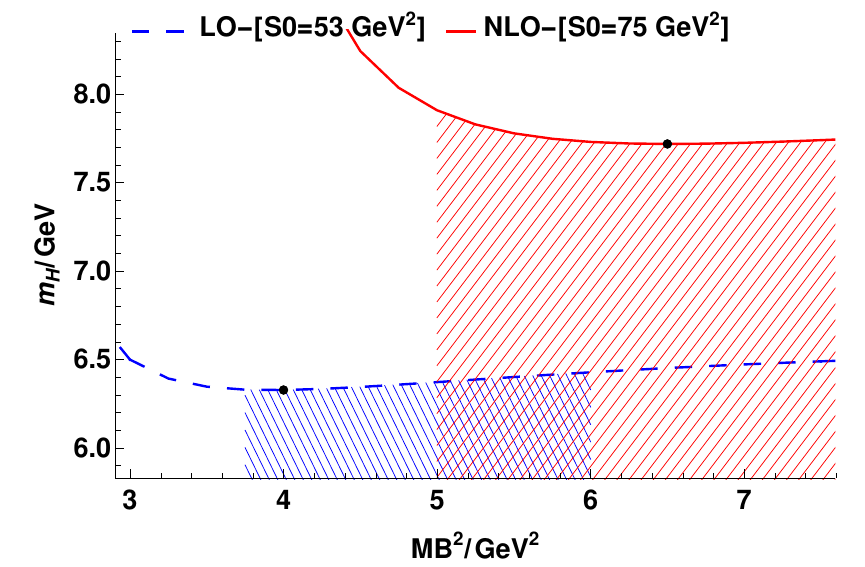}
		\includegraphics[scale=0.47]{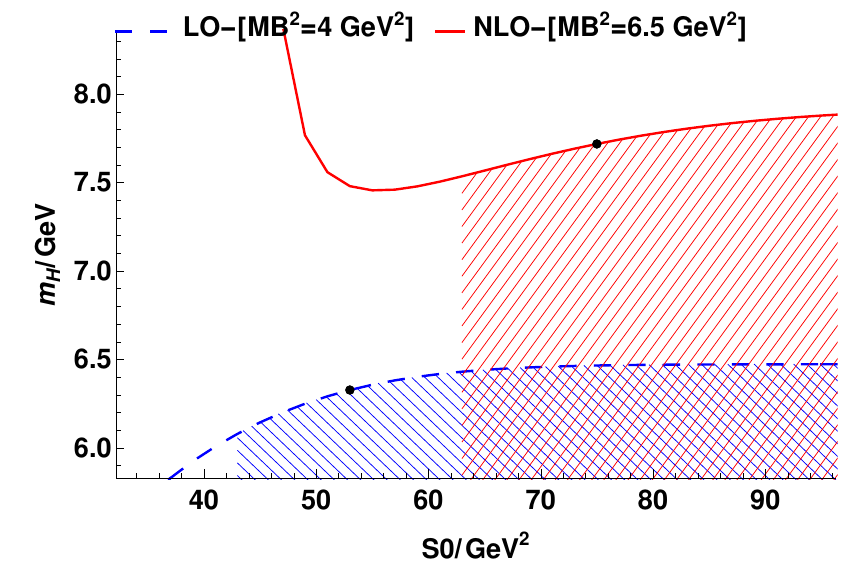}
	}\\
	\subfigure[OS]{
		\includegraphics[scale=0.47]{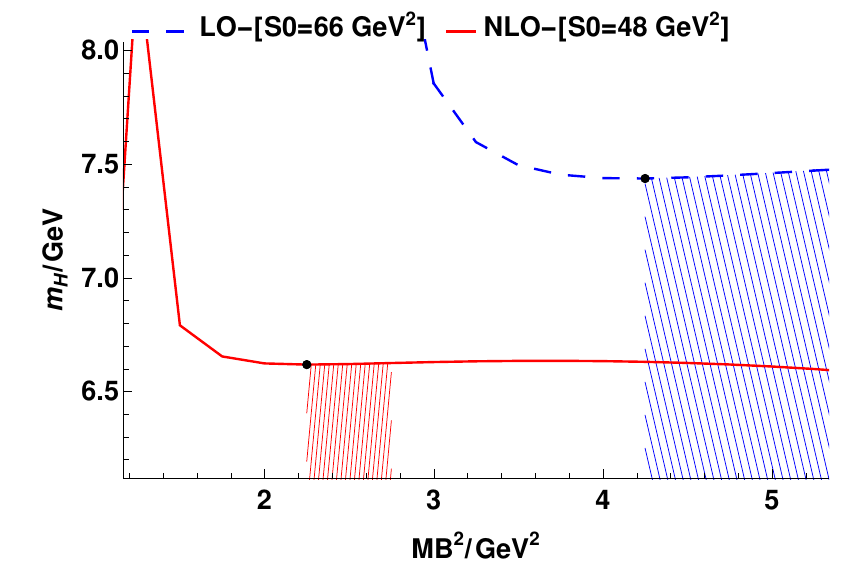}
		\includegraphics[scale=0.47]{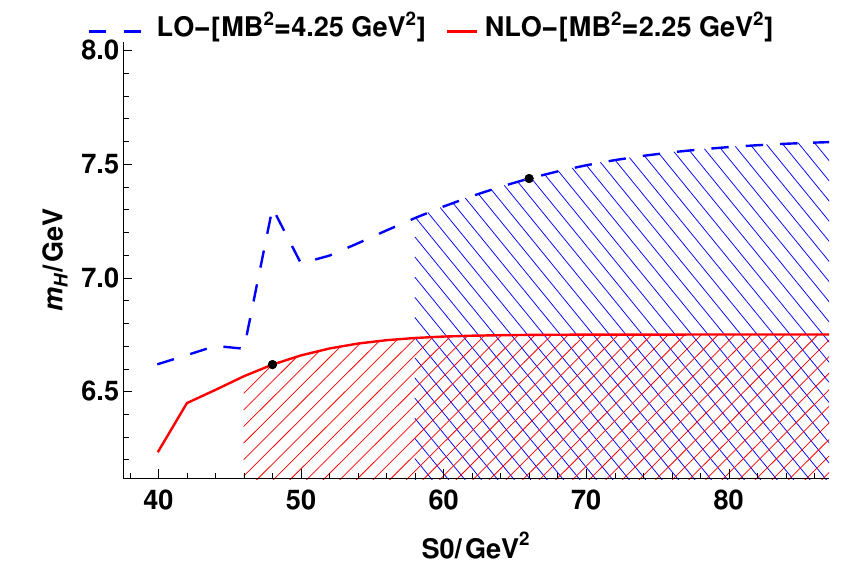}
	}
	\caption{\label{fig:0+-Mixed5-NLO-MSbar-OS}
		The Borel platform curves for $J_{S,5}^{\text{Dia}}$ with $J^{PC}=0^{++}$ in the $\overline{\text{MS}}$ and On-Shell schemes}
	\vspace{-0.2cm}
\end{figure}

\subsection{Numerical Results for $J^P=0^-$ states}

\begin{table}[H]
		\vspace{-0.5cm}
	\renewcommand\arraystretch{1.6}
	\setlength{\tabcolsep}{3 mm}
	\begin{center}
		\caption{The LO and NLO Results for $J^P=0^-$ with $\bar{c}c\bar{c}c$ system in the $\overline{\text{MS}}$ scheme}
		\begin{tabular}{cccc|@{*}|ccc}
			\hline\hline
			\multirow{2}{*}{Current} &
			\multicolumn{3}{c|@{*}|}{LO}& \multicolumn{3}{c}{NLO($\overline{\text{MS}}$)} \\ \cline{2-7}
			& \makecell{$M_H$ \\ (GeV)} & \makecell{$s_0$ \\ ($\text{GeV}^2$)} & \makecell{$M_B^2$ \\ ($\text{GeV}^2$)} &  \makecell{$M_H$ \\ (GeV)} & \makecell{$s_0$ \\ ($\text{GeV}^2$)} & \makecell{$M_B^2$ \\ ($\text{GeV}^2$)} \\ \hline
			$J_{P,1}^{\text{M-M}}$ &$6.55^{+0.12}_{-0.14}$ &$56.(\pm 10\%)$ &$4.25(\pm 10\%)$    &$8.43^{+0.17}_{-0.17}$ &$88.(\pm 10\%)$ &$9.50(\pm 10\%)$\\
			$J_{P,2}^{\text{M-M}}$ &$6.53^{+0.12}_{-0.14}$ &$56.(\pm 10\%)$ &$4.25(\pm 10\%)$    &$7.30^{+0.11}_{-0.13}$ &$68.(\pm 10\%)$ &$6.00(\pm 10\%)$\\
			$J_{P,3}^{\text{M-M}}$ &$6.56^{+0.12}_{-0.15}$ &$56.(\pm 10\%)$ &$4.25(\pm 10\%)$    &$8.53^{+0.15}_{-0.19}$ &$90.(\pm 10\%)$ &$9.00(\pm 10\%)$\\ \hline
			$J_{P,1}^{\text{Di-Di}}$ &$6.55^{+0.12}_{-0.14}$ &$56.(\pm 10\%)$ &$4.25(\pm 10\%)$    &$8.43^{+0.17}_{-0.17}$ &$88.(\pm 10\%)$ &$9.50(\pm 10\%)$\\
			$J_{P,2}^{\text{Di-Di}}$ &$6.55^{+0.12}_{-0.14}$ &$56.(\pm 10\%)$ &$4.25(\pm 10\%)$    &$8.08^{+0.15}_{-0.16}$ &$82.(\pm 10\%)$ &$8.00(\pm 10\%)$\\
			$J_{P,3}^{\text{Di-Di}}$ &$6.54^{+0.12}_{-0.14}$ &$56.(\pm 10\%)$ &$4.25(\pm 10\%)$    &$7.51^{+0.12}_{-0.16}$ &$72.(\pm 10\%)$ &$6.25(\pm 10\%)$\\ \hline
			$J_{P,1}^{\text{Dia}}$ &$6.55^{+0.12}_{-0.14}$ &$56.(\pm 10\%)$ &$4.25(\pm 10\%)$    &$8.43^{+0.17}_{-0.17}$ &$88.(\pm 10\%)$ &$9.50(\pm 10\%)$\\
			$J_{P,2}^{\text{Dia}}$ &$6.53^{+0.12}_{-0.14}$ &$56.(\pm 10\%)$ &$4.25(\pm 10\%)$    &$7.30^{+0.11}_{-0.13}$ &$68.(\pm 10\%)$ &$6.00(\pm 10\%)$\\
			$J_{P,3}^{\text{Dia}}$ &$6.56^{+0.12}_{-0.15}$ &$56.(\pm 10\%)$ &$4.25(\pm 10\%)$    &$8.59^{+0.15}_{-0.18}$ &$92.(\pm 10\%)$ &$9.00(\pm 10\%)$\\ \hline\hline
		\end{tabular}
		\label{tab:P-NLOresult-MSbar}
	\end{center}
	\vspace{-0.5cm}
\end{table}

\begin{table}[H]
		\vspace{-0.5cm}
	\renewcommand\arraystretch{1.6}
	\setlength{\tabcolsep}{3 mm}
	\begin{center}
		\caption{The LO and NLO Results for $J^P=0^-$ with $\bar{c}c\bar{c}c$ system in the On-Shell scheme}
		\begin{tabular}{cccc|@{*}|ccc}
			\hline\hline
			\multirow{2}{*}{Current} &
			\multicolumn{3}{c|@{*}|}{LO}& \multicolumn{3}{c}{NLO(OS)} \\ \cline{2-7}
			& \makecell{$M_H$ \\ (GeV)} & \makecell{$s_0$ \\ ($\text{GeV}^2$)} & \makecell{$M_B^2$ \\ ($\text{GeV}^2$)} &  \makecell{$M_H$ \\ (GeV)} & \makecell{$s_0$ \\ ($\text{GeV}^2$)} & \makecell{$M_B^2$ \\ ($\text{GeV}^2$)} \\ \hline
			$J_{P,1}^{\text{M-M}}$ &$7.74^{+0.14}_{-0.18}$ &$72.(\pm 10\%)$ &$4.50(\pm 10\%)$    &$6.87^{+0.15}_{-0.27}$ &$52.(\pm 10\%)$ &$2.75(\pm 10\%)$\\
			$J_{P,2}^{\text{M-M}}$ &$7.79^{+0.10}_{-0.15}$ &$74.(\pm 10\%)$ &$4.50(\pm 10\%)$    &$6.89^{+0.14}_{-0.23}$ &$52.(\pm 10\%)$ &$2.75(\pm 10\%)$\\
			$J_{P,3}^{\text{M-M}}$ &$7.70^{+0.12}_{-0.21}$ &$70.(\pm 10\%)$ &$4.50(\pm 10\%)$    &$6.84^{+0.13}_{-0.23}$ &$52.(\pm 10\%)$ &$2.50(\pm 10\%)$\\ \hline
			$J_{P,1}^{\text{Di-Di}}$ &$7.74^{+0.14}_{-0.18}$ &$72.(\pm 10\%)$ &$4.50(\pm 10\%)$    &$6.87^{+0.15}_{-0.27}$ &$52.(\pm 10\%)$ &$2.75(\pm 10\%)$\\
			$J_{P,2}^{\text{Di-Di}}$ &$7.74^{+0.14}_{-0.18}$ &$72.(\pm 10\%)$ &$4.50(\pm 10\%)$    &$6.86^{+0.15}_{-0.26}$ &$52.(\pm 10\%)$ &$2.75(\pm 10\%)$\\
			$J_{P,3}^{\text{Di-Di}}$ &$7.75^{+0.13}_{-0.18}$ &$72.(\pm 10\%)$ &$4.50(\pm 10\%)$    &$6.87^{+0.14}_{-0.25}$ &$52.(\pm 10\%)$ &$2.75(\pm 10\%)$\\ \hline
			$J_{P,1}^{\text{Dia}}$ &$7.74^{+0.14}_{-0.18}$ &$72.(\pm 10\%)$ &$4.50(\pm 10\%)$    &$6.87^{+0.15}_{-0.27}$ &$52.(\pm 10\%)$ &$2.75(\pm 10\%)$\\
			$J_{P,2}^{\text{Dia}}$ &$7.79^{+0.10}_{-0.15}$ &$74.(\pm 10\%)$ &$4.50(\pm 10\%)$    &$6.89^{+0.14}_{-0.23}$ &$52.(\pm 10\%)$ &$2.75(\pm 10\%)$\\
			$J_{P,3}^{\text{Dia}}$ &$7.70^{+0.12}_{-0.21}$ &$70.(\pm 10\%)$ &$4.50(\pm 10\%)$    &$6.84^{+0.13}_{-0.23}$ &$52.(\pm 10\%)$ &$2.50(\pm 10\%)$\\ \hline\hline
		\end{tabular}
		
		\label{tab:P-NLOresult-OS}
	\end{center}
	\vspace{-0.8cm}
\end{table}

\begin{figure}[H]
		\vspace{-0.8cm}
	\centering
	\subfigure[$\overline{\text{MS}}$]{
		\includegraphics[scale=0.47]{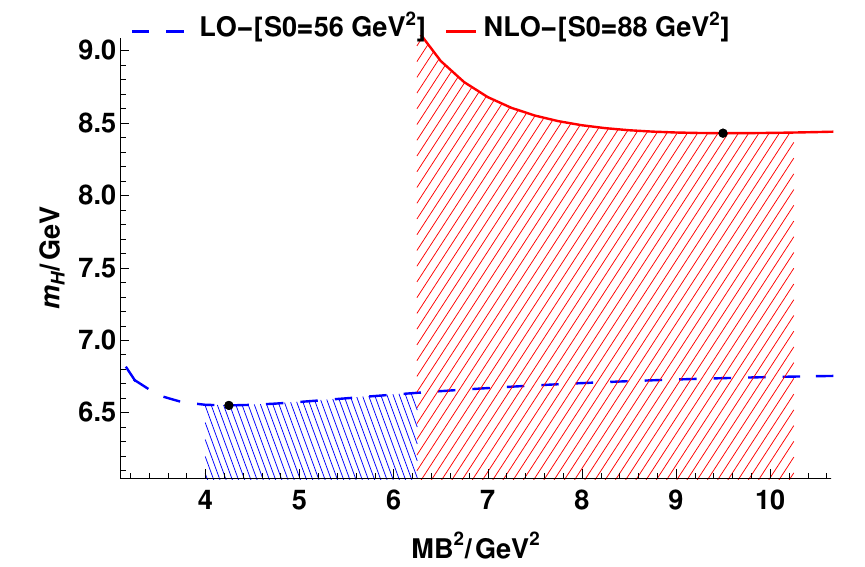}
		\includegraphics[scale=0.47]{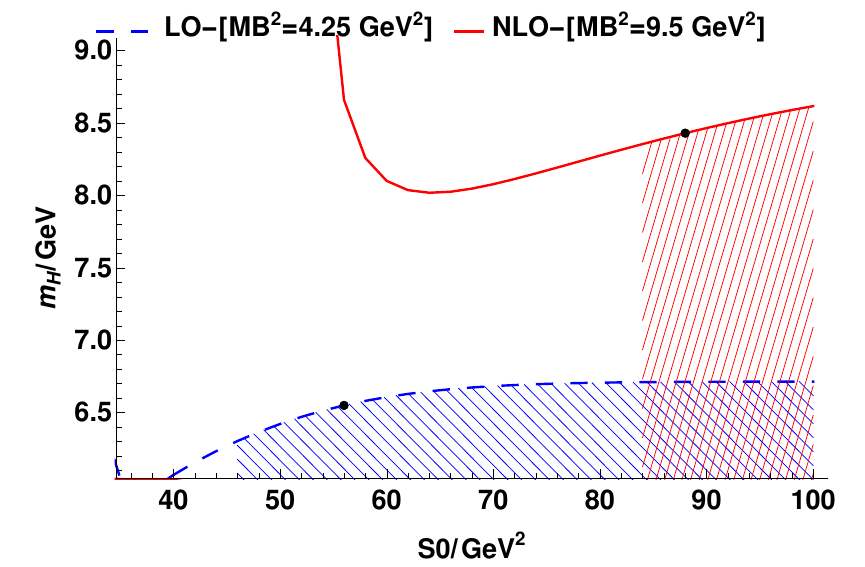}
	}\\
	\subfigure[OS]{
		\includegraphics[scale=0.47]{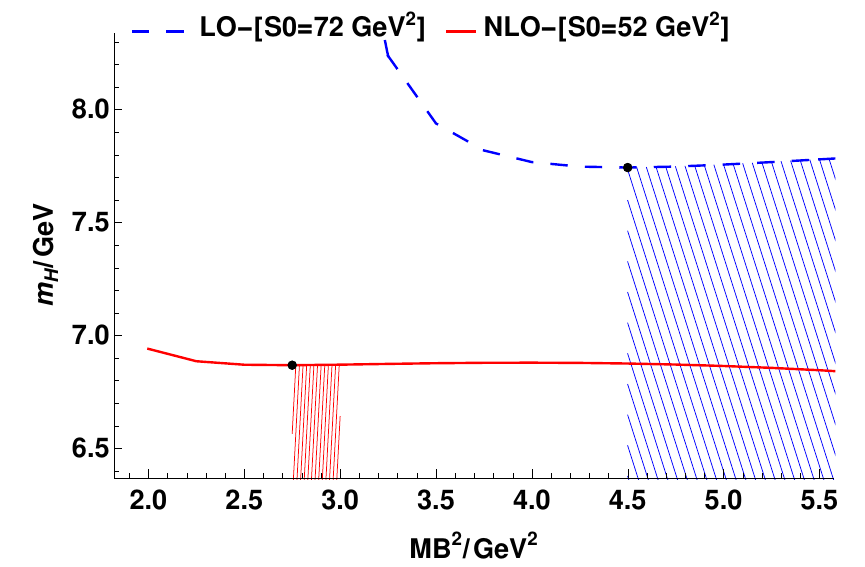}
		\includegraphics[scale=0.47]{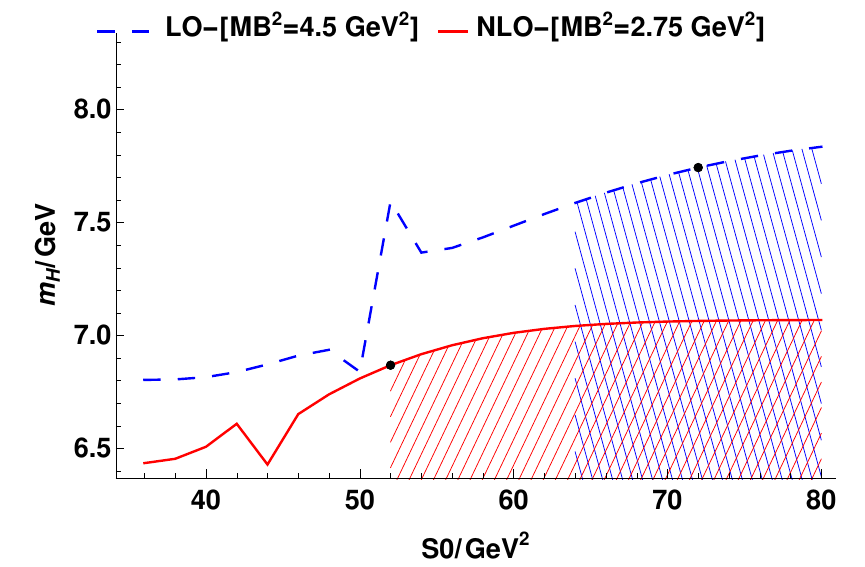}
	}
	\caption{\label{fig:0--Mixed1-NLO-MSbar-OS}
		The Borel platform curves for $J_{P,1}^{\text{Dia}}$ with $J^{PC}=0^{--}$ in the $\overline{\text{MS}}$ and On-Shell schemes}
		\vspace{-0.5cm}
\end{figure}
\begin{figure}[H]
		\vspace{-1cm}
	\centering
	\subfigure[$\overline{\text{MS}}$]{
		\includegraphics[scale=0.47]{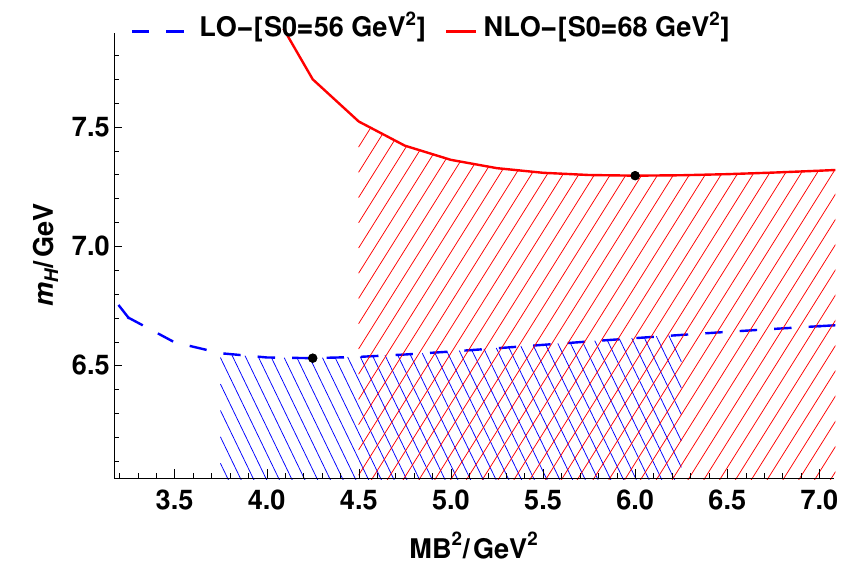}
		\includegraphics[scale=0.47]{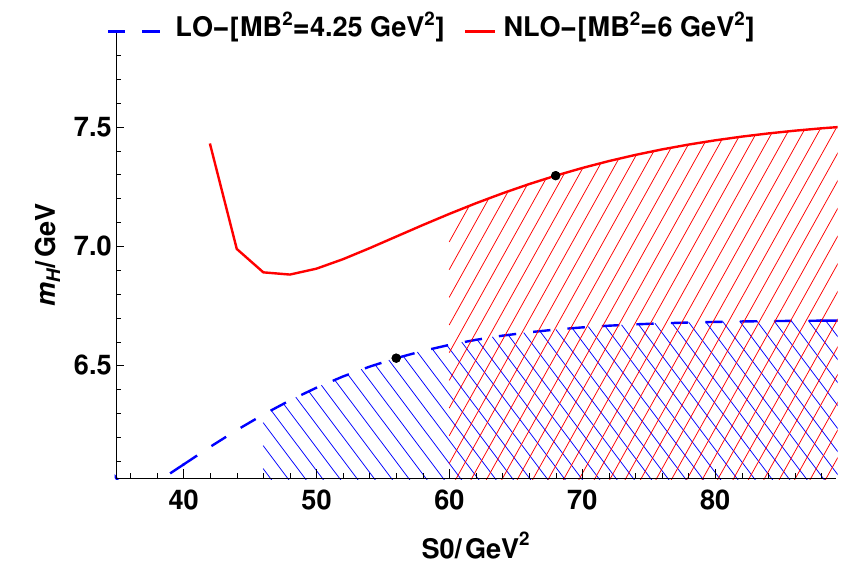}
	}\\
	\subfigure[OS]{
		\includegraphics[scale=0.47]{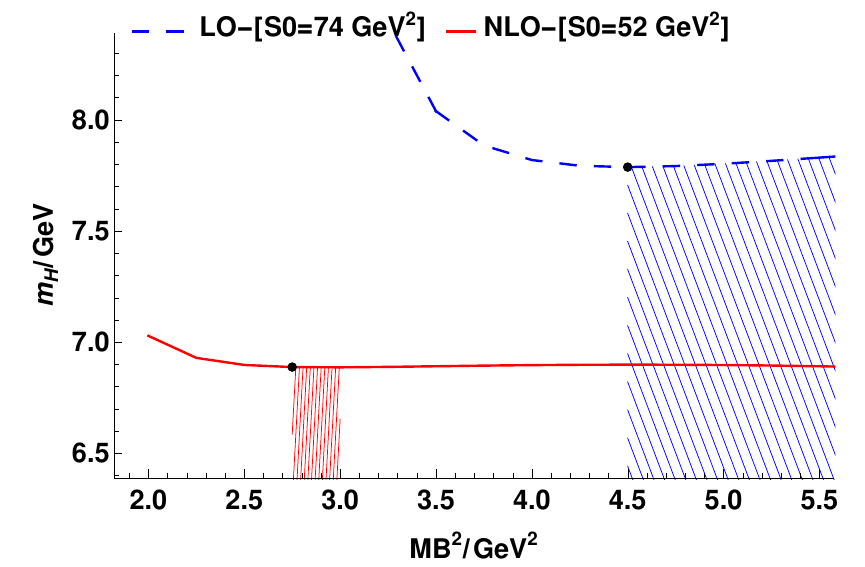}
		\includegraphics[scale=0.47]{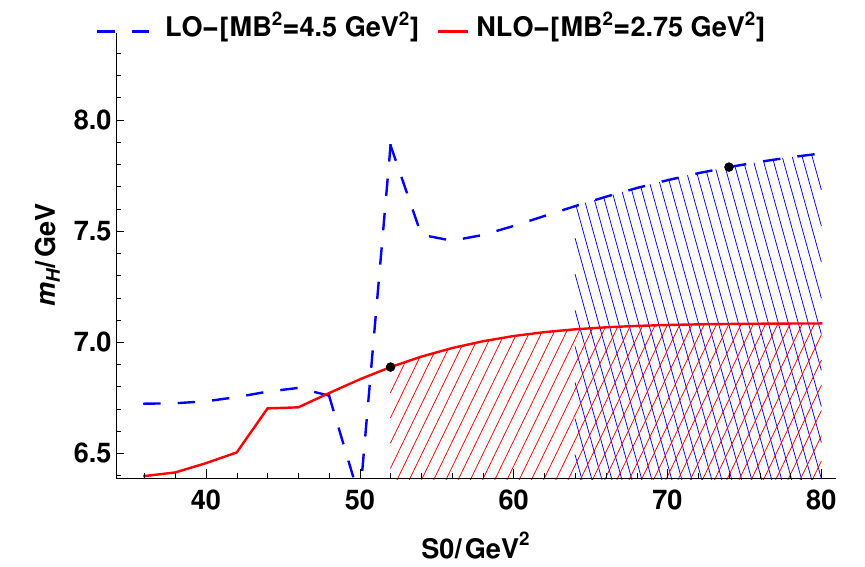}
	}
	\caption{\label{fig:0--Mixed2-NLO-MSbar-OS}
		The Borel platform curves for $J_{P,2}^{\text{Dia}}$ with $J^{PC}=0^{-+}$ in the $\overline{\text{MS}}$ and On-Shell schemes}
	\vspace{-0.2cm}
\end{figure}
\begin{figure}[H]
	\centering
	\subfigure[$\overline{\text{MS}}$]{
		\includegraphics[scale=0.47]{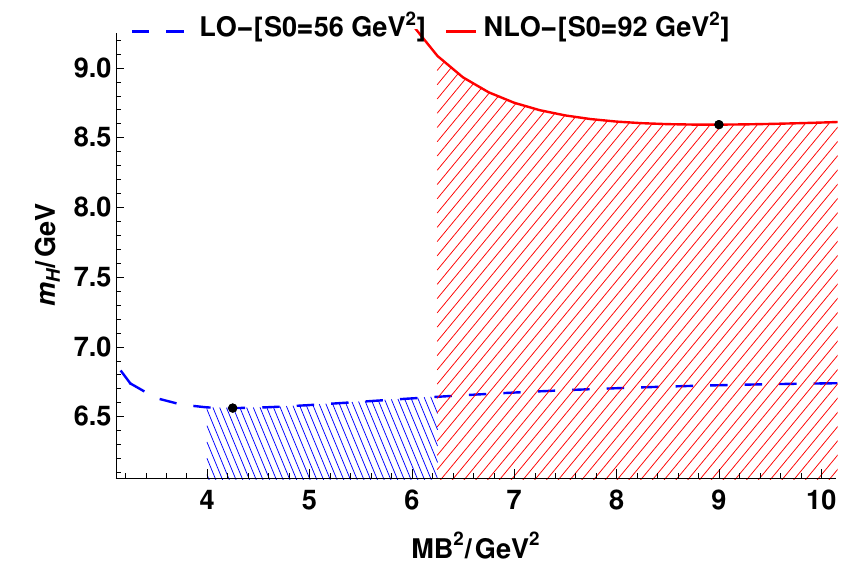}
		\includegraphics[scale=0.47]{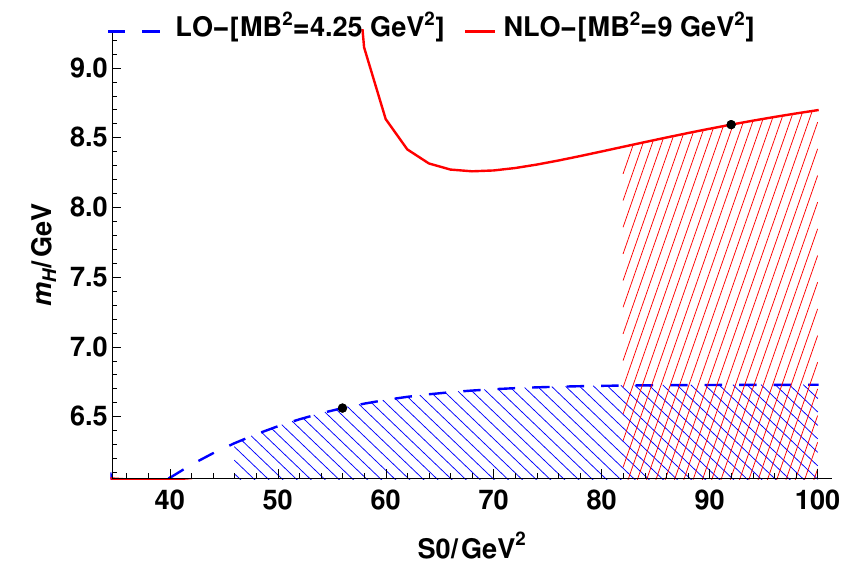}
	}\\
	\subfigure[OS]{
		\includegraphics[scale=0.47]{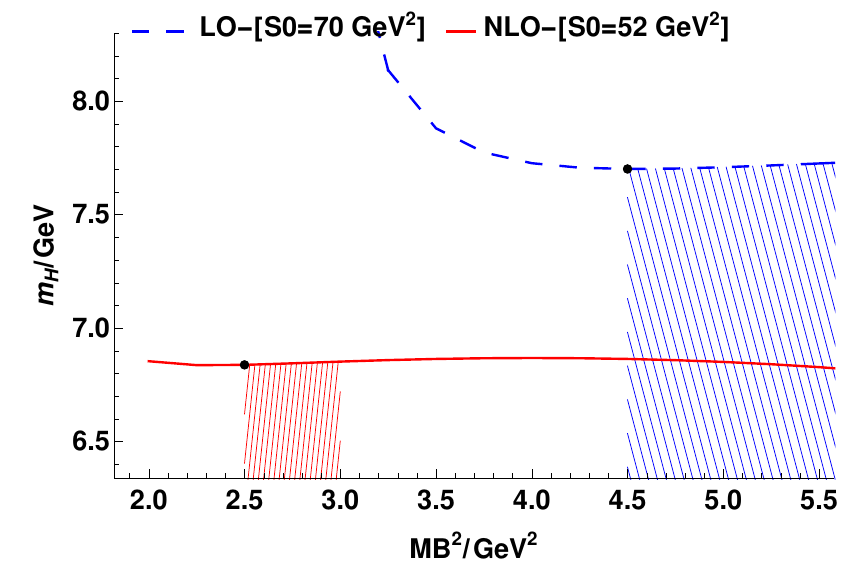}
		\includegraphics[scale=0.47]{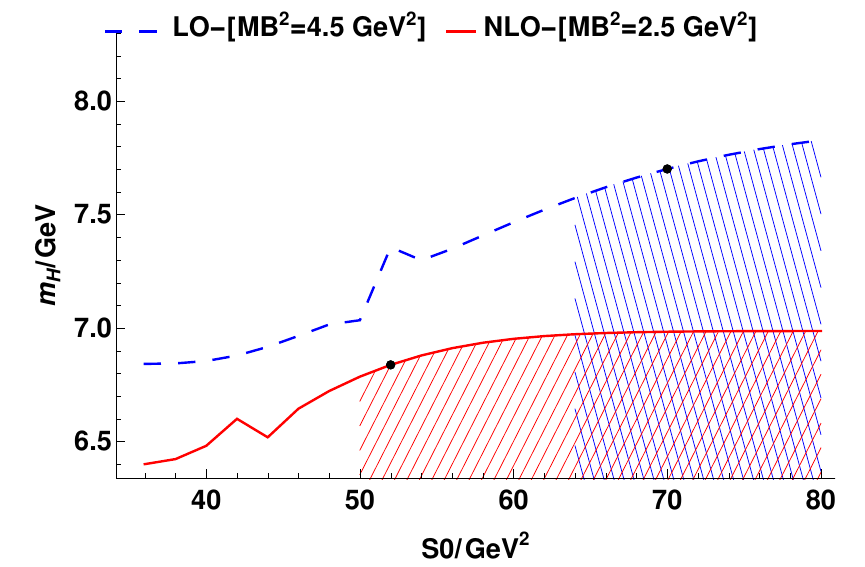}
	}
	\caption{\label{fig:0--Mixed3-NLO-MSbar-OS}
		The Borel platform curves for $J_{P,3}^{\text{Dia}}$ with $J^{PC}=0^{-+}$ in the $\overline{\text{MS}}$ and On-Shell schemes}
	\vspace{-0.8cm}
\end{figure}

\subsection{Numerical Results for $J^P=1^+$ states}

\begin{table}[H]
	\vspace{-0.4cm}
	\renewcommand\arraystretch{1.2}
	\setlength{\tabcolsep}{3 mm}
	\begin{center}
		\caption{The LO and NLO Results for $J^P=1^+$ with $\bar{c}c\bar{c}c$ system in the $\overline{\text{MS}}$ scheme}
		\begin{tabular}{cccc|@{*}|ccc}
			\hline\hline
			\multirow{2}{*}{Current} &
			\multicolumn{3}{c|@{*}|}{LO}& \multicolumn{3}{c}{NLO($\overline{\text{MS}}$)} \\ \cline{2-7}
			& \makecell{$M_H$ \\ (GeV)} & \makecell{$s_0$ \\ ($\text{GeV}^2$)} & \makecell{$M_B^2$ \\ ($\text{GeV}^2$)} &  \makecell{$M_H$ \\ (GeV)} & \makecell{$s_0$ \\ ($\text{GeV}^2$)} & \makecell{$M_B^2$ \\ ($\text{GeV}^2$)} \\ \hline
			$J_{A,1}^{\text{M-M}}$ &$7.07^{+0.14}_{-0.16}$ &$64.(\pm 10\%)$ &$5.50(\pm 10\%)$    &$8.32^{+0.18}_{-0.20}$ &$86.(\pm 10\%)$ &$9.00(\pm 10\%)$\\
			$J_{A,2}^{\text{M-M}}$ &$6.93^{+0.12}_{-0.15}$ &$62.(\pm 10\%)$ &$4.75(\pm 10\%)$    &$7.59^{+0.10}_{-0.06}$ &$62.(\pm 10\%)$ &$5.50(\pm 10\%)$\\
			$J_{A,3}^{\text{M-M}}$ &$6.04^{+0.06}_{-0.08}$ &$48.(\pm 10\%)$ &$3.25(\pm 10\%)$    &$6.65^{+0.09}_{-0.10}$ &$58.(\pm 10\%)$ &$4.25(\pm 10\%)$\\
			$J_{A,4}^{\text{M-M}}$ &$6.38^{+0.08}_{-0.13}$ &$54.(\pm 10\%)$ &$3.75(\pm 10\%)$    &$7.73^{+0.10}_{-0.12}$ &$76.(\pm 10\%)$ &$6.00(\pm 10\%)$\\ \hline
			$J_{A,1}^{\text{Di-Di}}$ &$7.00^{+0.12}_{-0.14}$ &$64.(\pm 10\%)$ &$5.00(\pm 10\%)$    &$8.84^{+0.09}_{-0.19}$ &$96.(\pm 10\%)$ &$10.00(\pm 10\%)$\\
			$J_{A,2}^{\text{Di-Di}}$ &$7.04^{+0.13}_{-0.15}$ &$64.(\pm 10\%)$ &$5.25(\pm 10\%)$    &$7.41^{+0.23}_{-0.30}$ &$70.(\pm 10\%)$ &$7.75(\pm 10\%)$\\
			$J_{A,3}^{\text{Di-Di}}$ &$6.95^{+0.13}_{-0.16}$ &$62.(\pm 10\%)$ &$5.00(\pm 10\%)$    &$8.81^{+0.08}_{-0.19}$ &$96.(\pm 10\%)$ &$10.00(\pm 10\%)$\\
			$J_{A,4}^{\text{Di-Di}}$ &$6.08^{+0.04}_{-0.10}$ &$50.(\pm 10\%)$ &$3.25(\pm 10\%)$    &$6.65^{+0.10}_{-0.13}$ &$56.(\pm 10\%)$ &$4.50(\pm 10\%)$\\ \hline
			$J_{A,1}^{\text{Dia}}$ &$6.92^{+0.12}_{-0.15}$ &$62.(\pm 10\%)$ &$4.75(\pm 10\%)$    &$7.49^{+0.23}_{-0.31}$ &$70.(\pm 10\%)$ &$7.50(\pm 10\%)$\\
			$J_{A,2}^{\text{Dia}}$ &$7.08^{+0.13}_{-0.16}$ &$64.(\pm 10\%)$ &$5.50(\pm 10\%)$    &$8.22^{+0.17}_{-0.19}$ &$84.(\pm 10\%)$ &$8.50(\pm 10\%)$\\
			$J_{A,3}^{\text{Dia}}$ &$6.21^{+0.07}_{-0.11}$ &$52.(\pm 10\%)$ &$3.50(\pm 10\%)$    &$7.07^{+0.09}_{-0.10}$ &$64.(\pm 10\%)$ &$5.00(\pm 10\%)$\\
			$J_{A,4}^{\text{Dia}}$ &$6.04^{+0.06}_{-0.08}$ &$48.(\pm 10\%)$ &$3.25(\pm 10\%)$    &$6.65^{+0.09}_{-0.10}$ &$58.(\pm 10\%)$ &$4.25(\pm 10\%)$\\ \hline\hline
		\end{tabular}
		
		\label{tab:A-NLOresult-MSbar}
	\end{center}
	\vspace{-0.4cm}
\end{table}

\begin{table}[H]
	\renewcommand\arraystretch{1.2}
	\setlength{\tabcolsep}{3 mm}
	\begin{center}
		\caption{The LO and NLO Results for $J^P=1^+$ with $\bar{c}c\bar{c}c$ system in the On-Shell scheme}
		\begin{tabular}{cccc|@{*}|ccc}
			\hline\hline
			\multirow{2}{*}{Current} &
			\multicolumn{3}{c|@{*}|}{LO}& \multicolumn{3}{c}{NLO(OS)} \\ \cline{2-7}
			& \makecell{$M_H$ \\ (GeV)} & \makecell{$s_0$ \\ ($\text{GeV}^2$)} & \makecell{$M_B^2$ \\ ($\text{GeV}^2$)} &  \makecell{$M_H$ \\ (GeV)} & \makecell{$s_0$ \\ ($\text{GeV}^2$)} & \makecell{$M_B^2$ \\ ($\text{GeV}^2$)} \\ \hline
			$J_{A,1}^{\text{M-M}}$ &$8.38^{+0.08}_{-0.18}$ &$85.(\pm 10\%)$ &$5.75(\pm 10\%)$    &$7.35^{+0.14}_{-0.21}$ &$61.(\pm 10\%)$ &$3.50(\pm 10\%)$\\
			$J_{A,2}^{\text{M-M}}$ &$8.10^{+0.17}_{-0.23}$ &$77.(\pm 10\%)$ &$5.25(\pm 10\%)$    &$6.74^{+0.21}_{-0.56}$ &$51.(\pm 10\%)$ &$2.75(\pm 10\%)$\\
			$J_{A,3}^{\text{M-M}}$ &$7.22^{+0.04}_{-0.05}$ &$67.(\pm 10\%)$ &$3.25(\pm 10\%)$    &$6.53^{+0.08}_{-0.11}$ &$47.(\pm 10\%)$ &$2.00(\pm 10\%)$\\
			$J_{A,4}^{\text{M-M}}$ &$7.46^{+0.11}_{-0.13}$ &$67.(\pm 10\%)$ &$4.00(\pm 10\%)$    &$6.58^{+0.11}_{-0.17}$ &$47.(\pm 10\%)$ &$2.25(\pm 10\%)$\\ \hline
			$J_{A,1}^{\text{Di-Di}}$ &$8.18^{+0.17}_{-0.22}$ &$79.(\pm 10\%)$ &$5.50(\pm 10\%)$    &$7.23^{+0.14}_{-0.23}$ &$59.(\pm 10\%)$ &$3.25(\pm 10\%)$\\
			$J_{A,2}^{\text{Di-Di}}$ &$8.27^{+0.16}_{-0.21}$ &$81.(\pm 10\%)$ &$5.75(\pm 10\%)$    &$6.63^{+0.16}_{-0.28}$ &$49.(\pm 10\%)$ &$2.50(\pm 10\%)$\\
			$J_{A,3}^{\text{Di-Di}}$ &$8.21^{+0.14}_{-0.18}$ &$81.(\pm 10\%)$ &$5.25(\pm 10\%)$    &$7.23^{+0.15}_{-0.23}$ &$59.(\pm 10\%)$ &$3.25(\pm 10\%)$\\
			$J_{A,4}^{\text{Di-Di}}$ &$7.22^{+0.07}_{-0.10}$ &$63.(\pm 10\%)$ &$3.50(\pm 10\%)$    &$6.54^{+0.08}_{-0.11}$ &$47.(\pm 10\%)$ &$2.00(\pm 10\%)$\\ \hline
			$J_{A,1}^{\text{Dia}}$ &$8.09^{+0.17}_{-0.23}$ &$77.(\pm 10\%)$ &$5.25(\pm 10\%)$    &$6.62^{+0.16}_{-0.24}$ &$49.(\pm 10\%)$ &$2.50(\pm 10\%)$\\
			$J_{A,2}^{\text{Dia}}$ &$8.39^{+0.08}_{-0.18}$ &$85.(\pm 10\%)$ &$5.75(\pm 10\%)$    &$7.25^{+0.16}_{-0.27}$ &$59.(\pm 10\%)$ &$3.50(\pm 10\%)$\\
			$J_{A,3}^{\text{Dia}}$ &$7.33^{+0.08}_{-0.11}$ &$65.(\pm 10\%)$ &$3.75(\pm 10\%)$    &$6.56^{+0.08}_{-0.12}$ &$47.(\pm 10\%)$ &$2.00(\pm 10\%)$\\
			$J_{A,4}^{\text{Dia}}$ &$7.23^{+0.04}_{-0.05}$ &$67.(\pm 10\%)$ &$3.25(\pm 10\%)$    &$6.53^{+0.08}_{-0.11}$ &$47.(\pm 10\%)$ &$2.00(\pm 10\%)$\\ \hline\hline
		\end{tabular}
		\label{tab:A-NLOresult-OS}
	\end{center}
	\vspace{-0.8cm}
\end{table}

\begin{figure}[H]
	\vspace{-1 cm}
	\centering
	\subfigure[$\overline{\text{MS}}$]{
		\includegraphics[scale=0.47]{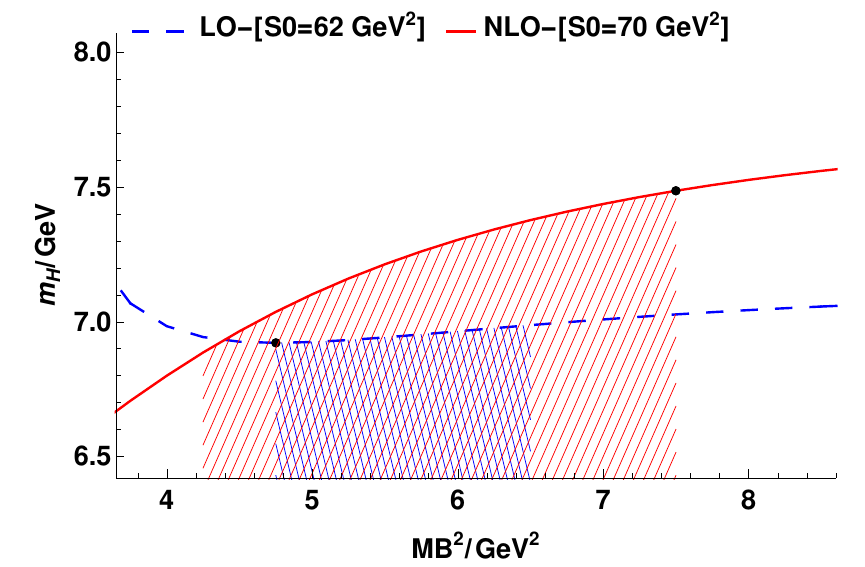}
		\includegraphics[scale=0.47]{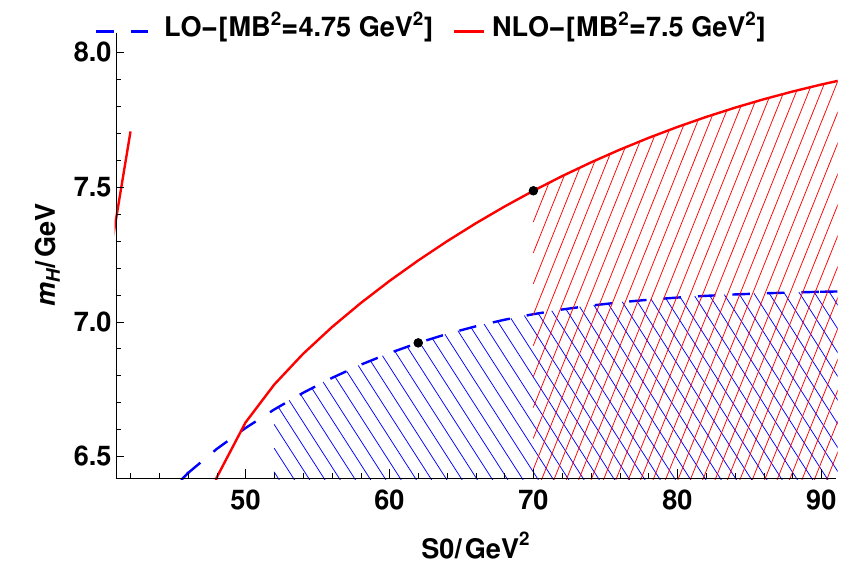}
	}\\
	\subfigure[OS]{
		\includegraphics[scale=0.47]{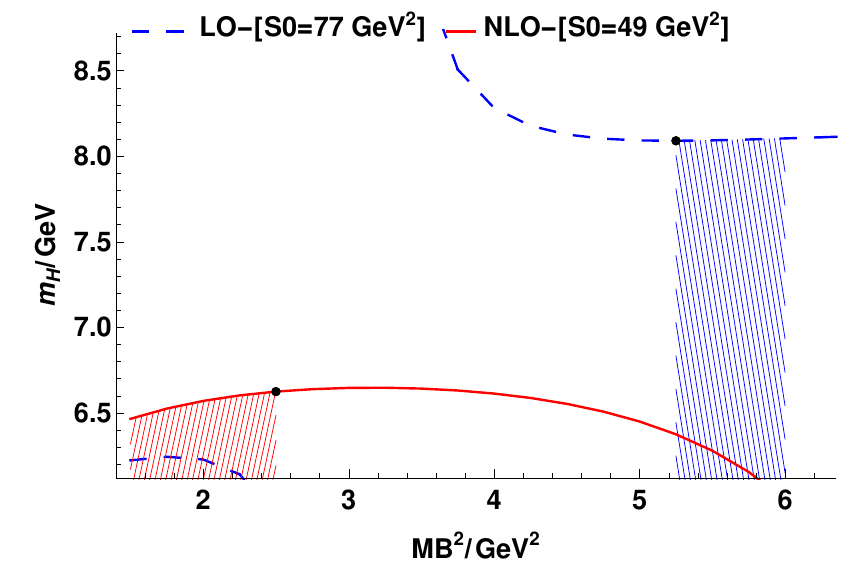}
		\includegraphics[scale=0.47]{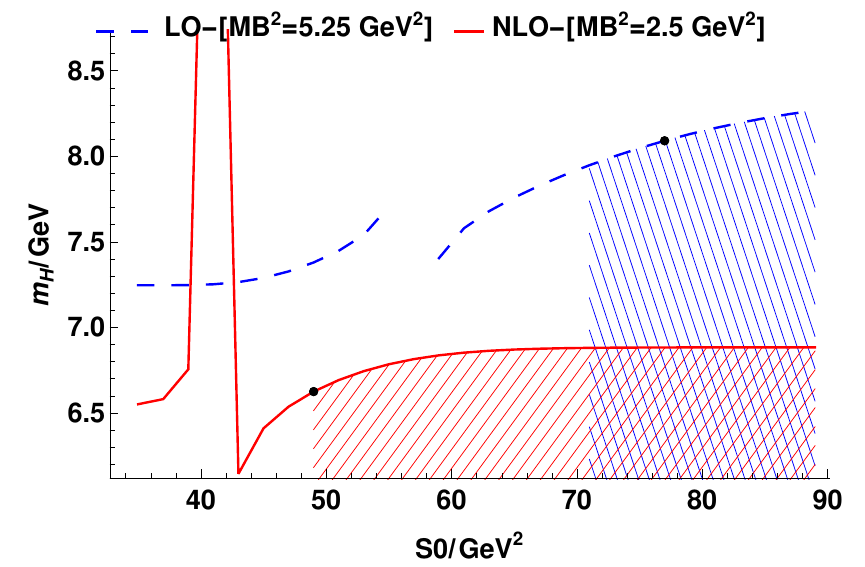}
	}
	\caption{\label{fig:1+-Mixed1-NLO-MSbar-OS}
		The Borel platform curves for $J_{A,1}^{\text{Dia}}$ with $J^{PC}=1^{++}$ in the $\overline{\text{MS}}$ and On-Shell schemes}
		\vspace{-0.3cm}
\end{figure}
\begin{figure}[H]
	\centering
	\subfigure[$\overline{\text{MS}}$]{
		\includegraphics[scale=0.47]{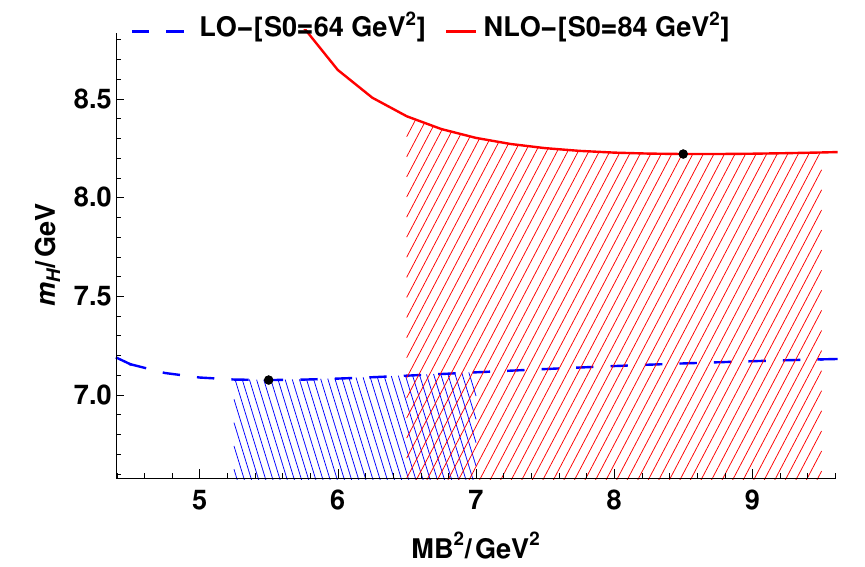}
		\includegraphics[scale=0.47]{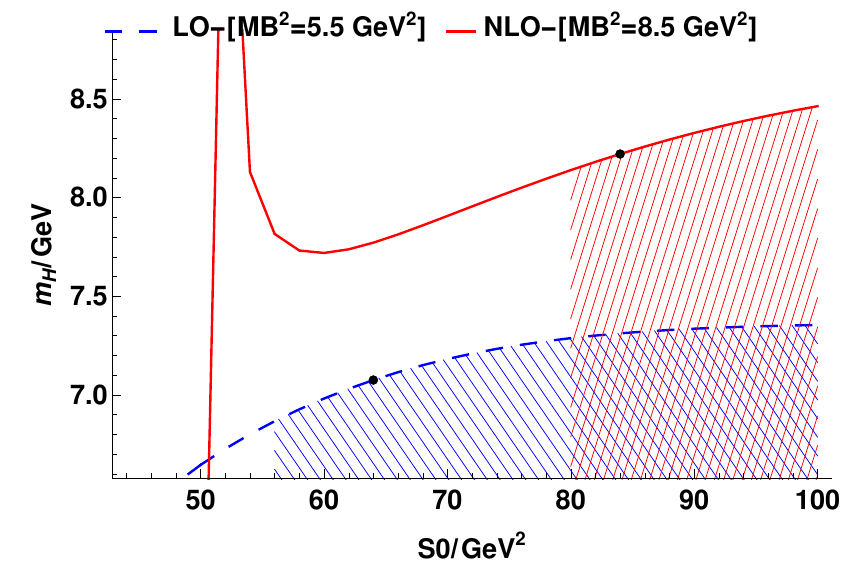}
	}\\
	\subfigure[OS]{
		\includegraphics[scale=0.47]{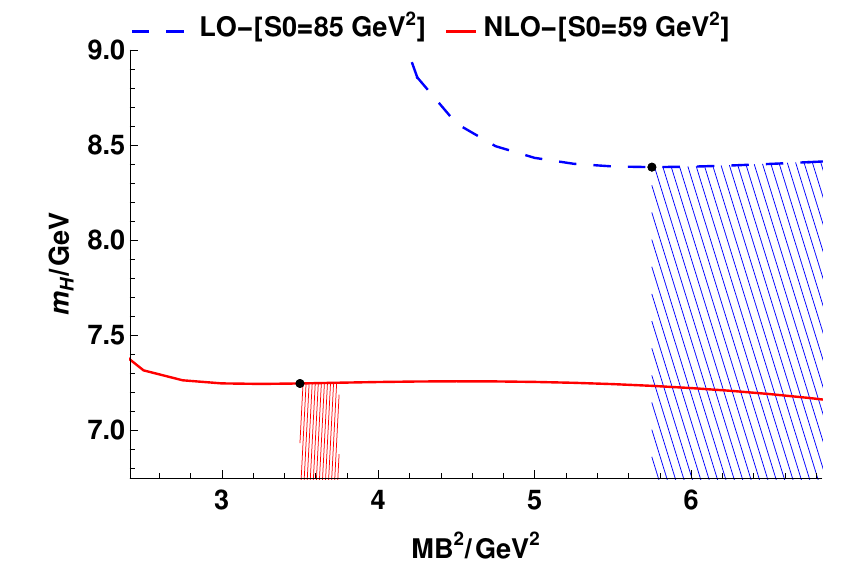}
		\includegraphics[scale=0.47]{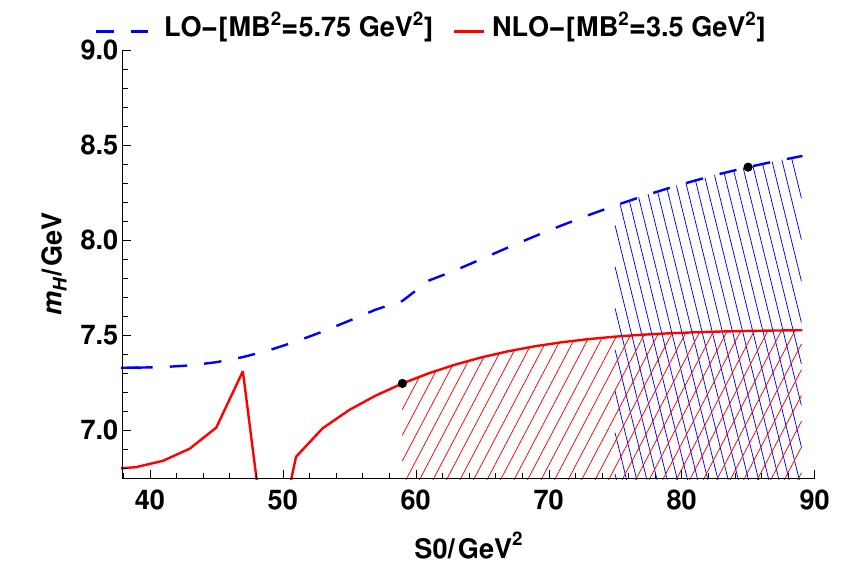}
	}
	\caption{\label{fig:1+-Mixed2-NLO-MSbar-OS}
		The Borel platform curves for $J_{A,2}^{\text{Dia}}$ with $J^{PC}=1^{++}$ in the $\overline{\text{MS}}$ and On-Shell schemes}
	\vspace{-0.8cm}
\end{figure}
\begin{figure}[H]
	\vspace{-0.8cm}
	\centering
	\subfigure[$\overline{\text{MS}}$]{
		\includegraphics[scale=0.47]{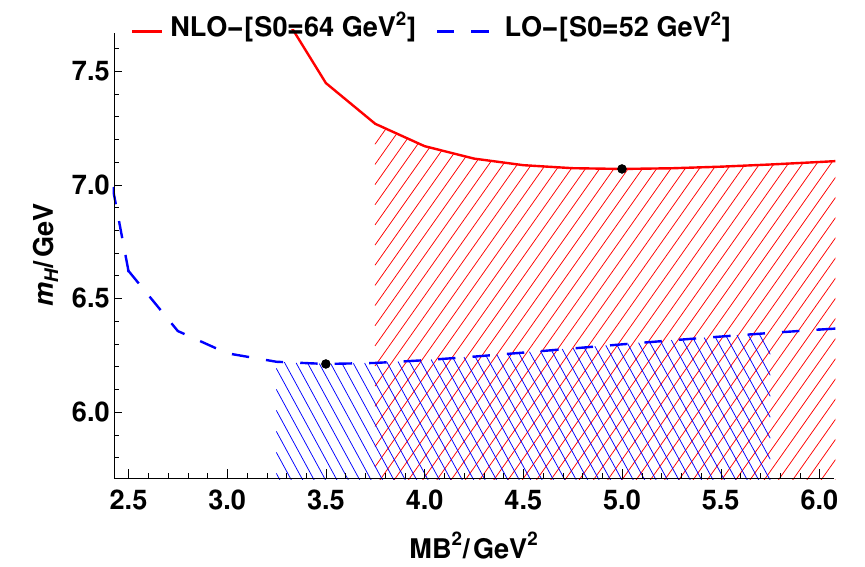}
		\includegraphics[scale=0.47]{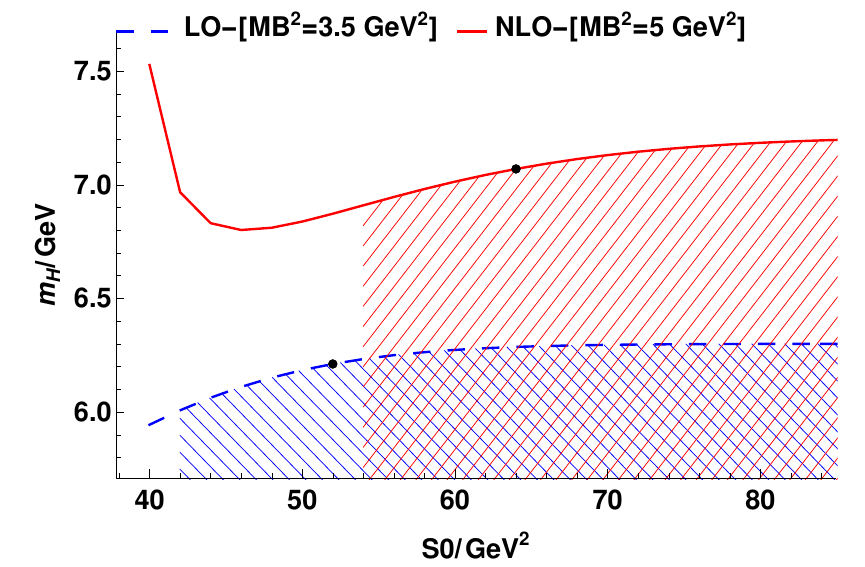}
	}\\
	\subfigure[OS]{
		\includegraphics[scale=0.47]{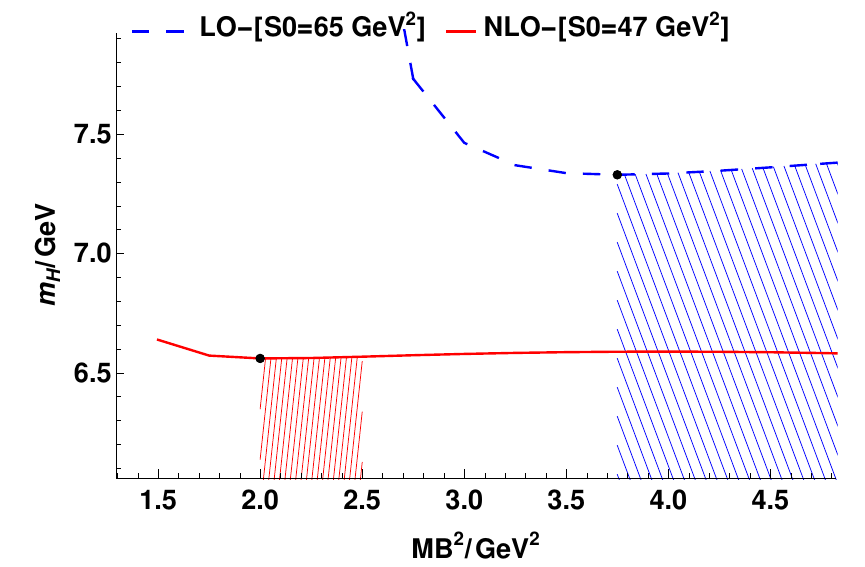}
		\includegraphics[scale=0.47]{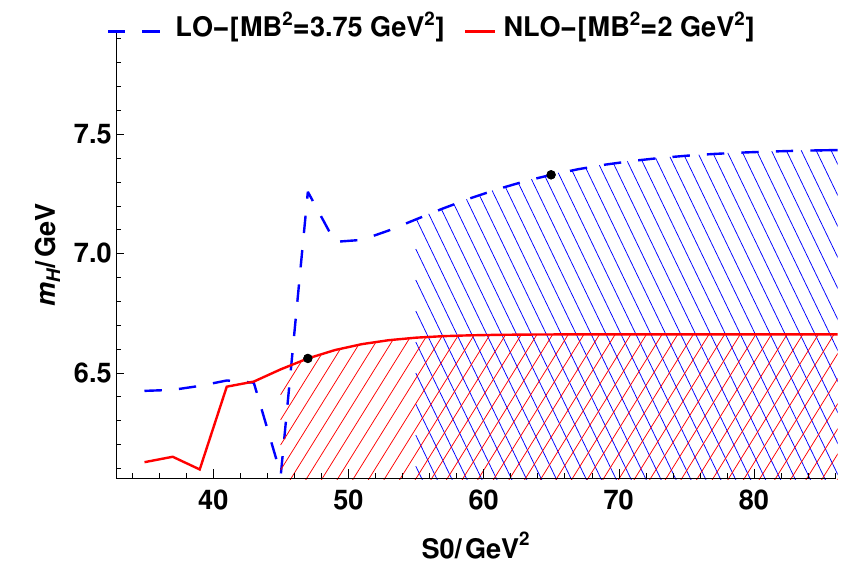}
	}
	\caption{\label{fig:1+-Mixed3-NLO-MSbar-OS}
		The Borel platform curves for $J_{A,3}^{\text{Dia}}$ with $J^{PC}=1^{+-}$ in the $\overline{\text{MS}}$ and On-Shell schemes}
		\vspace{-0.5cm}
\end{figure}
\begin{figure}[H]
	\centering
	\subfigure[$\overline{\text{MS}}$]{
		\includegraphics[scale=0.47]{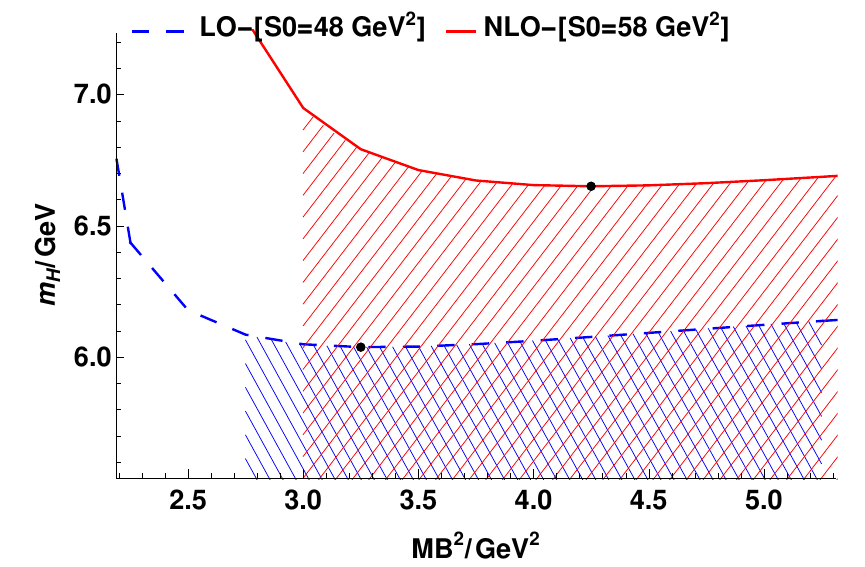}
		\includegraphics[scale=0.47]{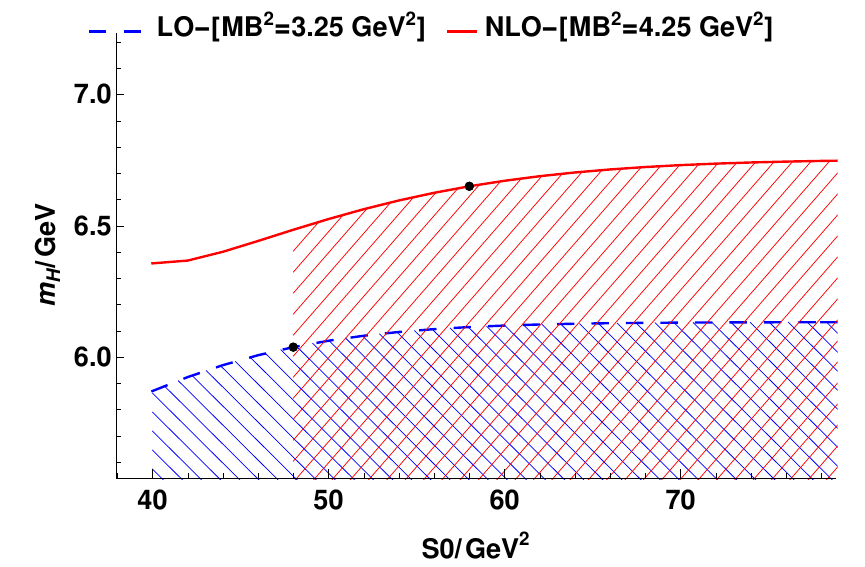}
	}\\
	\subfigure[OS]{
		\includegraphics[scale=0.47]{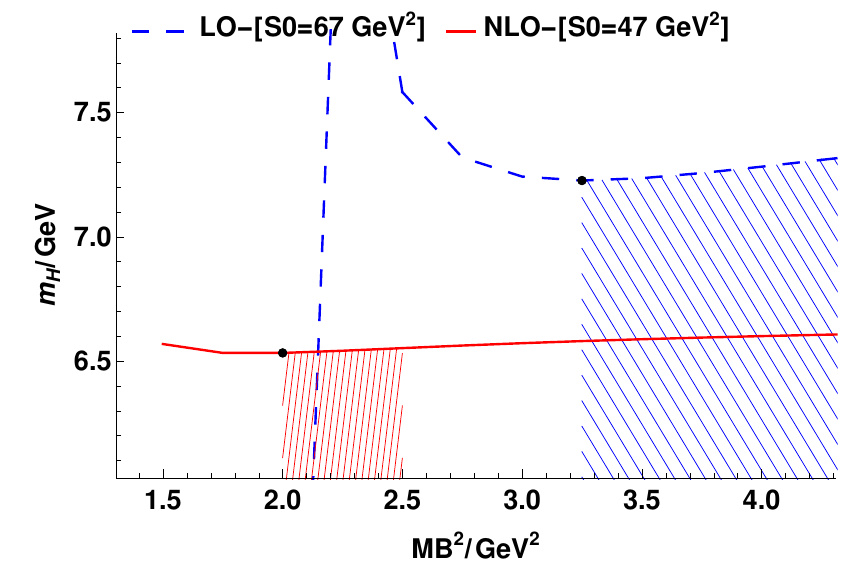}
		\includegraphics[scale=0.47]{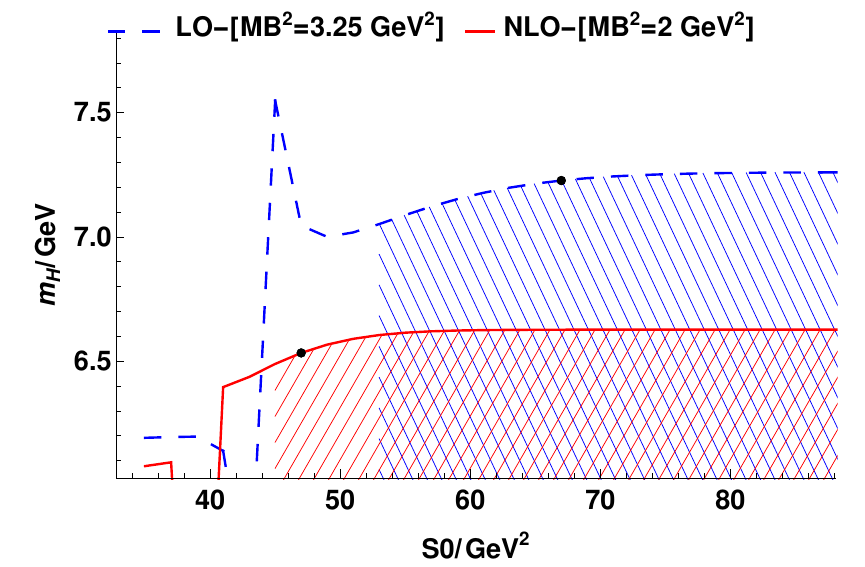}
	}
	\caption{\label{fig:1+-Mixed4-NLO-MSbar-OS}
		The Borel platform curves for $J_{A,4}^{\text{Dia}}$ with $J^{PC}=1^{+-}$ in the $\overline{\text{MS}}$ and On-Shell schemes}
	\vspace{-0.8cm}
\end{figure}

\subsection{Numerical Results for $J^P=1^-$ states}
\begin{table}[H]
	\vspace{-0.4cm}
	\renewcommand\arraystretch{1.2}
	\setlength{\tabcolsep}{3 mm}
	\begin{center}
		\caption{The LO and NLO Results for $J^P=1^-$ with $\bar{c}c\bar{c}c$ system in the $\overline{\text{MS}}$ scheme}
		\begin{tabular}{cccc|@{*}|ccc}
			\hline\hline
			\multirow{2}{*}{Current} &
			\multicolumn{3}{c|@{*}|}{LO}& \multicolumn{3}{c}{NLO($\overline{\text{MS}}$)} \\ \cline{2-7}
			& \makecell{$M_H$ \\ (GeV)} & \makecell{$s_0$ \\ ($\text{GeV}^2$)} & \makecell{$M_B^2$ \\ ($\text{GeV}^2$)} &  \makecell{$M_H$ \\ (GeV)} & \makecell{$s_0$ \\ ($\text{GeV}^2$)} & \makecell{$M_B^2$ \\ ($\text{GeV}^2$)} \\ \hline
			$J_{V,1}^{\text{M-M}}$ &$6.61^{+0.12}_{-0.14}$ &$57.(\pm 10\%)$ &$4.50(\pm 10\%)$    &$7.55^{+0.14}_{-0.15}$ &$73.(\pm 10\%)$ &$6.75(\pm 10\%)$\\
			$J_{V,2}^{\text{M-M}}$ &$6.55^{+0.11}_{-0.17}$ &$55.(\pm 10\%)$ &$4.50(\pm 10\%)$    &$7.98^{+0.10}_{-0.16}$ &$80.(\pm 10\%)$ &$7.75(\pm 10\%)$\\
			$J_{V,3}^{\text{M-M}}$ &$6.59^{+0.11}_{-0.14}$ &$57.(\pm 10\%)$ &$4.25(\pm 10\%)$    &$7.57^{+0.13}_{-0.15}$ &$73.(\pm 10\%)$ &$6.50(\pm 10\%)$\\
			$J_{V,4}^{\text{M-M}}$ &$6.53^{+0.11}_{-0.14}$ &$56.(\pm 10\%)$ &$4.00(\pm 10\%)$    &$8.09^{+0.06}_{-0.16}$ &$82.(\pm 10\%)$ &$7.75(\pm 10\%)$\\ \hline
			$J_{V,1}^{\text{Di-Di}}$ &$6.56^{+0.11}_{-0.13}$ &$57.(\pm 10\%)$ &$4.25(\pm 10\%)$    &$7.45^{+0.12}_{-0.14}$ &$71.(\pm 10\%)$ &$6.25(\pm 10\%)$\\
			$J_{V,2}^{\text{Di-Di}}$ &$6.61^{+0.12}_{-0.15}$ &$57.(\pm 10\%)$ &$4.50(\pm 10\%)$    &$7.97^{+0.10}_{-0.17}$ &$80.(\pm 10\%)$ &$8.00(\pm 10\%)$\\
			$J_{V,3}^{\text{Di-Di}}$ &$6.56^{+0.12}_{-0.15}$ &$56.(\pm 10\%)$ &$4.25(\pm 10\%)$    &$7.52^{+0.12}_{-0.14}$ &$72.(\pm 10\%)$ &$6.25(\pm 10\%)$\\
			$J_{V,4}^{\text{Di-Di}}$ &$6.53^{+0.11}_{-0.16}$ &$55.(\pm 10\%)$ &$4.25(\pm 10\%)$    &$8.02^{+0.08}_{-0.17}$ &$81.(\pm 10\%)$ &$7.75(\pm 10\%)$\\ \hline
			$J_{V,1}^{\text{Dia}}$ &$6.57^{+0.11}_{-0.13}$ &$57.(\pm 10\%)$ &$4.25(\pm 10\%)$    &$7.90^{+0.11}_{-0.15}$ &$79.(\pm 10\%)$ &$7.25(\pm 10\%)$\\
			$J_{V,2}^{\text{Dia}}$ &$6.60^{+0.12}_{-0.14}$ &$57.(\pm 10\%)$ &$4.50(\pm 10\%)$    &$7.52^{+0.14}_{-0.16}$ &$72.(\pm 10\%)$ &$6.75(\pm 10\%)$\\
			$J_{V,3}^{\text{Dia}}$ &$6.53^{+0.11}_{-0.13}$ &$56.(\pm 10\%)$ &$4.00(\pm 10\%)$    &$8.02^{+0.08}_{-0.15}$ &$81.(\pm 10\%)$ &$7.25(\pm 10\%)$\\
			$J_{V,4}^{\text{Dia}}$ &$6.59^{+0.11}_{-0.14}$ &$57.(\pm 10\%)$ &$4.25(\pm 10\%)$    &$7.56^{+0.14}_{-0.15}$ &$73.(\pm 10\%)$ &$6.50(\pm 10\%)$\\ \hline\hline
		\end{tabular}
			\vspace{-0.4cm}
		\label{tab:V-NLOresult-MSbar}
	\end{center}
\end{table}

\begin{table}[H]
	\renewcommand\arraystretch{1.2}
	\setlength{\tabcolsep}{3 mm}
	\begin{center}
		\caption{The LO and NLO Results for $J^P=1^-$ with $\bar{c}c\bar{c}c$ system in the On-Shell scheme}
		\begin{tabular}{cccc|@{*}|ccc}
			\hline\hline
			\multirow{2}{*}{Current} &
			\multicolumn{3}{c|@{*}|}{LO}& \multicolumn{3}{c}{NLO(OS)} \\ \cline{2-7}
			& \makecell{$M_H$ \\ (GeV)} & \makecell{$s_0$ \\ ($\text{GeV}^2$)} & \makecell{$M_B^2$ \\ ($\text{GeV}^2$)} &  \makecell{$M_H$ \\ (GeV)} & \makecell{$s_0$ \\ ($\text{GeV}^2$)} & \makecell{$M_B^2$ \\ ($\text{GeV}^2$)} \\ \hline
			$J_{V,1}^{\text{M-M}}$ &$7.87^{+0.12}_{-0.15}$ &$76.(\pm 10\%)$ &$4.75(\pm 10\%)$    &$6.96^{+0.12}_{-0.17}$ &$54.(\pm 10\%)$ &$3.00(\pm 10\%)$\\
			$J_{V,2}^{\text{M-M}}$ &$7.78^{+0.13}_{-0.17}$ &$72.(\pm 10\%)$ &$4.75(\pm 10\%)$    &$6.88^{+0.12}_{-0.20}$ &$52.(\pm 10\%)$ &$2.75(\pm 10\%)$\\
			$J_{V,3}^{\text{M-M}}$ &$7.86^{+0.11}_{-0.13}$ &$77.(\pm 10\%)$ &$4.50(\pm 10\%)$    &$6.95^{+0.11}_{-0.15}$ &$54.(\pm 10\%)$ &$2.75(\pm 10\%)$\\
			$J_{V,4}^{\text{M-M}}$ &$7.69^{+0.12}_{-0.15}$ &$71.(\pm 10\%)$ &$4.25(\pm 10\%)$    &$6.80^{+0.12}_{-0.20}$ &$51.(\pm 10\%)$ &$2.50(\pm 10\%)$\\ \hline
			$J_{V,1}^{\text{Di-Di}}$ &$7.79^{+0.11}_{-0.13}$ &$74.(\pm 10\%)$ &$4.50(\pm 10\%)$    &$6.87^{+0.13}_{-0.19}$ &$52.(\pm 10\%)$ &$2.75(\pm 10\%)$\\
			$J_{V,2}^{\text{Di-Di}}$ &$7.85^{+0.11}_{-0.16}$ &$75.(\pm 10\%)$ &$4.75(\pm 10\%)$    &$6.96^{+0.12}_{-0.18}$ &$54.(\pm 10\%)$ &$3.00(\pm 10\%)$\\
			$J_{V,3}^{\text{Di-Di}}$ &$7.78^{+0.12}_{-0.15}$ &$73.(\pm 10\%)$ &$4.50(\pm 10\%)$    &$6.87^{+0.13}_{-0.20}$ &$52.(\pm 10\%)$ &$2.75(\pm 10\%)$\\
			$J_{V,4}^{\text{Di-Di}}$ &$7.70^{+0.14}_{-0.18}$ &$70.(\pm 10\%)$ &$4.50(\pm 10\%)$    &$6.86^{+0.11}_{-0.17}$ &$52.(\pm 10\%)$ &$2.50(\pm 10\%)$\\ \hline
			$J_{V,1}^{\text{Dia}}$ &$7.78^{+0.11}_{-0.13}$ &$74.(\pm 10\%)$ &$4.50(\pm 10\%)$    &$6.87^{+0.13}_{-0.20}$ &$52.(\pm 10\%)$ &$2.75(\pm 10\%)$\\
			$J_{V,2}^{\text{Dia}}$ &$7.85^{+0.11}_{-0.16}$ &$75.(\pm 10\%)$ &$4.75(\pm 10\%)$    &$6.98^{+0.10}_{-0.17}$ &$55.(\pm 10\%)$ &$2.75(\pm 10\%)$\\
			$J_{V,3}^{\text{Dia}}$ &$7.68^{+0.12}_{-0.15}$ &$71.(\pm 10\%)$ &$4.25(\pm 10\%)$    &$6.80^{+0.13}_{-0.20}$ &$51.(\pm 10\%)$ &$2.50(\pm 10\%)$\\
			$J_{V,4}^{\text{Dia}}$ &$7.86^{+0.11}_{-0.13}$ &$77.(\pm 10\%)$ &$4.50(\pm 10\%)$    &$6.95^{+0.11}_{-0.15}$ &$54.(\pm 10\%)$ &$2.75(\pm 10\%)$\\ \hline\hline
		\end{tabular}
		
		\label{tab:V-NLOresult-OS}
	\end{center}
	\vspace{-0.8cm}
\end{table}

\begin{figure}[H]
	\vspace{-0.8cm}
	\centering
	\subfigure[$\overline{\text{MS}}$]{
		\includegraphics[scale=0.47]{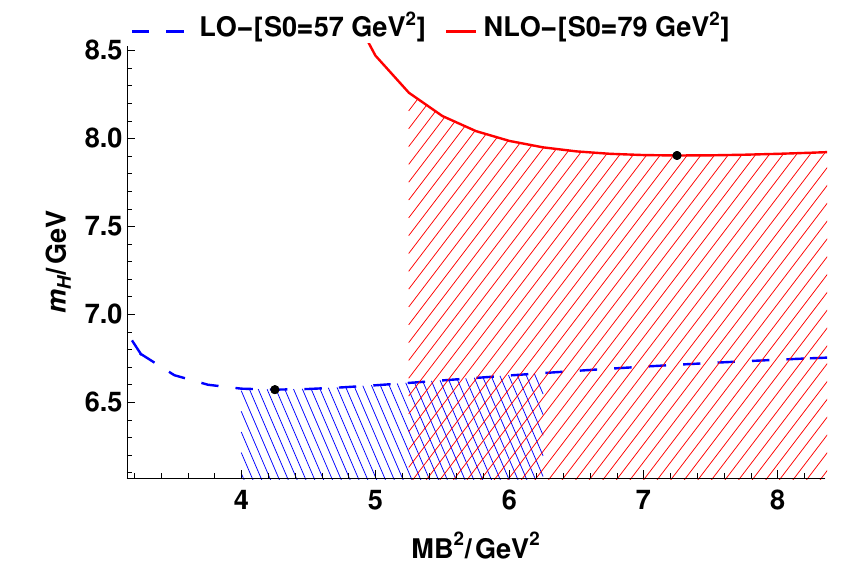}
		\includegraphics[scale=0.47]{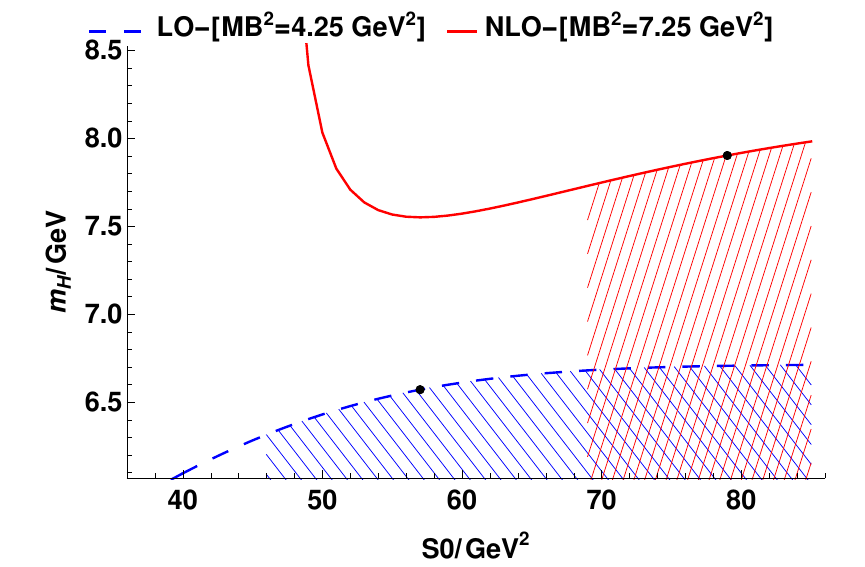}
	}\\
	\subfigure[OS]{
		\includegraphics[scale=0.47]{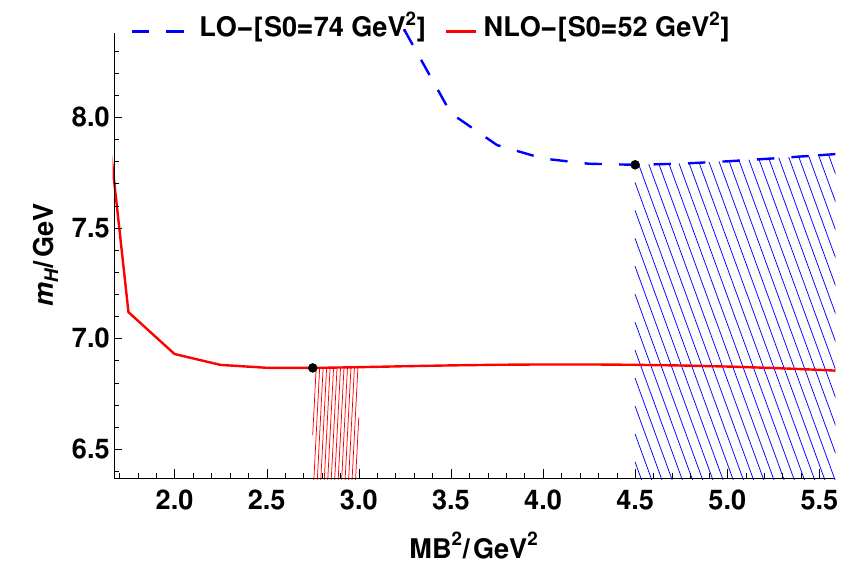}
		\includegraphics[scale=0.47]{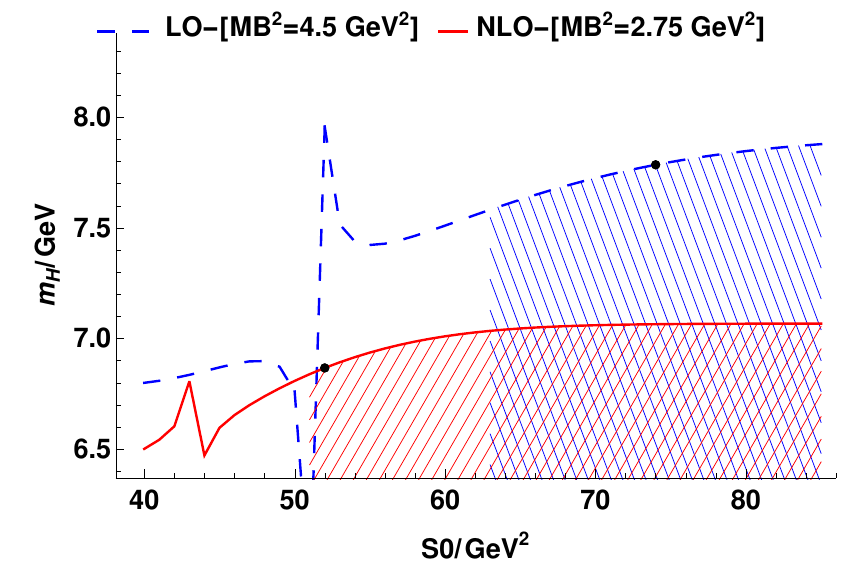}
	}
	\caption{\label{fig:1--Mixed1-NLO-MSbar-OS}
		The Borel platform curves for $J_{V,1}^{\text{Dia}}$ with $J^{PC}=1^{--}$ in the $\overline{\text{MS}}$ and On-Shell schemes}
		\vspace{-0.5cm}
\end{figure}

\begin{figure}[H]
	
	\centering
	\subfigure[$\overline{\text{MS}}$]{
		\includegraphics[scale=0.47]{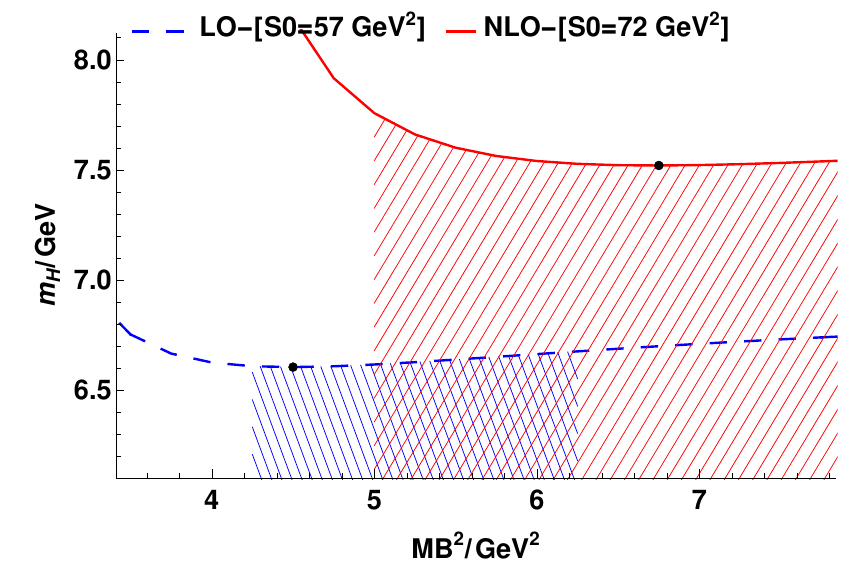}
		\includegraphics[scale=0.47]{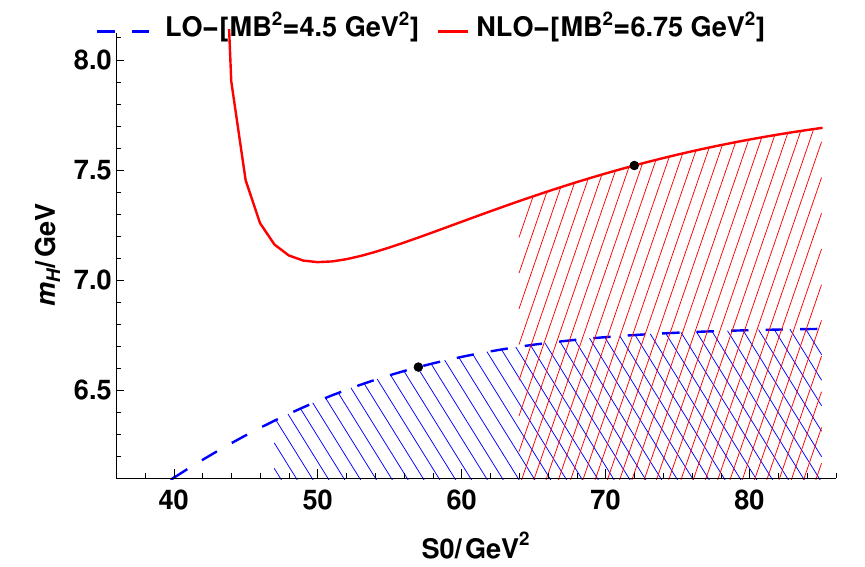}
	}\\
	\subfigure[OS]{
		\includegraphics[scale=0.47]{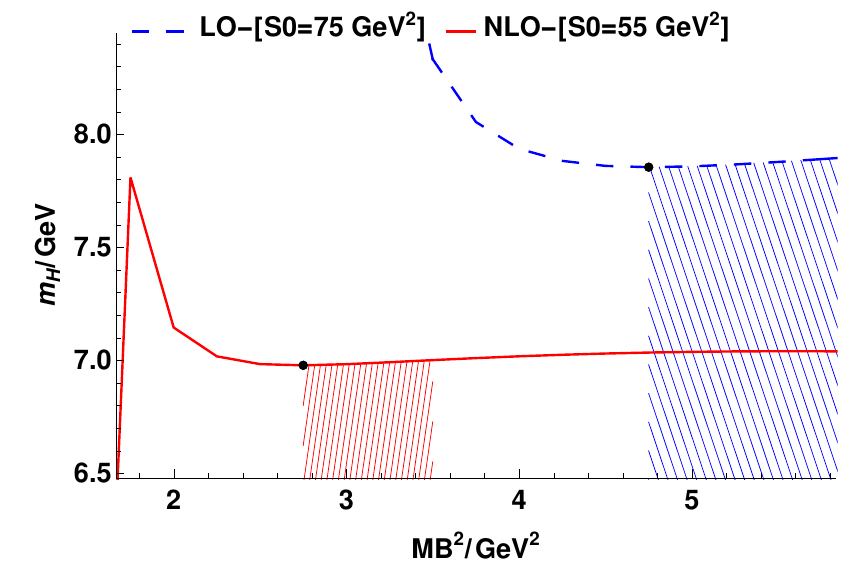}
		\includegraphics[scale=0.47]{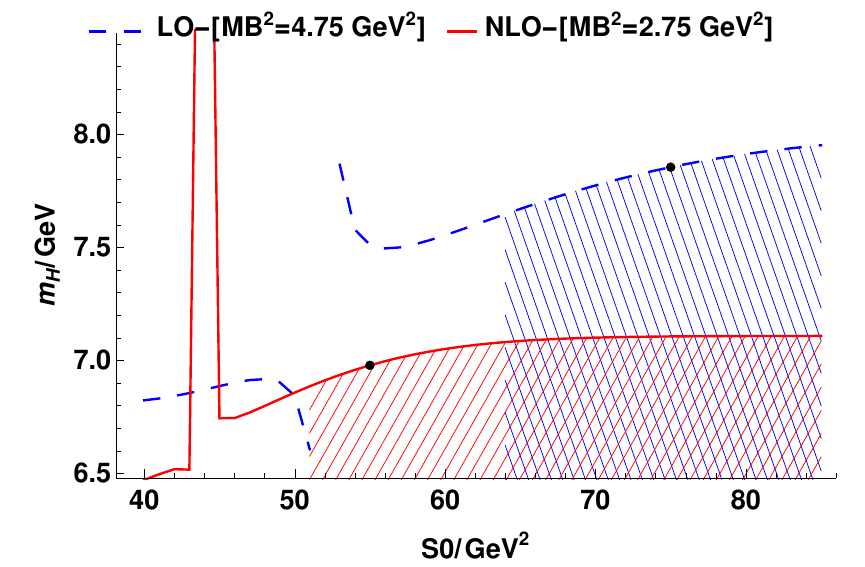}
	}
	\caption{\label{fig:1--Mixed2-NLO-MSbar-OS}
		The Borel platform curves for $J_{V,2}^{\text{Dia}}$ with $J^{PC}=1^{--}$ in the $\overline{\text{MS}}$ and On-Shell schemes}
	\vspace{-0.8cm}
\end{figure}

\begin{figure}[H]
	\vspace{-0.8cm}
	\centering
	\subfigure[$\overline{\text{MS}}$]{
		\includegraphics[scale=0.47]{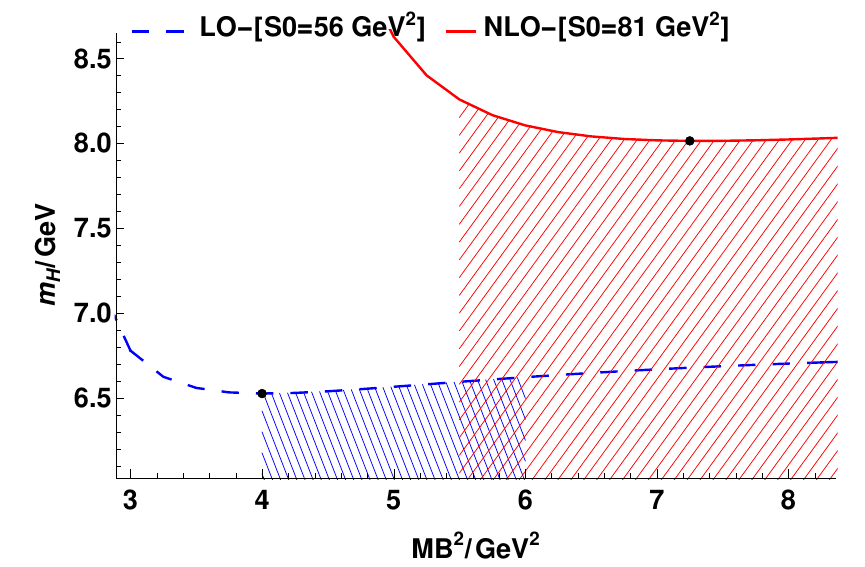}
		\includegraphics[scale=0.47]{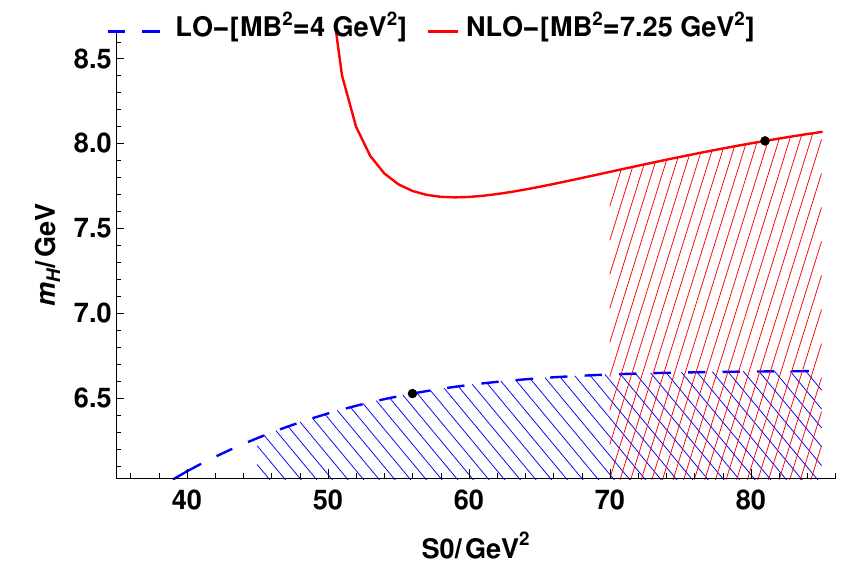}
	}\\
	\subfigure[OS]{
		\includegraphics[scale=0.47]{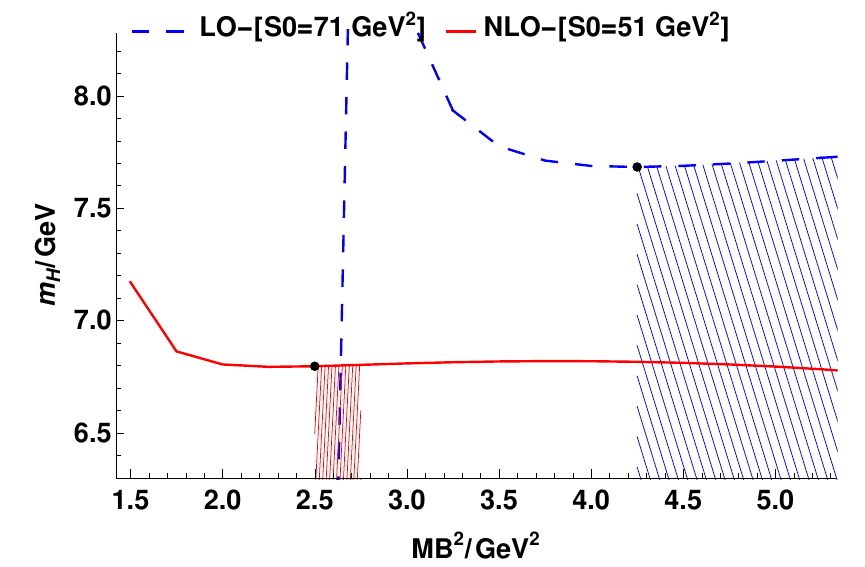}
		\includegraphics[scale=0.47]{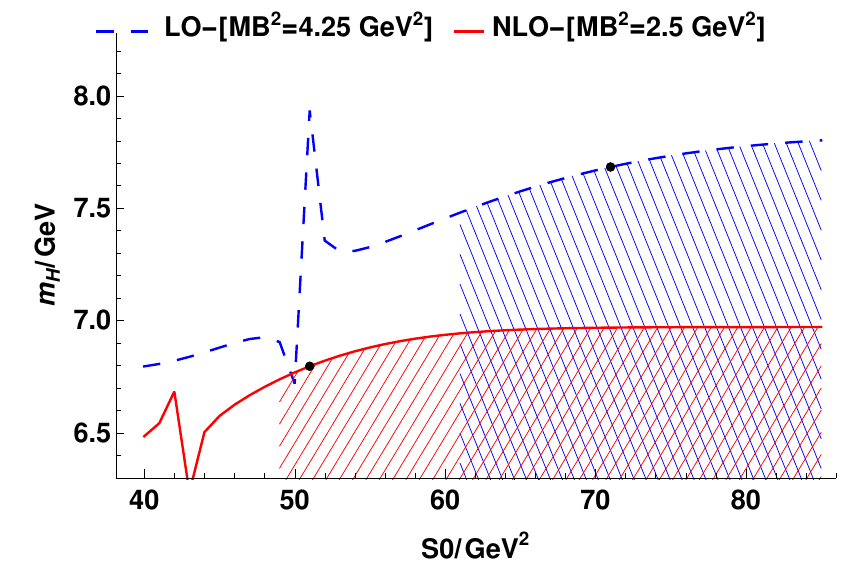}
	}
	\caption{\label{fig:1--Mixed3-NLO-MSbar-OS}
		The Borel platform curves for $J_{V,3}^{\text{Dia}}$ with $J^{PC}=1^{-+}$ in the $\overline{\text{MS}}$ and On-Shell schemes}
		\vspace{-0.5cm}
\end{figure}

\begin{figure}[H]
	
	\centering
	\subfigure[$\overline{\text{MS}}$]{
		\includegraphics[scale=0.47]{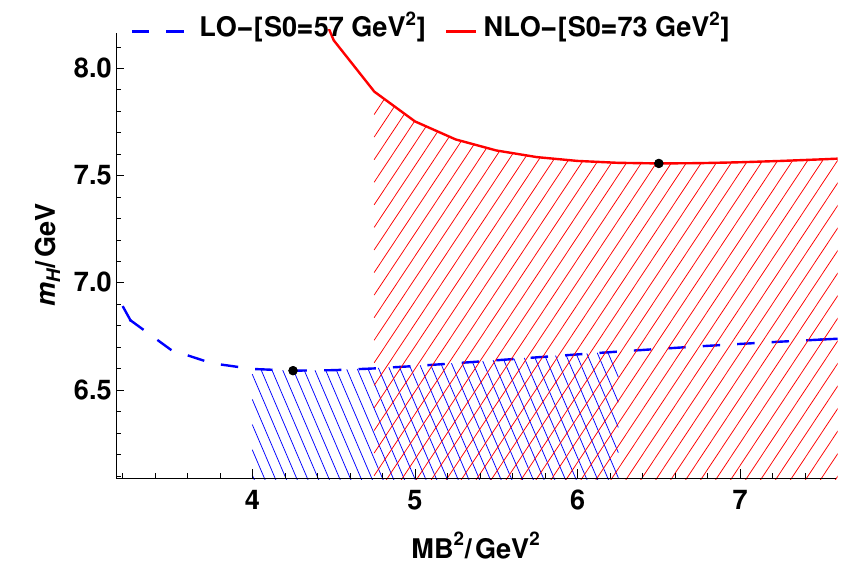}
		\includegraphics[scale=0.47]{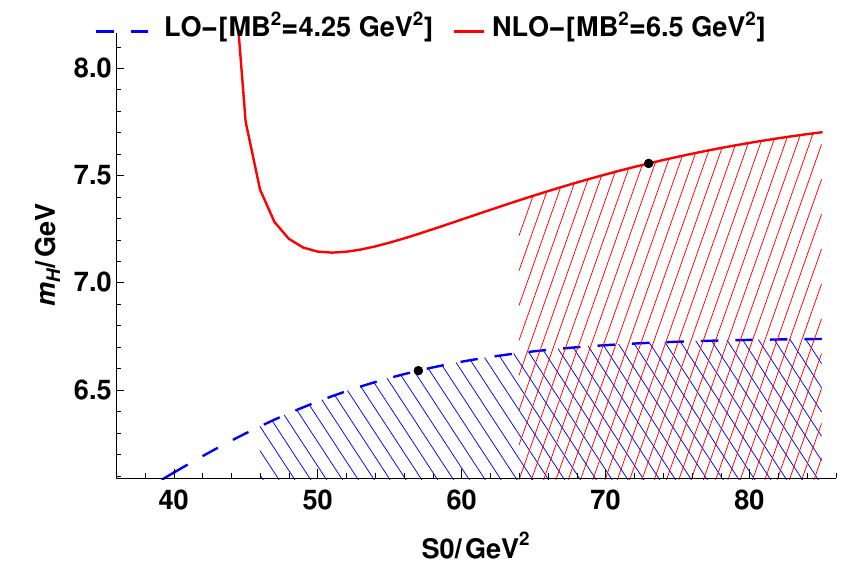}
	}\\
	\subfigure[OS]{
		\includegraphics[scale=0.47]{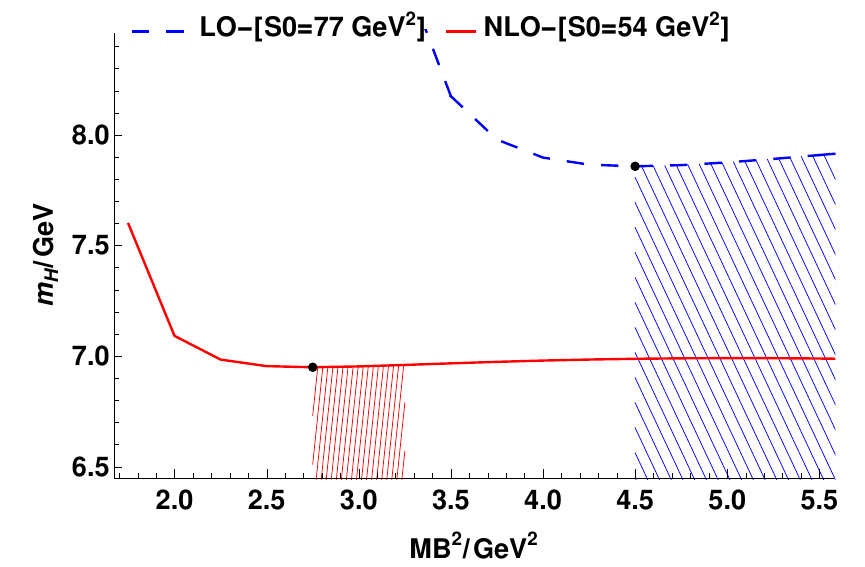}
		\includegraphics[scale=0.47]{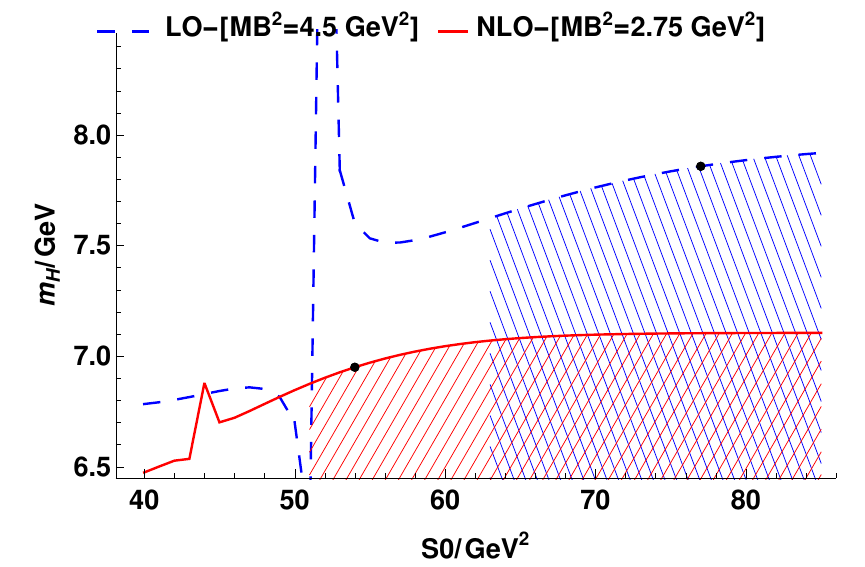}
	}
	\caption{\label{fig:1--Mixed4-NLO-MSbar-OS}
		The Borel platform curves for $J_{V,4}^{\text{Dia}}$ with $J^{PC}=1^{-+}$ in the $\overline{\text{MS}}$ and On-Shell schemes}
	\vspace{-0.8cm}
\end{figure}

\subsection{Numerical Results for $J^P=2^+$ states}
\begin{table}[H]
		\vspace{-0.6cm}
	\renewcommand\arraystretch{1.6}
	\setlength{\tabcolsep}{3 mm}
	\begin{center}
		\caption{The LO and NLO Results for $J^P=2^+$ with $\bar{c}c\bar{c}c$ system in the $\overline{\text{MS}}$ scheme}
		\begin{tabular}{cccc|@{*}|ccc}
			\hline\hline
			\multirow{2}{*}{Current} &
			\multicolumn{3}{c|@{*}|}{LO}& \multicolumn{3}{c}{NLO($\overline{\text{MS}}$)} \\ \cline{2-7}
			& \makecell{$M_H$ \\ (GeV)} & \makecell{$s_0$ \\ ($\text{GeV}^2$)} & \makecell{$M_B^2$ \\ ($\text{GeV}^2$)} &  \makecell{$M_H$ \\ (GeV)} & \makecell{$s_0$ \\ ($\text{GeV}^2$)} & \makecell{$M_B^2$ \\ ($\text{GeV}^2$)} \\ \hline
			$J_{T,1}^{\text{M-M}}$ &$6.11^{+0.06}_{-0.08}$ &$49.(\pm 10\%)$ &$3.50(\pm 10\%)$    &$7.03^{+0.10}_{-0.12}$ &$63.(\pm 10\%)$ &$5.50(\pm 10\%)$\\
			$J_{T,2}^{\text{M-M}}$ &$7.10^{+0.13}_{-0.15}$ &$65.(\pm 10\%)$ &$5.50(\pm 10\%)$    &$8.89^{+0.21}_{-0.24}$ &$97.(\pm 10\%)$ &$11.00(\pm 10\%)$\\
			$J_{T,3}^{\text{M-M}}$ &$6.23^{+0.10}_{-0.14}$ &$51.(\pm 10\%)$ &$3.75(\pm 10\%)$    &$7.35^{+0.10}_{-0.10}$ &$69.(\pm 10\%)$ &$5.75(\pm 10\%)$\\ \hline
			$J_{T,1}^{\text{Di-Di}}$ &$6.07^{+0.08}_{-0.10}$ &$47.(\pm 10\%)$ &$3.75(\pm 10\%)$    &$6.98^{+0.09}_{-0.11}$ &$63.(\pm 10\%)$ &$5.25(\pm 10\%)$\\
			$J_{T,2}^{\text{Di-Di}}$ &$7.02^{+0.13}_{-0.16}$ &$63.(\pm 10\%)$ &$5.25(\pm 10\%)$    &$9.00^{+0.21}_{-0.23}$ &$99.(\pm 10\%)$ &$11.25(\pm 10\%)$\\
			$J_{T,3}^{\text{Di-Di}}$ &$6.15^{+0.08}_{-0.10}$ &$49.(\pm 10\%)$ &$3.75(\pm 10\%)$    &$7.25^{+0.10}_{-0.11}$ &$67.(\pm 10\%)$ &$5.75(\pm 10\%)$\\ \hline
			$J_{T,1}^{\text{Dia}}$ &$6.14^{+0.07}_{-0.11}$ &$51.(\pm 10\%)$ &$3.50(\pm 10\%)$    &$7.03^{+0.11}_{-0.12}$ &$63.(\pm 10\%)$ &$5.50(\pm 10\%)$\\
			$J_{T,2}^{\text{Dia}}$ &$6.15^{+0.08}_{-0.10}$ &$49.(\pm 10\%)$ &$3.75(\pm 10\%)$    &$7.25^{+0.10}_{-0.11}$ &$67.(\pm 10\%)$ &$5.75(\pm 10\%)$\\
			$J_{T,3}^{\text{Dia}}$ &$6.23^{+0.10}_{-0.14}$ &$51.(\pm 10\%)$ &$3.75(\pm 10\%)$    &$7.35^{+0.10}_{-0.10}$ &$69.(\pm 10\%)$ &$5.75(\pm 10\%)$\\ \hline\hline
		\end{tabular}
		\label{tab:T-NLOresult-MSbar}
	\end{center}
	\vspace{-0.4cm}
\end{table}

\begin{table}[H]
	\renewcommand\arraystretch{1.6}
	\setlength{\tabcolsep}{3 mm}
	\begin{center}
		\caption{The LO and NLO Results for $J^P=2^+$ with $\bar{c}c\bar{c}c$ system in the On-Shell scheme}
		
		\begin{tabular}{cccc|@{*}|ccc}
			\hline\hline
			\multirow{2}{*}{Current} &
			\multicolumn{3}{c|@{*}|}{LO}& \multicolumn{3}{c}{NLO(OS)} \\ \cline{2-7}
			& \makecell{$M_H$ \\ (GeV)} & \makecell{$s_0$ \\ ($\text{GeV}^2$)} & \makecell{$M_B^2$ \\ ($\text{GeV}^2$)} &  \makecell{$M_H$ \\ (GeV)} & \makecell{$s_0$ \\ ($\text{GeV}^2$)} & \makecell{$M_B^2$ \\ ($\text{GeV}^2$)} \\ \hline
			$J_{T,1}^{\text{M-M}}$ &$7.31^{+0.05}_{-0.08}$ &$68.(\pm 10\%)$ &$3.50(\pm 10\%)$    &$6.56^{+0.09}_{-0.14}$ &$47.(\pm 10\%)$ &$2.00(\pm 10\%)$\\
			$J_{T,2}^{\text{M-M}}$ &$8.38^{+0.14}_{-0.20}$ &$85.(\pm 10\%)$ &$5.75(\pm 10\%)$    &$7.37^{+0.16}_{-0.25}$ &$61.(\pm 10\%)$ &$3.75(\pm 10\%)$\\
			$J_{T,3}^{\text{M-M}}$ &$7.39^{+0.08}_{-0.11}$ &$68.(\pm 10\%)$ &$3.75(\pm 10\%)$    &$6.57^{+0.12}_{-0.19}$ &$47.(\pm 10\%)$ &$2.25(\pm 10\%)$\\ \hline
			$J_{T,1}^{\text{Di-Di}}$ &$7.32^{+0.04}_{-0.07}$ &$69.(\pm 10\%)$ &$3.50(\pm 10\%)$    &$6.56^{+0.09}_{-0.14}$ &$47.(\pm 10\%)$ &$2.00(\pm 10\%)$\\
			$J_{T,2}^{\text{Di-Di}}$ &$8.27^{+0.16}_{-0.22}$ &$81.(\pm 10\%)$ &$5.75(\pm 10\%)$    &$7.30^{+0.15}_{-0.24}$ &$60.(\pm 10\%)$ &$3.50(\pm 10\%)$\\
			$J_{T,3}^{\text{Di-Di}}$ &$7.32^{+0.07}_{-0.13}$ &$65.(\pm 10\%)$ &$3.75(\pm 10\%)$    &$6.57^{+0.12}_{-0.18}$ &$47.(\pm 10\%)$ &$2.25(\pm 10\%)$\\ \hline
			$J_{T,1}^{\text{Dia}}$ &$7.31^{+0.05}_{-0.08}$ &$68.(\pm 10\%)$ &$3.50(\pm 10\%)$    &$6.56^{+0.09}_{-0.15}$ &$47.(\pm 10\%)$ &$2.00(\pm 10\%)$\\
			$J_{T,2}^{\text{Dia}}$ &$7.32^{+0.07}_{-0.13}$ &$65.(\pm 10\%)$ &$3.75(\pm 10\%)$    &$6.57^{+0.12}_{-0.18}$ &$47.(\pm 10\%)$ &$2.25(\pm 10\%)$\\
			$J_{T,3}^{\text{Dia}}$ &$7.39^{+0.08}_{-0.11}$ &$68.(\pm 10\%)$ &$3.75(\pm 10\%)$    &$6.57^{+0.12}_{-0.19}$ &$47.(\pm 10\%)$ &$2.25(\pm 10\%)$\\ \hline\hline
		\end{tabular}
		\label{tab:T-NLOresult-OS}
	\end{center}
	\vspace{-1.2 cm}
\end{table}

\begin{figure}[H]
	\vspace{-0.8cm}
	\centering
	\subfigure[$\overline{\text{MS}}$]{
		\includegraphics[scale=0.47]{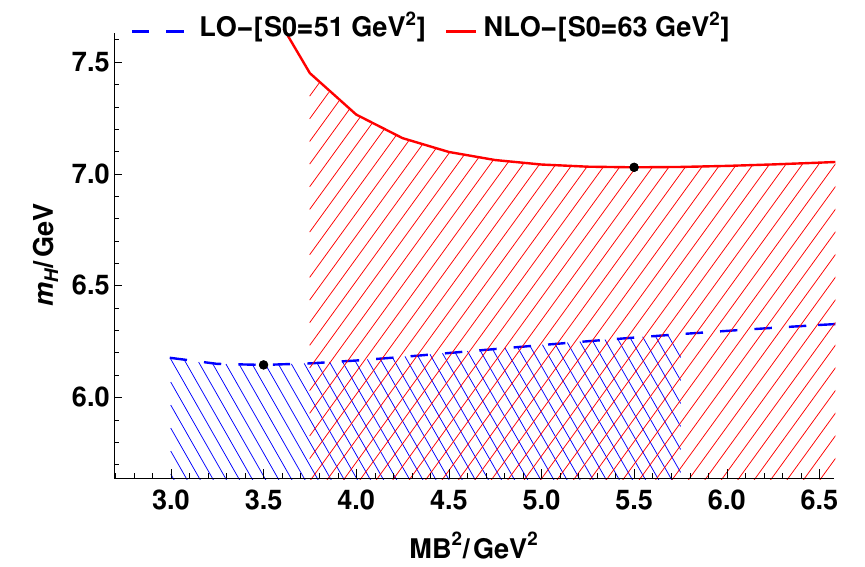}
		\includegraphics[scale=0.47]{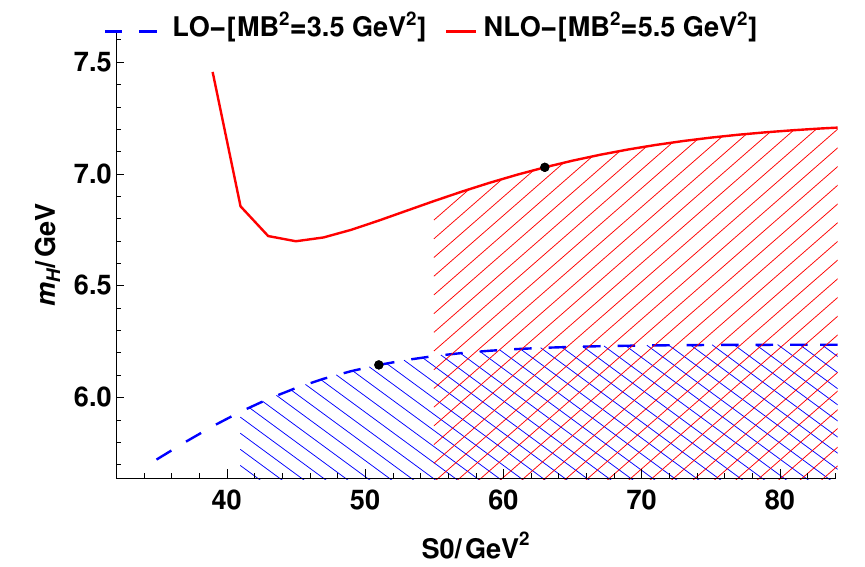}
	}\\
	\subfigure[OS]{
		\includegraphics[scale=0.47]{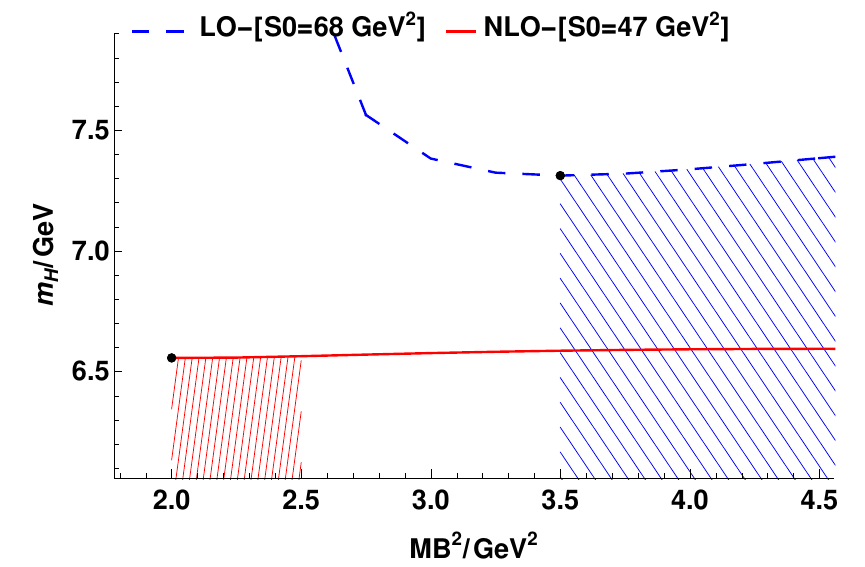}
		\includegraphics[scale=0.47]{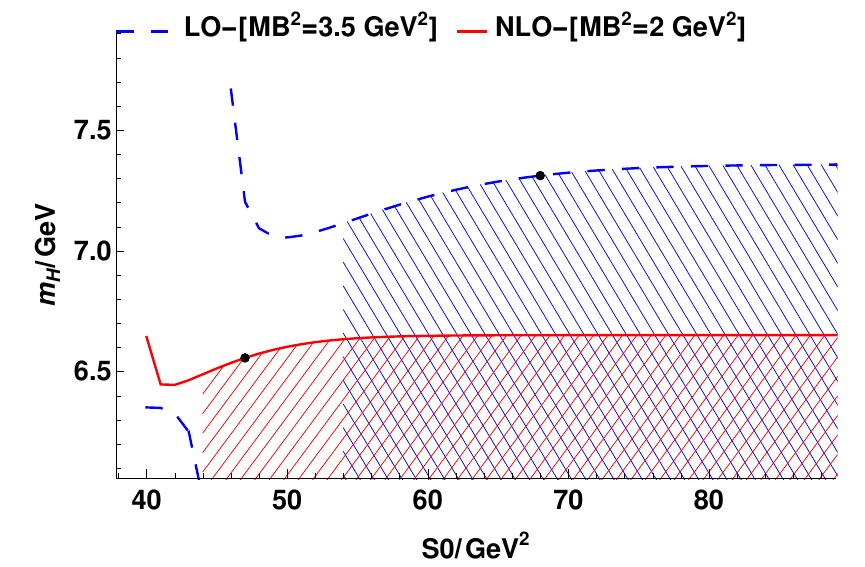}
	}
	\caption{\label{fig:2+-Mixed1-NLO-MSbar-OS}
		The Borel platform curves for $J_{T,1}^{\text{Dia}}$ with $J^{PC}=2^{++}$ in the $\overline{\text{MS}}$ and On-Shell schemes}
		\vspace{-0.5cm}
\end{figure}

\begin{figure}[H]
	\centering
	\subfigure[$\overline{\text{MS}}$]{
		\includegraphics[scale=0.47]{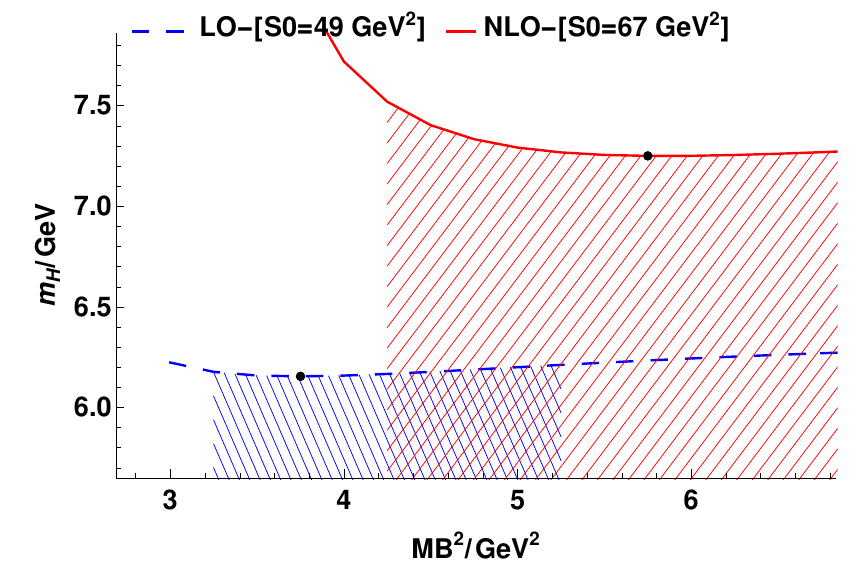}
		\includegraphics[scale=0.47]{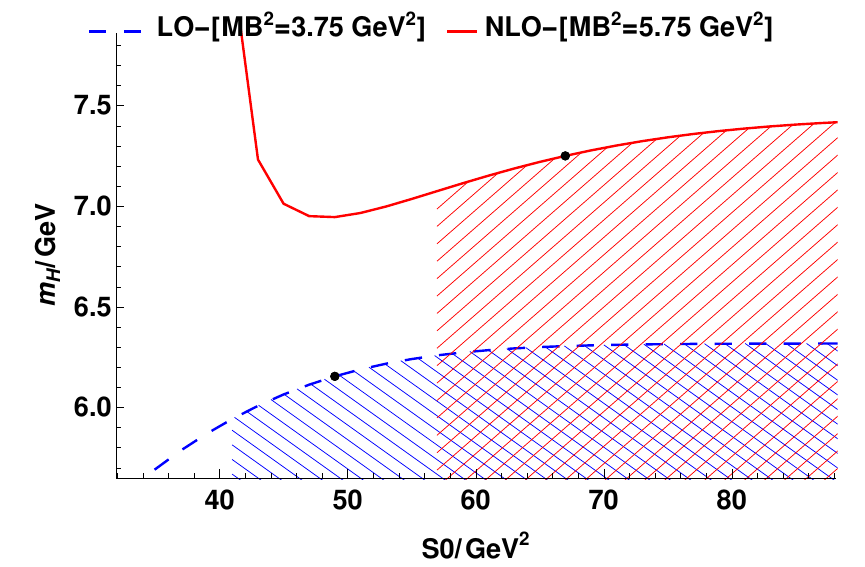}
	}\\
	\subfigure[OS]{
		\includegraphics[scale=0.47]{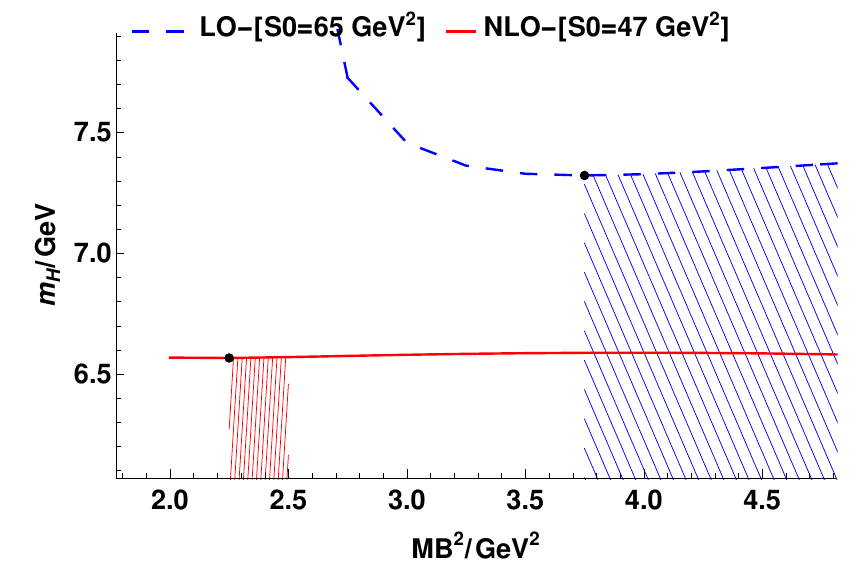}
		\includegraphics[scale=0.47]{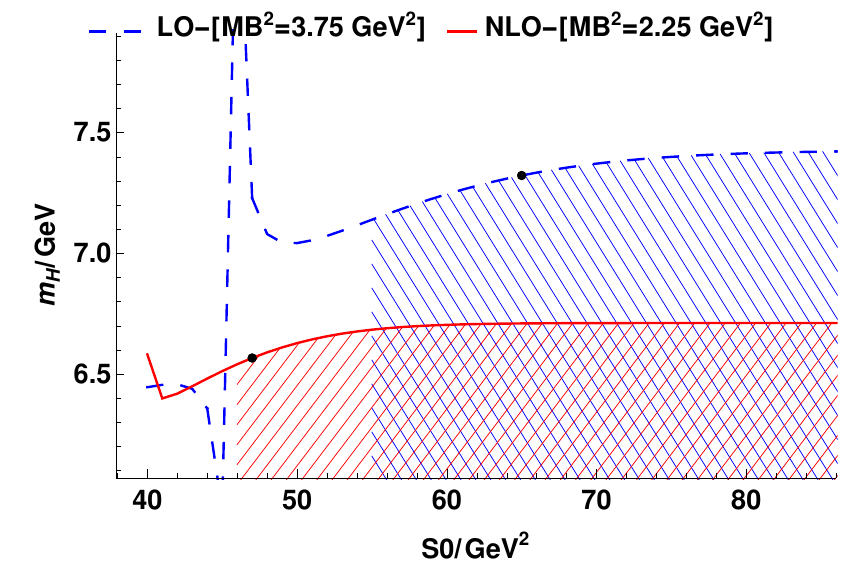}
	}
	\caption{\label{fig:2+-Mixed2-NLO-MSbar-OS}
		The Borel platform curves for $J_{T,2}^{\text{Dia}}$ with $J^{PC}=2^{++}$ in the $\overline{\text{MS}}$ and On-Shell schemes}
	\vspace{-0.8cm}
\end{figure}

\begin{figure}[H]
		\vspace{-0.8cm}
	\centering
	\subfigure[$\overline{\text{MS}}$]{
		\includegraphics[scale=0.4]{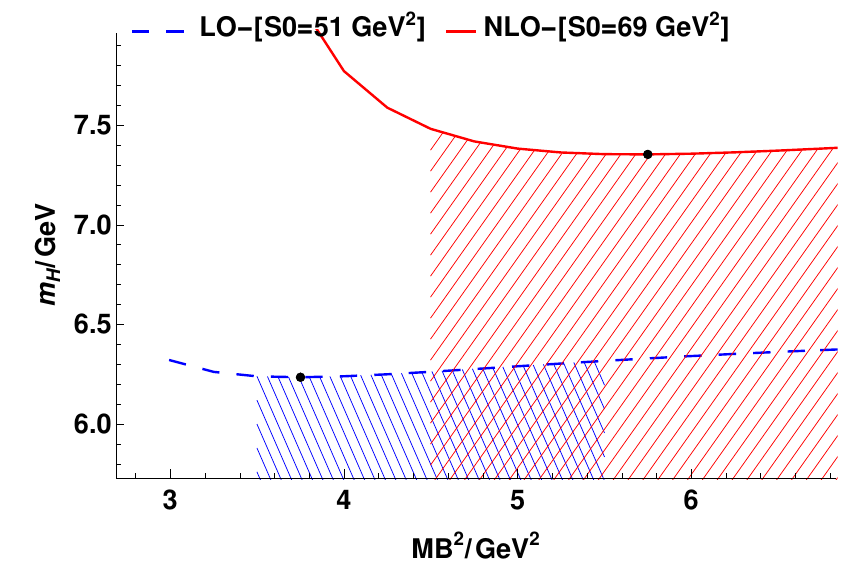}
		\includegraphics[scale=0.4]{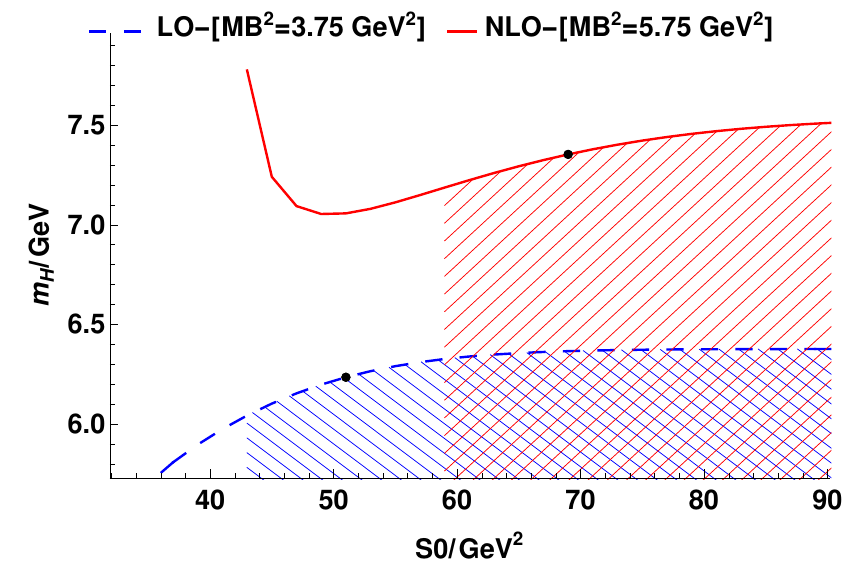}
	}\\
	\subfigure[OS]{
		\includegraphics[scale=0.4]{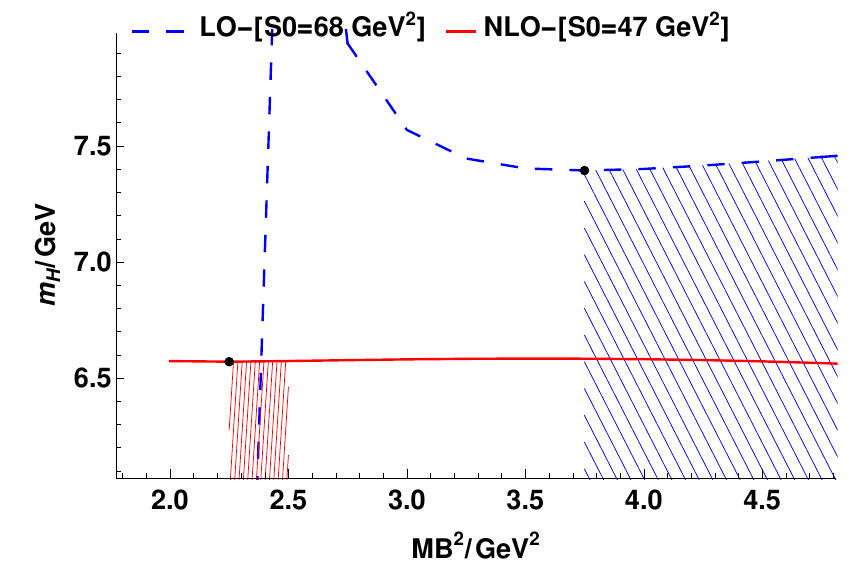}
		\includegraphics[scale=0.4]{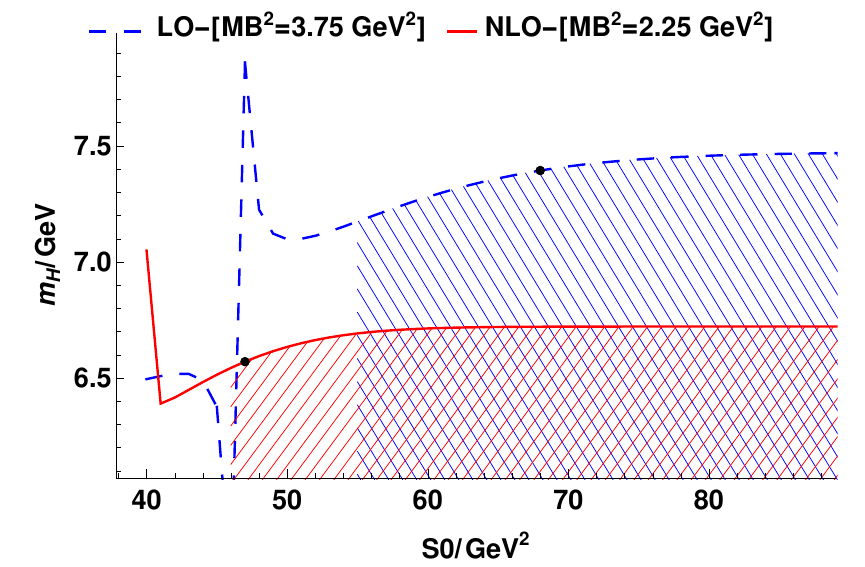}
	}
	\caption{\label{fig:2+-Mixed3-NLO-MSbar-OS}
		The Borel platform curves for $J_{T,3}^{\text{Dia}}$ with $J^{PC}=2^{++}$ in the $\overline{\text{MS}}$ and On-Shell schemes}
\end{figure}

\subsection{Renormalization scale dependence }

\begin{figure}[H]

	\centering
	\includegraphics[scale=0.53]{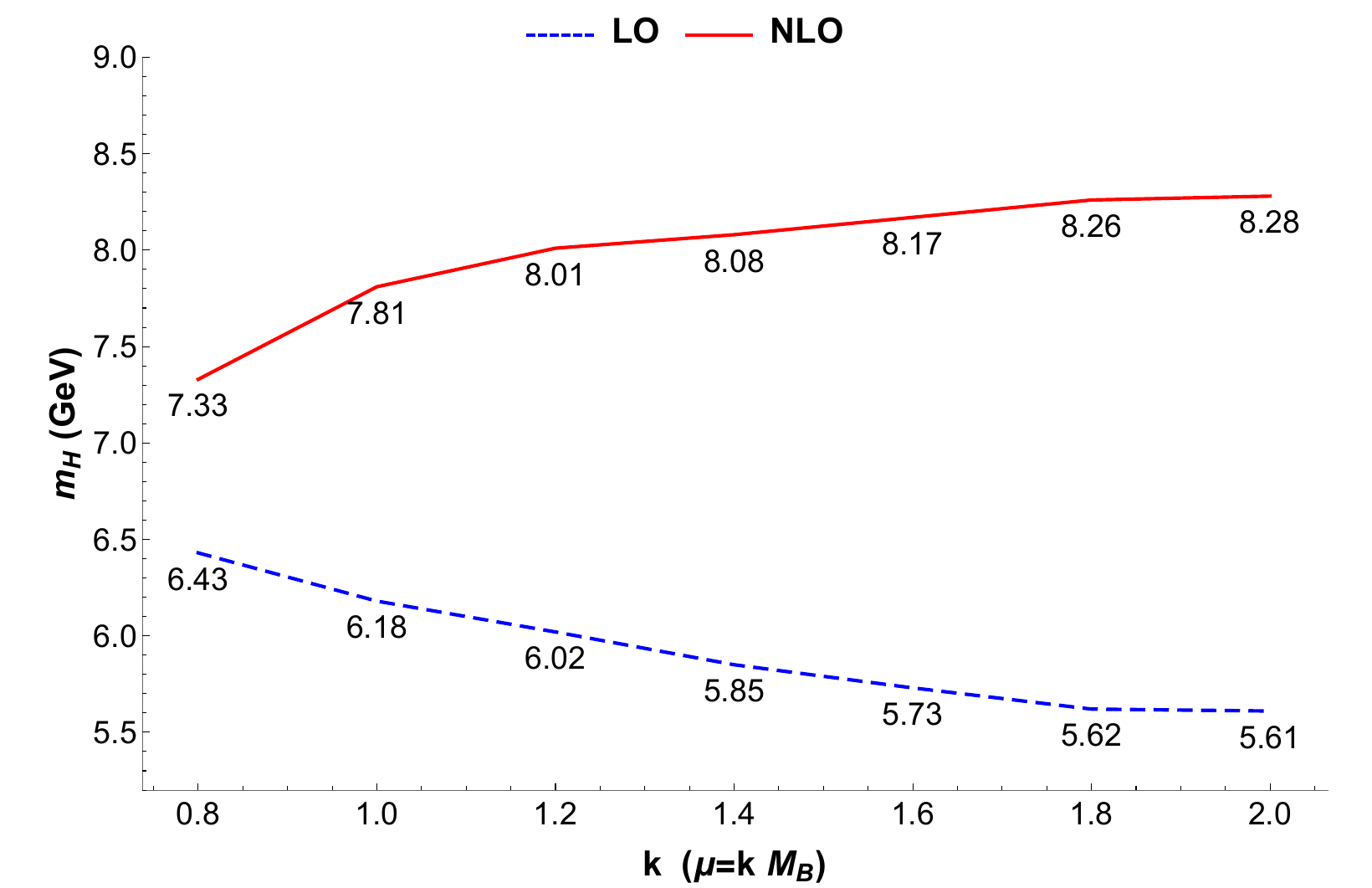}
	\caption{
		The renormalization scale $\mu$ dependence of the LO and NLO results of $J_{S,1}^{\text{Dia}}$ in $\overline{\text{MS}}$ scheme }
	\label{4c-mu-dependence-S-1}
	\vspace{-0.13cm}
\end{figure}

\begin{figure}[H]

	\centering
	\includegraphics[scale=0.53]{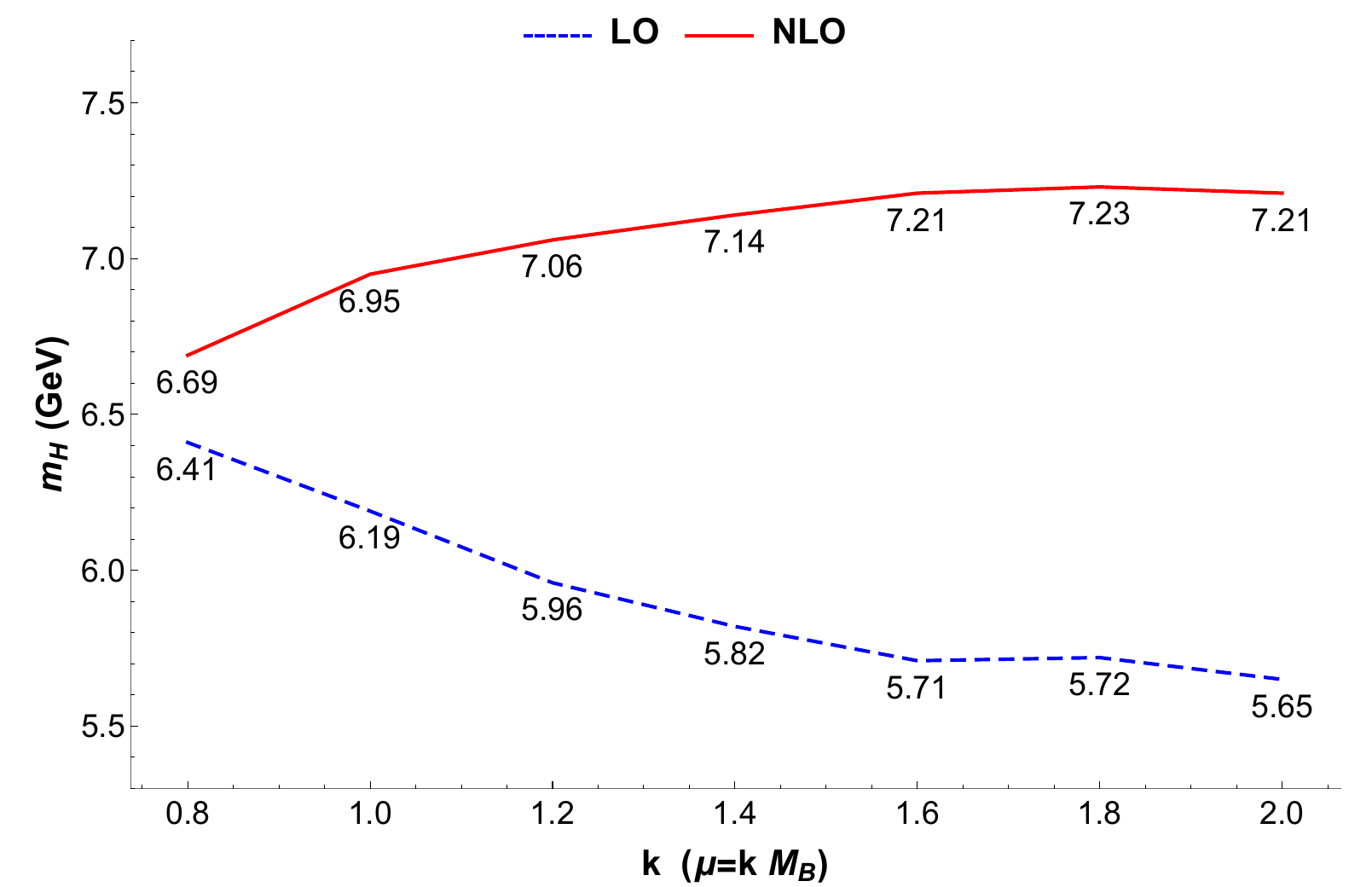}
	\caption{
		The renormalization scale $\mu$ dependence of the LO and NLO results of $J_{S,2}^{\text{Dia}}$ in $\overline{\text{MS}}$ scheme }
	\label{4c-mu-dependence-S-2}

\end{figure}

\begin{figure}[H]

	\centering
	\includegraphics[scale=0.5]{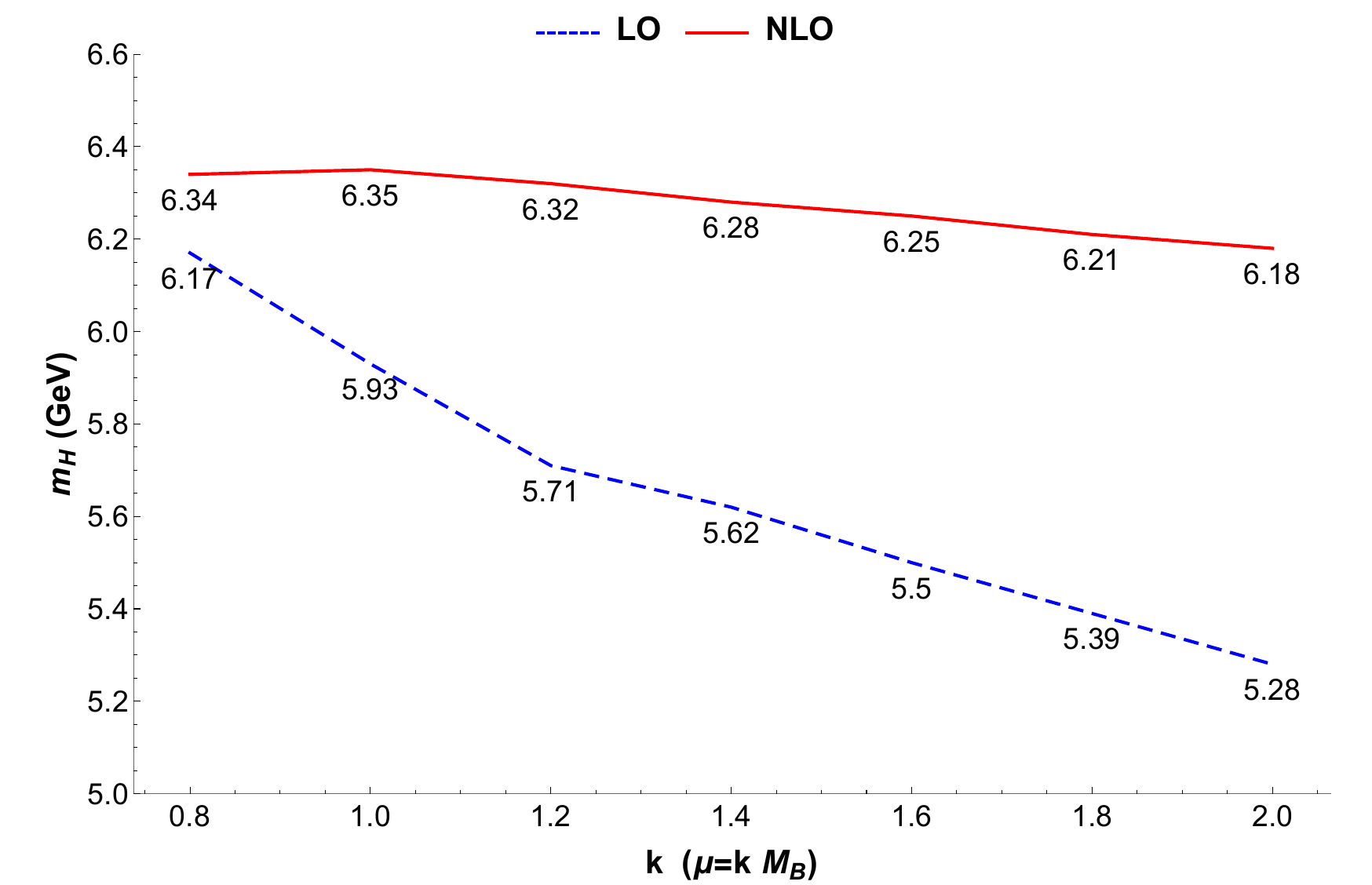}
	\caption{
		The renormalization scale $\mu$ dependence of the LO and NLO results of $J_{S,3}^{\text{Dia}}$ in $\overline{\text{MS}}$ scheme }
	\label{4c-mu-dependence-S-3}

\end{figure}

\begin{figure}[H]
	\vspace{-0.4cm}
	\centering
	\includegraphics[scale=0.47]{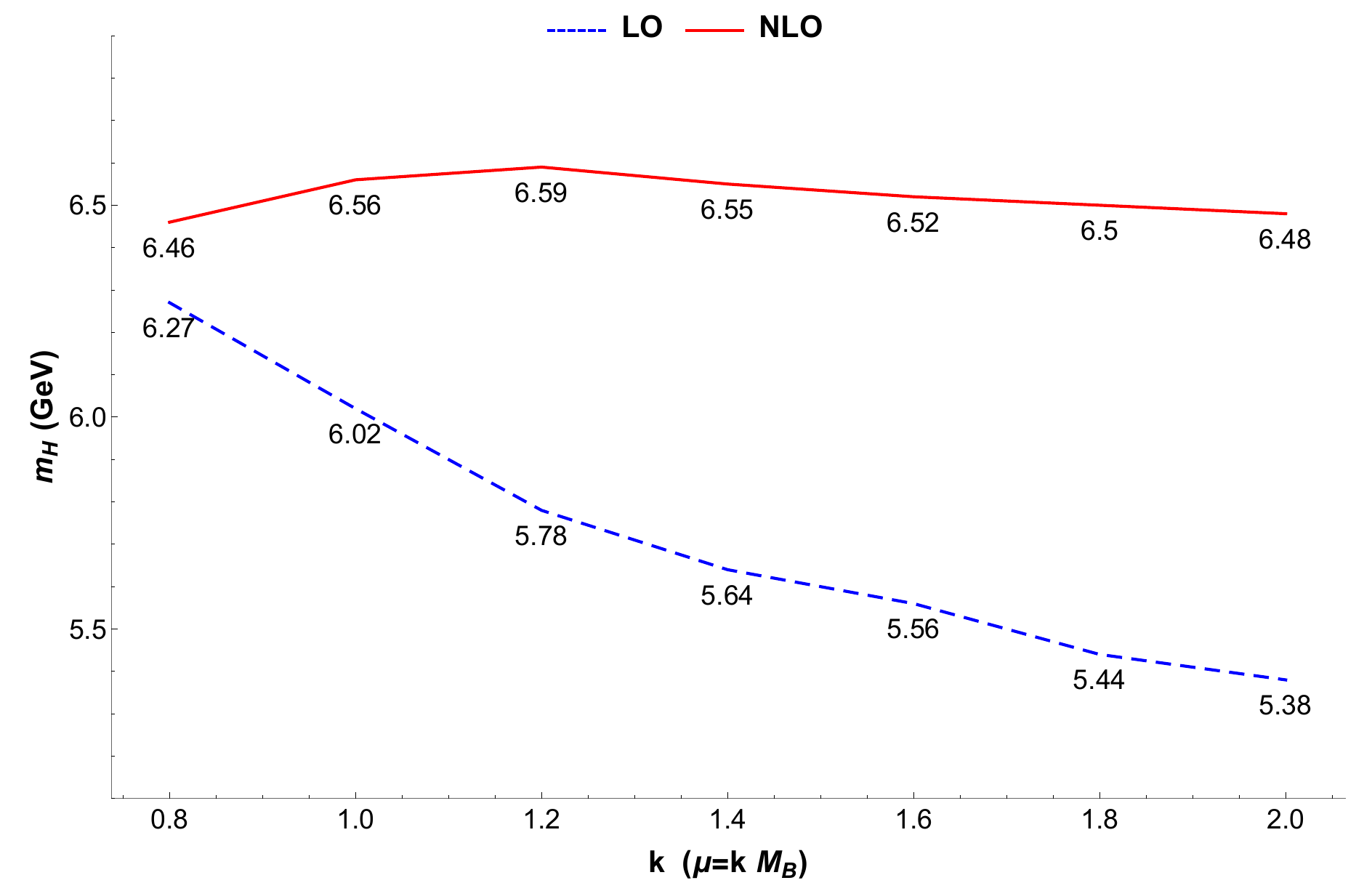}
	\caption{
		The renormalization scale $\mu$ dependence of the LO and NLO results of $J_{S,4}^{\text{Dia}}$ in $\overline{\text{MS}}$ scheme }
	\label{4c-mu-dependence-S-4}

\end{figure}

\begin{figure}[H]
	\vspace{-0.4cm}
	\centering
	\includegraphics[scale=0.41]{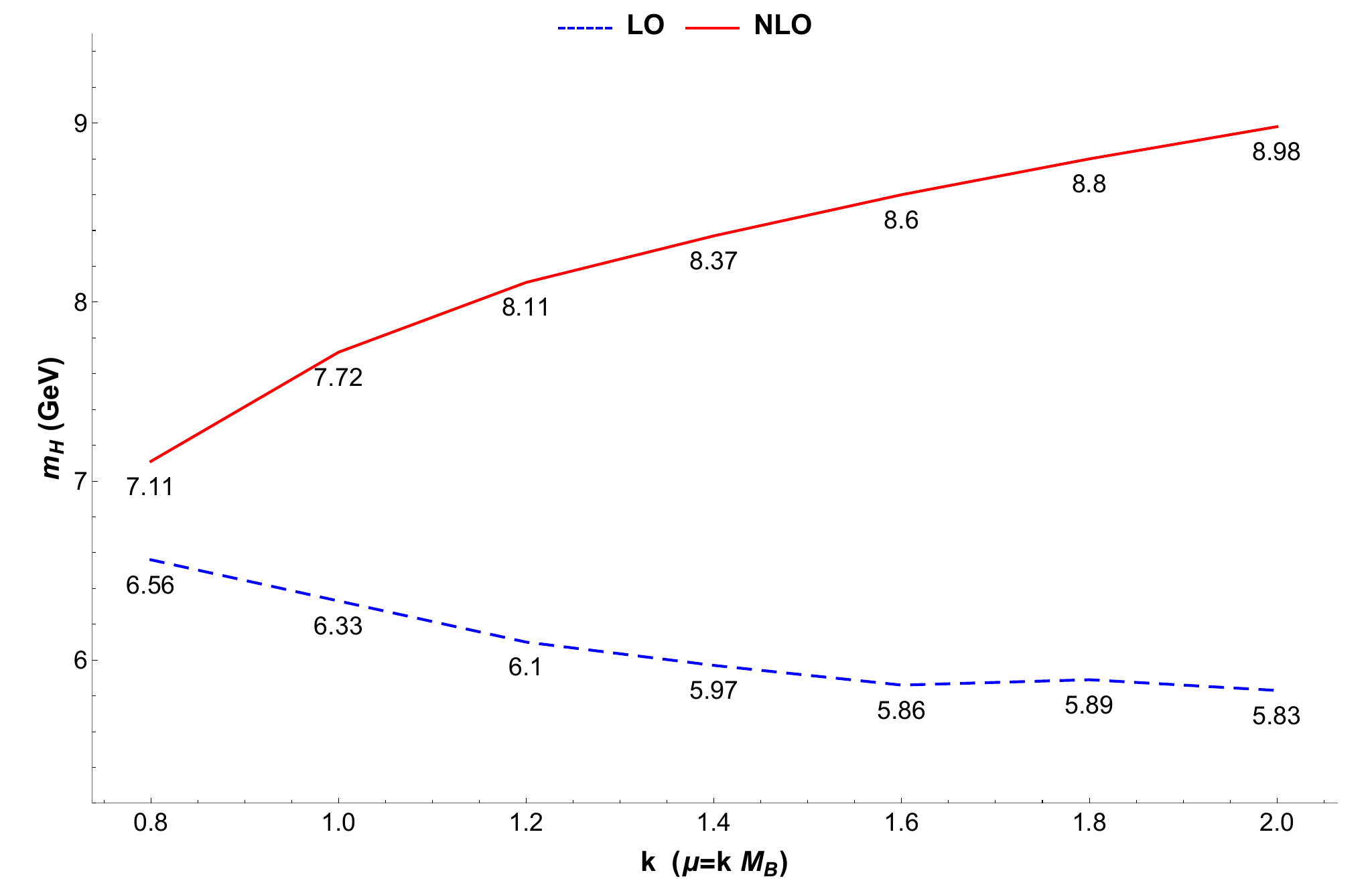}
	\caption{
		The renormalization scale $\mu$ dependence of the LO and NLO results of $J_{S,5}^{\text{Dia}}$ in $\overline{\text{MS}}$ scheme }
	\label{4c-mu-dependence-S-5}

\end{figure}

\begin{figure}[H]

	\centering
	\includegraphics[scale=0.52]{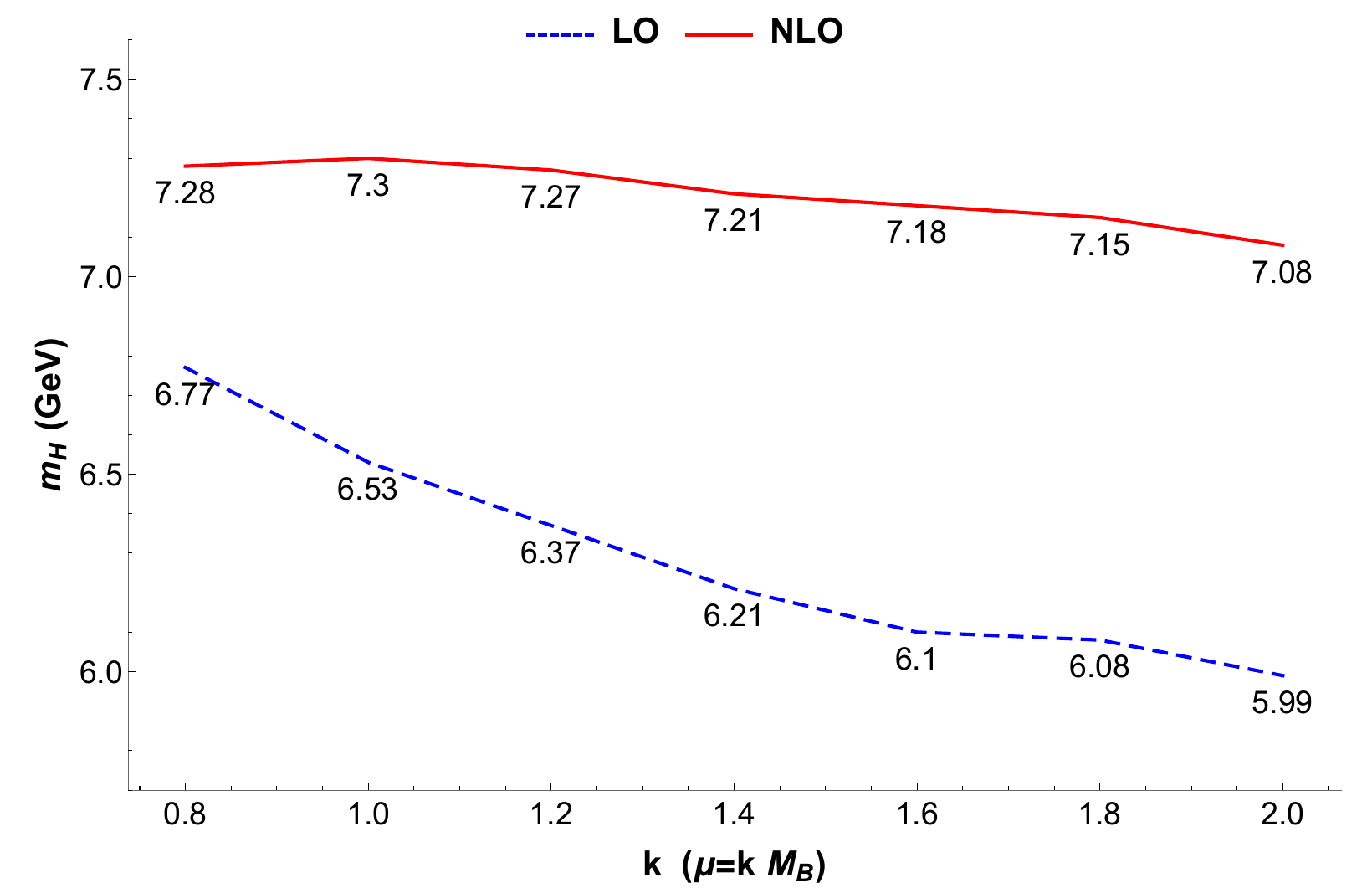}
	\caption{
		The renormalization scale $\mu$ dependence of the LO and NLO results of $J_{P,2}^{\text{Dia}}$ in $\overline{\text{MS}}$ scheme }
	\label{4c-mu-dependence-P-2}

\end{figure}

\begin{figure}[H]

	\centering
	\includegraphics[scale=0.5]{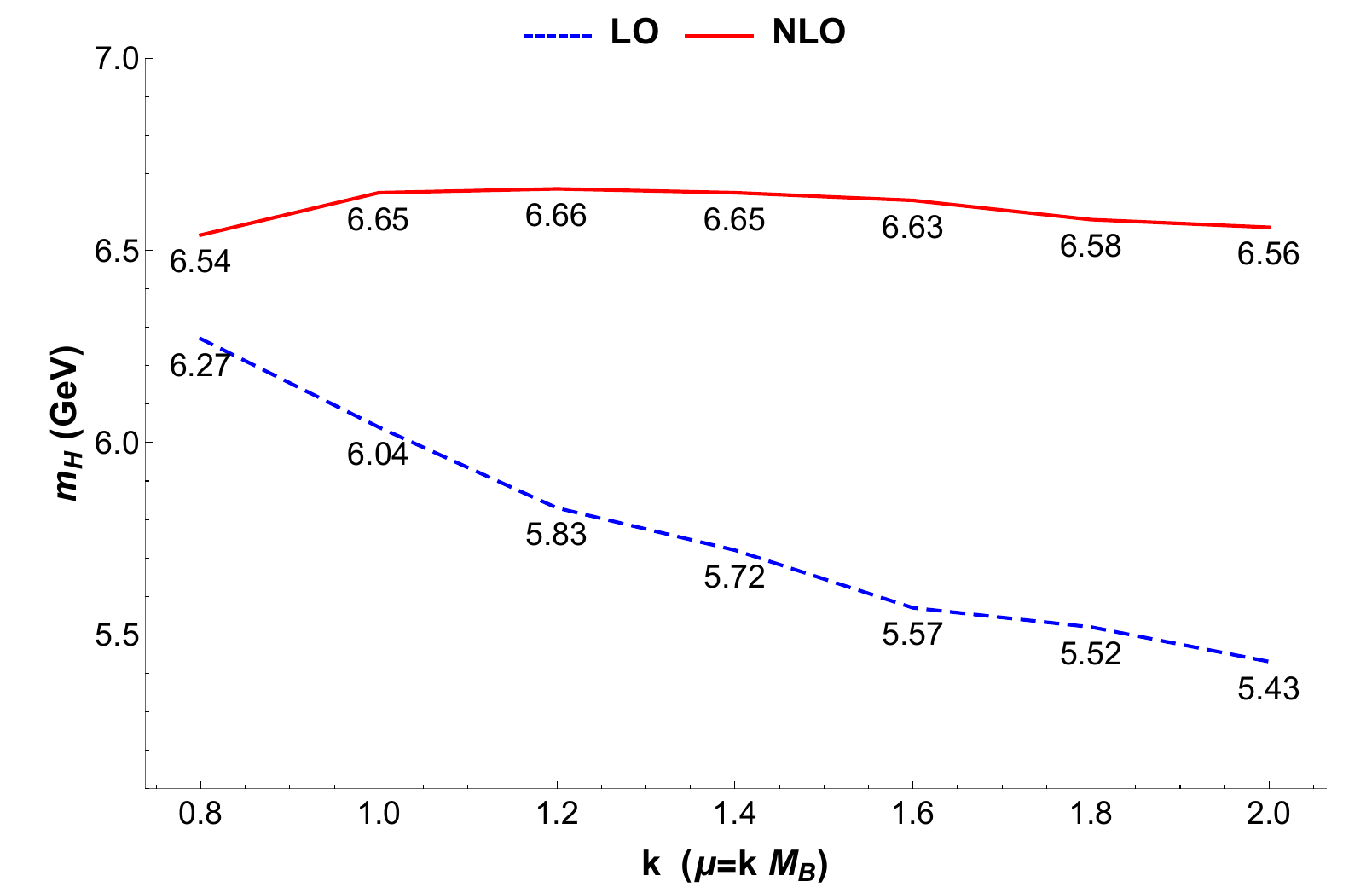}
	\caption{
		The renormalization scale $\mu$ dependence of the LO and NLO results of $J_{A,4}^{\text{Dia}}$ in $\overline{\text{MS}}$ scheme }
	\label{4c-mu-dependence-A-4}

\end{figure}

\begin{figure}[H]

	\centering
	\includegraphics[scale=0.5]{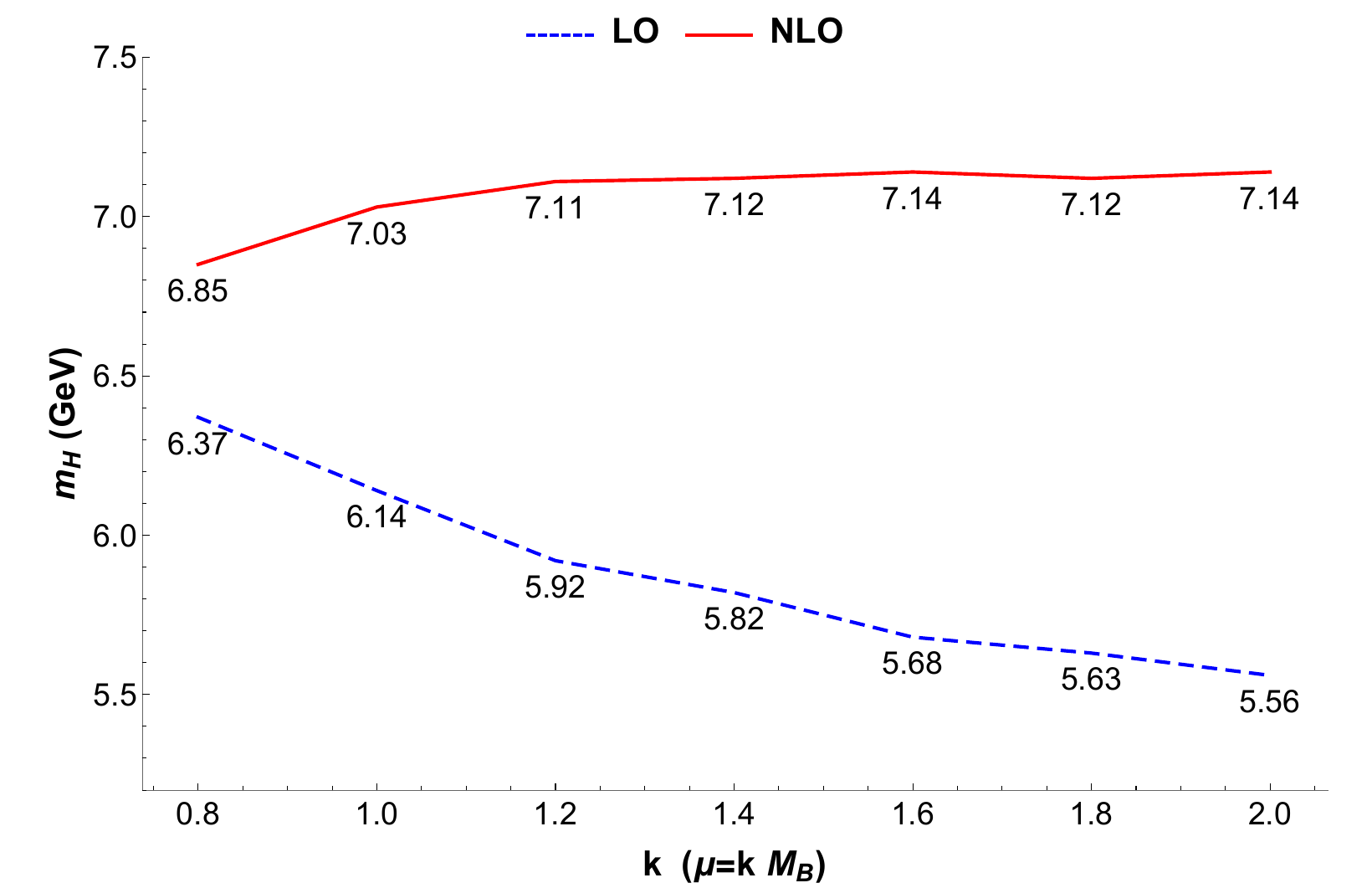}
	\caption{
		The renormalization scale $\mu$ dependence of the LO and NLO results of $J_{T,1}^{\text{Dia}}$ in $\overline{\text{MS}}$ scheme }
	\label{4c-mu-dependence-T-1}

\end{figure}

\begin{figure}[H]

	\centering
	\includegraphics[scale=0.5]{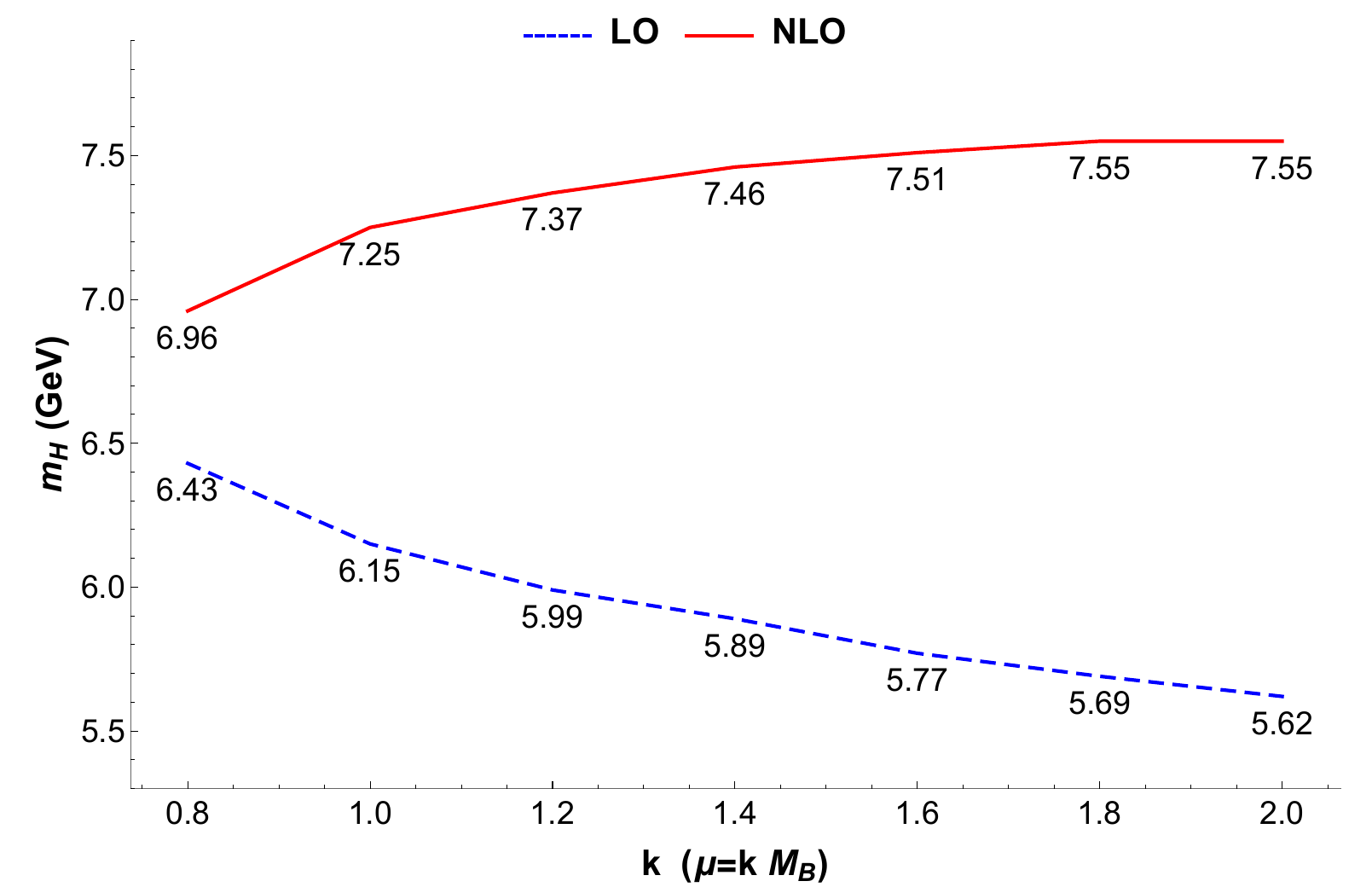}
	\caption{
		The renormalization scale $\mu$ dependence of the LO and NLO results of $J_{T,2}^{\text{Dia}}$ in $\overline{\text{MS}}$ scheme }
	\label{4c-mu-dependence-T-2}

\end{figure}

\section{Details for $\bar{b}b\bar{b}b$ system}
\label{sec:bottom}

\subsection{Numerical Results with $J^P=0^+$}

\begin{table}[H]

	\renewcommand\arraystretch{1.8}
	\begin{center}
		\setlength{\tabcolsep}{3 mm}
		\caption{The LO and NLO Results for $J^P=0^+$ with $\bar{b}b\bar{b}b$ system in the $\overline{\text{MS}}$ scheme. ($\Delta$ denotes this result is got in Borel condition ($\leq 40\%$))}
		\begin{tabular}{cccc|@{*}|ccc}
			\hline\hline
			\multirow{2}{*}{Current} &
			\multicolumn{3}{c|@{*}|}{LO}& \multicolumn{3}{c}{NLO($\overline{\text{MS}}$)} \\ \cline{2-7}
			& \makecell{$M_H$ \\ (GeV)} & \makecell{$S_0$ \\ ($\text{GeV}^2$)} & \makecell{$M_B^2$ \\ ($\text{GeV}^2$)} &  \makecell{$M_H$ \\ (GeV)} & \makecell{$S_0$ \\ ($\text{GeV}^2$)} & \makecell{$M_B^2$ \\ ($\text{GeV}^2$)} \\ \hline
			$J_{S,1}^{\text{M-M}}$ &$18.51^{+0.17}_{-0.26}$ &$380.(\pm 5\%)$ &$19.00(\pm 5\%)$    &$19.00^{+0.05}_{-0.10}$ &$400.(\pm 5\%)$ &$9.00(\pm 5\%)$\\
			$J_{S,2}^{\text{M-M}}$ &$18.55^{+0.19}_{-0.26}$ &$382.(\pm 5\%)$ &$18.00(\pm 5\%)$    &$18.92^{+0.10}_{-0.17}$ &$384.(\pm 5\%)$ &$9.50(\pm 5\%)$\\
			$J_{S,3}^{\text{M-M}}$ &$19.21^{+0.20}_{-0.26}$ &$408.(\pm 5\%)$ &$18.00(\pm 5\%)$    &($\Delta$)$19.66^{+0.05}_{-0.10}$ &$420.(\pm 5\%)$ &$7.00(\pm 5\%)$\\
			$J_{S,4}^{\text{M-M}}$ &$18.50^{+0.17}_{-0.26}$ &$380.(\pm 5\%)$ &$19.00(\pm 5\%)$    &$18.97^{+0.05}_{-0.11}$ &$398.(\pm 5\%)$ &$9.50(\pm 5\%)$\\
			$J_{S,5}^{\text{M-M}}$ &$18.51^{+0.17}_{-0.26}$ &$380.(\pm 5\%)$ &$19.00(\pm 5\%)$    &$18.93^{+0.09}_{-0.16}$ &$386.(\pm 5\%)$ &$9.50(\pm 5\%)$\\ \hline
			$J_{S,1}^{\text{Di-Di}}$ &$18.50^{+0.17}_{-0.26}$ &$380.(\pm 5\%)$ &$19.00(\pm 5\%)$    &$18.97^{+0.05}_{-0.11}$ &$398.(\pm 5\%)$ &$9.50(\pm 5\%)$\\
			$J_{S,2}^{\text{Di-Di}}$ &$18.52^{+0.17}_{-0.26}$ &$380.(\pm 5\%)$ &$18.00(\pm 5\%)$    &$18.95^{+0.08}_{-0.14}$ &$390.(\pm 5\%)$ &$9.50(\pm 5\%)$\\
			$J_{S,3}^{\text{Di-Di}}$ &$19.17^{+0.20}_{-0.26}$ &$406.(\pm 5\%)$ &$17.50(\pm 5\%)$    &$19.42^{+0.10}_{-0.17}$ &$404.(\pm 5\%)$ &$8.00(\pm 5\%)$\\
			$J_{S,4}^{\text{Di-Di}}$ &$18.50^{+0.17}_{-0.26}$ &$380.(\pm 5\%)$ &$19.00(\pm 5\%)$    &$19.00^{+0.05}_{-0.10}$ &$400.(\pm 5\%)$ &$9.00(\pm 5\%)$\\
			$J_{S,5}^{\text{Di-Di}}$ &$18.51^{+0.17}_{-0.26}$ &$380.(\pm 5\%)$ &$19.00(\pm 5\%)$    &$18.97^{+0.05}_{-0.11}$ &$398.(\pm 5\%)$ &$9.50(\pm 5\%)$\\ \hline
			$J_{S,1}^{\text{Dia}}$ &$18.51^{+0.17}_{-0.26}$ &$380.(\pm 5\%)$ &$19.00(\pm 5\%)$    &$19.01^{+0.05}_{-0.10}$ &$400.(\pm 5\%)$ &$9.00(\pm 5\%)$\\
			$J_{S,2}^{\text{Dia}}$ &$18.51^{+0.17}_{-0.26}$ &$380.(\pm 5\%)$ &$19.00(\pm 5\%)$    &$18.97^{+0.06}_{-0.11}$ &$398.(\pm 5\%)$ &$9.50(\pm 5\%)$\\
			$J_{S,3}^{\text{Dia}}$ &$18.50^{+0.18}_{-0.26}$ &$380.(\pm 5\%)$ &$19.00(\pm 5\%)$    &$18.96^{+0.05}_{-0.11}$ &$398.(\pm 5\%)$ &$9.50(\pm 5\%)$\\
			$J_{S,4}^{\text{Dia}}$ &$18.50^{+0.17}_{-0.26}$ &$380.(\pm 5\%)$ &$19.00(\pm 5\%)$    &$18.97^{+0.06}_{-0.11}$ &$398.(\pm 5\%)$ &$9.50(\pm 5\%)$\\
			$J_{S,5}^{\text{Dia}}$ &$18.51^{+0.17}_{-0.26}$ &$380.(\pm 5\%)$ &$19.00(\pm 5\%)$    &$18.95^{+0.08}_{-0.14}$ &$390.(\pm 5\%)$ &$9.50(\pm 5\%)$\\ \hline\hline
		\end{tabular}

		\label{tab:4b-S-NLOresult-MSbar}
	\end{center}
\end{table}
\begin{table}[H]
	\renewcommand\arraystretch{1.8}
	\setlength{\tabcolsep}{3 mm}
	\begin{center}
		\caption{The LO and NLO Results for $J^P=0^+$ with $\bar{b}b\bar{b}b$ system in the On-Shell scheme}
		\begin{tabular}{cccc|@{*}|ccc}
			\hline\hline
			\multirow{2}{*}{Current} &
			\multicolumn{3}{c|@{*}|}{LO}& \multicolumn{3}{c}{NLO(OS)} \\ \cline{2-7}
			& \makecell{$M_H$ \\ (GeV)} & \makecell{$S_0$ \\ ($\text{GeV}^2$)} & \makecell{$M_B^2$ \\ ($\text{GeV}^2$)} &  \makecell{$M_H$ \\ (GeV)} & \makecell{$S_0$ \\ ($\text{GeV}^2$)} & \makecell{$M_B^2$ \\ ($\text{GeV}^2$)} \\ \hline
			
			$J_{S,1}^{\text{M-M}}$ &$19.68^{+0.04}_{-0.10}$ &$420.(\pm 5\%)$ &$7.50(\pm 5\%)$    &$18.98^{+0.07}_{-0.28}$ &$366.(\pm 5\%)$ &$3.50(\pm 5\%)$\\
			$J_{S,2}^{\text{M-M}}$ &$19.68^{+0.04}_{-0.10}$ &$420.(\pm 5\%)$ &$7.50(\pm 5\%)$    &$18.98^{+0.07}_{-0.28}$ &$366.(\pm 5\%)$ &$3.50(\pm 5\%)$\\
			$J_{S,3}^{\text{M-M}}$ &$20.51^{+0.05}_{-0.18}$ &$452.(\pm 5\%)$ &$11.00(\pm 5\%)$    &$19.51^{+0.72}_{-1.55}$ &$392.(\pm 5\%)$ &$6.00(\pm 5\%)$\\
			$J_{S,4}^{\text{M-M}}$ &$19.64^{+0.02}_{-0.06}$ &$426.(\pm 5\%)$ &$7.00(\pm 5\%)$    &$18.98^{+0.07}_{-0.35}$ &$366.(\pm 5\%)$ &$3.50(\pm 5\%)$\\
			$J_{S,5}^{\text{M-M}}$ &$19.71^{+0.03}_{-0.08}$ &$426.(\pm 5\%)$ &$7.50(\pm 5\%)$    &$18.98^{+0.07}_{-0.27}$ &$366.(\pm 5\%)$ &$3.50(\pm 5\%)$\\ \hline
			$J_{S,1}^{\text{Di-Di}}$ &$19.64^{+0.06}_{-0.12}$ &$412.(\pm 5\%)$ &$7.50(\pm 5\%)$    &$18.98^{+0.07}_{-0.31}$ &$366.(\pm 5\%)$ &$3.50(\pm 5\%)$\\
			$J_{S,2}^{\text{Di-Di}}$ &$19.64^{+0.06}_{-0.12}$ &$412.(\pm 5\%)$ &$7.50(\pm 5\%)$    &$18.98^{+0.07}_{-0.30}$ &$366.(\pm 5\%)$ &$3.50(\pm 5\%)$\\
			$J_{S,3}^{\text{Di-Di}}$ &$20.25^{+0.12}_{-0.20}$ &$436.(\pm 5\%)$ &$9.50(\pm 5\%)$    &$19.31^{+0.10}_{-1.33}$ &$382.(\pm 5\%)$ &$4.50(\pm 5\%)$\\
			$J_{S,4}^{\text{Di-Di}}$ &$19.68^{+0.04}_{-0.10}$ &$420.(\pm 5\%)$ &$7.50(\pm 5\%)$    &$18.98^{+0.07}_{-0.28}$ &$366.(\pm 5\%)$ &$3.50(\pm 5\%)$\\
			$J_{S,5}^{\text{Di-Di}}$ &$19.65^{+0.05}_{-0.11}$ &$414.(\pm 5\%)$ &$7.50(\pm 5\%)$    &$18.98^{+0.07}_{-0.31}$ &$366.(\pm 5\%)$ &$3.50(\pm 5\%)$\\ \hline
			$J_{S,1}^{\text{Dia}}$ &$19.68^{+0.04}_{-0.10}$ &$420.(\pm 5\%)$ &$7.50(\pm 5\%)$    &$18.98^{+0.07}_{-0.28}$ &$366.(\pm 5\%)$ &$3.50(\pm 5\%)$\\
			$J_{S,2}^{\text{Dia}}$ &$19.67^{+0.04}_{-0.10}$ &$418.(\pm 5\%)$ &$7.50(\pm 5\%)$    &$18.98^{+0.07}_{-0.28}$ &$366.(\pm 5\%)$ &$3.50(\pm 5\%)$\\
			$J_{S,3}^{\text{Dia}}$ &$19.64^{+0.02}_{-0.06}$ &$426.(\pm 5\%)$ &$7.00(\pm 5\%)$    &$18.98^{+0.07}_{-0.36}$ &$366.(\pm 5\%)$ &$3.50(\pm 5\%)$\\
			$J_{S,4}^{\text{Dia}}$ &$19.61^{+0.07}_{-0.14}$ &$408.(\pm 5\%)$ &$7.50(\pm 5\%)$    &$18.98^{+0.07}_{-0.33}$ &$366.(\pm 5\%)$ &$3.50(\pm 5\%)$\\
			$J_{S,5}^{\text{Dia}}$ &$19.66^{+0.08}_{-0.15}$ &$410.(\pm 5\%)$ &$8.00(\pm 5\%)$    &$18.98^{+0.07}_{-0.26}$ &$366.(\pm 5\%)$ &$3.50(\pm 5\%)$\\ \hline\hline
		\end{tabular}
		
		\label{tab:4b-S-NLOresult-OS}
	\end{center}
\end{table}

\begin{figure}[H]
	\centering
	\subfigure[$\overline{\text{MS}}$]{
		\includegraphics[scale=0.4]{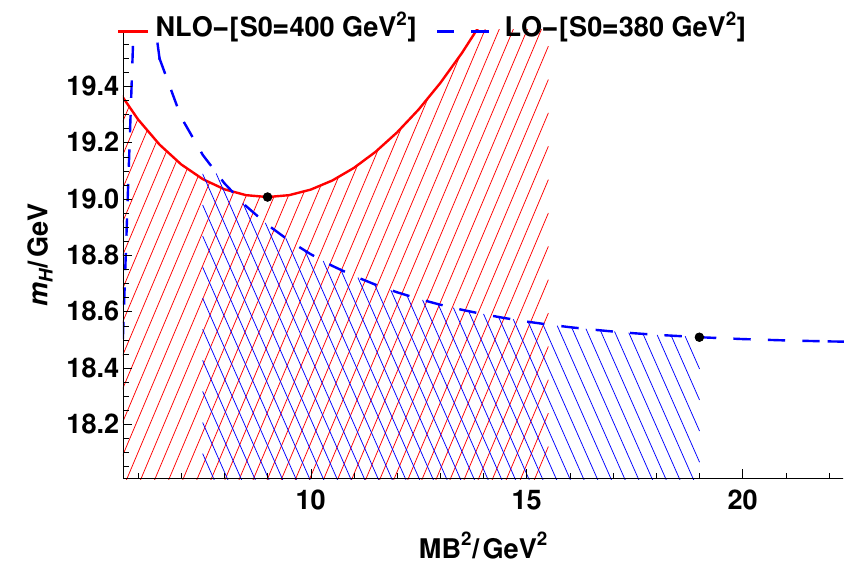}
		\includegraphics[scale=0.4]{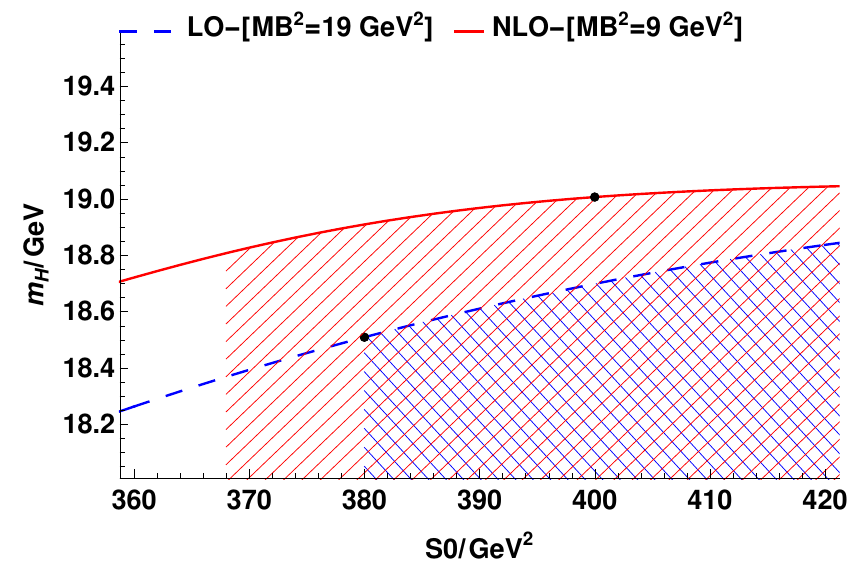}
	}\\
	\subfigure[OS]{
		\includegraphics[scale=0.4]{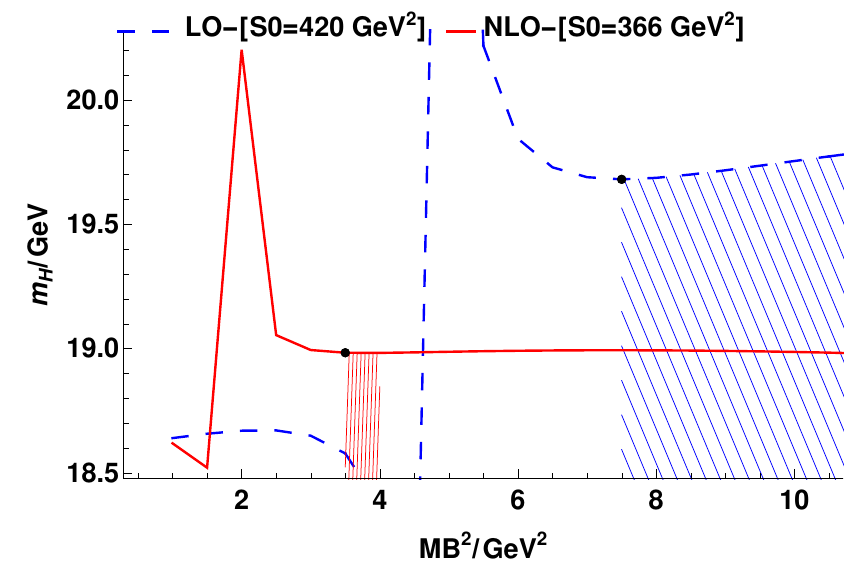}
		\includegraphics[scale=0.4]{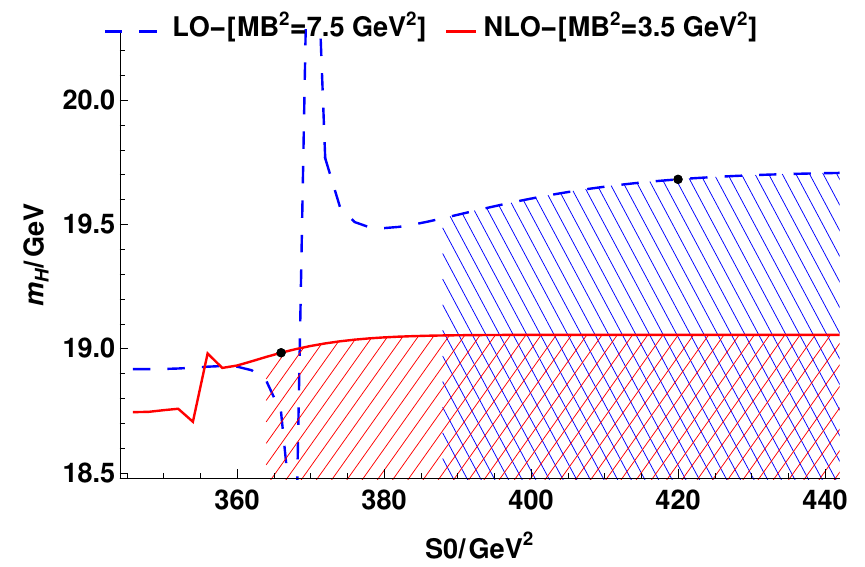}
	}
	\caption{\label{fig:4b-0+-Mixed1-NLO-MSbar-OS}
		The Borel platform curves for $J_{S,1}^{\text{Dia}}$ with $J^{PC}=0^{++}$ in the $\overline{\text{MS}}$ and On-Shell schemes}
\end{figure}
\begin{figure}[H]
	\centering
	\subfigure[$\overline{\text{MS}}$]{
		\includegraphics[scale=0.4]{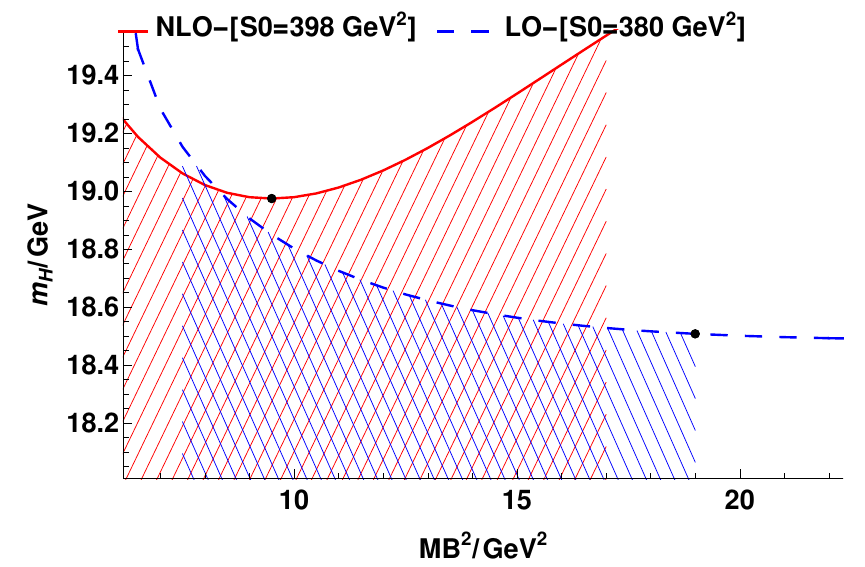}
		\includegraphics[scale=0.4]{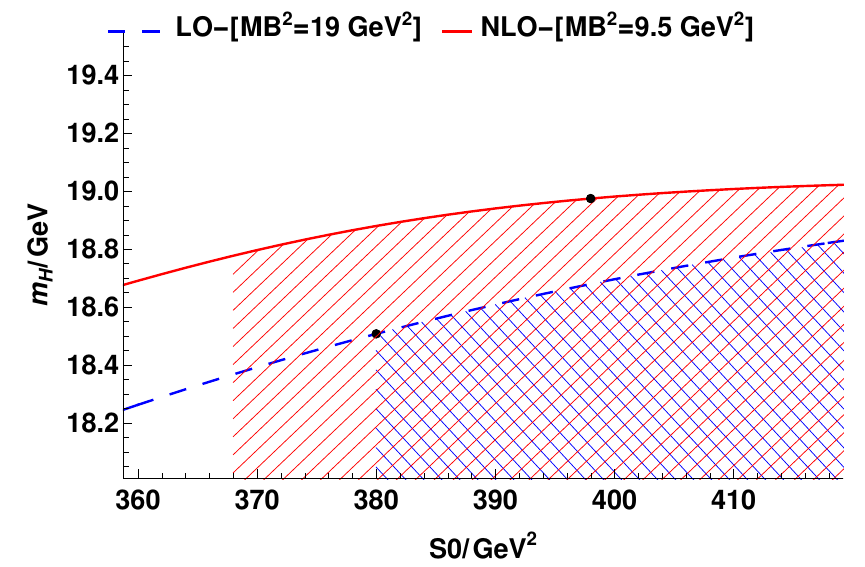}
	}\\
	\subfigure[OS]{
		\includegraphics[scale=0.4]{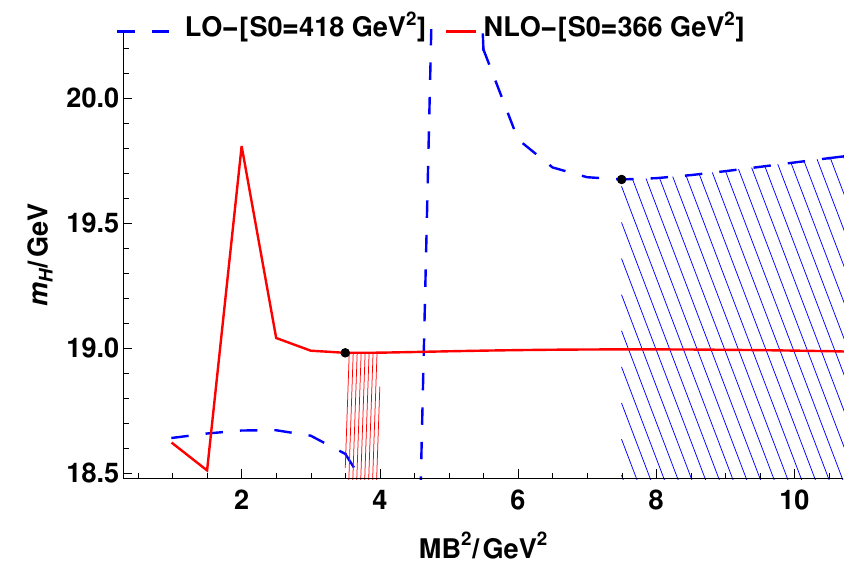}
		\includegraphics[scale=0.4]{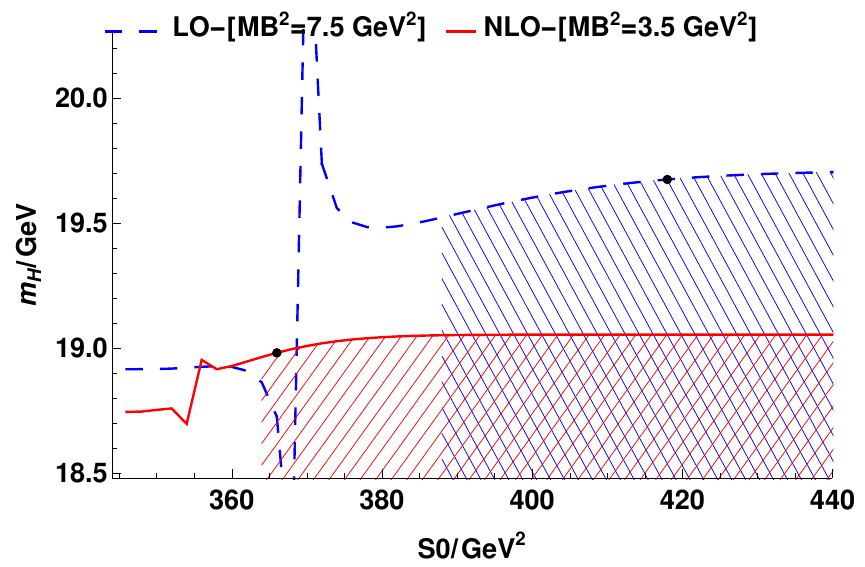}
	}
	\caption{\label{fig:4b-0+-Mixed2-NLO-MSbar-OS}
		The Borel platform curves for $J_{S,2}^{\text{Dia}}$ with $J^{PC}=0^{++}$ in the $\overline{\text{MS}}$ and On-Shell schemes}
\end{figure}
\begin{figure}[H]
	\centering
	\subfigure[$\overline{\text{MS}}$]{
		\includegraphics[scale=0.4]{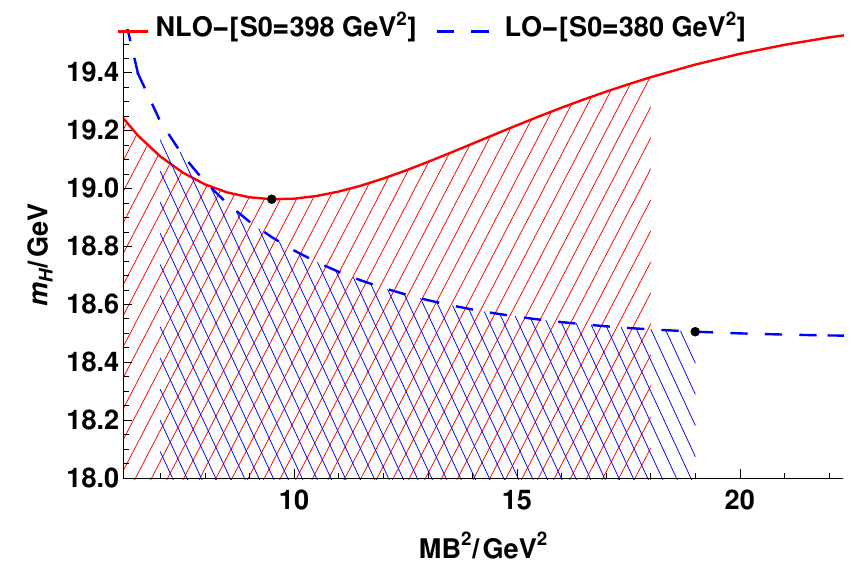}
		\includegraphics[scale=0.4]{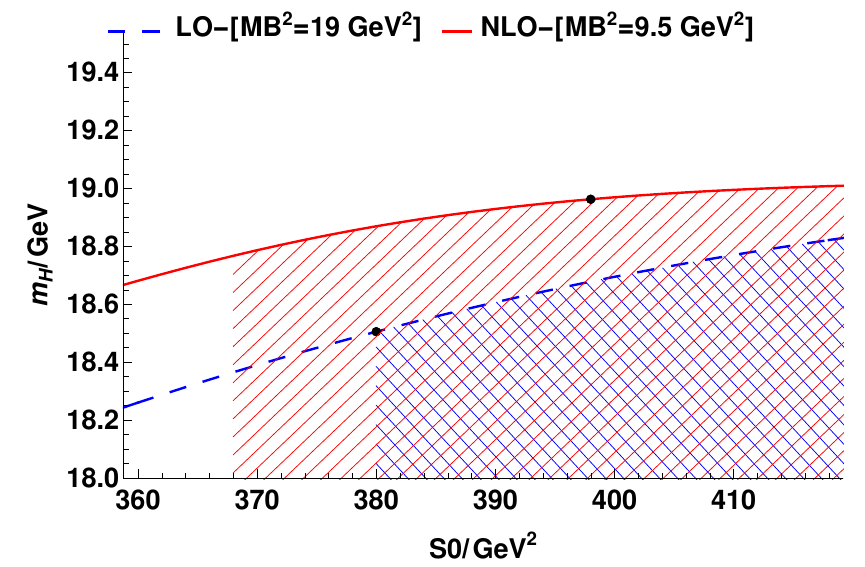}
	}\\
	\subfigure[OS]{
		\includegraphics[scale=0.4]{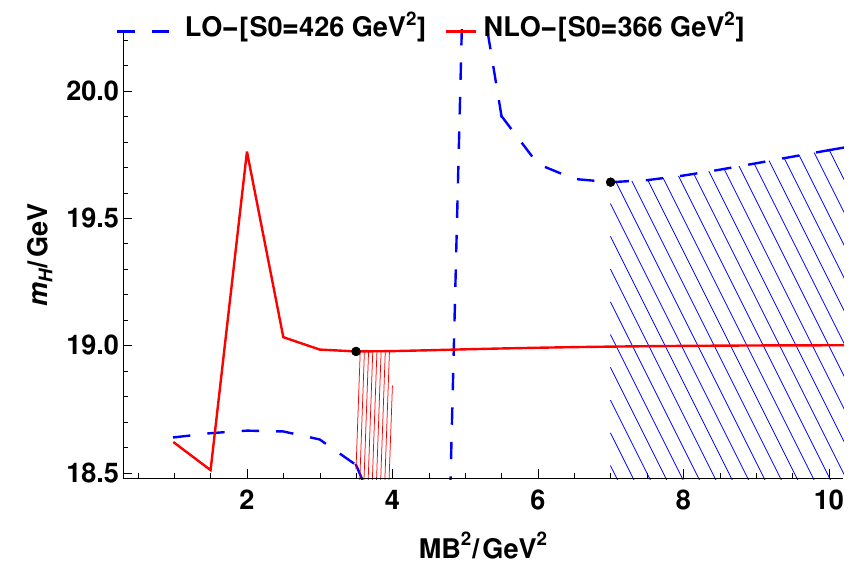}
		\includegraphics[scale=0.4]{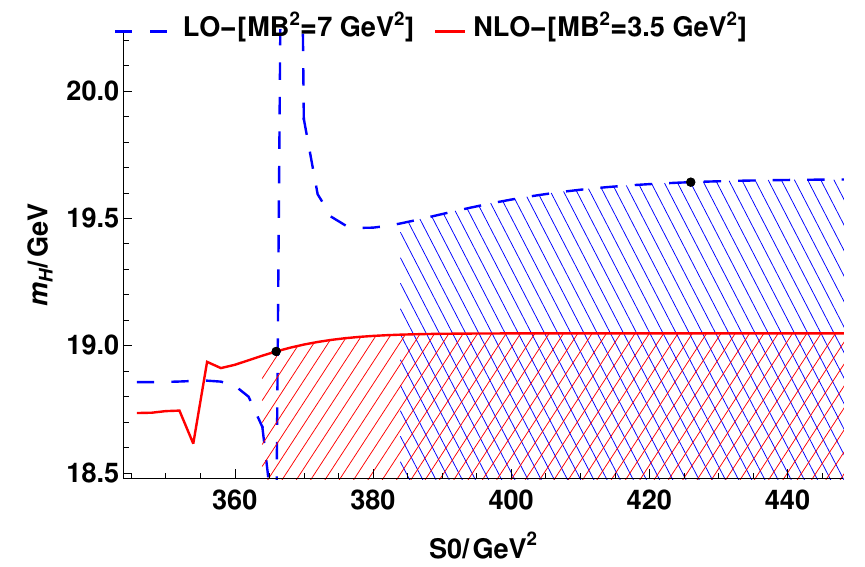}
	}
	\caption{\label{fig:4b-0+-Mixed3-NLO-MSbar-OS}
		The Borel platform curves for $J_{S,3}^{\text{Dia}}$ with $J^{PC}=0^{++}$ in the $\overline{\text{MS}}$ and On-Shell schemes}
\end{figure}
\begin{figure}[H]
	\centering
	\subfigure[$\overline{\text{MS}}$]{
		\includegraphics[scale=0.4]{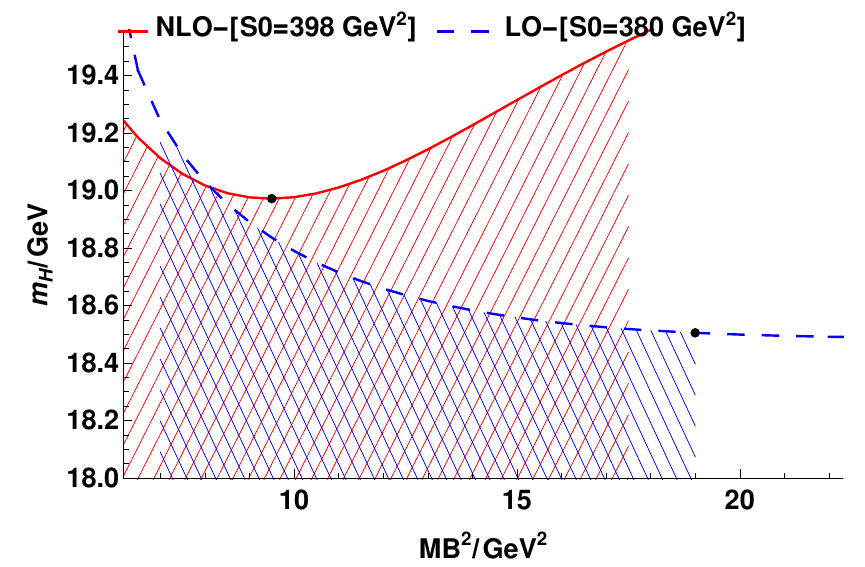}
		\includegraphics[scale=0.4]{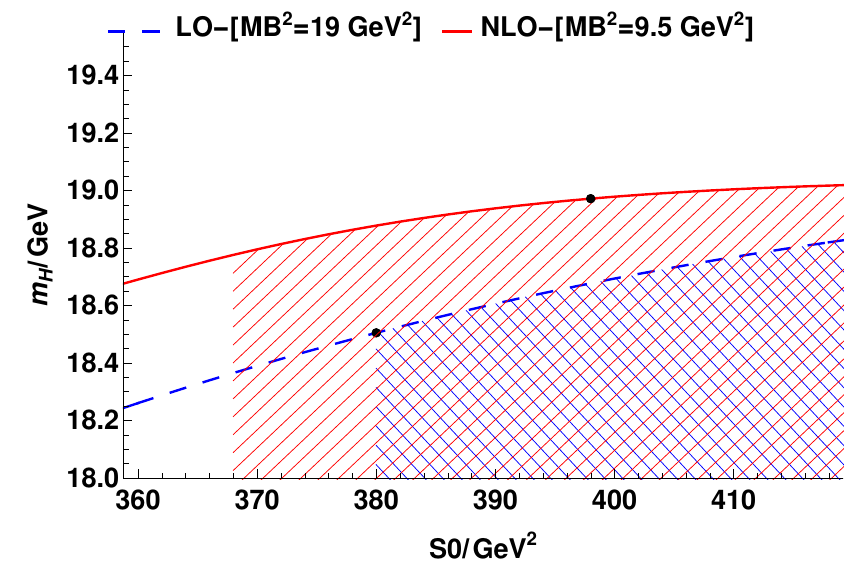}
	}\\
	\subfigure[OS]{
		\includegraphics[scale=0.4]{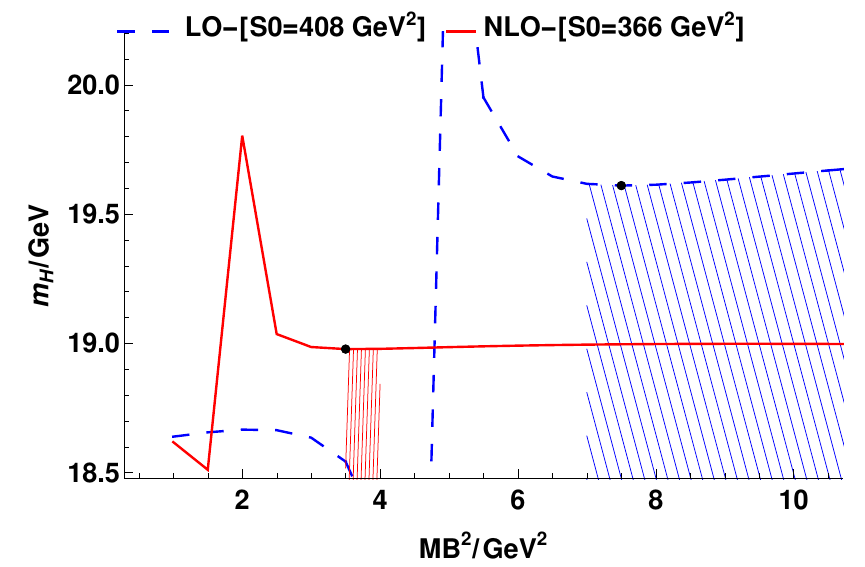}
		\includegraphics[scale=0.4]{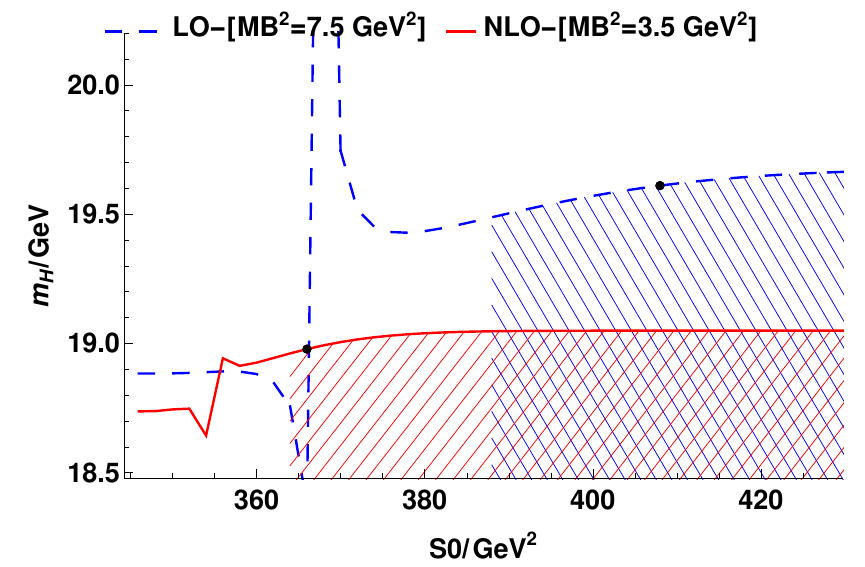}
	}
	\caption{\label{fig:4b-0+-Mixed4-NLO-MSbar-OS}
		The Borel platform curves for $J_{S,4}^{\text{Dia}}$ with $J^{PC}=0^{++}$ in the $\overline{\text{MS}}$ and On-Shell schemes}
\end{figure}
\begin{figure}[H]
	\centering
	\subfigure[$\overline{\text{MS}}$]{
		\includegraphics[scale=0.4]{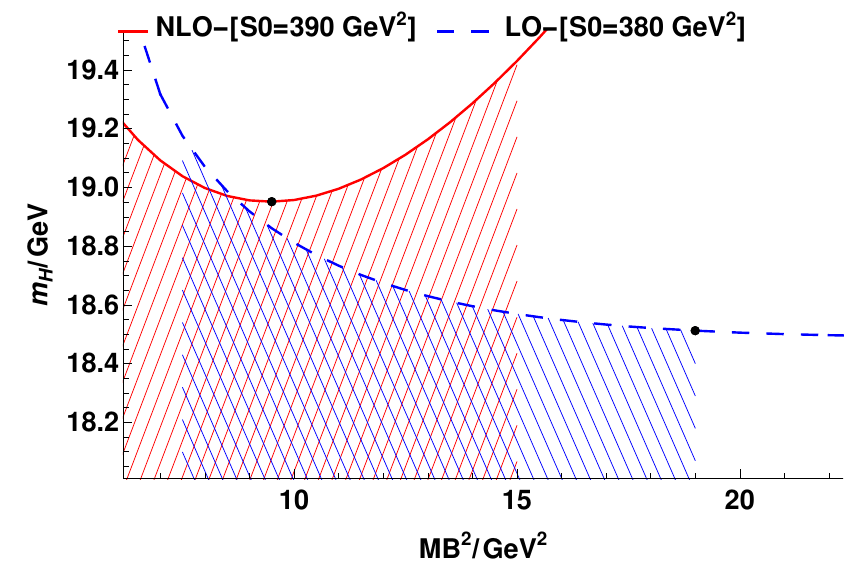}
		\includegraphics[scale=0.4]{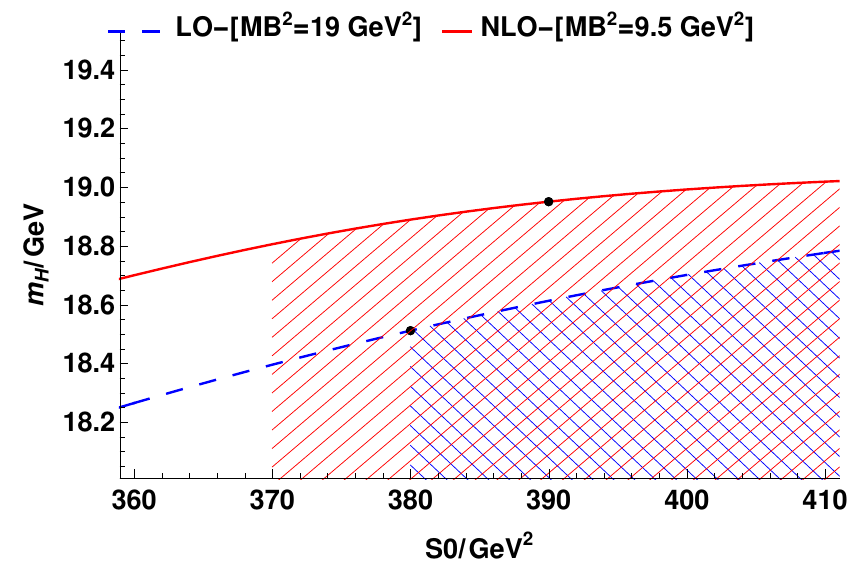}
	}\\
	\subfigure[OS]{
		\includegraphics[scale=0.4]{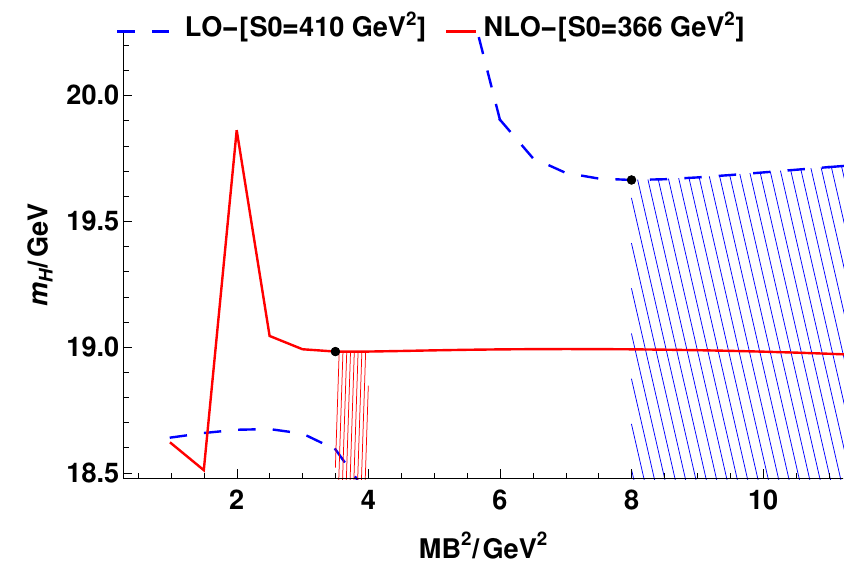}
		\includegraphics[scale=0.4]{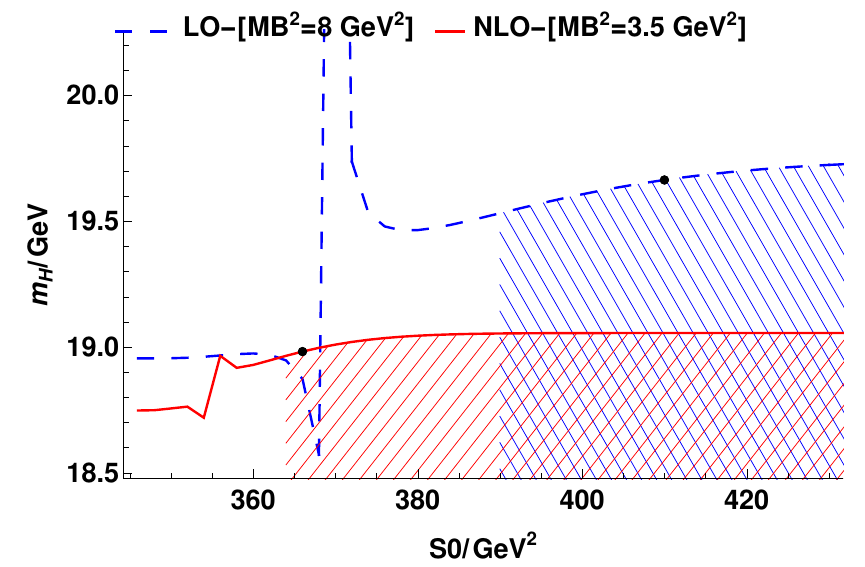}
	}
	\caption{\label{fig:4b-0+-Mixed5-NLO-MSbar-OS}
		The Borel platform curves for $J_{S,5}^{\text{Dia}}$ with $J^{PC}=0^{++}$ in the $\overline{\text{MS}}$ and On-Shell schemes}
\end{figure}

\subsection{Numerical Results with $J^P=0^-$}
\begin{table}[H]
	\vspace{-0.5cm}
	\renewcommand\arraystretch{1.5}
	\setlength{\tabcolsep}{3 mm}
	\begin{center}
		\caption{The LO and NLO Results for $J^P=0^-$ with $\bar{b}b\bar{b}b$ system in the $\overline{\text{MS}}$ scheme}
		\begin{tabular}{cccc|@{*}|ccc}
			\hline\hline
			\multirow{2}{*}{Current} &
			\multicolumn{3}{c|@{*}|}{LO}& \multicolumn{3}{c}{NLO($\overline{\text{MS}}$)} \\ \cline{2-7}
			& \makecell{$M_H$ \\ (GeV)} & \makecell{$S_0$ \\ ($\text{GeV}^2$)} & \makecell{$M_B^2$ \\ ($\text{GeV}^2$)} &  \makecell{$M_H$ \\ (GeV)} & \makecell{$S_0$ \\ ($\text{GeV}^2$)} & \makecell{$M_B^2$ \\ ($\text{GeV}^2$)} \\ \hline
			
			$J_{P,1}^{\text{M-M}}$ &$18.85^{+0.19}_{-0.26}$ &$394.(\pm 5\%)$ &$18.00(\pm 5\%)$    &$19.18^{+0.11}_{-0.18}$ &$392.(\pm 5\%)$ &$8.50(\pm 5\%)$\\
			$J_{P,2}^{\text{M-M}}$ &$18.86^{+0.19}_{-0.26}$ &$394.(\pm 5\%)$ &$18.00(\pm 5\%)$    &$19.31^{+0.04}_{-0.09}$ &$412.(\pm 5\%)$ &$8.00(\pm 5\%)$\\
			$J_{P,3}^{\text{M-M}}$ &$18.85^{+0.19}_{-0.26}$ &$394.(\pm 5\%)$ &$18.00(\pm 5\%)$    &$19.23^{+0.05}_{-0.11}$ &$408.(\pm 5\%)$ &$8.50(\pm 5\%)$\\ \hline
			$J_{P,1}^{\text{Di-Di}}$ &$18.85^{+0.19}_{-0.24}$ &$394.(\pm 5\%)$ &$18.00(\pm 5\%)$    &$19.18^{+0.11}_{-0.18}$ &$392.(\pm 5\%)$ &$8.50(\pm 5\%)$\\
			$J_{P,2}^{\text{Di-Di}}$ &$18.85^{+0.19}_{-0.24}$ &$394.(\pm 5\%)$ &$18.00(\pm 5\%)$    &$19.24^{+0.06}_{-0.12}$ &$406.(\pm 5\%)$ &$8.50(\pm 5\%)$\\
			$J_{P,3}^{\text{Di-Di}}$ &$18.86^{+0.19}_{-0.24}$ &$394.(\pm 5\%)$ &$18.00(\pm 5\%)$    &$19.23^{+0.08}_{-0.14}$ &$400.(\pm 5\%)$ &$8.50(\pm 5\%)$\\ \hline
			$J_{P,1}^{\text{Dia}}$ &$18.85^{+0.19}_{-0.26}$ &$394.(\pm 5\%)$ &$18.00(\pm 5\%)$    &$19.18^{+0.11}_{-0.18}$ &$392.(\pm 5\%)$ &$8.50(\pm 5\%)$\\
			$J_{P,2}^{\text{Dia}}$ &$18.86^{+0.19}_{-0.26}$ &$394.(\pm 5\%)$ &$18.00(\pm 5\%)$    &$19.31^{+0.04}_{-0.09}$ &$412.(\pm 5\%)$ &$8.00(\pm 5\%)$\\
			$J_{P,3}^{\text{Dia}}$ &$18.85^{+0.19}_{-0.26}$ &$394.(\pm 5\%)$ &$18.00(\pm 5\%)$    &$19.22^{+0.05}_{-0.11}$ &$408.(\pm 5\%)$ &$8.50(\pm 5\%)$\\ \hline\hline
		\end{tabular}
		\label{tab:4b-P-NLOresult-MSbar}
	\end{center}
\end{table}

\begin{table}[H]
	\renewcommand\arraystretch{1.5}
	\setlength{\tabcolsep}{3 mm}
	\begin{center}
		\caption{The LO and NLO Results for $J^P=0^-$ with $\bar{b}b\bar{b}b$ system in the On-Shell scheme}
		\begin{tabular}{cccc|@{*}|ccc}
			\hline\hline
			\multirow{2}{*}{Current} &
			\multicolumn{3}{c|@{*}|}{LO}& \multicolumn{3}{c}{NLO(OS)} \\ \cline{2-7}
			& \makecell{$M_H$ \\ (GeV)} & \makecell{$S_0$ \\ ($\text{GeV}^2$)} & \makecell{$M_B^2$ \\ ($\text{GeV}^2$)} &  \makecell{$M_H$ \\ (GeV)} & \makecell{$S_0$ \\ ($\text{GeV}^2$)} & \makecell{$M_B^2$ \\ ($\text{GeV}^2$)} \\ \hline
			
			$J_{P,1}^{\text{M-M}}$ &$19.97^{+0.08}_{-0.15}$ &$428.(\pm 5\%)$ &$8.50(\pm 5\%)$    &$19.14^{+0.08}_{-0.55}$ &$374.(\pm 5\%)$ &$4.00(\pm 5\%)$\\
			$J_{P,2}^{\text{M-M}}$ &$20.08^{+0.06}_{-0.12}$ &$438.(\pm 5\%)$ &$9.00(\pm 5\%)$    &$19.24^{+0.07}_{-1.29}$ &$380.(\pm 5\%)$ &$4.50(\pm 5\%)$\\
			$J_{P,3}^{\text{M-M}}$ &$19.89^{+0.10}_{-0.18}$ &$418.(\pm 5\%)$ &$8.50(\pm 5\%)$    &$19.12^{+0.08}_{-0.21}$ &$374.(\pm 5\%)$ &$4.00(\pm 5\%)$\\ \hline
			$J_{P,1}^{\text{Di-Di}}$ &$19.97^{+0.08}_{-0.15}$ &$428.(\pm 5\%)$ &$8.50(\pm 5\%)$    &$19.14^{+0.08}_{-0.55}$ &$374.(\pm 5\%)$ &$4.00(\pm 5\%)$\\
			$J_{P,2}^{\text{Di-Di}}$ &$19.97^{+0.08}_{-0.15}$ &$428.(\pm 5\%)$ &$8.50(\pm 5\%)$    &$19.14^{+0.08}_{-0.93}$ &$374.(\pm 5\%)$ &$4.00(\pm 5\%)$\\
			$J_{P,3}^{\text{Di-Di}}$ &$20.01^{+0.09}_{-0.17}$ &$428.(\pm 5\%)$ &$9.00(\pm 5\%)$    &$19.18^{+0.09}_{-0.21}$ &$376.(\pm 5\%)$ &$4.50(\pm 5\%)$\\ \hline
			$J_{P,1}^{\text{Dia}}$ &$19.97^{+0.08}_{-0.15}$ &$428.(\pm 5\%)$ &$8.50(\pm 5\%)$    &$19.14^{+0.08}_{-0.55}$ &$374.(\pm 5\%)$ &$4.00(\pm 5\%)$\\
			$J_{P,2}^{\text{Dia}}$ &$20.09^{+0.06}_{-0.12}$ &$438.(\pm 5\%)$ &$9.00(\pm 5\%)$    &$19.24^{+0.07}_{-1.24}$ &$380.(\pm 5\%)$ &$4.50(\pm 5\%)$\\
			$J_{P,3}^{\text{Dia}}$ &$19.92^{+0.07}_{-0.14}$ &$426.(\pm 5\%)$ &$8.00(\pm 5\%)$    &$19.12^{+0.08}_{-0.22}$ &$374.(\pm 5\%)$ &$4.00(\pm 5\%)$\\ \hline\hline
		\end{tabular}
		\label{tab:4b-P-NLOresult-OS}
	\end{center}
\end{table}

\begin{figure}[H]
	\centering
	\subfigure[$\overline{\text{MS}}$]{
		\includegraphics[scale=0.4]{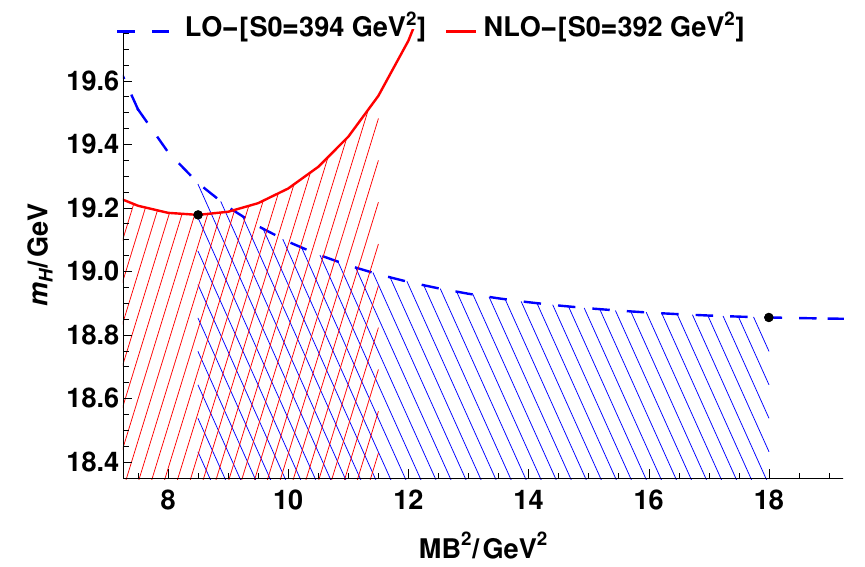}
		\includegraphics[scale=0.4]{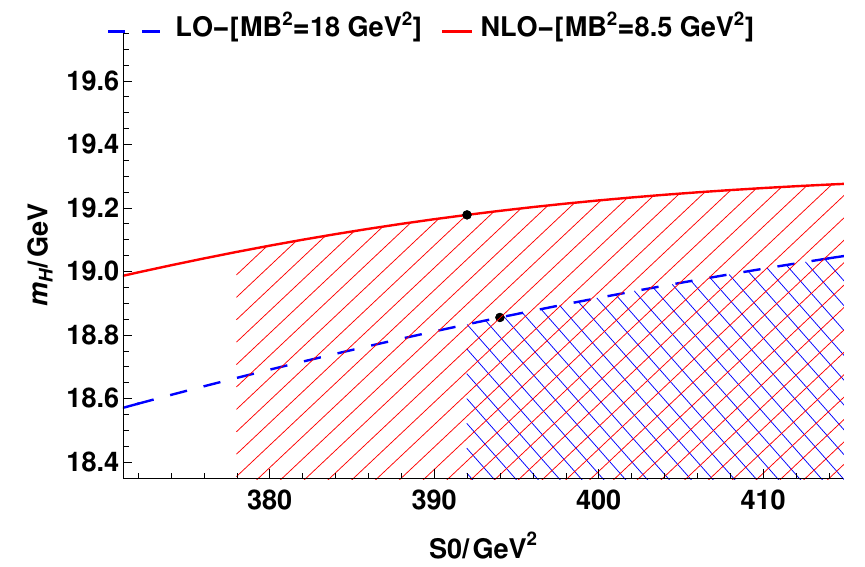}
	}\\
	\subfigure[OS]{
		\includegraphics[scale=0.4]{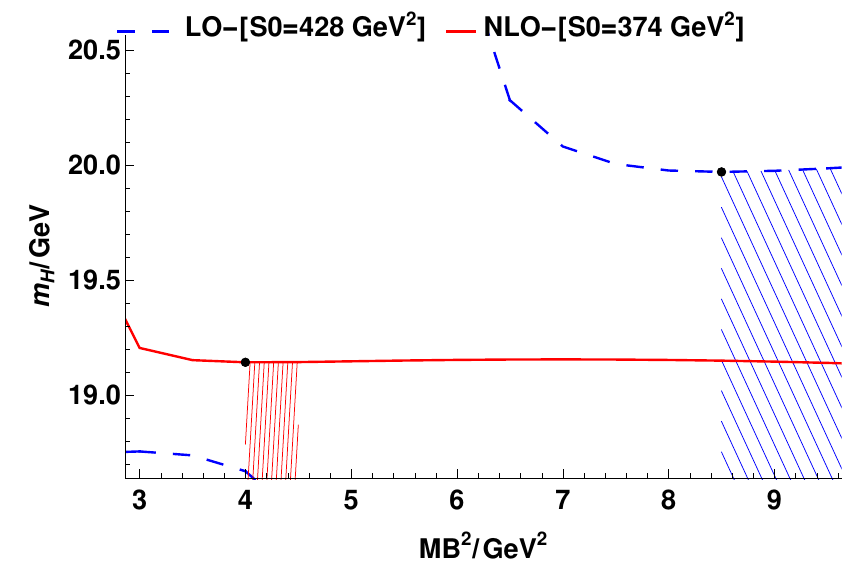}
		\includegraphics[scale=0.4]{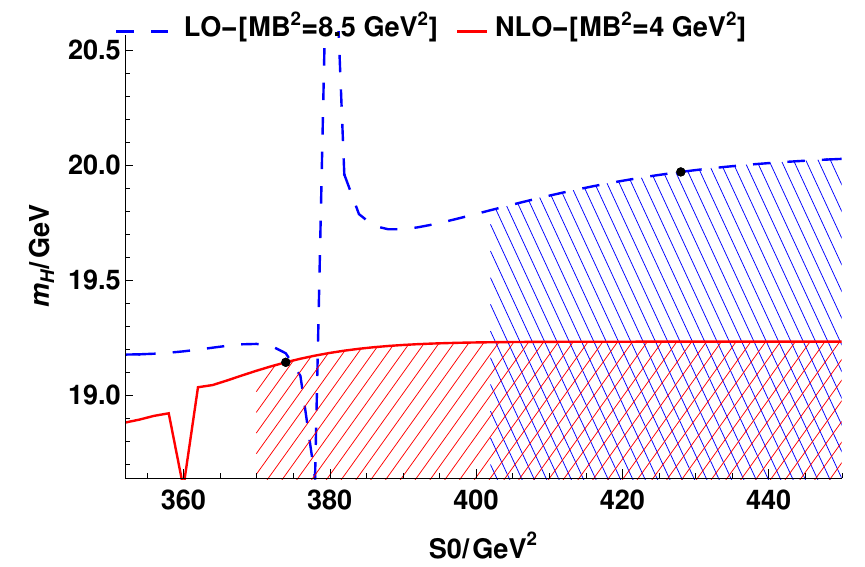}
	}
	\caption{\label{fig:4b-0--Mixed1-NLO-MSbar-OS}
		The Borel platform curves for $J_{P,1}^{\text{Dia}}$ with $J^{PC}=0^{--}$ in the $\overline{\text{MS}}$ and On-Shell schemes}
\end{figure}
\begin{figure}[H]
	\centering
	\subfigure[$\overline{\text{MS}}$]{
		\includegraphics[scale=0.4]{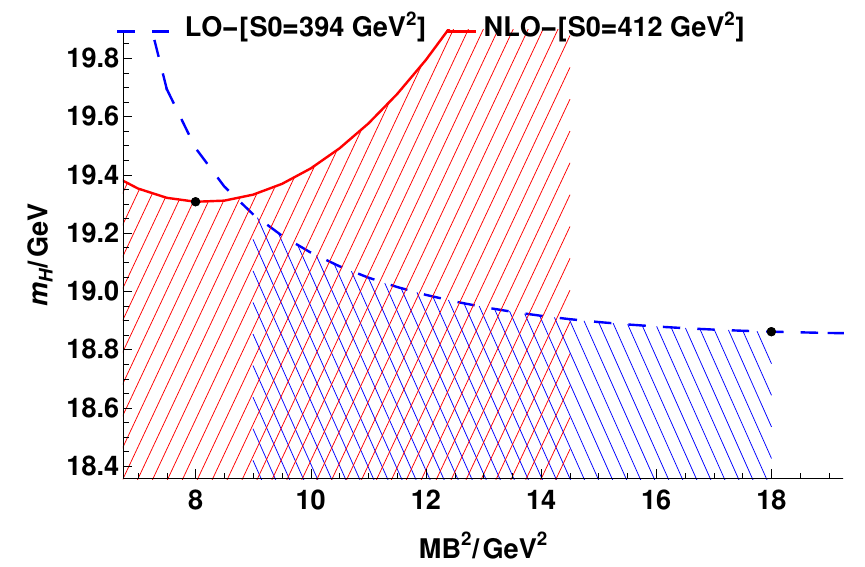}
		\includegraphics[scale=0.4]{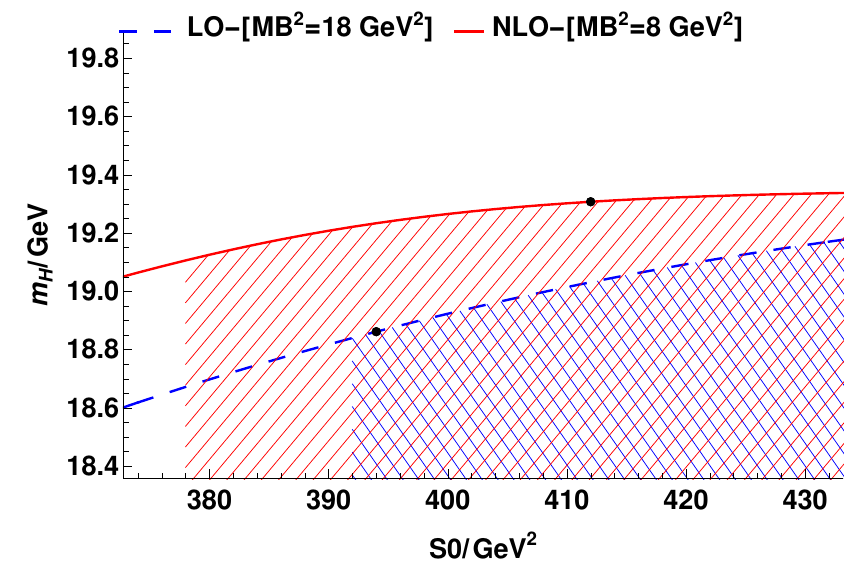}
	}\\
	\subfigure[OS]{
		\includegraphics[scale=0.4]{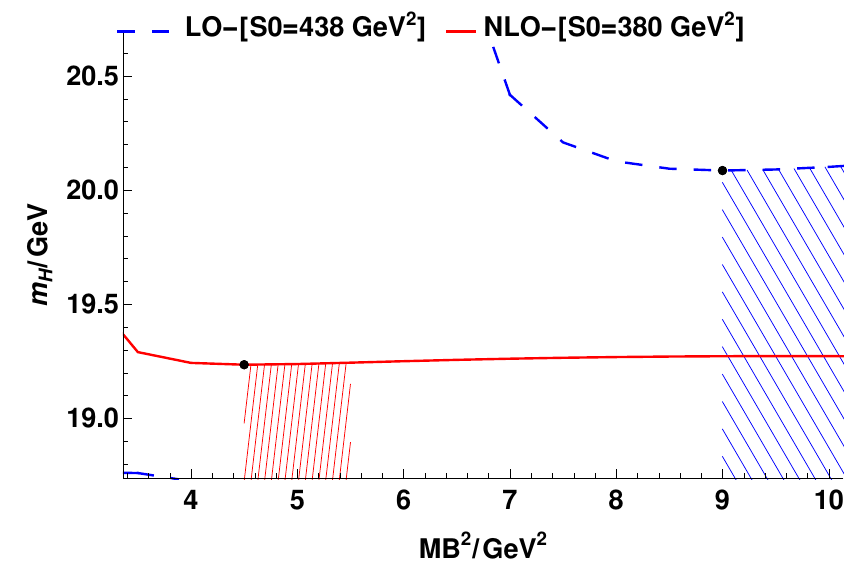}
		\includegraphics[scale=0.4]{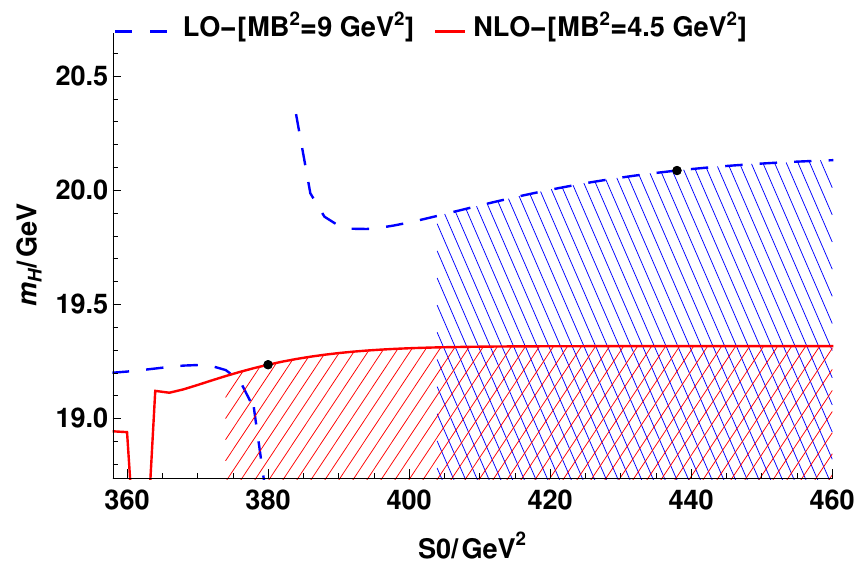}
	}
	\caption{\label{fig:4b-0--Mixed2-NLO-MSbar-OS}
		The Borel platform curves for $J_{P,2}^{\text{Dia}}$ with $J^{PC}=0^{-+}$ in the $\overline{\text{MS}}$ and On-Shell schemes}
\end{figure}
\begin{figure}[H]
	\centering
	\subfigure[$\overline{\text{MS}}$]{
		\includegraphics[scale=0.4]{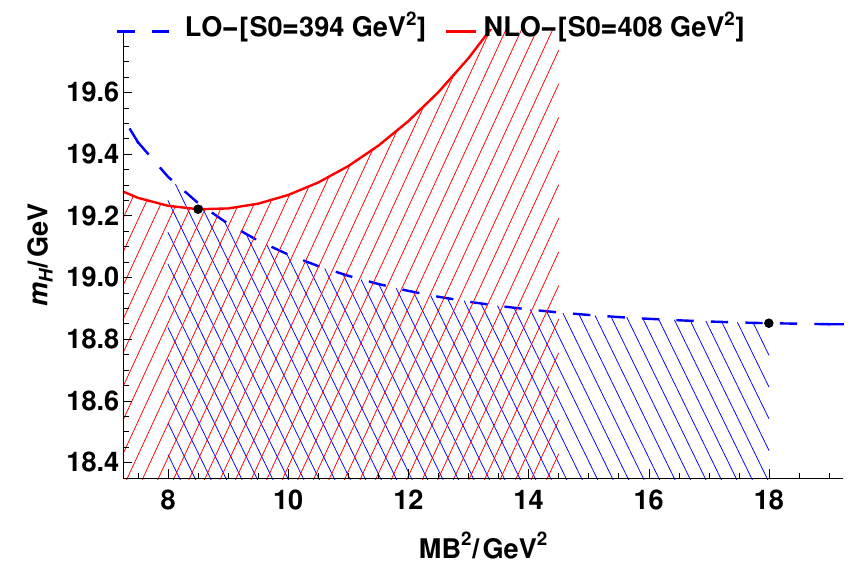}
		\includegraphics[scale=0.4]{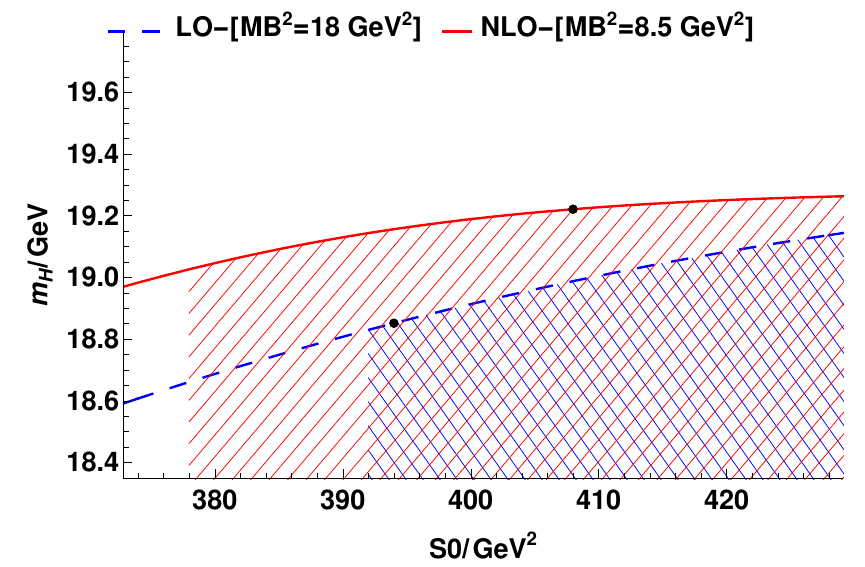}
	}\\
	\subfigure[OS]{
		\includegraphics[scale=0.4]{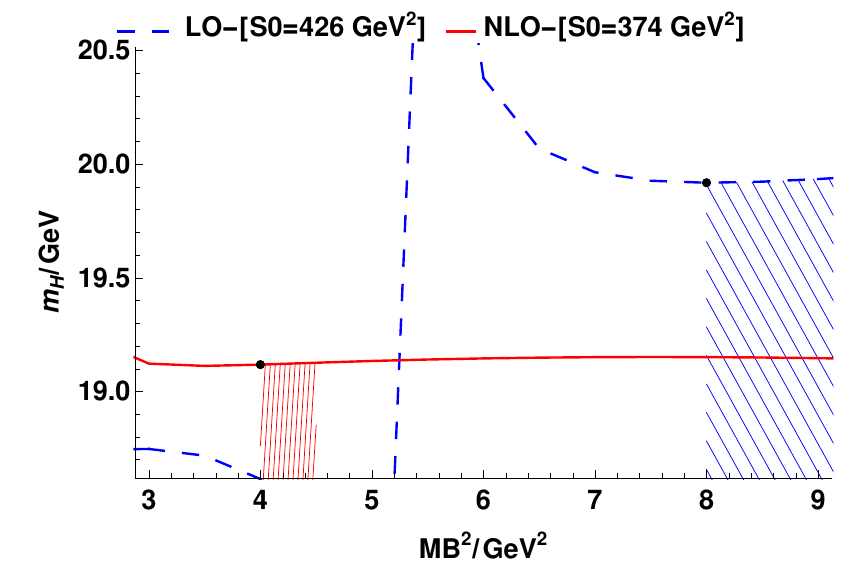}
		\includegraphics[scale=0.4]{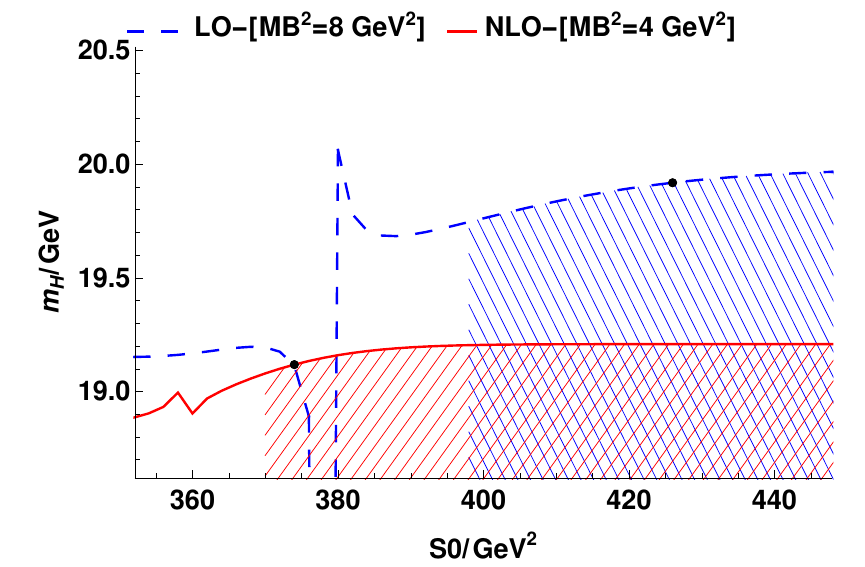}
	}
	\caption{\label{fig:4b-0--Mixed3-NLO-MSbar-OS}
		The Borel platform curves for $J_{P,3}^{\text{Dia}}$ with $J^{PC}=0^{-+}$ system in the $\overline{\text{MS}}$ and On-Shell schemes}
\end{figure}

\subsection{Numerical Results with $J^P=1^+$}
\begin{table}[H]
	\vspace{-0.4cm}
	\renewcommand\arraystretch{1.2}
	\setlength{\tabcolsep}{3 mm}
	\begin{center}
		\caption{The LO and NLO Results for $J^P=1^+$ with $\bar{b}b\bar{b}b$ system in the $\overline{\text{MS}}$ scheme}
		\begin{tabular}{cccc|@{*}|ccc}
			\hline\hline
			\multirow{2}{*}{Current} &
			\multicolumn{3}{c|@{*}|}{LO}& \multicolumn{3}{c}{NLO($\overline{\text{MS}}$)} \\ \cline{2-7}
			& \makecell{$M_H$ \\ (GeV)} & \makecell{$S_0$ \\ ($\text{GeV}^2$)} & \makecell{$M_B^2$ \\ ($\text{GeV}^2$)} &  \makecell{$M_H$ \\ (GeV)} & \makecell{$S_0$ \\ ($\text{GeV}^2$)} & \makecell{$M_B^2$ \\ ($\text{GeV}^2$)} \\ \hline
			
			$J_{A,1}^{\text{M-M}}$ &$19.21^{+0.20}_{-0.26}$ &$408.(\pm 5\%)$ &$18.00(\pm 5\%)$    &$19.53^{+0.11}_{-0.17}$ &$402.(\pm 5\%)$ &$7.50(\pm 5\%)$\\
			$J_{A,2}^{\text{M-M}}$ &$19.17^{+0.20}_{-0.26}$ &$406.(\pm 5\%)$ &$17.50(\pm 5\%)$    &$18.60^{+0.19}_{-0.26}$ &$384.(\pm 5\%)$ &$16.50(\pm 5\%)$\\
			$J_{A,3}^{\text{M-M}}$ &$18.50^{+0.17}_{-0.26}$ &$380.(\pm 5\%)$ &$19.00(\pm 5\%)$    &$18.97^{+0.06}_{-0.12}$ &$396.(\pm 5\%)$ &$9.50(\pm 5\%)$\\
			$J_{A,4}^{\text{M-M}}$ &$18.55^{+0.18}_{-0.25}$ &$382.(\pm 5\%)$ &$18.00(\pm 5\%)$    &$18.97^{+0.06}_{-0.11}$ &$398.(\pm 5\%)$ &$9.50(\pm 5\%)$\\ \hline
			$J_{A,1}^{\text{Di-Di}}$ &$19.17^{+0.20}_{-0.26}$ &$406.(\pm 5\%)$ &$17.50(\pm 5\%)$    &$19.40^{+0.12}_{-0.20}$ &$398.(\pm 5\%)$ &$8.00(\pm 5\%)$\\
			$J_{A,2}^{\text{Di-Di}}$ &$19.20^{+0.20}_{-0.26}$ &$408.(\pm 5\%)$ &$18.00(\pm 5\%)$    &$18.55^{+0.19}_{-0.26}$ &$382.(\pm 5\%)$ &$17.00(\pm 5\%)$\\
			$J_{A,3}^{\text{Di-Di}}$ &$19.17^{+0.20}_{-0.26}$ &$406.(\pm 5\%)$ &$17.50(\pm 5\%)$    &$19.40^{+0.12}_{-0.20}$ &$398.(\pm 5\%)$ &$8.00(\pm 5\%)$\\
			$J_{A,4}^{\text{Di-Di}}$ &$18.50^{+0.17}_{-0.25}$ &$380.(\pm 5\%)$ &$19.00(\pm 5\%)$    &$18.97^{+0.06}_{-0.11}$ &$398.(\pm 5\%)$ &$9.50(\pm 5\%)$\\ \hline
			$J_{A,1}^{\text{Dia}}$ &$19.17^{+0.20}_{-0.26}$ &$406.(\pm 5\%)$ &$17.00(\pm 5\%)$    &$18.55^{+0.20}_{-0.26}$ &$382.(\pm 5\%)$ &$17.50(\pm 5\%)$\\
			$J_{A,2}^{\text{Dia}}$ &$19.21^{+0.20}_{-0.26}$ &$408.(\pm 5\%)$ &$18.00(\pm 5\%)$    &$19.11^{+0.10}_{-0.18}$ &$404.(\pm 5\%)$ &$10.00(\pm 5\%)$\\
			$J_{A,3}^{\text{Dia}}$ &$18.51^{+0.17}_{-0.26}$ &$380.(\pm 5\%)$ &$19.00(\pm 5\%)$    &$18.97^{+0.06}_{-0.11}$ &$398.(\pm 5\%)$ &$9.50(\pm 5\%)$\\
			$J_{A,4}^{\text{Dia}}$ &$18.50^{+0.17}_{-0.26}$ &$380.(\pm 5\%)$ &$19.00(\pm 5\%)$    &$18.97^{+0.06}_{-0.12}$ &$396.(\pm 5\%)$ &$9.50(\pm 5\%)$\\ \hline\hline
		\end{tabular}
		
		\label{tab:4b-A-NLOresult-MSbar}
	\end{center}
\end{table}
\begin{table}[H]
	\vspace{-0.4cm}
	\renewcommand\arraystretch{1.2}
	\setlength{\tabcolsep}{3 mm}
	\begin{center}
		\caption{The LO and NLO Results for $J^P=1^+$ with $\bar{b}b\bar{b}b$ system in the On-Shell scheme}
		\begin{tabular}{cccc|@{*}|ccc}
			\hline\hline
			\multirow{2}{*}{Current} &
			\multicolumn{3}{c|@{*}|}{LO}& \multicolumn{3}{c}{NLO(OS)} \\ \cline{2-7}
			& \makecell{$M_H$ \\ (GeV)} & \makecell{$S_0$ \\ ($\text{GeV}^2$)} & \makecell{$M_B^2$ \\ ($\text{GeV}^2$)} &  \makecell{$M_H$ \\ (GeV)} & \makecell{$S_0$ \\ ($\text{GeV}^2$)} & \makecell{$M_B^2$ \\ ($\text{GeV}^2$)} \\ \hline
			
			$J_{A,1}^{\text{M-M}}$ &$20.46^{+0.11}_{-0.18}$ &$450.(\pm 5\%)$ &$10.50(\pm 5\%)$    &$19.47^{+0.32}_{-0.82}$ &$390.(\pm 5\%)$ &$5.50(\pm 5\%)$\\
			$J_{A,2}^{\text{M-M}}$ &$20.21^{+0.14}_{-0.22}$ &$432.(\pm 5\%)$ &$9.50(\pm 5\%)$    &$18.97^{+0.12}_{-1.47}$ &$368.(\pm 5\%)$ &$4.00(\pm 5\%)$\\
			$J_{A,3}^{\text{M-M}}$ &$19.64^{+0.06}_{-0.12}$ &$412.(\pm 5\%)$ &$7.50(\pm 5\%)$    &$18.98^{+0.07}_{-0.31}$ &$366.(\pm 5\%)$ &$3.50(\pm 5\%)$\\
			$J_{A,4}^{\text{M-M}}$ &$19.70^{+0.04}_{-0.09}$ &$424.(\pm 5\%)$ &$7.50(\pm 5\%)$    &$18.98^{+0.07}_{-0.27}$ &$366.(\pm 5\%)$ &$3.50(\pm 5\%)$\\ \hline
			$J_{A,1}^{\text{Di-Di}}$ &$20.28^{+0.14}_{-0.22}$ &$436.(\pm 5\%)$ &$10.00(\pm 5\%)$    &$19.36^{+0.11}_{-0.47}$ &$386.(\pm 5\%)$ &$5.00(\pm 5\%)$\\
			$J_{A,2}^{\text{Di-Di}}$ &$20.40^{+0.10}_{-0.17}$ &$448.(\pm 5\%)$ &$10.00(\pm 5\%)$    &$19.20^{+0.18}_{-0.32}$ &$382.(\pm 5\%)$ &$6.50(\pm 5\%)$\\
			$J_{A,3}^{\text{Di-Di}}$ &$20.32^{+0.09}_{-0.17}$ &$444.(\pm 5\%)$ &$9.50(\pm 5\%)$    &$19.36^{+0.11}_{-0.46}$ &$386.(\pm 5\%)$ &$5.00(\pm 5\%)$\\
			$J_{A,4}^{\text{Di-Di}}$ &$19.65^{+0.05}_{-0.11}$ &$414.(\pm 5\%)$ &$7.50(\pm 5\%)$    &$18.98^{+0.07}_{-0.30}$ &$366.(\pm 5\%)$ &$3.50(\pm 5\%)$\\ \hline
			$J_{A,1}^{\text{Dia}}$ &$20.18^{+0.14}_{-0.23}$ &$430.(\pm 5\%)$ &$9.50(\pm 5\%)$    &$19.20^{+0.18}_{-0.31}$ &$382.(\pm 5\%)$ &$6.50(\pm 5\%)$\\
			$J_{A,2}^{\text{Dia}}$ &$20.47^{+0.10}_{-0.17}$ &$452.(\pm 5\%)$ &$10.50(\pm 5\%)$    &$19.02^{+2.93}_{-0.42}$ &$370.(\pm 5\%)$ &$4.00(\pm 5\%)$\\
			$J_{A,3}^{\text{Dia}}$ &$19.68^{+0.04}_{-0.10}$ &$420.(\pm 5\%)$ &$7.50(\pm 5\%)$    &$18.98^{+0.07}_{-0.29}$ &$366.(\pm 5\%)$ &$3.50(\pm 5\%)$\\
			$J_{A,4}^{\text{Dia}}$ &$19.64^{+0.05}_{-0.11}$ &$414.(\pm 5\%)$ &$7.50(\pm 5\%)$    &$18.98^{+0.07}_{-0.31}$ &$366.(\pm 5\%)$ &$3.50(\pm 5\%)$\\ \hline\hline
		\end{tabular}
			\vspace{-0.8cm}
		\label{tab:4b-A-NLOresult-OS}
	\end{center}
\end{table}

\begin{figure}[H]
	\centering
	\subfigure[$\overline{\text{MS}}$]{
		\includegraphics[scale=0.4]{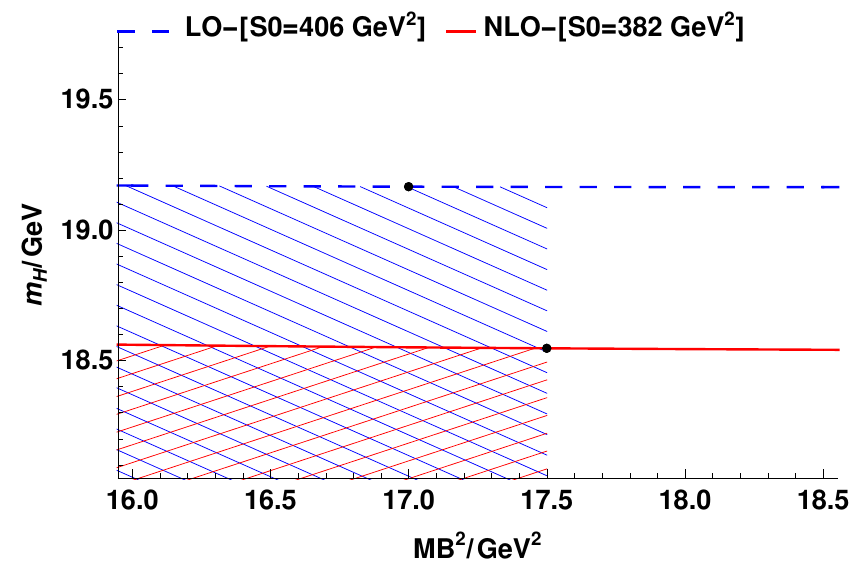}
		\includegraphics[scale=0.4]{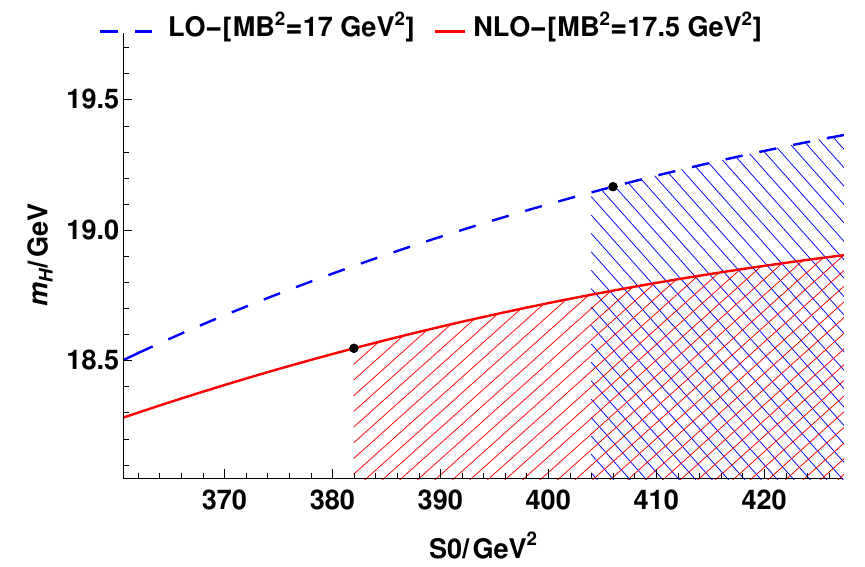}
	}\\
	\subfigure[OS]{
		\includegraphics[scale=0.4]{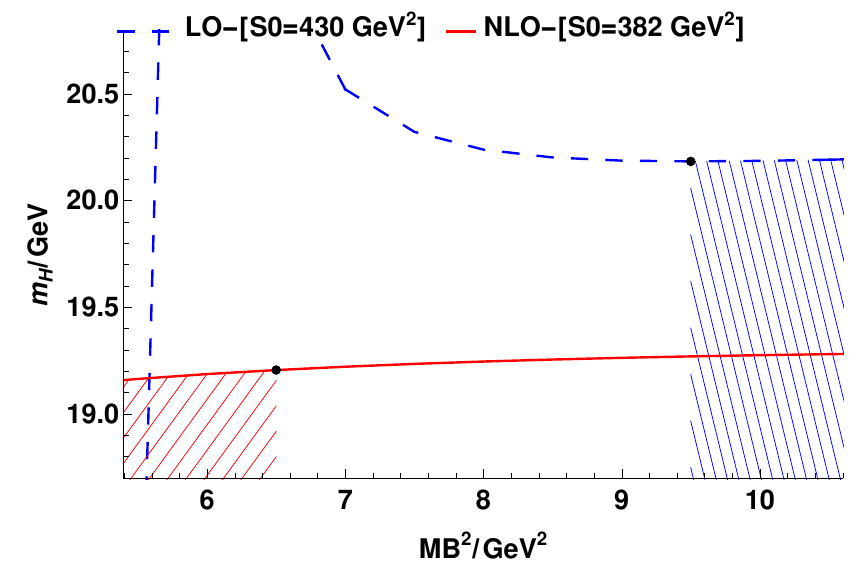}
		\includegraphics[scale=0.4]{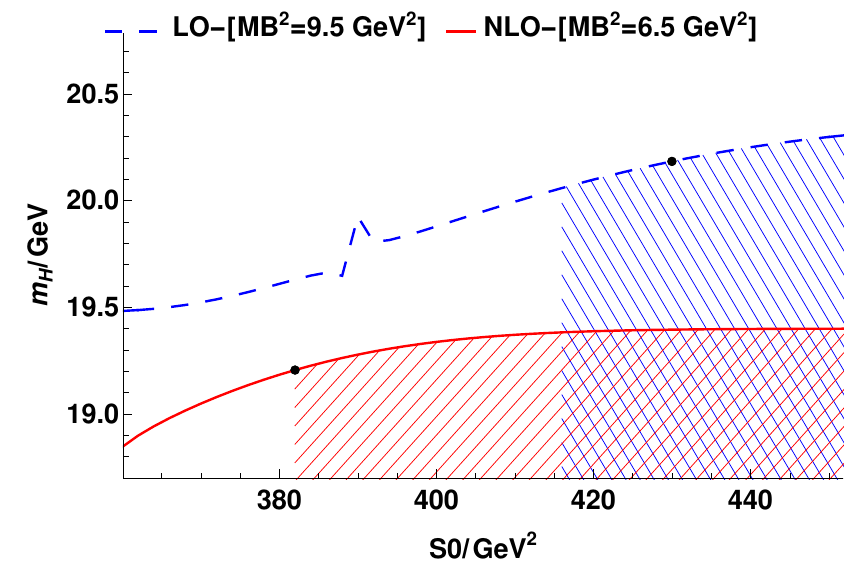}
	}
	\caption{\label{fig:4b-1+-Mixed1-NLO-MSbar-OS}
		The Borel platform curves for $J_{A,1}^{\text{Dia}}$ with $J^{PC}=1^{++}$ system in the $\overline{\text{MS}}$ and On-Shell schemes}
\end{figure}
\begin{figure}[H]
	\centering
	\subfigure[$\overline{\text{MS}}$]{
		\includegraphics[scale=0.4]{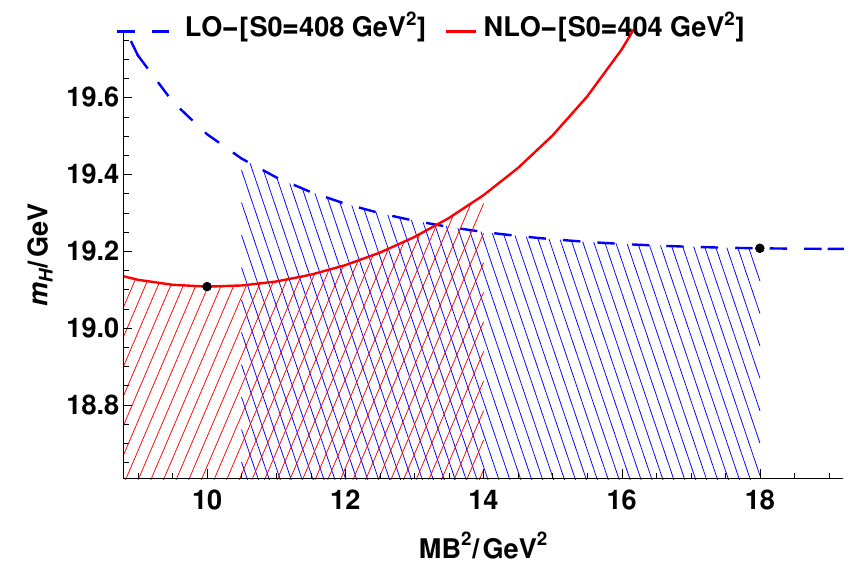}
		\includegraphics[scale=0.4]{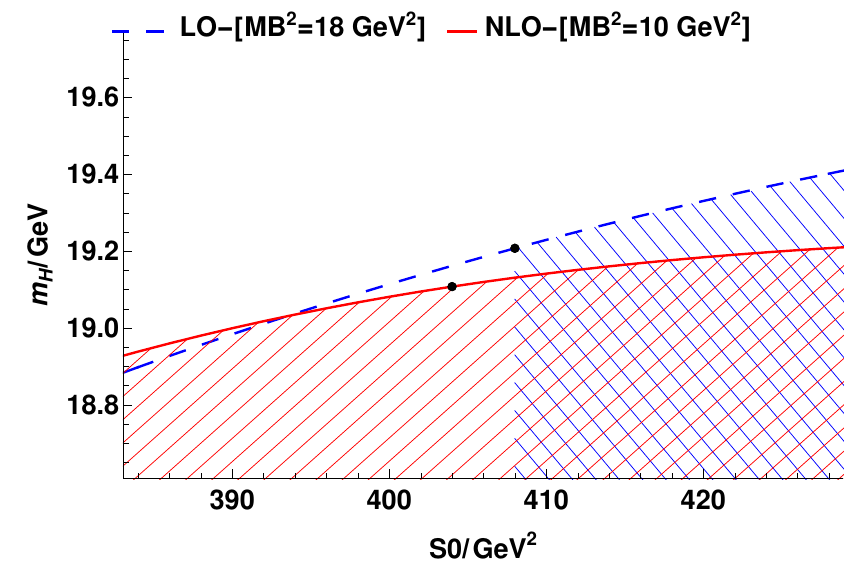}
	}\\
	\subfigure[OS]{
		\includegraphics[scale=0.4]{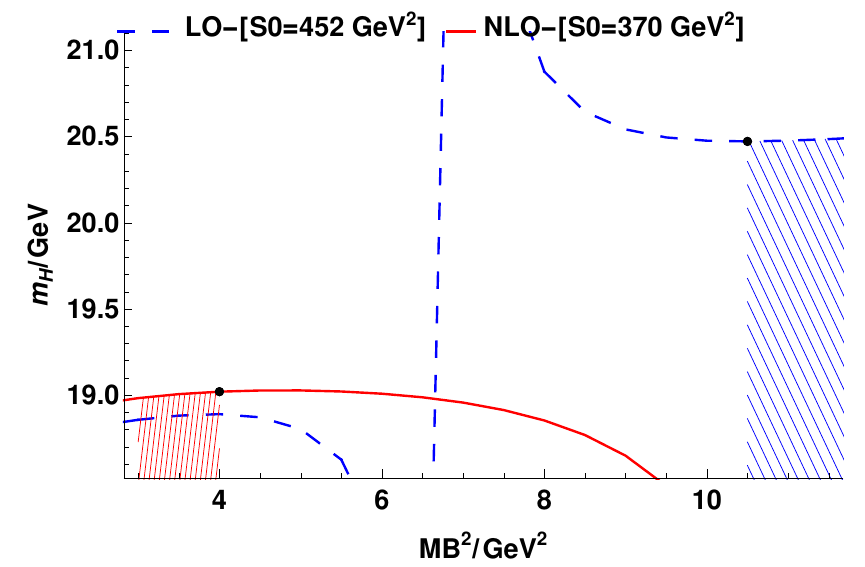}
		\includegraphics[scale=0.4]{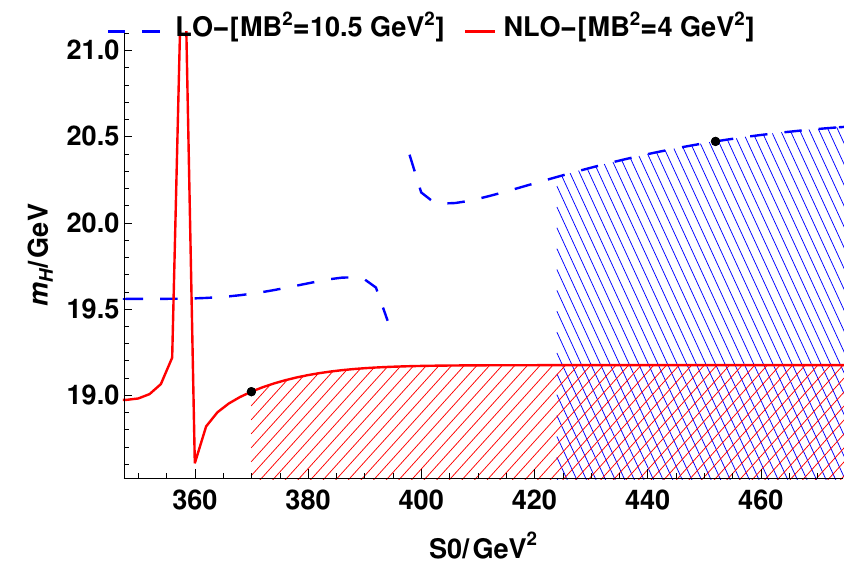}
	}
	\caption{\label{fig:4b-1+-Mixed2-NLO-MSbar-OS}
		The Borel platform curves for $J_{A,2}^{\text{Dia}}$ with $J^{PC}=1^{++}$ in the $\overline{\text{MS}}$ and On-Shell schemes}
\end{figure}
\begin{figure}[H]
	\centering
	\subfigure[$\overline{\text{MS}}$]{
		\includegraphics[scale=0.4]{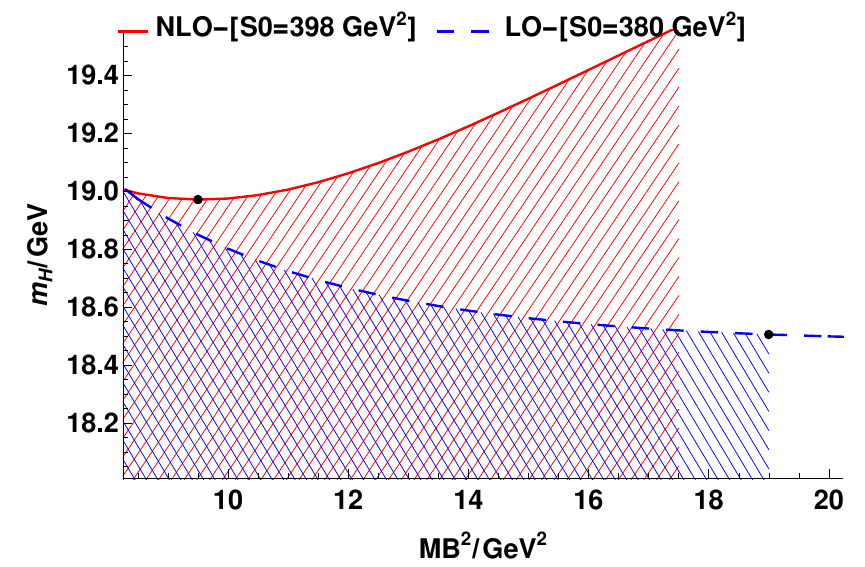}
		\includegraphics[scale=0.4]{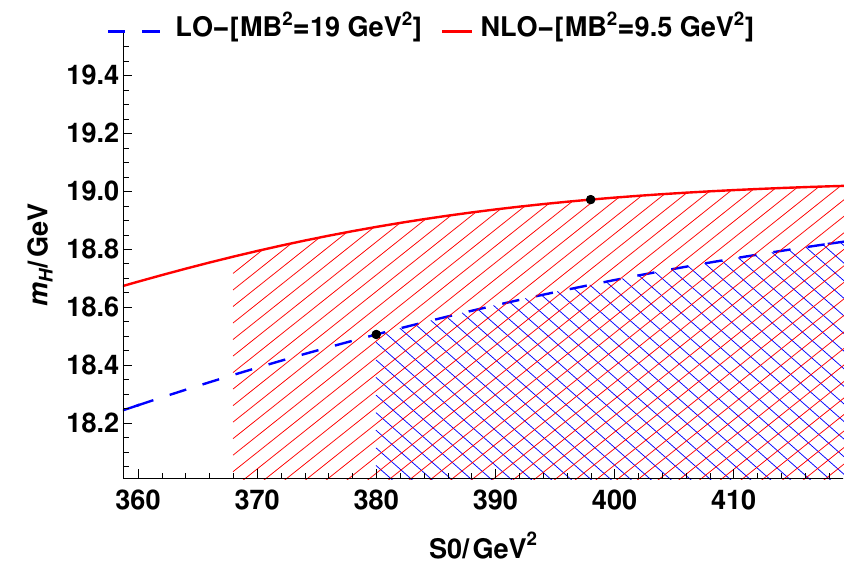}
	}\\
	\subfigure[OS]{
		\includegraphics[scale=0.4]{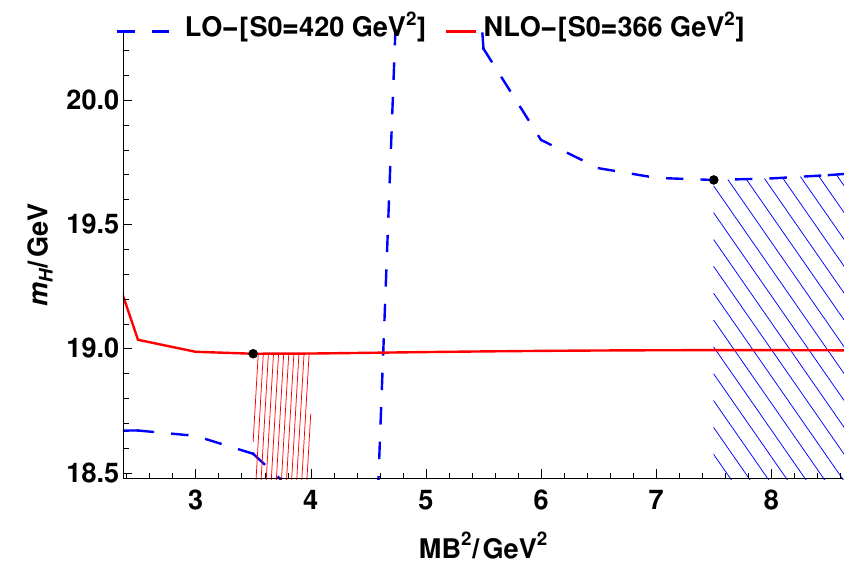}
		\includegraphics[scale=0.4]{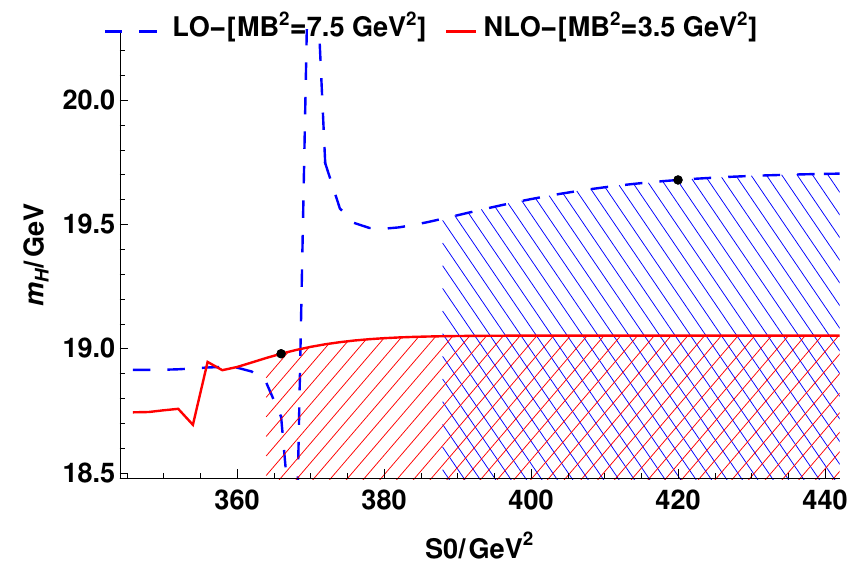}
	}
	\caption{\label{fig:4b-1+-Mixed3-NLO-MSbar-OS}
		The Borel platform curves for $J_{A,3}^{\text{Dia}}$ with $J^{PC}=1^{+-}$ in the $\overline{\text{MS}}$ and On-Shell schemes}
\end{figure}
\begin{figure}[H]
	\centering
	\subfigure[$\overline{\text{MS}}$]{
		\includegraphics[scale=0.4]{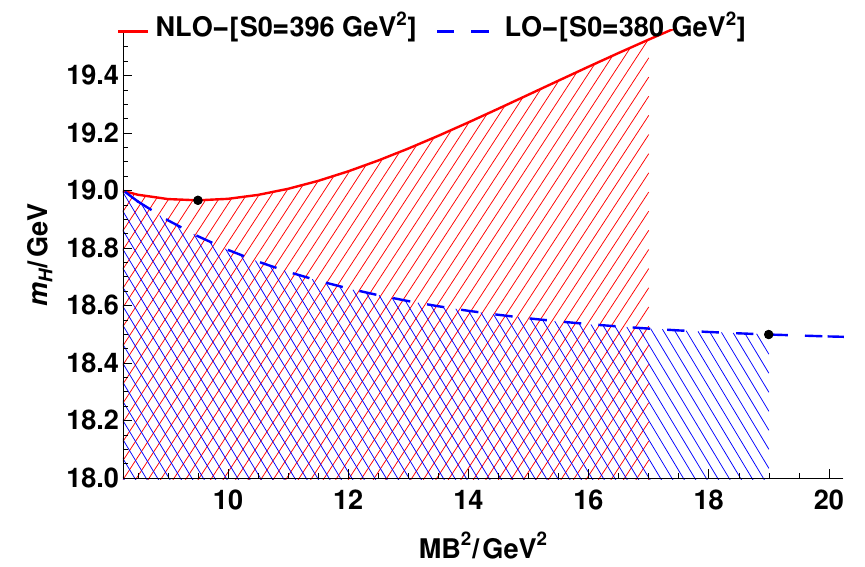}
		\includegraphics[scale=0.4]{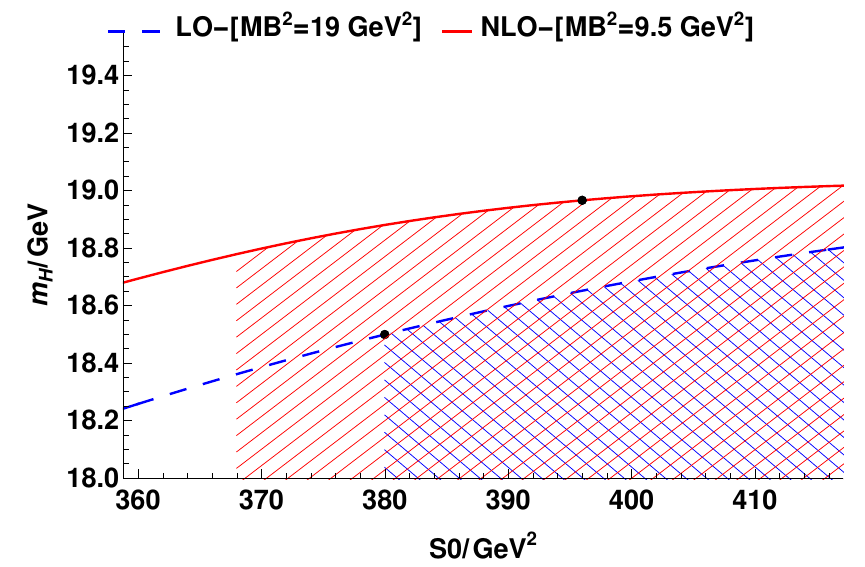}
	}\\
	\subfigure[OS]{
		\includegraphics[scale=0.4]{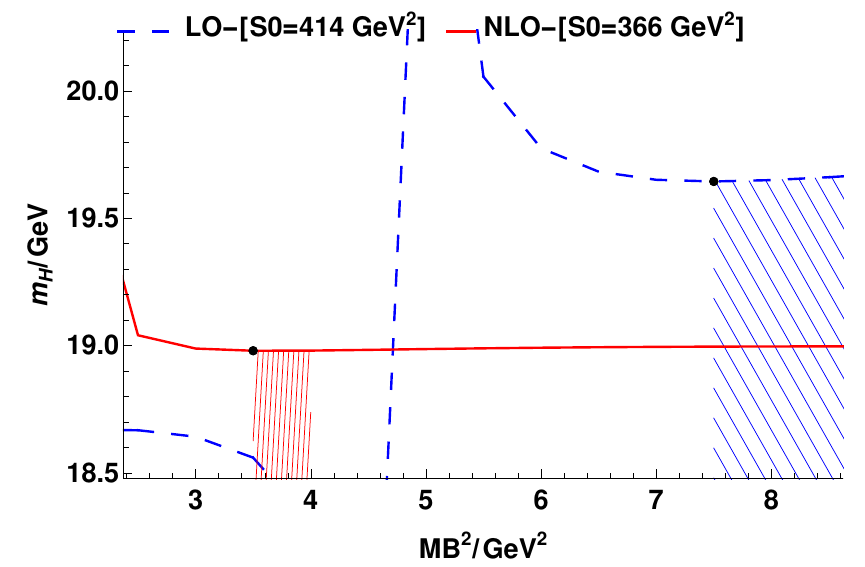}
		\includegraphics[scale=0.4]{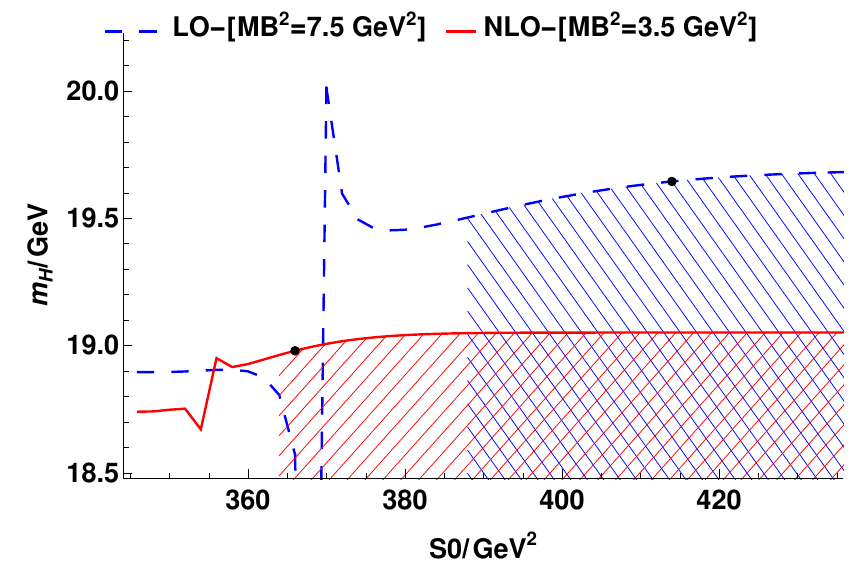}
	}
	\caption{\label{fig:4b-1+-Mixed4-NLO-MSbar-OS}
		The Borel platform curves for $J_{A,4}^{\text{Dia}}$ with $J^{PC}=1^{+-}$ in the $\overline{\text{MS}}$ and On-Shell schemes}
\end{figure}

\subsection{Numerical Results with $J^P=1^-$}
\begin{table}[H]
	\vspace{-0.4cm}
	\renewcommand\arraystretch{1.2}
	\setlength{\tabcolsep}{3 mm}
	\begin{center}
		\caption{The LO and NLO Results for $J^P=1^-$ with $\bar{b}b\bar{b}b$ system in the $\overline{\text{MS}}$ scheme}
		\begin{tabular}{cccc|@{*}|ccc}
			\hline\hline
			\multirow{2}{*}{Current} &
			\multicolumn{3}{c|@{*}|}{LO}& \multicolumn{3}{c}{NLO($\overline{\text{MS}}$)} \\ \cline{2-7}
			& \makecell{$M_H$ \\ (GeV)} & \makecell{$S_0$ \\ ($\text{GeV}^2$)} & \makecell{$M_B^2$ \\ ($\text{GeV}^2$)} &  \makecell{$M_H$ \\ (GeV)} & \makecell{$S_0$ \\ ($\text{GeV}^2$)} & \makecell{$M_B^2$ \\ ($\text{GeV}^2$)} \\ \hline
			
			$J_{V,1}^{\text{M-M}}$ &$18.86^{+0.19}_{-0.25}$ &$394.(\pm 5\%)$ &$18.50(\pm 5\%)$    &$19.31^{+0.04}_{-0.09}$ &$412.(\pm 5\%)$ &$8.00(\pm 5\%)$\\
			$J_{V,2}^{\text{M-M}}$ &$18.85^{+0.19}_{-0.26}$ &$394.(\pm 5\%)$ &$18.00(\pm 5\%)$    &$19.23^{+0.07}_{-0.13}$ &$402.(\pm 5\%)$ &$8.50(\pm 5\%)$\\
			$J_{V,3}^{\text{M-M}}$ &$18.87^{+0.19}_{-0.26}$ &$394.(\pm 5\%)$ &$18.00(\pm 5\%)$    &$19.31^{+0.04}_{-0.09}$ &$412.(\pm 5\%)$ &$8.00(\pm 5\%)$\\
			$J_{V,4}^{\text{M-M}}$ &$18.85^{+0.20}_{-0.26}$ &$394.(\pm 5\%)$ &$18.00(\pm 5\%)$    &$19.22^{+0.05}_{-0.11}$ &$408.(\pm 5\%)$ &$8.50(\pm 5\%)$\\ \hline
			$J_{V,1}^{\text{Di-Di}}$ &$18.85^{+0.19}_{-0.25}$ &$394.(\pm 5\%)$ &$18.00(\pm 5\%)$    &$19.26^{+0.05}_{-0.11}$ &$408.(\pm 5\%)$ &$8.50(\pm 5\%)$\\
			$J_{V,2}^{\text{Di-Di}}$ &$18.86^{+0.19}_{-0.26}$ &$394.(\pm 5\%)$ &$18.00(\pm 5\%)$    &$19.18^{+0.13}_{-0.20}$ &$388.(\pm 5\%)$ &$8.50(\pm 5\%)$\\
			$J_{V,3}^{\text{Di-Di}}$ &$18.86^{+0.19}_{-0.26}$ &$394.(\pm 5\%)$ &$18.00(\pm 5\%)$    &$19.25^{+0.07}_{-0.13}$ &$404.(\pm 5\%)$ &$8.50(\pm 5\%)$\\
			$J_{V,4}^{\text{Di-Di}}$ &$18.86^{+0.19}_{-0.26}$ &$394.(\pm 5\%)$ &$18.00(\pm 5\%)$    &$19.23^{+0.07}_{-0.13}$ &$402.(\pm 5\%)$ &$8.50(\pm 5\%)$\\ \hline
			$J_{V,1}^{\text{Dia}}$ &$18.85^{+0.19}_{-0.25}$ &$394.(\pm 5\%)$ &$18.00(\pm 5\%)$    &$19.25^{+0.05}_{-0.10}$ &$410.(\pm 5\%)$ &$8.50(\pm 5\%)$\\
			$J_{V,2}^{\text{Dia}}$ &$18.86^{+0.19}_{-0.25}$ &$394.(\pm 5\%)$ &$18.50(\pm 5\%)$    &$19.31^{+0.04}_{-0.09}$ &$412.(\pm 5\%)$ &$8.00(\pm 5\%)$\\
			$J_{V,3}^{\text{Dia}}$ &$18.85^{+0.20}_{-0.26}$ &$394.(\pm 5\%)$ &$18.00(\pm 5\%)$    &$19.12^{+0.12}_{-0.19}$ &$390.(\pm 5\%)$ &$9.00(\pm 5\%)$\\
			$J_{V,4}^{\text{Dia}}$ &$18.87^{+0.19}_{-0.26}$ &$394.(\pm 5\%)$ &$18.00(\pm 5\%)$    &$19.31^{+0.04}_{-0.09}$ &$412.(\pm 5\%)$ &$8.00(\pm 5\%)$\\ \hline\hline
		\end{tabular}
		
		\label{tab:4b-V-NLOresult-MSbar}
	\end{center}
\end{table}

\begin{table}[H]
	\vspace{-0.4cm}
	\renewcommand\arraystretch{1.2}
	\setlength{\tabcolsep}{3 mm}
	\begin{center}
		\caption{The LO and NLO Results for $J^P=1^-$ with $\bar{b}b\bar{b}b$ system in the On-Shell scheme}
		\begin{tabular}{cccc|@{*}|ccc}
			\hline\hline
			\multirow{2}{*}{Current} &
			\multicolumn{3}{c|@{*}|}{LO}& \multicolumn{3}{c}{NLO(OS)} \\ \cline{2-7}
			& \makecell{$M_H$ \\ (GeV)} & \makecell{$S_0$ \\ ($\text{GeV}^2$)} & \makecell{$M_B^2$ \\ ($\text{GeV}^2$)} &  \makecell{$M_H$ \\ (GeV)} & \makecell{$S_0$ \\ ($\text{GeV}^2$)} & \makecell{$M_B^2$ \\ ($\text{GeV}^2$)} \\ \hline
			
			$J_{V,1}^{\text{M-M}}$ &$20.06^{+0.10}_{-0.17}$ &$430.(\pm 5\%)$ &$9.50(\pm 5\%)$    &$19.24^{+0.07}_{-0.77}$ &$380.(\pm 5\%)$ &$4.50(\pm 5\%)$\\
			$J_{V,2}^{\text{M-M}}$ &$19.99^{+0.10}_{-0.18}$ &$426.(\pm 5\%)$ &$9.00(\pm 5\%)$    &$19.17^{+0.09}_{-0.20}$ &$376.(\pm 5\%)$ &$4.50(\pm 5\%)$\\
			$J_{V,3}^{\text{M-M}}$ &$20.10^{+0.06}_{-0.12}$ &$440.(\pm 5\%)$ &$9.00(\pm 5\%)$    &$19.24^{+0.07}_{-1.03}$ &$380.(\pm 5\%)$ &$4.50(\pm 5\%)$\\
			$J_{V,4}^{\text{M-M}}$ &$19.92^{+0.07}_{-0.14}$ &$426.(\pm 5\%)$ &$8.00(\pm 5\%)$    &$19.12^{+0.08}_{-0.20}$ &$374.(\pm 5\%)$ &$4.00(\pm 5\%)$\\ \hline
			$J_{V,1}^{\text{Di-Di}}$ &$20.00^{+0.10}_{-0.17}$ &$426.(\pm 5\%)$ &$9.00(\pm 5\%)$    &$19.18^{+0.09}_{-0.21}$ &$376.(\pm 5\%)$ &$4.50(\pm 5\%)$\\
			$J_{V,2}^{\text{Di-Di}}$ &$20.06^{+0.07}_{-0.14}$ &$434.(\pm 5\%)$ &$9.00(\pm 5\%)$    &$19.19^{+0.09}_{-0.23}$ &$376.(\pm 5\%)$ &$4.50(\pm 5\%)$\\
			$J_{V,3}^{\text{Di-Di}}$ &$20.01^{+0.09}_{-0.17}$ &$428.(\pm 5\%)$ &$9.00(\pm 5\%)$    &$19.18^{+0.10}_{-0.21}$ &$376.(\pm 5\%)$ &$4.50(\pm 5\%)$\\
			$J_{V,4}^{\text{Di-Di}}$ &$19.96^{+0.08}_{-0.16}$ &$426.(\pm 5\%)$ &$8.50(\pm 5\%)$    &$19.14^{+0.08}_{-0.92}$ &$374.(\pm 5\%)$ &$4.00(\pm 5\%)$\\ \hline
			$J_{V,1}^{\text{Dia}}$ &$20.00^{+0.06}_{-0.13}$ &$432.(\pm 5\%)$ &$8.50(\pm 5\%)$    &$19.15^{+0.08}_{-0.44}$ &$374.(\pm 5\%)$ &$4.00(\pm 5\%)$\\
			$J_{V,2}^{\text{Dia}}$ &$20.05^{+0.10}_{-0.18}$ &$428.(\pm 5\%)$ &$9.50(\pm 5\%)$    &$19.23^{+0.07}_{-1.02}$ &$380.(\pm 5\%)$ &$4.50(\pm 5\%)$\\
			$J_{V,3}^{\text{Dia}}$ &$19.90^{+0.07}_{-0.15}$ &$424.(\pm 5\%)$ &$8.00(\pm 5\%)$    &$19.12^{+0.08}_{-0.21}$ &$374.(\pm 5\%)$ &$4.00(\pm 5\%)$\\
			$J_{V,4}^{\text{Dia}}$ &$20.10^{+0.06}_{-0.12}$ &$440.(\pm 5\%)$ &$9.00(\pm 5\%)$    &$19.24^{+0.07}_{-1.03}$ &$380.(\pm 5\%)$ &$4.50(\pm 5\%)$\\ \hline\hline
		\end{tabular}
			\vspace{-0.8cm}
		\label{tab:4b-V-NLOresult-OS}
	\end{center}

\end{table}

\begin{figure}[H]
	\centering
	\subfigure[$\overline{\text{MS}}$]{
		\includegraphics[scale=0.4]{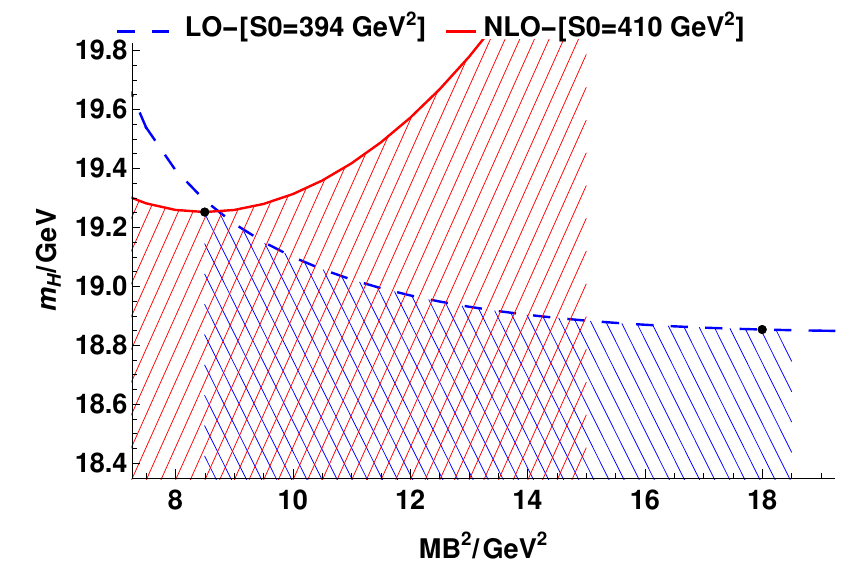}
		\includegraphics[scale=0.4]{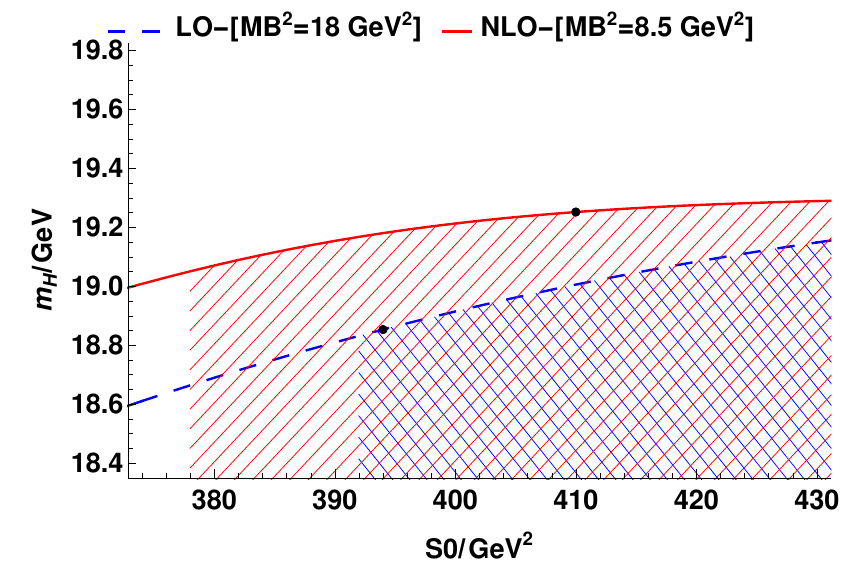}
	}\\
	\subfigure[OS]{
		\includegraphics[scale=0.4]{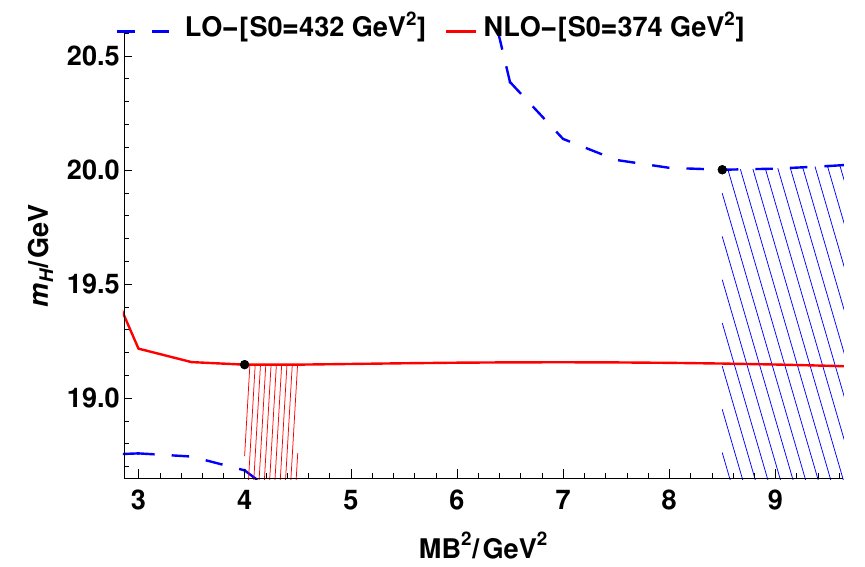}
		\includegraphics[scale=0.4]{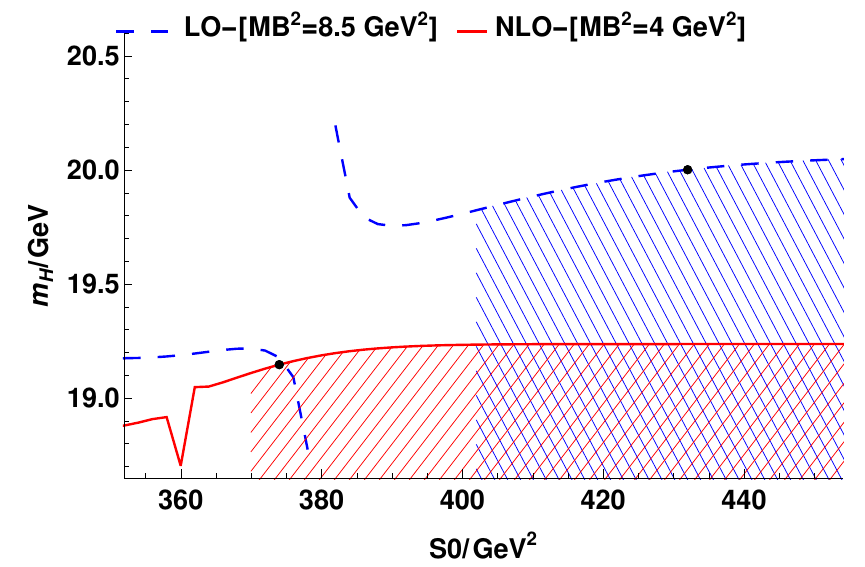}
	}
	\caption{\label{fig:4b-1--Mixed1-NLO-MSbar-OS}
		The Borel platform curves for $J_{V,1}^{\text{Dia}}$ with $J^{PC}=1^{--}$ in the $\overline{\text{MS}}$ and On-Shell schemes}
\end{figure}
\begin{figure}[H]
	\centering
	\subfigure[$\overline{\text{MS}}$]{
		\includegraphics[scale=0.4]{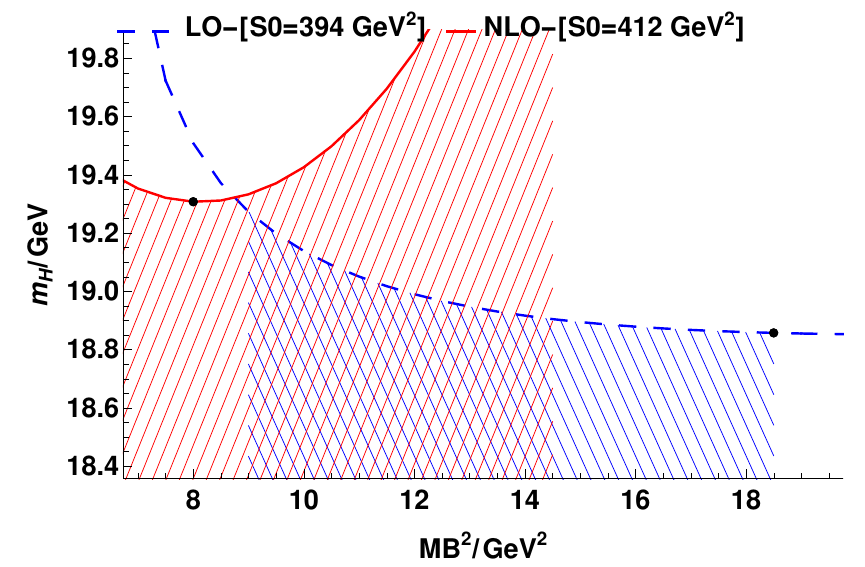}
		\includegraphics[scale=0.4]{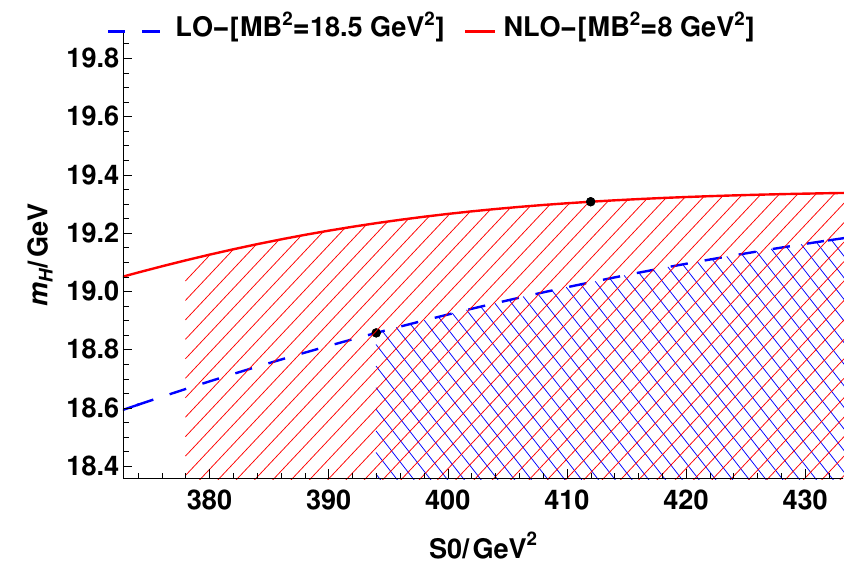}
	}\\
	\subfigure[OS]{
		\includegraphics[scale=0.4]{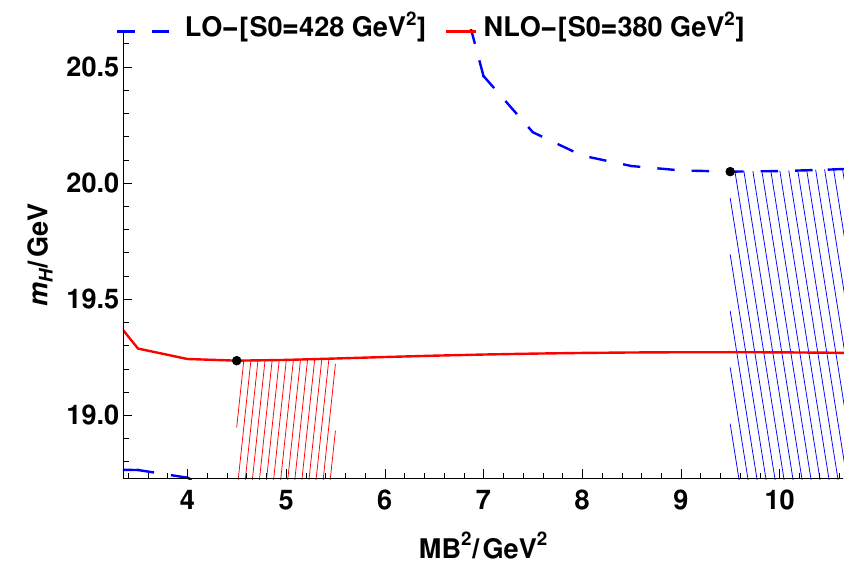}
		\includegraphics[scale=0.4]{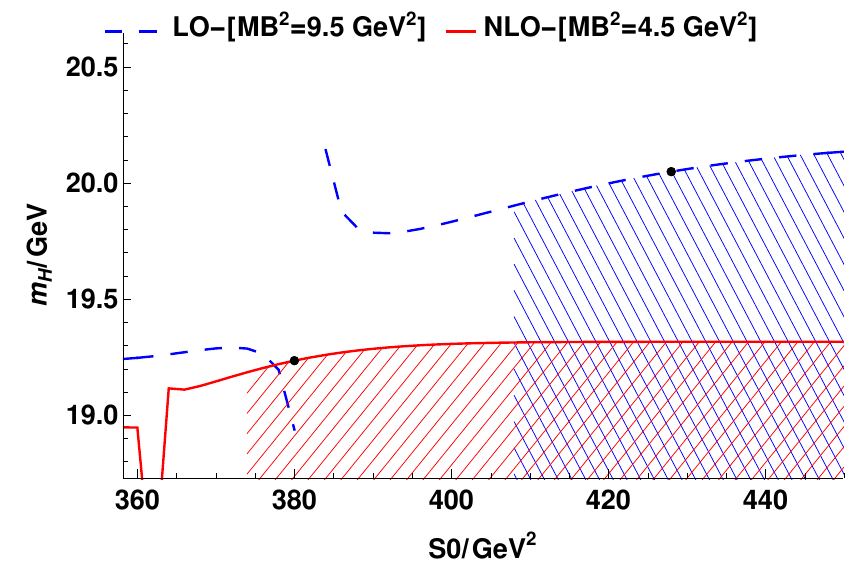}
	}
	\caption{\label{fig:4b-1--Mixed2-NLO-MSbar-OS}
		The Borel platform curves for $J_{V,2}^{\text{Dia}}$ with $J^{PC}=1^{--}$ in the $\overline{\text{MS}}$ and On-Shell schemes}
\end{figure}
\begin{figure}[H]
	\centering
	\subfigure[$\overline{\text{MS}}$]{
		\includegraphics[scale=0.4]{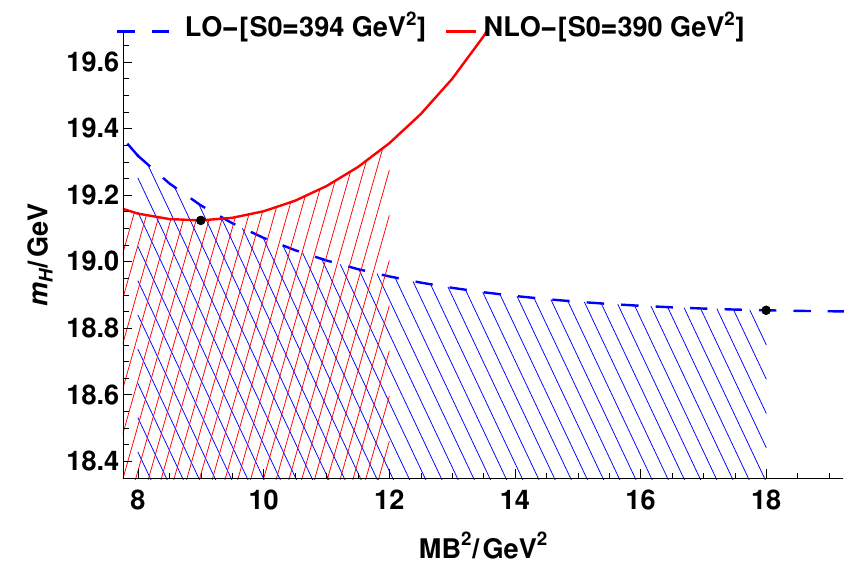}
		\includegraphics[scale=0.4]{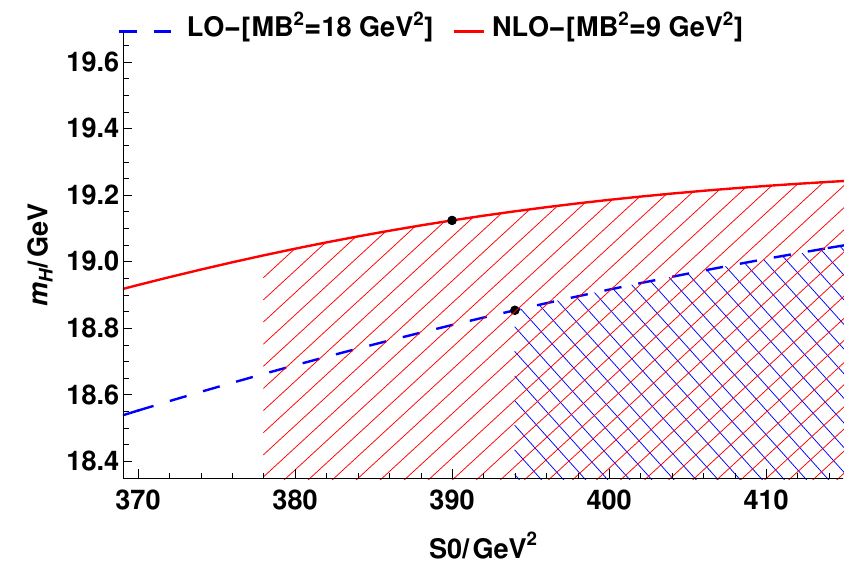}
	}\\
	\subfigure[OS]{
		\includegraphics[scale=0.4]{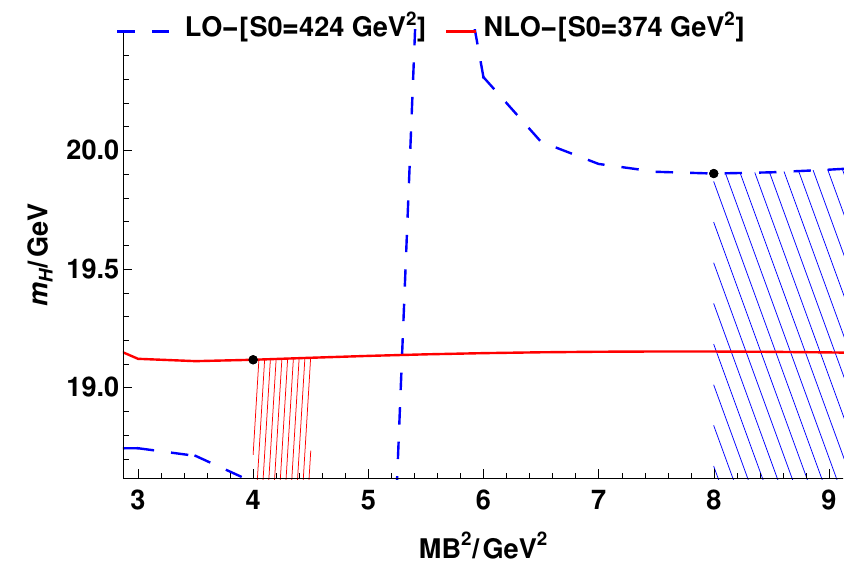}
		\includegraphics[scale=0.4]{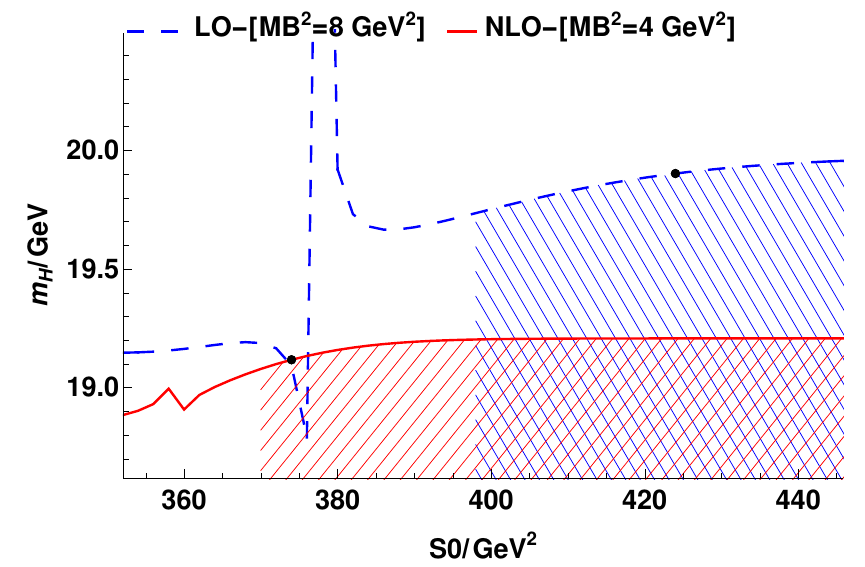}
	}
	\caption{\label{fig:4b-1--Mixed3-NLO-MSbar-OS}
		The Borel platform curves for $J_{V,3}^{\text{Dia}}$ with $J^{PC}=1^{-+}$ in the $\overline{\text{MS}}$ and On-Shell schemes}
\end{figure}
\begin{figure}[H]
	\centering
	\subfigure[$\overline{\text{MS}}$]{
		\includegraphics[scale=0.4]{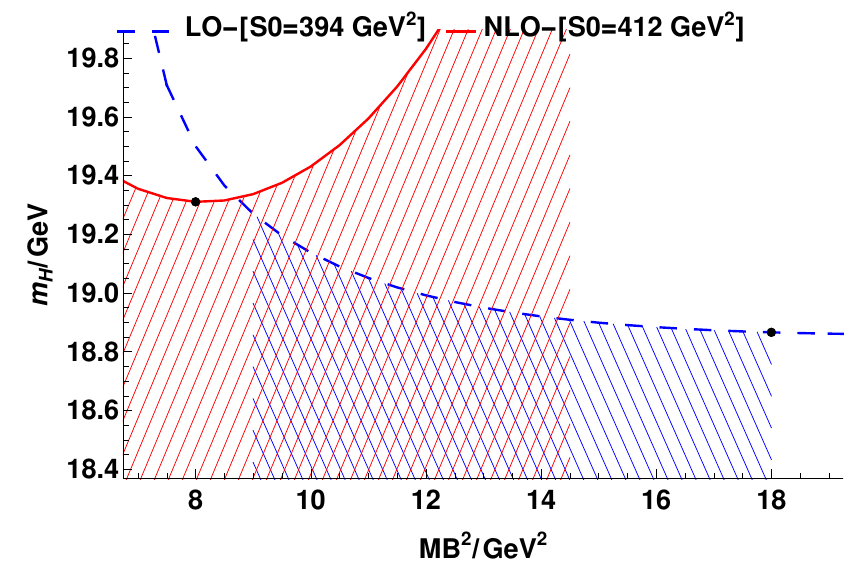}
		\includegraphics[scale=0.4]{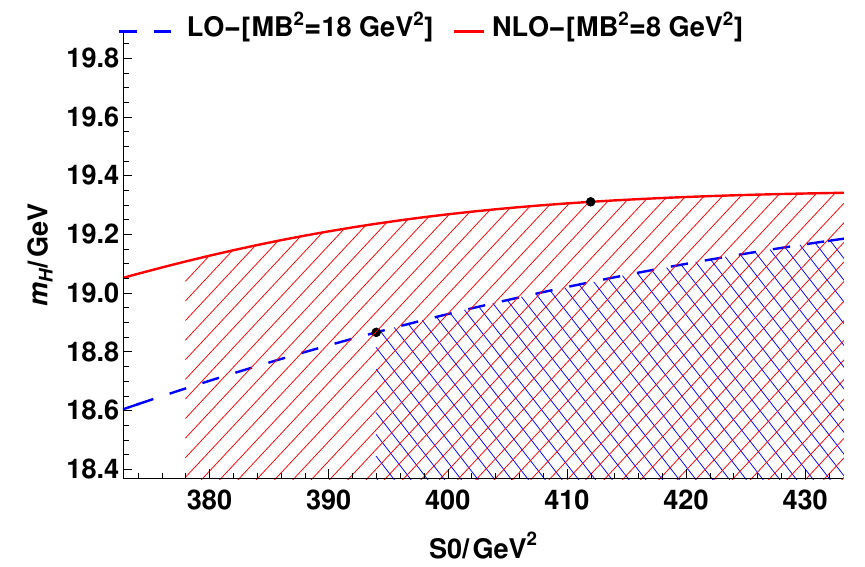}
	}\\
	\subfigure[OS]{
		\includegraphics[scale=0.4]{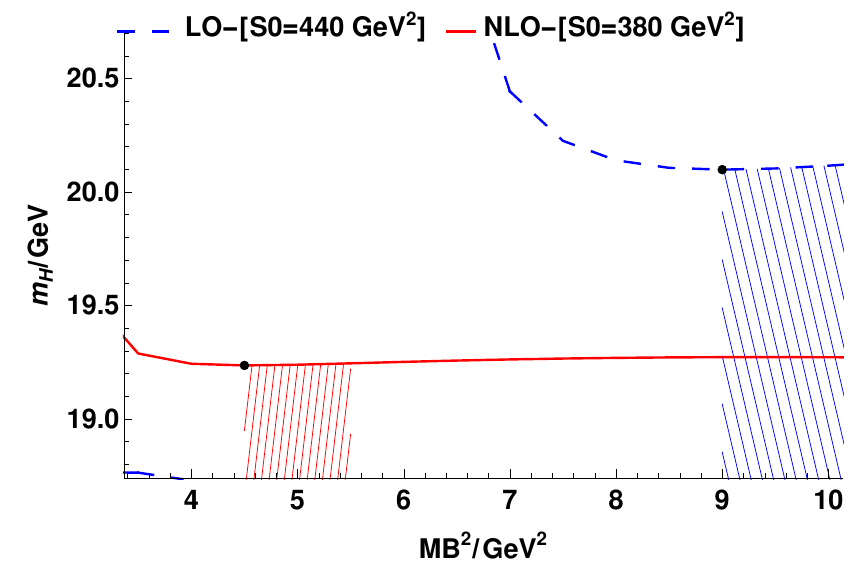}
		\includegraphics[scale=0.4]{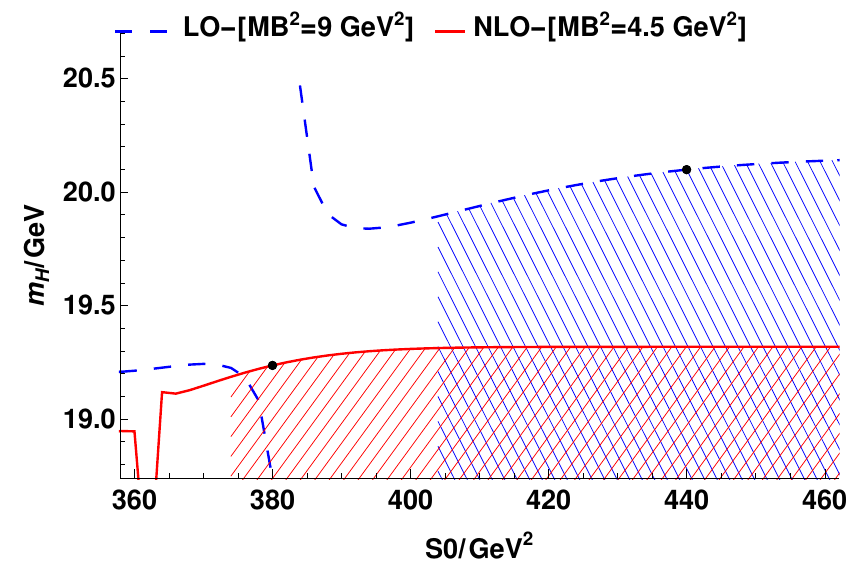}
	}
	\caption{\label{fig:4b-1--Mixed4-NLO-MSbar-OS}
		The Borel platform curves for $J_{V,4}^{\text{Dia}}$ with $J^{PC}=1^{-+}$ in the $\overline{\text{MS}}$ and On-Shell schemes}
\end{figure}

\subsection{Numerical Results with $J^P=2^+$}

\begin{table}[H]
	\vspace{-0.4cm}
	\renewcommand\arraystretch{1.5}
	\setlength{\tabcolsep}{3 mm}
	\begin{center}
		\caption{The LO and NLO Results for $J^P=2^+$ with $\bar{b}b\bar{b}b$ system in the $\overline{\text{MS}}$ scheme}
		\begin{tabular}{cccc|@{*}|ccc}
			\hline\hline
			\multirow{2}{*}{Current} &
			\multicolumn{3}{c|@{*}|}{LO}& \multicolumn{3}{c}{NLO($\overline{\text{MS}}$)} \\ \cline{2-7}
			& \makecell{$M_H$ \\ (GeV)} & \makecell{$S_0$ \\ ($\text{GeV}^2$)} & \makecell{$M_B^2$ \\ ($\text{GeV}^2$)} &  \makecell{$M_H$ \\ (GeV)} & \makecell{$S_0$ \\ ($\text{GeV}^2$)} & \makecell{$M_B^2$ \\ ($\text{GeV}^2$)} \\ \hline
			
			$J_{T,1}^{\text{M-M}}$ &$18.50^{+0.17}_{-0.25}$ &$380.(\pm 5\%)$ &$19.00(\pm 5\%)$    &$18.89^{+0.11}_{-0.18}$ &$380.(\pm 5\%)$ &$9.50(\pm 5\%)$\\
			$J_{T,2}^{\text{M-M}}$ &$19.21^{+0.20}_{-0.26}$ &$408.(\pm 5\%)$ &$18.00(\pm 5\%)$    &$19.62^{+0.04}_{-0.08}$ &$424.(\pm 5\%)$ &$7.00(\pm 5\%)$\\
			$J_{T,3}^{\text{M-M}}$ &$18.50^{+0.17}_{-0.26}$ &$380.(\pm 5\%)$ &$19.00(\pm 5\%)$    &$18.95^{+0.07}_{-0.13}$ &$392.(\pm 5\%)$ &$9.50(\pm 5\%)$\\ \hline
			$J_{T,1}^{\text{Di-Di}}$ &$18.50^{+0.17}_{-0.25}$ &$380.(\pm 5\%)$ &$19.00(\pm 5\%)$    &$18.93^{+0.09}_{-0.16}$ &$386.(\pm 5\%)$ &$9.50(\pm 5\%)$\\
			$J_{T,2}^{\text{Di-Di}}$ &$19.20^{+0.21}_{-0.26}$ &$408.(\pm 5\%)$ &$18.00(\pm 5\%)$    &$19.55^{+0.04}_{-0.10}$ &$422.(\pm 5\%)$ &$7.50(\pm 5\%)$\\
			$J_{T,3}^{\text{Di-Di}}$ &$18.50^{+0.17}_{-0.26}$ &$380.(\pm 5\%)$ &$19.00(\pm 5\%)$    &$18.91^{+0.11}_{-0.18}$ &$382.(\pm 5\%)$ &$9.50(\pm 5\%)$\\ \hline
			$J_{T,1}^{\text{Dia}}$ &$18.50^{+0.17}_{-0.26}$ &$380.(\pm 5\%)$ &$19.00(\pm 5\%)$    &$18.91^{+0.11}_{-0.18}$ &$382.(\pm 5\%)$ &$9.50(\pm 5\%)$\\
			$J_{T,2}^{\text{Dia}}$ &$18.50^{+0.17}_{-0.26}$ &$380.(\pm 5\%)$ &$19.00(\pm 5\%)$    &$18.91^{+0.11}_{-0.18}$ &$382.(\pm 5\%)$ &$9.50(\pm 5\%)$\\
			$J_{T,3}^{\text{Dia}}$ &$18.50^{+0.17}_{-0.26}$ &$380.(\pm 5\%)$ &$19.00(\pm 5\%)$    &$18.95^{+0.07}_{-0.13}$ &$392.(\pm 5\%)$ &$9.50(\pm 5\%)$\\ \hline\hline
		\end{tabular}
		
		\label{tab:4b-T-NLOresult-MSbar}
	\end{center}
\end{table}
\begin{table}[H]
	\vspace{-0.8cm}
	\renewcommand\arraystretch{1.5}
	\setlength{\tabcolsep}{3 mm}
	\begin{center}
		\caption{The LO and NLO Results for $J^P=2^+$ with $\bar{b}b\bar{b}b$ system in the On-Shell scheme}
		\begin{tabular}{cccc|@{*}|ccc}
			\hline\hline
			\multirow{2}{*}{Current} &
			\multicolumn{3}{c|@{*}|}{LO}& \multicolumn{3}{c}{NLO(OS)} \\ \cline{2-7}
			& \makecell{$M_H$ \\ (GeV)} & \makecell{$S_0$ \\ ($\text{GeV}^2$)} & \makecell{$M_B^2$ \\ ($\text{GeV}^2$)} &  \makecell{$M_H$ \\ (GeV)} & \makecell{$S_0$ \\ ($\text{GeV}^2$)} & \makecell{$M_B^2$ \\ ($\text{GeV}^2$)} \\ \hline
			
			$J_{T,1}^{\text{M-M}}$ &$19.67^{+0.04}_{-0.09}$ &$418.(\pm 5\%)$ &$7.50(\pm 5\%)$    &$18.98^{+0.07}_{-0.29}$ &$366.(\pm 5\%)$ &$3.50(\pm 5\%)$\\
			$J_{T,2}^{\text{M-M}}$ &$20.44^{+0.11}_{-0.18}$ &$448.(\pm 5\%)$ &$10.50(\pm 5\%)$    &$19.44^{+0.54}_{-1.79}$ &$388.(\pm 5\%)$ &$5.50(\pm 5\%)$\\
			$J_{T,3}^{\text{M-M}}$ &$19.68^{+0.04}_{-0.10}$ &$420.(\pm 5\%)$ &$7.50(\pm 5\%)$    &$18.98^{+0.07}_{-0.28}$ &$366.(\pm 5\%)$ &$3.50(\pm 5\%)$\\ \hline
			$J_{T,1}^{\text{Di-Di}}$ &$19.67^{+0.04}_{-0.10}$ &$418.(\pm 5\%)$ &$7.50(\pm 5\%)$    &$18.98^{+0.07}_{-0.29}$ &$366.(\pm 5\%)$ &$3.50(\pm 5\%)$\\
			$J_{T,2}^{\text{Di-Di}}$ &$20.38^{+0.11}_{-0.18}$ &$446.(\pm 5\%)$ &$10.00(\pm 5\%)$    &$19.39^{+0.23}_{-1.89}$ &$386.(\pm 5\%)$ &$5.00(\pm 5\%)$\\
			$J_{T,3}^{\text{Di-Di}}$ &$19.68^{+0.04}_{-0.10}$ &$420.(\pm 5\%)$ &$7.50(\pm 5\%)$    &$18.98^{+0.07}_{-0.28}$ &$366.(\pm 5\%)$ &$3.50(\pm 5\%)$\\ \hline
			$J_{T,1}^{\text{Dia}}$ &$19.67^{+0.04}_{-0.09}$ &$418.(\pm 5\%)$ &$7.50(\pm 5\%)$    &$18.98^{+0.07}_{-0.29}$ &$366.(\pm 5\%)$ &$3.50(\pm 5\%)$\\
			$J_{T,2}^{\text{Dia}}$ &$19.68^{+0.04}_{-0.10}$ &$420.(\pm 5\%)$ &$7.50(\pm 5\%)$    &$18.98^{+0.07}_{-0.28}$ &$366.(\pm 5\%)$ &$3.50(\pm 5\%)$\\
			$J_{T,3}^{\text{Dia}}$ &$19.68^{+0.04}_{-0.10}$ &$420.(\pm 5\%)$ &$7.50(\pm 5\%)$    &$18.98^{+0.07}_{-0.28}$ &$366.(\pm 5\%)$ &$3.50(\pm 5\%)$\\ \hline\hline
		\end{tabular}
		
		\label{tab:4b-T-NLOresult-OS}
	\end{center}
\end{table}
\begin{figure}[H]
	\centering
	\subfigure[$\overline{\text{MS}}$]{
		\includegraphics[scale=0.4]{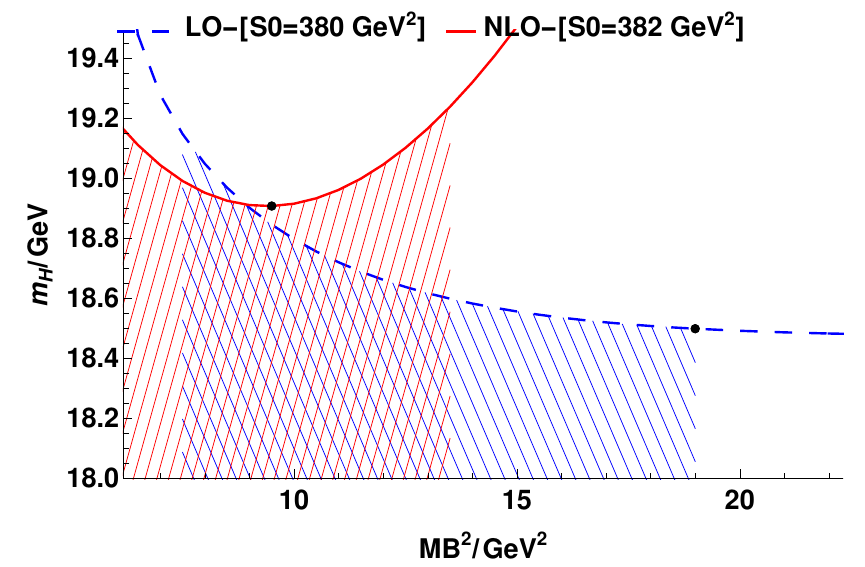}
		\includegraphics[scale=0.4]{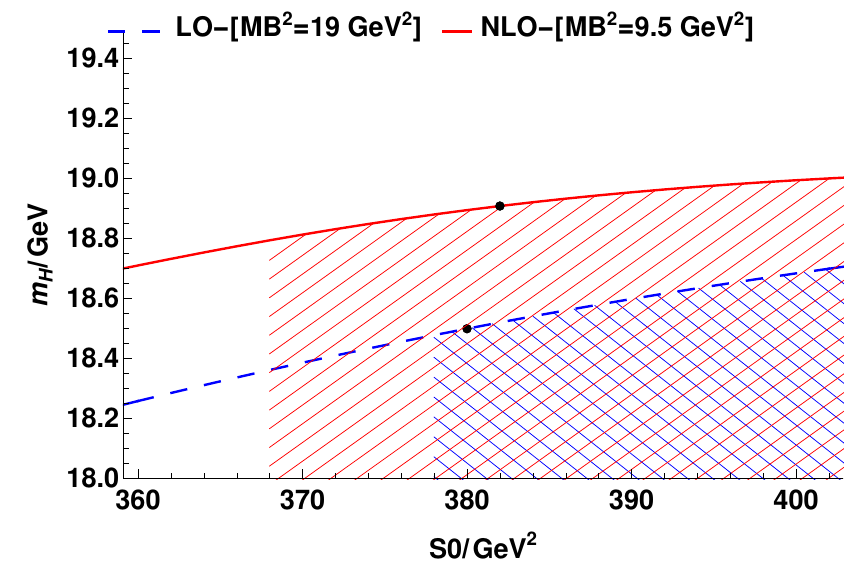}
	}\\
	\subfigure[OS]{
		\includegraphics[scale=0.4]{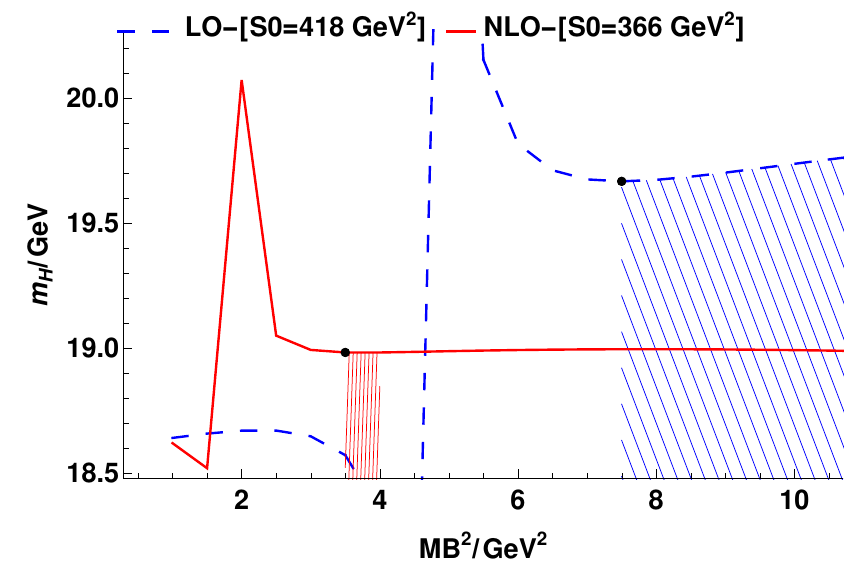}
		\includegraphics[scale=0.4]{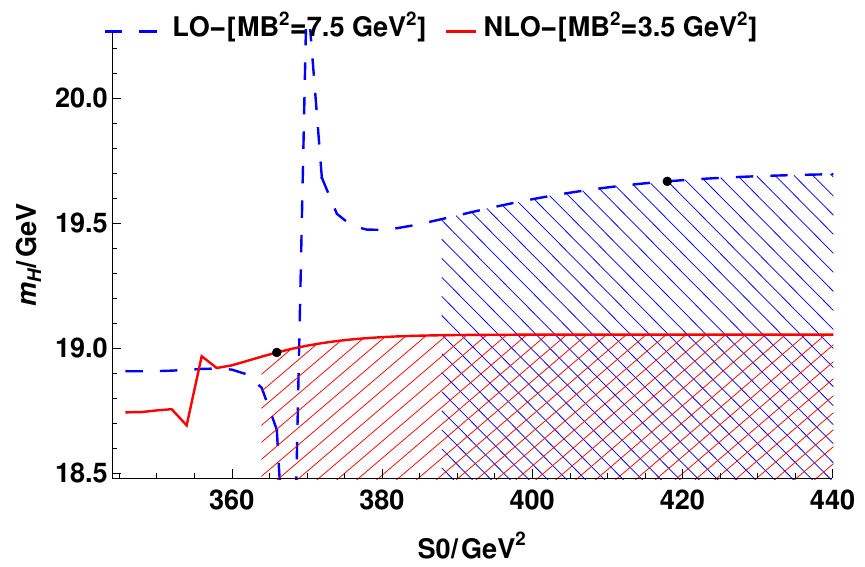}
	}
	\caption{\label{fig:4b-2+-Mixed1-NLO-MSbar-OS}
		The Borel platform curves for $J_{T,1}^{\text{Dia}}$ with $J^{PC}=2^{++}$ in the $\overline{\text{MS}}$ and On-Shell schemes}
\end{figure}
\begin{figure}[H]
	\centering
	\subfigure[$\overline{\text{MS}}$]{
		\includegraphics[scale=0.4]{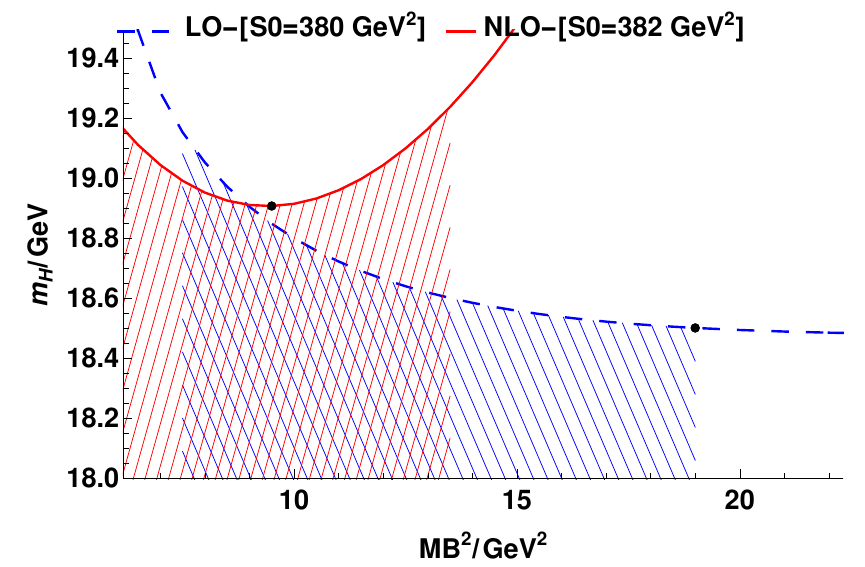}
		\includegraphics[scale=0.4]{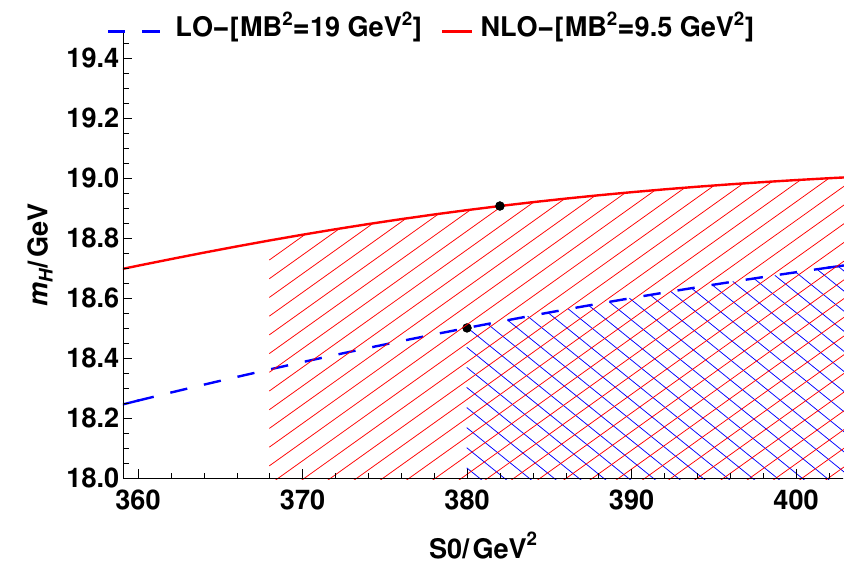}
	}\\
	\subfigure[OS]{
		\includegraphics[scale=0.4]{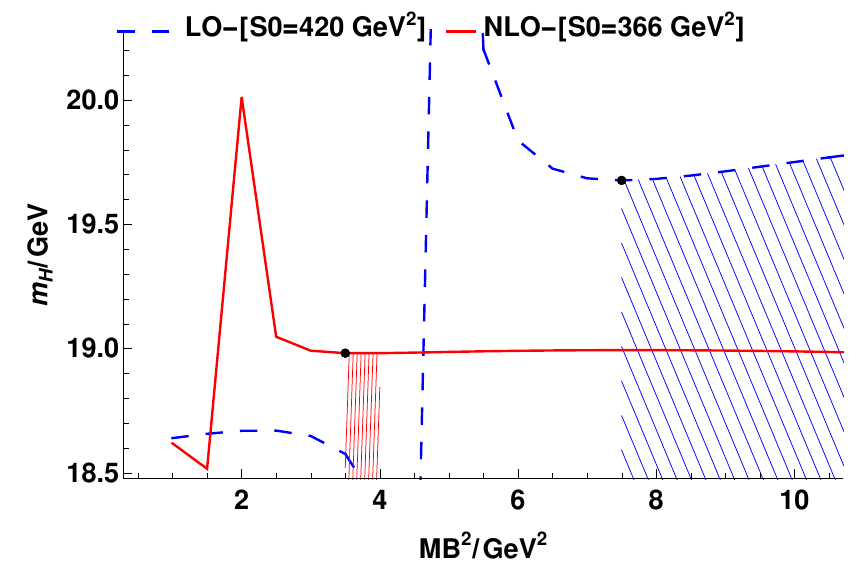}
		\includegraphics[scale=0.4]{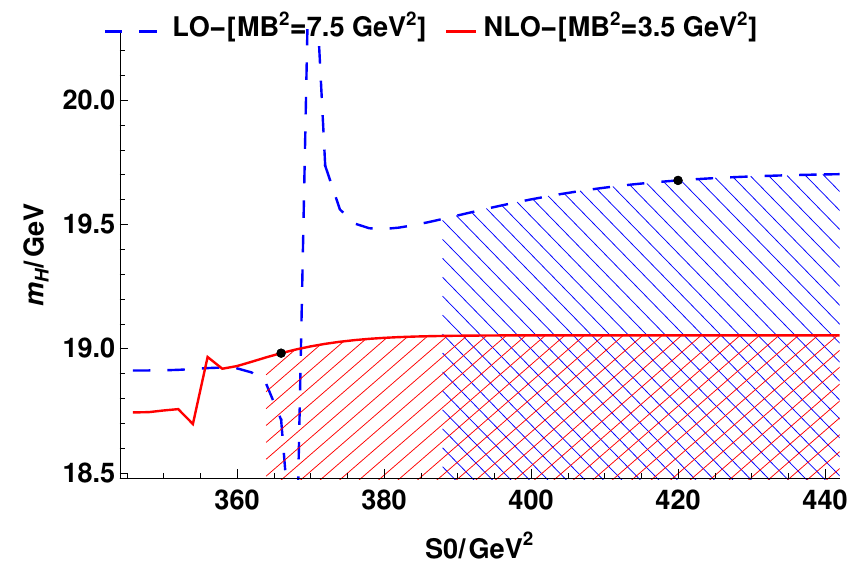}
	}
	\caption{\label{fig:4b-2+-Mixed2-NLO-MSbar-OS}
		The Borel platform curves for $J_{T,2}^{\text{Dia}}$ with $J^{PC}=2^{++}$ in the $\overline{\text{MS}}$ and On-Shell schemes}
\end{figure}
\begin{figure}[H]
	\centering
	\subfigure[$\overline{\text{MS}}$]{
		\includegraphics[scale=0.4]{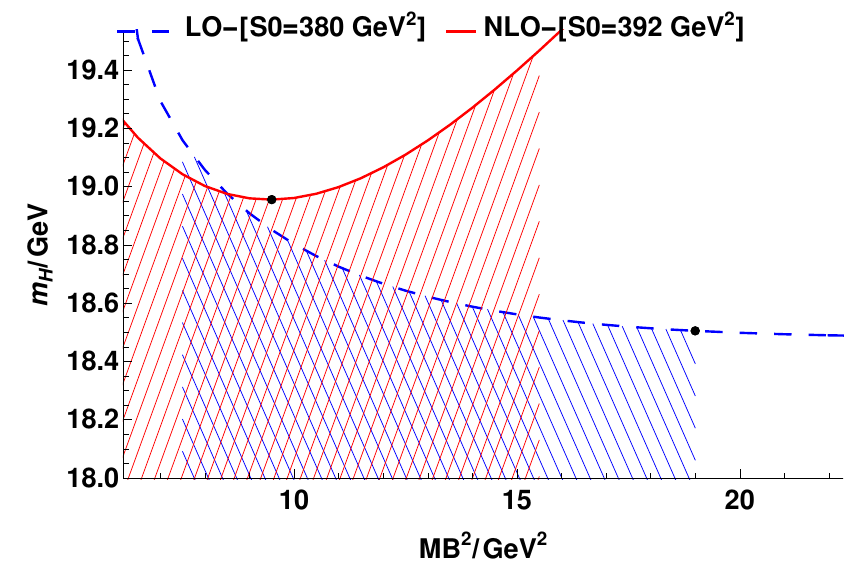}
		\includegraphics[scale=0.4]{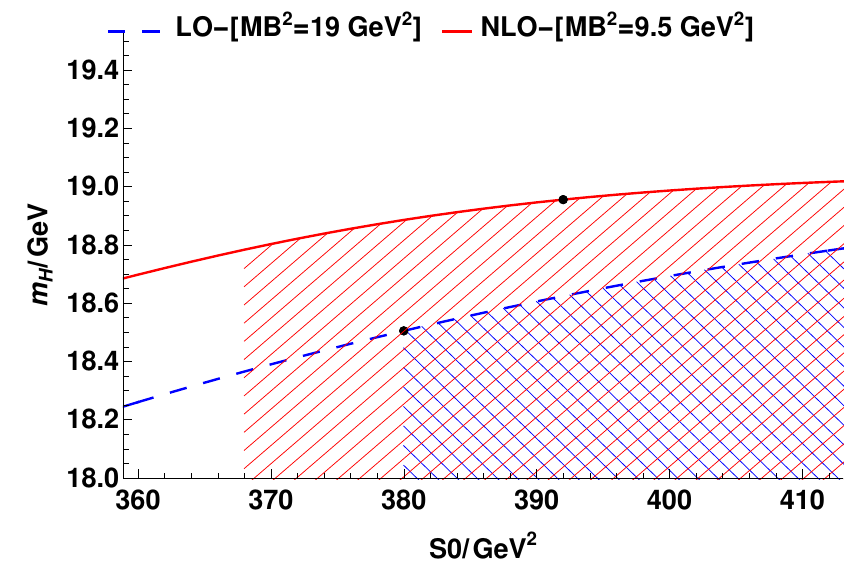}
	}\\
	\subfigure[OS]{
		\includegraphics[scale=0.4]{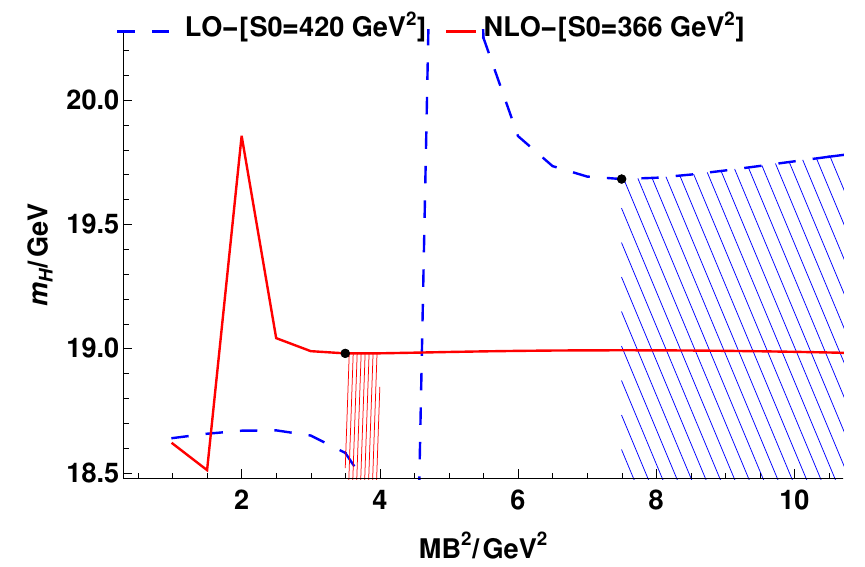}
		\includegraphics[scale=0.4]{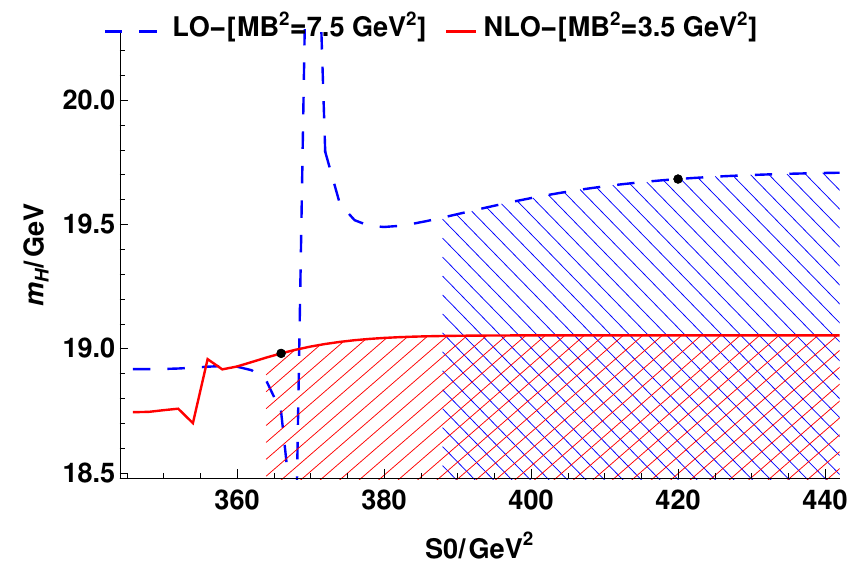}
	}
	\caption{\label{fig:4b-2+-Mixed3-NLO-MSbar-OS}
		The Borel platform curves for $J_{T,3}^{\text{Dia}}$ with $J^{PC}=2^{++}$ in the $\overline{\text{MS}}$ and On-Shell schemes}
\end{figure}

\subsection{Renormalization scale dependence}

\begin{figure}[H]
	\vspace{0.5cm}
	\centering
	\includegraphics[scale=0.5]{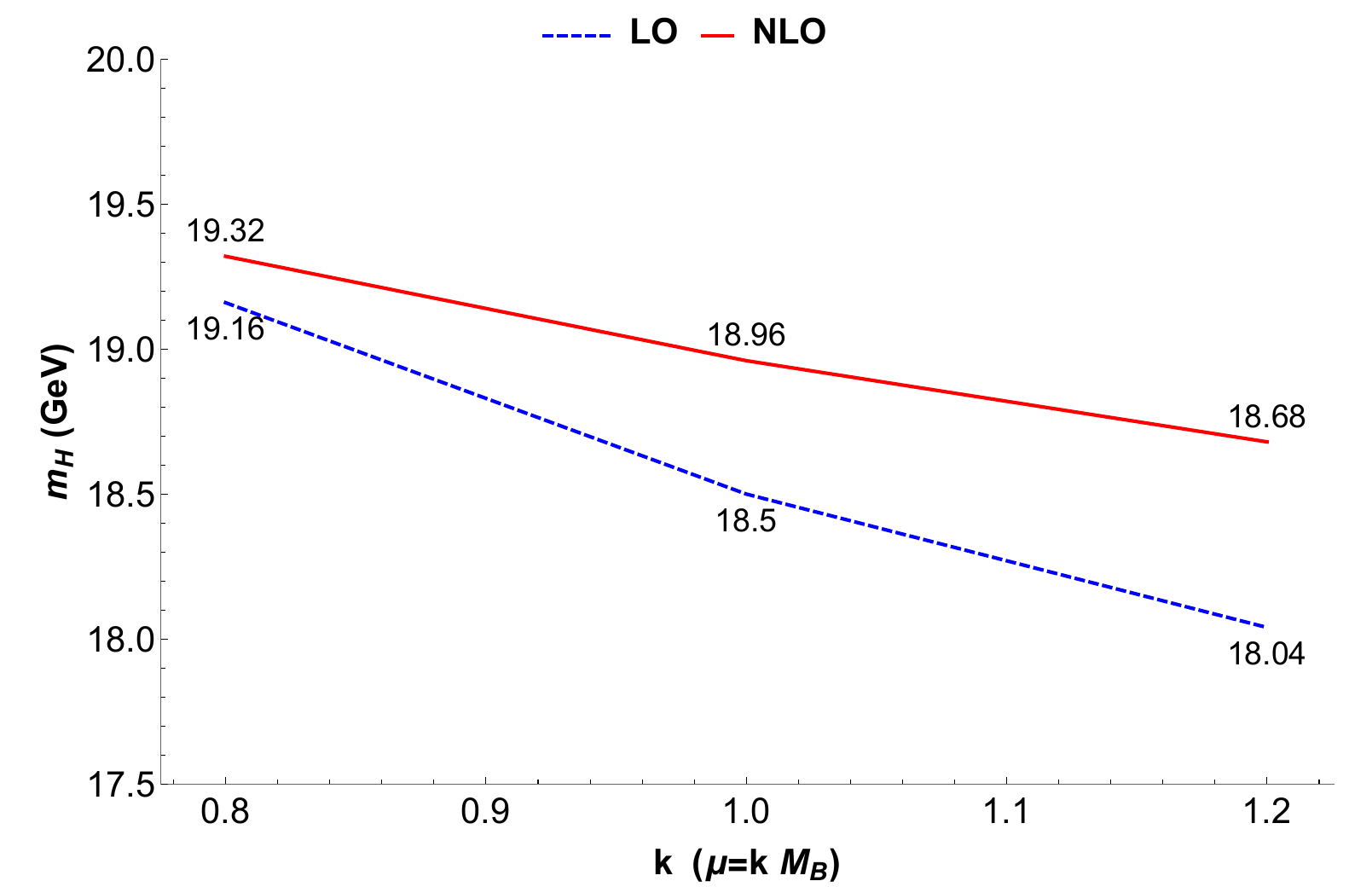}
	
	\caption{\label{fig:4b-mu-dependence-S-3}
		The renormalization scale $\mu$ dependence of the LO and NLO results of $J_{S,3}^{\text{Dia}}$ in $\overline{\text{MS}}$ scheme }
	
	\vspace{0.2cm}
\end{figure}

\begin{figure}[H]
	\vspace{-0.4cm}
	\centering
	\includegraphics[scale=0.5]{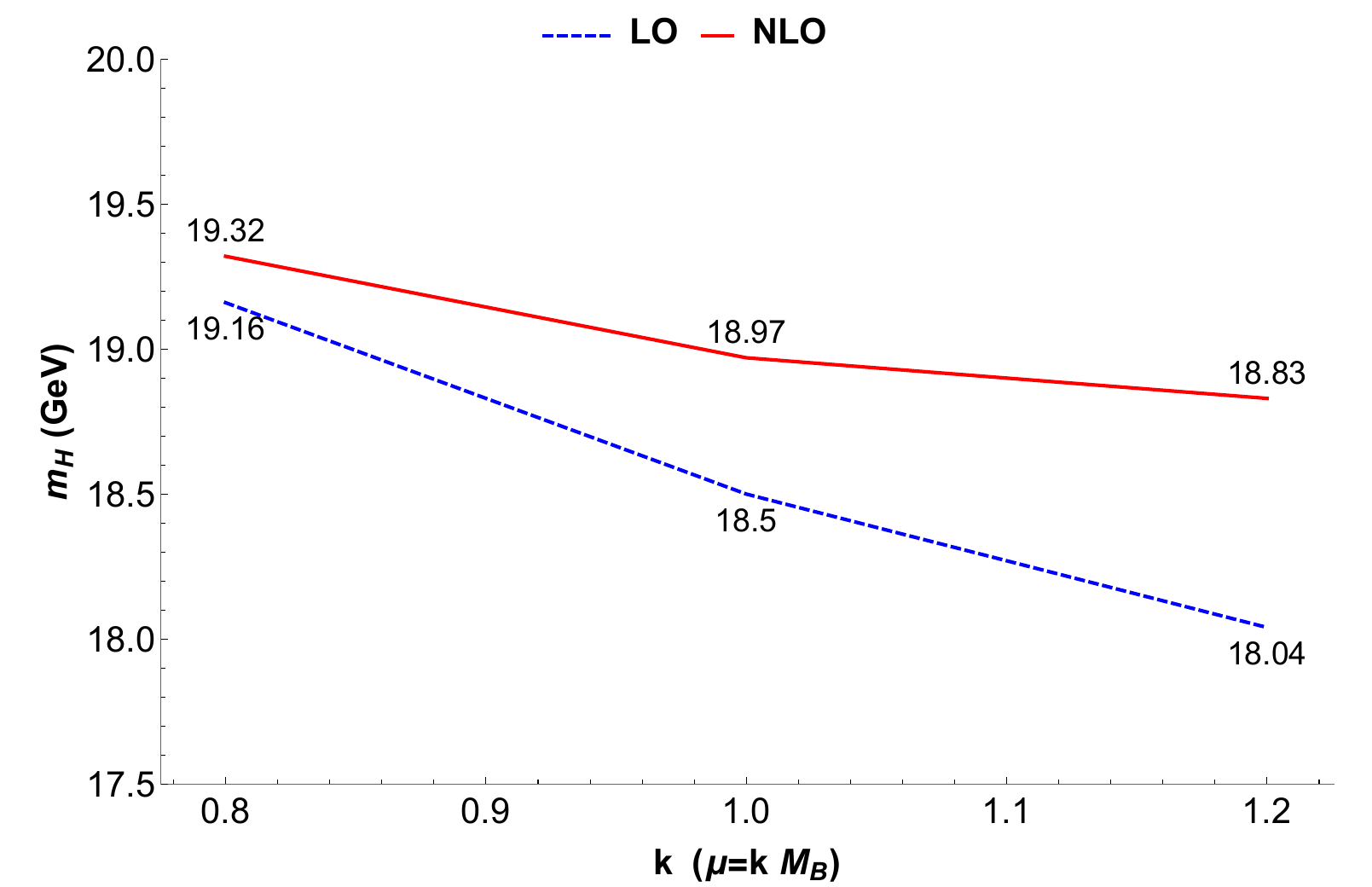}
	
	\caption{\label{fig:4b-mu-dependence-S-4}
		The renormalization scale $\mu$ dependence of the LO and NLO results of $J_{S,4}^{\text{Dia}}$ in $\overline{\text{MS}}$ scheme }
	
	\vspace{-0.5cm}
\end{figure}

%%%%%%%%%%%%%%%%%%%%%%%%%%%%%%%%%%%%%%%%%%%%%%%%%
%%%%%%%%%%%%%%%%%%%%%%%%%%%%%%%%%%%%%%%%%%%%%%%%%
% references
\newpage
\bibliographystyle{utphysMa}
\bibliography{SumRule}

%\bibliography{4Q}

\end{document}